\documentclass[a4paper,11pt]{article}
\pdfoutput=1
\usepackage{jheppub}
\usepackage{tikz}
\usepackage{slashed}
\usepackage{enumitem}
\usepackage[T1]{fontenc} 
\usepackage{floatrow}
\usepackage{appendix}
\usepackage{simpler-wick}
\usepackage{tabularx}
\usepackage[normalem]{ulem}
\usepackage{soul}

\allowdisplaybreaks

\usepackage{epsf}
\usepackage{amsmath}
\usepackage{amsfonts}
\usepackage{amssymb}
\usepackage{psfrag,epsfig,graphicx,graphics}
\usepackage{empheq}

\newcommand\numberthis[1][]{%
    \refstepcounter{equation}%
    \ifx#1\empty\else\label{eq:#1}\fi%
    \tag{\theequation}%
}

\usepackage{xargs} 
\usepackage[colorinlistoftodos,prependcaption,textsize=tiny]{todonotes}
\newcommandx{\SW}[2][1=]{\todo[linecolor=OliveGreen,backgroundcolor=OliveGreen!25,bordercolor=OliveGreen,#1]{#2}}

\newcommandx{\LS}[2][1=]{\todo[linecolor=Plum,backgroundcolor=Plum!25,bordercolor=Plum,#1]{#2}}

\newcommandx{\RBO}[2][1=]{\todo[linecolor=orange,backgroundcolor=orange!25,bordercolor=orange,#1]{#2}}

\newcommandx{\MF}[2][1=]{\todo[linecolor=blue,backgroundcolor=blue!25,bordercolor=blue,#1]{#2}}

\providecommand{\U}[1]{\protect\rule{.1in}{.1in}}

\def\slashchar#1{\setbox0=\hbox{$#1$}
   \dimen0=\wd0
   \setbox1=\hbox{/} \dimen1=\wd1
   \ifdim\dimen0>\dimen1
      \rlap{\hbox to \dimen0{\hfil/\hfil}}
      #1
   \else
      \rlap{\hbox to \dimen1{\hfil$#1$\hfil}}
      /
   \fi}

\def\bei{\begin{itemize}}
\def\ei{\end{itemize}}

\def\beeq{\begin{eqnarray}} 
\def\beqa{\begin{eqnarray}}
\def\bea{\begin{eqnarray}}

\def\eea{\end{eqnarray}}
\def\eqa{\end{eqnarray}}
\def\eeeq{\end{eqnarray}}

\def\eqar{\end{array}}
\def\beqar{\begin{array}}

\def\beas{\begin{eqnarray*}}
\def\beqas{\begin{eqnarray*}}

\def\eqas{\end{eqnarray*}}
\def\eeas{\end{eqnarray*}}

\def\beq{\begin{equation}} 
\def\be{\begin{equation}}

\def\ee{\end{equation}}
\def\eq{\end{equation}}
\def\eeq{\end{equation}}

\def\beqd{\begin{displaymath}}
\def\eeqd{\end{displaymath}}
\def\eqd{\end{displaymath}}

\def\beeq{\begin{eqnarray}} \def\eeeq{\end{eqnarray}}

\newcommand{\fin}{\end{document}}

\title{\boldmath Twist corrections to exclusive vector meson production in a saturation framework}

\author[a]{Renaud Boussarie,} 
\author[b]{Michael Fucilla,}
\author[c]{Lech Szymanowski,}
\author[b]{Samuel Wallon}

\affiliation[a]{CPHT, CNRS, Ecole Polytechnique, Institut Polytechnique de Paris, 91128 Palaiseau, France}
\affiliation[b]{Université Paris-Saclay, CNRS/IN2P3, IJCLab, 91405, Orsay, France}
\affiliation[c]{National Centre for Nuclear Research (NCBJ),Pasteura 7, 02-093 Warsaw,  Poland}

\emailAdd{Michael.Fucilla@ijclab.in2p3.fr}
\emailAdd{Renaud.Boussarie@polytechnique.edu}
\emailAdd{Lech.Szymanowski@ncbj.gov.pl}
\emailAdd{Samuel.Wallon@ijclab.in2p3.fr}

\abstract{We develop a framework combining the higher-twist formalism of exclusive processes in the $s$ channel with the semi-classical effective description of small-$x$ physics in the $t$ channel. We apply it to transversely polarized light vector meson production, $\gamma^{*} p \rightarrow V (\rho, \varphi ,\omega) \; p$, which starts at the next-to-leading power and for which a purely collinear treatment leads to end-point singularities. The result is obtained in the most general kinematics, including both forward and non-forward cases by preserving the full impact parameter dependence in the non-perturbative correlators, in both momentum and coordinate space representations. A systematic expansion of the Wilson lines in terms of Reggeized gluon fields is performed in order to obtain the results in the weak-field BFKL approximation. These new results will allow for investigating the dilute-to-dense regime transition of QCD for a wide class of observables.}

\begin{document} 
\maketitle
\flushbottom

\newpage

\section{Introduction}
\label{sec:intro}

The HERA experiments revealed, based on $e^\pm p$ deep inelastic scattering (DIS), several important features of QCD at very large energies.
First, 
it was realized that diffractive events represent a fraction of up to 10\% of the total $e^\pm p$ cross-section for DIS~\cite{Ahmed:1995ns,Adloff:1997sc,Aktas:2006hx,Aktas:2006hy,Aaron:2010aa,Aaron:2012ad,Derrick:1995wv,Breitweg:1997aa,Breitweg:1998gc,Chekanov:2005vv,Chekanov:2008fh,Chekanov:2004hy,Aaron:2012hua}. In such a situation, the proton remains intact, with a color singlet exchange in the $t$ channel. Second, the behaviour of several experimental measurements done in the kinematical domain where the photon virtuality $Q^2$ is moderate (more precisely, below the saturation scale $Q_s$) and the Bjorken $x$ variable is asymptotically small could be interpreted as a sign of gluonic saturation inside the proton, for both  inclusive and diffractive deep inelastic scattering, as first shown by 
Golec-Biernat and W\"usthoff~\cite{GolecBiernat:1998js,GolecBiernat:1999qd}. 
It has been further realized that exclusive diffractive processes
could give an excellent lever arm to scrutinize the proton's internal structure at asymptotic energies. Among various diffractive states, the exclusive diffractive production of a light vector meson $M$ ($\rho, \phi, \omega$)~\cite{Ivanov:1998gk,Ivanov:2000uq,Munier:2001nr,Forshaw:2001pf,Enberg:2003jw,Poludniowski:2003yk} 
\begin{equation}
\label{process}
\gamma^{(*)} p \to M (\rho, \phi, \omega) \, p
\end{equation}
is of particular interest, and
was studied at HERA both for forward electroproduction~\cite{Chekanov:2007zr,Aaron:2009xp} and large $t$ photoproduction~\cite{Breitweg:1999jy,Chekanov:2002rm,Aktas:2006qs,Aaron:2009xp} kinematics.
This process indeed provides multidimensional information: the photon virtuality is related to the transverse size of the probed partons and the transverse momentum
exchanged in the $t$ channel gives access to the impact parameter distribution
of partons inside the proton after Fourier transformation. 

Understanding gluonic saturation is one of the most important and longstanding problems
of QCD. Indeed, at large center of mass energy $\sqrt{s}$, the proton is a dense system with high field strengths, very high gluonic occupation numbers, but still in the weak-coupling regime. This provides a very unique system, the color glass condensate (CGC), in which the collective effect of saturation can be treated relying on perturbative methods.
Several equivalent frameworks have been developed in order to deal with such a system, including the large-$N_c$ dipole model~\cite{Mueller:1993rr,Mueller:1994jq,Mueller:1994gb,Chen:1995pa,Kovchegov:1999yj,Kovchegov:1999ua}, a semi-classical projectile-oriented approach ~\cite{Balitsky:1995ub,Balitsky:1998kc,Balitsky:1998ya,Balitsky:2001re}, and a semi-classical target-oriented approach~\cite{JalilianMarian:1997jx,JalilianMarian:1997gr,JalilianMarian:1997dw,JalilianMarian:1998cb,Kovner:2000pt,Weigert:2000gi,Iancu:2000hn,Iancu:2001ad,Ferreiro:2001qy}. 

Still, in order to get clear evidence of the CGC regime and to study its features, one should reach the frontier of precision. In the case of exclusive diffractive meson production, one can measure the various possible 
transition amplitudes between an initial polarized photon and a produced polarized meson, and describe these various observables with the highest possible precision. This means that both Next-to-Leading Order (NLO) corrections and power corrections should be included. These latter include both powers of the hard scale $Q^2$ (i.e. higher-twist corrections) and of the center-of-mass energy $\sqrt{s}$ (i.e. subeikonal corrections)~\cite{Altinoluk:2014oxa,Altinoluk:2015gia,Altinoluk:2015xuy,Agostini:2019avp,Agostini:2019hkj,Altinoluk:2020oyd,Altinoluk:2021lvu,Altinoluk:2022jkk,Altinoluk:2023qfr,Agostini:2024xqs,Chirilli:2018kkw,Chirilli:2021lif}, including spin effects~\cite{Kovchegov:2015pbl,Kovchegov:2019rrz,Cougoulic:2022gbk}.

A first step towards NLO corrections was performed at the level of the evolution kernel for CGC, including the running coupling effects~\cite{Kovchegov:2006vj} and then the whole NLO corrections to the kernel~\cite{Balitsky:2008zza,Balitsky:2013fea,Grabovsky:2013mba,Balitsky:2014mca,Kovner:2013ona,Lublinsky:2016meo,Caron-Huot:2015bja}. 
First steps have been made concerning the corrections to
the coupling to a probe in inclusive and semi-inclusive processes, called impact factors. The  NLO impact factor has been obtained 
for the inclusive coupling to a $\gamma^\ast$ \cite{Balitsky:2010ze,Balitsky:2012bs} and for semi-inclusive hadron production, involving the coupling to a parton, in view of studying $p_\perp$-broadening effects~\cite{Chirilli:2011km,Ivanov:2012iv,Iancu:2016vyg}.
Finally, the first computation of an exclusive NLO impact factor in the CGC framework was performed for exclusive dijets~\cite{Boussarie:2014lxa,Boussarie:2016ogo,Boussarie:2019ero}. Other results were obtained at NLO, for inclusive DIS~\cite{Beuf:2022ndu}, inclusive photoproduction of dijets~\cite{Altinoluk:2020qet,Taels:2022tza}, photon-dijet production in DIS~\cite{Roy:2019hwr}, dijets in DIS~\cite{Caucal:2021ent,Caucal:2022ulg,Caucal:2023fsf}, single hadron~\cite{Bergabo:2022zhe}, dihadrons production in DIS~\cite{Bergabo:2022tcu,Iancu:2022gpw}, exclusive quarkonium production~\cite{Mantysaari:2021ryb,Mantysaari:2022kdm}, inclusive DDIS~\cite{Beuf:2022kyp}, diffractive single~\cite{Fucilla:2023mkl} and di-hadron~\cite{Fucilla:2022wcg} production, forward production of a Drell-Yan pair and a jet~\cite{Taels:2023czt}, single jet in DIS~\cite{Caucal:2024cdq}.

In this article, we focus on the exclusive production of a neutral  vector meson of arbitrary polarization.
It is known that among DIS events, one can
describe the exclusive production of a meson from the open production a $q \bar{q}$ pair,
relying on
the collinear QCD factorization approach~\cite{Brodsky:1994kf,Frankfurt:1995jw}. At moderate energies, the scattering amplitude is given as  a convolution of quark or gluon generalized parton
distributions (GPDs) inside the nucleon, the  distribution amplitude (DA) describing the partonic content of the 
light meson,
and a 
hard scattering amplitude which is perturbatively calculable~\cite{Collins:1996fb, Radyushkin:1997ki}. The DAs and GPDs are subject to specific QCD evolution equations~\cite{Farrar:1979aw,Lepage:1979zb,Efremov:1979qk,Dittes:1988xz,Muller:1994ses}. It turns out that this collinear factorization is proven only for the twist-2 dominated transition between a longitudinally polarized photon and a longitudinally polarized vector meson~\cite{Collins:1996fb}. Furthermore, explicit breaking of collinear factorization can be exhibited at twist 3, through the presence of end-point singularities, i.e. when the momentum of one of the parton vanishes, in exclusive electroproduction of transversely polarized vector mesons~\cite{Mankiewicz:1999tt}. In order to evade from these divergences, an improved collinear approximation scheme~\cite{Li:1992nu} has been proposed and
 applied to $\rho$ electroproduction~\cite{Vanderhaeghen:1999xj,Goloskokov:2005sd,Goloskokov:2006hr,Goloskokov:2007nt}. At high energies, the end-point singularities are naturally regularized by non-zero transverse
momenta of $t-$channel gluons~\cite{Ivanov:1998gk, Anikin:2009hk,Anikin:2009bf}, and models for the target can be used to describe HERA data~\cite{Anikin:2011sa,Besse:2012ia,Besse:2013muy,Bolognino:2018rhb,Bolognino:2019pba,Bolognino:2021niq}.

Describing exclusive diffractive meson production in a saturated regime, furthermore with the aim of precision physiscs, is a highly challenging task. A first step in this direction was made by some of us a few years ago~\cite{Boussarie:2016bkq}. There, we described for the first time the exclusive diffractive production of a longitudinally polarized vector meson including NLO corrections in a saturated regime, combining a saturation approach with leading twist collinear factorization to describe the transition between a quark-antiquark final state and the meson. In this description, the final $q\bar{q}$ pair is forced to fly collinearly to produce the meson, resulting in infrared (IR) divergences which are absorbed in the Efremov-Radyushkin-Brodsky-Lepage (ERBL) evolution equation.

As previously mentioned, including higher twist is a prerequisite in order to describe many transitions like $T \to T$ which simply vanish at leading twist (here, twist 2), but doing this consistently in a saturated framework is also highly non trivial.
At twist 3, one should include not only kinematic twist 3 contributions (i.e. with a slightly off-collinear $q\bar{q}$ pair in the final state), but also genuine twist 3 contributions (i.e. with a $q\bar{q} g$ collinear final state, where the gluon is physically polarized). It is the purpose of the present article to make such a step at twist 3. We will present
a complete LO calculation for the scattering amplitude of
exclusive diffractive meson production in $\gamma^{(*)} p$
or $\gamma^{(*)} A$ collisions, with completely general kinematics, combining higher-twist collinear factorization and high-energy small-$x$ factorization techniques. 
Contrary to previous formulations of effective approaches to saturation, in order to accommodate for a systematic $s$-channel higher-twist expansion, we have to introduce effective quark, antiquark and gluon fields in the shockwave background instead of the standard and more convenient effective propagators and effective fermion or gluon lines.
This allows us in a first place to perform the summation of small-$x$ logarithms within the exclusive higher-twist framework, while accounting for full kinematic twist effects, i.e. at any order up to genuine twist 4. Those results can be used for LO numerical predictions as they are using phenomenological models for light cone wave functions or transverse momentum dependent DAs, but our purpose being to build a consistent approach with less model dependence and with a possibility to extend computations beyond leading order we later expand it in kinematic twists. \\

Our results are in the most general kinematics, so that they cover both forward and non-forward cases, preserving the full impact parameter dependence of the dipole operators. In contrast to previous studies, since our computation has completely general kinematics, we are not restricted to the $s$-channel helicity conserving amplitudes, being able, for the first time, to describe the production of a transversely polarized meson from a longitudinally polarized incoming photon within full twist-3 accuracy. Moreover our results are given both in momentum and coordinate space representations, which is particularly convenient in view of future phenomenological applications. Furthermore, a systematic expansion of the Wilson lines in terms of Reggeized gluon fields is performed in order to also obtain the LO BFKL~\cite{Fadin:1975cb, Kuraev:1976ge, Kuraev:1977fs, Balitsky:1978ic} impact factors, in the non saturated regime. As a consistency check, we extract the forward result, which is in full agreement with the result of some of us in a pure BFKL approach~\cite{Anikin:2009hk,Anikin:2009bf}.

From the phenomenological point of view, our present results provides a vast class of observables to deal with exclusive processes.
Indeed, the complete generality of the kinematics allows it to be applied to a wide range of experimental conditions, including the electroproduction of vector mesons with general kinematics, as well as their photoproduction at large transfered momentum. It thus
can be used both at $ep$ and $eA$ colliders, like the future EIC~\cite{Boer:2011fh} or LHeC~\cite{AbelleiraFernandez:2012cc} and in ultraperipheral collisions at RHIC or at the LHC~\cite{Baltz:2007kq,N.Cartiglia:2015gve}, to describe all spin density matrix elements. Since we provide expressions both in the linear and non-linear framework, our results can be used to investigate the transition from a dilute to a saturated regime. 

This article is organized as follows. In Section~\ref{Sec:theoretical-framework} we set up the theoretical framework before the twist expansion, going through the semi-classical formalism for high-energy $t$-channel scatterings, in particular introducing effective operators. In Section~\ref{Sec:exclusive-meson} we explicitly compute the amplitude for diffractive exclusive light vector-meson production in our approach. This result contains an infinite tower of kinematic twist effects. In Section~\ref{Sec:CovColl} we introduce the ingredients which are necessary for a fixed order twist expansion, which we use in Section~\ref{Sec:impact-factor-twist3}, where we provide the full twist-3 result. In Section~\ref{Sec:dilute-limit} we consider the BFKL dilute limit and show that we get full agreement with the previously known result in the forward case. 
The conclusions of our studies are presented in Section~\ref{Sec:Conc}.

\section{Theoretical framework}
\label{Sec:theoretical-framework}
In the present article, we focus on the computation of exclusive diffractive vector meson production (EDMP)~\footnote{The current description may be used to describe photoproduction at large-$t$, which means we are not specifically studying standard deeply virtual meson production (DVMP) which is why we will refer to our process as EDMP rather than DVMP.}, for a meson $M$, namely
\begin{equation}
\gamma_{\lambda_{\gamma}}^{(*)} (p_{\gamma}) \; P (p) \to \; {M}_{\lambda_{ M}} (p_{{ M}}) \, P(p')
\end{equation}
where $P$ is a nucleon or a nucleus target, $\lambda_{\gamma}$ is the incoming photon polarization (it can be either longitudinal or transverse) and $\lambda_{M}$ is the outgoing meson polarization (it can also be either longitudinal or transverse). We stress here that in contrast to previous studies, since our computation has completely general kinematics, we can describe the non-forward EDMP process. Moreover, we are not restricted to the $s$-channel helicity conserving amplitudes which allows us to describe for the first time the production of a transversely polarized meson from a longitudinally polarized incoming photon within full twist-3 accuracy. We also insist on the fact that our study includes both linear small-$x$ effects (à la BFKL) and non-linear effects responsible for saturation dynamics. We will discuss how to properly isolate the linear effects in Section~\ref{Sec:dilute-limit}. The initial photon (projectile) plays the role of a probe. Our computation can be applied both to the photoproduction case at large transferred momenta squared $t$, and to the electroproduction case. 



\subsubsection*{Kinematics}

We introduce a light-cone basis composed of $n_1 $ and $n_2$, with $n_1 \cdot n_2 = 1$ defining the $+/-$ direction.
We write the Sudakov decomposition for any vector $k$ as 
\begin{equation}
k^\mu = k^+ n_1^\mu + k^- n_2^\mu + k_\perp^\mu
\end{equation}
and the scalar product of two vectors as\footnote{Any transverse momentum in Euclidean space will be denoted as a bold character, while a $\perp$ index will be used in Minkowski space.} 

\begin{equation}
    k \cdot q = k^+ q^- + k^- q^+ + k_\perp \cdot q_\perp  
    = k^+ q^- + k^- q^+ -\boldsymbol{k} \cdot \boldsymbol{q}\,.
\end{equation}
We consider the semi-hard kinematics characterized by the scale hierachy $s = (p_\gamma + p)^2 \gg Q^2=-p_\gamma^2 \gg \Lambda_{\text{QCD}}^2$. We work in a reference frame, called projectile frame,
such that the target moves towards the projectile ultrarelativistically. Particles on the projectile side are moving in the $n_1$ (i.e. $+$) direction while particles on the target side have a large component along $n_2$ (i.e. $-$ direction).

\subsection{Semi-classical small $x$ formalism}
For convenience of the reader we will shortly review the effective semi-classical effective theory description for small $x$ physics we will use to account for saturation effects. Since our main focus is the projectile transition (here, photon-to-meson) rather than the target dynamics, we will use what we will refer to as the shockwave formalism, the projectile-oriented description from~\cite{Balitsky:1995ub,Balitsky:1998kc,Balitsky:1998ya,Balitsky:2001re}, companion to the target-oriented CGC effective theory~\cite{McLerran:1994vd}. In the shockwave approach the gluonic field $\mathcal{A}$ is separated into external background fields $b$ (resp. internal fields $A$) depending on whether their $+$-momentum is below (resp. above) the arbitrary rapidity cut-off $e^\eta p_\gamma^+$, with $\eta < 0$. After being highly boosted from the target rest frame to the projectile frame, the external field takes the form 
\begin{equation}
    b^\mu (x) = b^-(x^+, x_\perp) n_2^\mu = \delta (x^+) \mathbf{B} (\boldsymbol{x})  n_2^\mu \,,
\end{equation}
often referred to as the (eikonal) shockwave approximation. We now start from the QCD Lagrangian, which is split between the interacting and the free parts,
\begin{gather} 
\mathcal{L} =-\frac{1}{4} \mathcal{F}_{a \mu \nu} \mathcal{F}^{a \mu \nu}+i \bar{\psi} \slashed{D} \psi=\mathcal{L}_{\text {free }}+\mathcal{L}_{\text {int }} \nonumber \; , \\ \mathcal{L}_{\text {int }}  =-g f^{a b c} \mathcal{A}_\mu^b \mathcal{A}_\nu^c \partial^\mu \mathcal{A}^{\nu a}-\frac{1}{4} g^2 f^{a b c} f^{a d e} \mathcal{A}_\mu^b \mathcal{A}_\nu^c \mathcal{A}^{\mu d} \mathcal{A}^{\nu e}+i \bar{\psi}\left(-i g t^a \slashed{\mathcal{A}^a} \right) \psi \; .
\end{gather}
We further split the gluon field as 
\begin{equation}
    \mathcal{A}_\mu^a (k) = A_\mu^a ( k^+ > e^\eta p_\gamma^+, k^-, k_{\perp}) + b_\mu^a ( k^+ < e^\eta p_\gamma^+, k^-, k_{\perp}) \; .
\end{equation}
Taking into account that $b^2 \propto n_2^2 = 0$, we get
\begin{gather} 
\mathcal{L}_{\rm i n t} =-g f^{a b c}\left[\left(A_\mu^b \partial^\mu A^{\nu a}+b_\mu^b \partial^\mu A^{\nu a}\right)\left(A_\nu^c+b_\nu^c\right)+A_\mu^b \partial^\mu b^{\nu a} A_\nu^c+b_\mu^b \partial^\mu b^{\nu a} A_\nu^c\right] \nonumber \\ -\frac{1}{4} g^2 f^{a b c} f^{a d e}\left[\left(A_\mu^b A^{\mu d}+A_\mu^b b^{\mu d}\right)\left(A^{\nu e} A_\nu^c+A^{\nu e} b_\nu^c+b^{\nu e} A_\nu^c\right)\right. \nonumber \\ \left.+b_\mu^b A^{\mu d}\left(A^{\nu e} A_{\nu }^{c} + A^{\nu e} b_{\nu}^{c} + b^{\nu e} A_{\nu}^{c} \right)\right]+i \bar{\psi}\left[-i g t^a\left(\slashed{A}^a+\slashed{b}^a\right)\right] \psi \; .
\label{Eq:Effective_Lagran_inter}
\end{gather}
From eq.~(\ref{Eq:Effective_Lagran_inter}) it is clear that, in the $n_2 \cdot A = 0$ light-cone gauge, this effective field theory drastically simplifies. Adopting this choice, we get
\begin{gather}
\mathcal{L}_{\text {int }}=-g f^{a b c} A_\mu^b\left(\partial^\mu A^{\nu a}\right) A_\nu^c-\frac{1}{4} g^2 f^{a b c} f^{a d e} A_\mu^b A^{\mu d} A^{\nu e} A_\nu^c+\bar{\psi}\left(g t^a \slashed{A}^a\right) \psi \nonumber \\
-g f^{a b c} b_\mu^b\left(\partial^\mu A^{\nu a}\right) A_\nu^c+\bar{\psi}\left(g t^a \slashed{b}^a\right) \psi \; .
\end{gather}
The first line contains the usual interaction terms of the standard QCD Lagrangian, while the last line corresponds to the interaction with the shockwave field and takes the form
\begin{gather}
\mathcal{L}_{\text {int }}^S =-g f^{a c b} b^{-c} g^{\alpha \beta}\left(\frac{\partial A_\beta^a}{\partial x^{-}}\right) A_\alpha^b+\bar{\psi}\left(g t^a \slashed{b}^a\right) \psi  =-g f^{a c b} b^{-c} g^{\alpha \beta}\left(\frac{\partial A_\beta^a}{\partial x^{-}}\right) A_\alpha^b+\bar{\psi} g b^{-} \gamma^{+} \psi \nonumber \\ =g f^{c a b} b^{-c} g_{\perp}^{\alpha \beta}\left(\frac{\partial A_\beta^a}{\partial x^{-}}\right) A_\alpha^b+\bar{\psi} g b^{-} \gamma^{+} \psi =i g T_{a b}^c b^{-c} g_{\perp}^{\alpha \beta}\left(\frac{\partial A_\beta^a}{\partial x^{-}}\right) A_\alpha^b+\bar{\psi} g b^{-} \gamma^{+} \psi \; ,
\label{Eq:Effective_Lagran_inter_Shockwave}
\end{gather}
where $b^{-}(x) = t^a b^{-a} (x), \; T_{ab}^{c} = - i f^{abc}$. \\

In our approach, the resummation of all order interactions with the background field leads to a Wilson line which represents the shockwave and is located at $z^- =0$. For instance, in the fundamental representation, one gets
\begin{equation}
    V_{\boldsymbol{z}} = \mathcal{P} \exp \left(i g \int d z^+ b^-(z)\right)\,,
\end{equation}
where $\mathcal{P}$ is the usual path ordering operator for the $+$ direction. The scattering amplitude can then be written as a small-$x$-factorized form i.e. as the convolution of the projectile impact factor with the non-perturbative matrix element of operators built from Wilson lines on the target states.

\subsection{Effective operators}
As explained in the introduction, we now introduce the effective background field operators, which allow the resummation of small-$x$ logarithms within the exclusive higher-twist framework. Using the shockwave approach, we construct the following effective field operators: \\

\textbf{Fermionic field effective operators}
\begin{equation}
    \left[\psi_{\text {eff }}\left(z_0\right)\right]_{z_0^{+}<0}=\psi\left(z_0\right)-\int \mathrm{d}^D z_2 G_0\left(z_{02}\right)\left(V_{\boldsymbol{z}_2}^{\dagger}-1\right) \gamma^{+} \psi\left(z_2\right) \delta\left(z_2^{+}\right) \; ,
    \label{Eq:EffecQBarOpe}
\end{equation}
\begin{equation}
    \left[\bar{\psi}_{\text{eff }}\left(z_0\right)\right]_{z_0^{+}<0}=\bar{\psi}\left(z_0\right)+\int \mathrm{d}^D z_1 \bar{\psi}\left(z_1\right) \gamma^{+}\left(V_{\boldsymbol{z}_1}-1\right) G_0\left(z_{10}\right) \delta\left(z_1^{+}\right) \; ,
    \label{Eq:EffecQOpe}
\end{equation}

\textbf{Gluon field effective operator}
\begin{equation}
   \left[ A_{\text {eff }}^{\mu a}\left(z_0\right) \right]_{z_0^{+}<0} =A^{\mu a}\left(z_0\right)+2 i \int \mathrm{d}^D z_3 \delta\left(z_3^{+}\right) F_{-\sigma}^b\left(z_3\right) G^{\mu \sigma_{\perp}}\left(z_{30}\right)\left(U_{\boldsymbol{z}_3}^{a b}-\delta^{a b}\right) \; , \vspace{0.2 cm}
   \label{Eq:EffecGOpe}
\end{equation}
the proof of which will be given in section \ref{Derivation_of_the_effective_operators}. Here and throughout this article, we denote for two coordinates (and only for coordinates) $z_{ij}\equiv z_i -z_j$. For the sake of generality, most equations in this article are written in arbitrary dimension $D=d+2$, although it is not necessary for the specific observable we will compute. For the time being, we use these operators to reproduce the effective shockwave Feynman rules, which are commonly used for standard CGC calculations. The action of the effective operators on single particle antiquark, quark and gluon states, i.e.
\begin{gather}
  \left[ v_{\alpha}^{ij} \left( p_{\bar{q}}, z_0 \right) \right]_{z_0^{+}<0} \equiv  \left[\psi_{\text{eff},\alpha}^{j} \left(z_0\right)\right]_{z_0^{+}<0} |i, p_{\bar{q}} \rangle =  - \frac{(-i)^{d/2}}{2 (2 \pi)^{d/2}} \left( \frac{p_{\bar{q}}^{+}}{-z_0^{+}} \right)^{d/2} \theta \left( p_{\bar{q}}^{+} \right) \theta \left( - z_0^+ \right) \nonumber \\ \times
    \int d^d z_2 V^{ij \dagger}_{\boldsymbol{z}_2} \frac{-z_0^{+} \gamma^{-} + \slashed{z}_{20 \perp}}{-z_0^{+}} \gamma^{+} \frac{v (p_{\bar{q}})}{\sqrt{2 p_{\bar{q}}^{+}}} \;  \exp \left \{ i p_{\bar{q}}^{+} \left( z_0^{-} - \frac{\boldsymbol{z}_{20}^{\; 2}}{2 z_0^+} + i 0 \right) - i \boldsymbol{p}_{\bar{q}} \cdot \boldsymbol{z}_{20} \right \} \; ,
\end{gather}
\begin{gather}
 \left[ \bar{u}_{\alpha}^{ij} \left( p_{q}, z_0 \right) \right]_{z_0^{+}<0} \equiv \langle i, p_q | \left[\bar{\psi}_{\text{eff},\alpha}\left(z_0\right)\right]_{z_0^{+}<0} =  \frac{(-i)^{d/2}}{2 (2 \pi)^{d/2}} \left( \frac{p_{q}^{+}}{-z_0^{+}} \right)^{d/2} \theta \left( p_{q}^{+} \right) \theta \left( - z_0^+ \right) \nonumber \\ \times
    \int d^d z_1 V^{ij}_{\boldsymbol{z}_1} \frac{\bar{u} (p_{q})}{\sqrt{2 p_{q}^{+}}} \gamma^{+} \frac{-z_0^{+} \gamma^{-} + \slashed{z}_{10 \perp}}{-z_0^{+}}  \;  \exp \left \{ i p_{q}^{+} \left( z_0^{-} - \frac{\boldsymbol{z}_{10}^{ 2}}{2 z_0^+} + i 0 \right) - i \boldsymbol{p}_{q} \cdot \boldsymbol{z}_{10} \right \} 
\end{gather}
and
\begin{gather}
  \left[ \varepsilon^{* \; ab}_{\nu} (p_g) \right]_{z_0^{+}<0} \equiv  \left[ A_{\text {eff }}^{\mu a}\left(z_0\right) \right]_{z_0^{+}<0} | p_g, b \rangle = \frac{(-i)^{d/2}}{ (2 \pi)^{d/2}} \left( \frac{p_{g}^{+}}{-z_0^{+}} \right)^{d/2} \theta \left( p_{g}^{+} \right) \theta \left( - z_0^+ \right) \nonumber \\
  \times
    \int d^d z_3\frac{\varepsilon^{*}_{\sigma} (p_g)}{\sqrt{2 p_g^+}} U^{ab}_{\boldsymbol{z}_3}  \frac{-z_0^{+} g_{\perp}^{\mu \sigma} - z_{30 \perp}^{\sigma} n_2^{\mu}}{-z_0^{+}} \;  \exp \left \{ i p_{g}^{+} \left( z_0^{-} - \frac{\boldsymbol{z}_{30}^{ 2}}{2 z_0^+} + i 0 \right) - i \boldsymbol{p}_{g} \cdot \boldsymbol{z}_{30} \right \} \; ,
\end{gather}
give the Feynman rules of external outgoing antiquark, quark and gluon external line passing through the shockwave at light cone time $z^+=0$.\footnote{See e.g. \cite{Boussarie:2016txb,Li:2023ihv}} \\

In a similar fashion, the effective propagators can be built as Wick contraction between a standard QCD field located at a light-cone time $z_2^+ > 0$ and an effective field located at $z_0^+ < 0$, i.e.
\begin{gather}
 \left.G_{i j}\left(z_2, z_0\right)\right|_{z_2^{+}>0>z_0^{+}} \equiv \wick{
        \c2 \psi_{i} (z_2) 
         \left[ \; \overline{\c2 \psi}_{\text{eff}, j} \left(z_0\right) \right]_{z_0^{+}<0}} \nonumber \\ =\frac{i \Gamma(d+1)}{4(2 \pi)^{d+1}} \int d^d \boldsymbol{z}_1 V^{i j}_{\boldsymbol{z}_1} \frac{\left(z_2^{+} \gamma^{-}+\hat{z}_{21 \perp}\right) \gamma^{+}\left(-z_0^{+} \gamma^{-}+\hat{z}_{10 \perp}\right)}{\left(-z_0^{+} z_2^{+}\right)^{\frac{D}{2}}\left(-z_{20}^{-}+\frac{\boldsymbol{z}_{21}^{\;2}}{2 z_2^{+}}-\frac{\boldsymbol{z}_{10}^{\;2}}{2 z_0^{+}}+i \varepsilon\right)^{d+1}} \theta (z_2^{+} ) \theta (-z_0^{+} ) \; ,
         \label{Eq:Effective_Quark_Prop}
\end{gather}
\begin{gather}
   G^{ab}_{\mu \nu} (z_2, z_0) |_{z_2^{+}>0>z_0^{+}} \equiv \wick{ \c2 A_{\nu}^{b} (z_2)
         \left[ \c2 A_{\text{eff},\mu }^{a}\left(z_0\right) \right]_{z_0^{+}<0}  } \nonumber \\ = - \frac{\Gamma (d)}{2 (2 \pi)^{1+d}} \int d^d \boldsymbol{z}_{1} \frac{ \left[ z_2^+ g_{\perp \mu}^{\alpha} - z_{21 \perp}^{\alpha} n_{2 \mu} \right] U_{z_1}^{ab} \left[ -z_0^+ g^{\perp}_{\alpha \nu} - z_{10 \perp \alpha} n_{2 \nu} \right] }{\left(-z_0^{+} z_2^{+}\right)^{\frac{D}{2}}\left(-z_{20}^{-}+\frac{\boldsymbol{z}_{21}^{\;2}}{2 z_2^{+}}-\frac{\boldsymbol{z}_{10}^{\;2}}{2 z_0^{+}}+i \varepsilon\right)^{d}} \; .
\end{gather}
Now, there is an important point to stress. In eqs.~(\ref{Eq:EffecQBarOpe}, \ref{Eq:EffecQOpe}, \ref{Eq:EffecGOpe}), contributions without Wilson lines, that we will name from now on as \textit{monopole terms}, appear. These terms cancel each other at the level of propagators and external states, as can be seen above. However at the level of field operators, their cancellation is non trivial. The nature of these terms will be further discussed in section \ref{Monopole_terms}.
\subsection{Derivation of the effective operators}
\label{Derivation_of_the_effective_operators}
\begin{figure} \includegraphics[scale=0.50]{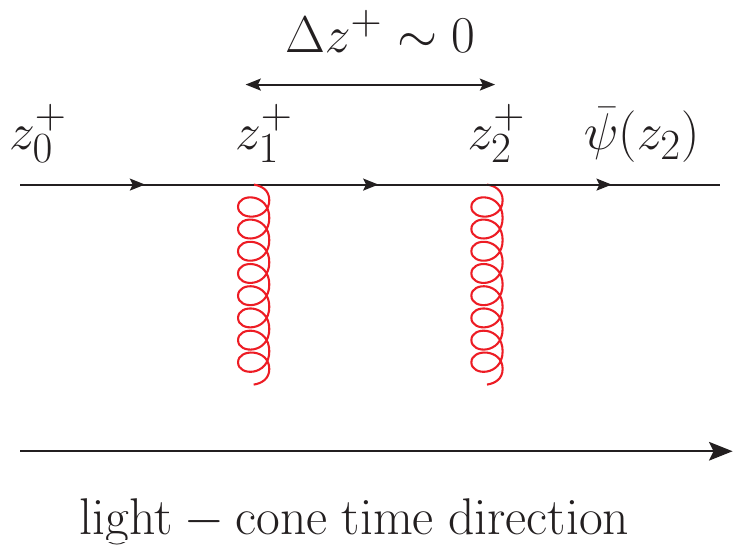}
\caption{Pictorial representation of a double interaction with the shockwave field.}
  \label{fig:DoubleInterShock}
\end{figure}
A easygoing derivation of these results can be obtain by mathematical induction. Let us consider the antiquark field within the background field of eq.~(\ref{Eq:Effective_Lagran_inter_Shockwave}). We first consider the case of two interactions, pictorially represented in fig.~\ref{fig:DoubleInterShock}:
\begin{gather}
    \left[ \bar{\psi}_{\rm eff} (z_0) \right]^{(2)} = \int d^D z_1 \int d^D z_2 \; \bar{\psi} (z_2) \left[ i g b^{-} (z_2) \gamma^+ \right] G_0 (z_{21}) \left[ i g b^{-} (z_1) \gamma^+ \right] G_0 (z_{10}) \theta (-z_{0}^{+}) \nonumber \\
    = (ig)^2 \int d z_1^+ d z_1^{-} d^d \boldsymbol{z}_1 \int d z_2^+ d z_2^{-} d^d \boldsymbol{z}_2 b^{-} (z_2) b^{-} (z_1)  \theta (-z_0^+) \int \frac{d p_1^+ d p_1^- d^d \boldsymbol{p}_1}{(2 \pi)^D} \int \frac{d p_2^+ d p_2^- d^d \boldsymbol{p}_2}{(2 \pi)^D} \nonumber \\ \times \int \frac{d p_3^+ d p_3^- d^d \boldsymbol{p}_3}{(2 \pi)^D}  \bar{\psi} (p_3) \gamma^+ \frac{i p_2^+ \gamma^{-}}{2 p_2^+ p_2^{-} + p_{2 \perp}^2 + i0 } \gamma^+ \frac{i ( p_1^+ \gamma^{-} + \slashed{p}_{1 \perp} )}{2 p_1^+ p_1^{-} + p_{1 \perp}^2 + i0 } e^{i p_3 \cdot z_2 - i p_2 z_{21} - i p_1 \cdot z_{10}} \; .
\end{gather}
We can now perform the integration over $z_2^{-}$ and $z_1^{-}$ exploiting the fact that within the eikonal approximation the field $b(x)$ is independent of the variable $x^{-}$. In this way, we obtain the conservation of momentum along the longitudinal direction, i.e. $p_1^+ = p_2^+ = p_3^+ \equiv p^+$. Then, we can immediately integrate with respect to $p_1^{-}$ and $p_2^{-}$ to get  
\begin{gather}
    \left[ \bar{\psi}_{\rm eff} (z_0) \right]^{(2)} = (ig)^2 \int d z_1^+ d^d \boldsymbol{z}_1 \int d z_2^+ d^d \boldsymbol{z}_2 b^{-} (z_2) b^{-} (z_1)  \theta (-z_0^+) \theta (z_{10}^+) \theta (z_{21}^+) \int \frac{d p^+ d^d \boldsymbol{p}_1}{(2 \pi)^{D-4}} \nonumber \\ \times  \frac{\theta (p^+)}{2 p^+}  \int \frac{ d^d \boldsymbol{p}_2 }{(2 \pi)^D}  \int \frac{d p_3^- d^d \boldsymbol{p}_3}{(2 \pi)^D} \bar{\psi} (p^+, p_3^-, \boldsymbol{p}_3) \gamma^+ (p^+ \gamma^{-} + \slashed{p}_{1 \perp}) e^{i p^+ z_0^{-}} e^{i p_3^- z_2^+} \nonumber \\ \times e^{- i \boldsymbol{p}_3 \cdot \boldsymbol{z}_2 } e^{ i \boldsymbol{p}_2 \cdot \boldsymbol{z}_{12} } e^{ i \boldsymbol{p}_1 \cdot \boldsymbol{z}_{10} } e^{-i \frac{\boldsymbol{p}_2^2 - i0}{2 p^+} z_{21}^+} e^{-i \frac{\boldsymbol{p}_1^2 - i0}{2 p^+} z_{10}^+} \; .
\end{gather}
Now, we exploit again the shockwave approximation, i.e. the fact that $z_1^+ \approx z_2^+ \approx 0$, to integrate over $\boldsymbol{p}_2$ and get
$\delta (\boldsymbol{z}_{12})$, which tell us that the interaction with the shockwave occurs at a single transverse coordinate. Then, the shockwave approximation allows us to write
\begin{gather}
    \int \frac{d p_3^- d^d \boldsymbol{p}_3}{(2 \pi)^D} \bar{\psi} (p^+, p_3^-, \boldsymbol{p}_3) e^{- i \boldsymbol{p}_3 \cdot \boldsymbol{z}_1 }  e^{i p_3^- z_2^+} \approx \int d z_1^- \int \frac{d p_3^- d^d \boldsymbol{p}_3}{(2 \pi)^D} \tilde{\bar{\psi}} (z_1^-, p_3^-, \boldsymbol{p}_3) e^{- i \boldsymbol{p}_3 \cdot \boldsymbol{z}_1 }  e^{i p^+ z_1^-} \; ,
\end{gather}
where the $\tilde{\bar{\psi}}$ is a quark field in a mixed momentum/coordinate representation and we exploited the $z_2^+ \approx 0$ approximation. Then, we can conveniently write
\begin{gather}
    \int d z_1^- \int \frac{d p_3^- d^d \boldsymbol{p}_3}{(2 \pi)^D} \tilde{\bar{\psi}} (z_1^-, p_3^-, \boldsymbol{p}_3) e^{- i \boldsymbol{p}_3 \cdot \boldsymbol{z}_1 } e^{i p^+ z_1^-} = \frac{1}{2 \pi} \int d z_1^- \bar{\psi} (z_1) |_{z_1^+ = 0} \; e^{i p^+ z_1^-} \; ,
\end{gather}
to obtain
\begin{gather}
    \left[ \bar{\psi}_{\rm eff} (z_0) \right]^{(2)} \hspace{-0.15 cm} = \hspace{-0.15 cm} \int \hspace{-0.15 cm} d z_1^{-}  d^d \boldsymbol{z}_1 \bar{\psi} (z_1) |_{z_1^+ = 0} \gamma^+ \hspace{-0.15 cm}  \left[ (ig)^2 \hspace{-0.15 cm} \int d z_1^+ \hspace{-0.15 cm}  \int \hspace{-0.15 cm} d z_2^+ b^{-} (z_1^+, \boldsymbol{z}_1) b^{-} (z_2^+, \boldsymbol{z}_2) \theta (z_{10}^+) \theta (z_{21}^+) \right] \nonumber \\
    \times \int \frac{ d^d \boldsymbol{p}_1 d p^+}{( 2 \pi)^{D-1}} \frac{\theta (p^+)}{2 p^+} e^{-i p^+ z_{10}^-} e^{-i ( - \boldsymbol{p}_1 \cdot \boldsymbol{z}_{10} - \frac{\boldsymbol{p}_1^2-i0}{2 p^+} z_0^+ )} (p^+ \gamma^- + \slashed{p}_{1 \perp}) \theta (-z_0^+) \; .
\end{gather}
Integrating first over $\boldsymbol{p}_1$ and then over $p^+$, we finally find\footnote{Please note that in the final step we reintroduce a variables that for 
aesthetics we call $z_1^+$ again, it should not be confused with the one that has the same name inside the 2-order expansion of the Wilson line.}
\begin{gather}
    \left[ \bar{\psi}_{\rm eff} (z_0) \right]^{(2)} = \int d z_1^{-}  d^d \boldsymbol{z}_1 \bar{\psi} (z_1) |_{z_1^+ = 0} \gamma^+ V_{\boldsymbol{z}_1}^{(2)} \frac{\Gamma (D/2)}{2 \pi^{D/2}} \frac{i (-z_0^+ \gamma^{-} + \slashed{z}_{10 \perp})}{\left( 2 z_0^+ z_{10}^{-} - z_{10 \perp}^2 + i0\right)^{D/2}} \theta (- z_0^+) \nonumber \\
    = \int d^D z_1 \bar{\psi} (z_1) \gamma^+ V_{\boldsymbol{z}_1}^{(2)} G_0 (z_{10}) \delta (z_1^+) \theta (- z_0^+) \; ,
    \label{Eq:Double_ShockWave_Int}
\end{gather}
where
\begin{gather}
    [V_{\boldsymbol{z}_1}^{ij}]^{(2)} = (ig)^2 \int d z_1^+ \int d z_2^+ b^{-} (z_1^+, \boldsymbol{z}_1) b^{-} (z_2^+, \boldsymbol{z}_2) \theta (z_{10}^+) \theta (z_{21}^+) \; .
\end{gather}
It is clear that eq.~(\ref{Eq:Double_ShockWave_Int}) is the second order expansion of the term containing the Wilson line in eq.~(\ref{Eq:EffecQOpe}). More generally, we define $V_{\boldsymbol{z}_1}^{(n)}$ as the $n$-th order in the expansion of the Wilson line in the fundamental representation. We observe that the case of a single interaction ($n=1$) is trivial:
\begin{gather}
    \left[ \bar{\psi}_{\rm eff} (z_0) \right]^{(1)} = \int d^D z_1  \bar{\psi} (z_1) \left[ i g b^{-} (z_1) \gamma^+ \right] G_0 (z_{10}) \theta (-z_0^+) \nonumber \\ = \int d^D z_1 \bar{\psi} (z_1) \gamma^+ V_{\boldsymbol{z}_1}^{(1)} G_0 (z_{10}) \delta (z_1^+) \theta (- z_0^+) \; .
\end{gather}
Now, we show that if we assume that the result is correct for $n$ interactions, i.e.
\begin{gather}
    \left[ \bar{\psi}_{\rm eff} (z_0) \right]^{(n)} = \int d^D z_1 \bar{\psi} (z_1) \gamma^+ V_{\boldsymbol{z}_1}^{(n)} G_0 (z_{10}) \delta (z_1^+) \theta (- z_0^+) \; ,
    \label{Eq:Induction_n}
\end{gather}
then, the $n+1$ case follows. The $n+1$-interaction case reads
\begin{gather}
    \left[ \bar{\psi}_{\rm eff} (z_0) \right]^{(n+1)} = \int d^D z_{n+1} \int d^D z_{n} \; ... \int d^D z_{1} \nonumber \\ \times \bar{\psi} (z_{n+1}) \left[ i g b^{-} (z_{n+1}) \gamma^+ \right] G_0 (z_{(n+1) n}) \; ... \; G_0 (z_{21}) 
    \left[ i g b^{-} (z_1) \gamma^+ \right] G_0 (z_{10}) \theta (-z_0^+) \; .
\end{gather}
We can immediately rewrite this expression as
\begin{equation}
   \left[ \bar{\psi}_{\rm eff} (z_0) \right]^{(n+1)} = \int d^D z_{n+1} \bar{\psi} (z_{n+1}) \gamma^+ i g b^{-} (z_{n+1}) G^{(n)} \left(z_{n+1}, z_0\right) \; ,
\end{equation}
where  
\begin{equation}
     G^{(n)} \left(z_{n+1}, z_0\right) = \int d^D z_n \; ... \int d^D z_1 G_0 (z_{(n+1) n}) \; ... \; G_0 (z_{21}) 
    \left[ i g b^{-} (z_1) \gamma^+ \right] G_0 (z_{10}) \theta (-z_0^+)
\end{equation}
is the $n$-order expansion of the effective propagator in eq.~(\ref{Eq:Effective_Quark_Prop}) that can be easily expressed as
\begin{gather}
    G^{(n)} \left(z_{n+1}, z_0\right)  \equiv \wick{
        \c2 \psi  (z_{n+1}) 
         \left[ \; \overline{\c2 \psi}_{\text{eff}} \left(z_0\right) \right]^{(n)} } \; .
\end{gather}
Using expression (\ref{Eq:Induction_n}), we get
\begin{gather}
    \left[ \bar{\psi}_{\rm eff} (z_0) \right]^{(n+1)} = \int d z_{n+1}^{-} d^d \boldsymbol{z}_{n+1} \bar{\psi} (z_{n+1}) |_{z_{n+1}^+ = 0} \nonumber \\
    \times \left[ \gamma^+ \int d^D z_1 G_0 (z_{(n+1)1}) \gamma^+ G_0 (z_{10}) i g \mathbf{B} (\boldsymbol{z}_{n+1}) V_{\boldsymbol{z}_{1}}^{(n)} \delta (z_1^+) \theta (-z_0^+) \right]_{z_{n+1}^+=0} \; .
\end{gather}
Performing algebraic manipulations similar to those used for the two interactions case, we obtain
\begin{gather}
    \left[ \bar{\psi}_{\rm eff} (z_0) \right]^{(n+1)} = \int d z_{n+1}^{-} d^d \boldsymbol{z}_{n+1} \bar{\psi} (z_{n+1}) |_{z_{n+1}^+ = 0} \gamma^+ G_{0} (z_{(n+1) 0}) |_{z_{n+1}^+ = 0} V_{\boldsymbol{z}_{n+1}}^{(n+1)} \nonumber \\
    = \int \mathrm{d}^D z_{n+1} \bar{\psi}\left(z_{n+1} \right) \gamma^{+} V_{\boldsymbol{z}_{n+1}}^{(n+1)} G_0\left(z_{(n+1) 0}\right) \theta \left( - z_{0}^{+}\right) \; ,
\end{gather}
which is exactly in the expected form. We have demonstrated that the effective operator for $n \geq 1$ is the one described by the term with the Wilson line in eq.~(\ref{Eq:EffecQOpe}). If we subtract the term corresponding to the $n=0$ case and we add the standard QCD antiquark field, we end up with the effective operators in eq.~(\ref{Eq:EffecQOpe}). Similarly, one obtains the two effective operators in eqs.~(\ref{Eq:EffecQBarOpe}, \ref{Eq:EffecGOpe}).

\subsection{Monopole terms}
\label{Monopole_terms}
The direct use of the effective operators (\ref{Eq:EffecQBarOpe}, \ref{Eq:EffecQOpe}, \ref{Eq:EffecGOpe}) would naturally lead to the appearance of monopole terms, which are always associated with the external "legs" of the impact factor. In this section, we will derive a general expression for the sum of these terms in eq.~(\ref{Eq:EffecQBarOpe}, \ref{Eq:EffecQOpe}, \ref{Eq:EffecGOpe}), at the operator level, and show that they never contributes to the coefficient function (the impact factor in the present case). \\

\noindent \textbf{Antiquark and quark monopoles} \\

\noindent We start with the antiquark case, considering the sum of the non-interacting terms (i.e. the first and the third term) in eq.~(\ref{Eq:EffecQBarOpe}):
\begin{equation}
    \left[\bar{\psi}_{\text{eff }}\left(z_0\right)\right]^{\text{n.i}}_{z_0^+ < 0}  = \bar{\psi}\left(z_0\right) - \int \mathrm{d}^D z_1 \bar{\psi}\left(z_1\right) \gamma^{+} G_0\left(z_{10}\right) \delta\left(z_1^{+}\right)  \; .
\end{equation}
Writing the first term as\footnote{We stress here that $z_0^+ < 0 $ and therefore the $\theta (-z_1^+)$ can be safely introduced.}
\begin{equation}
    \bar{\psi}\left(z_0\right)  =  \int d^D z_1 \int \frac{d^D k}{(2 \pi)^D} e^{-i k \cdot z_{10}} \bar{\psi} (z_1) \theta (-z_1^{+})
    \label{Eq:DeltaTrick}
\end{equation}
and using the explicit expression for the quark propagator, we get
\begin{gather}
   \left[\bar{\psi}_{\text{eff }} \left(z_0\right)\right]^{\text{n.i}} \nonumber \\  =  \int d^D z_1 \int \frac{d^D k}{(2 \pi)^D} \frac{\bar{\psi} (z_1)}{k^2 + i0} \left[ k^2 \theta (-z_1^+) - i (2 k^+ - \slashed{k} \gamma^+) \delta (z_1^+) \right] e^{-i k \cdot z_{10}} \; .
\end{gather}
We can rewrite the momenta in the square bracket as a derivative acting on the exponential, i.e.
\begin{gather}
    \left[\bar{\psi}_{\text{eff }}\left(z_0\right)\right]^{\text{n.i}} =   \int d^D z_1 \int \frac{d^D k}{(2 \pi)^D} \frac{\bar{\psi} (z_1)}{k^2 + i0} \nonumber \\ \times \left[ - \theta (-z_1^+) \left( 2 \frac{\partial}{\partial z_1^{-}} \frac{\partial}{\partial z_1^{+}} + \frac{\partial}{\partial z_{1 \perp}^{\mu}} \frac{\partial}{\partial z_{1\perp \mu}} \right) + \left( 2 \frac{\partial}{\partial z_1^{-}} - \frac{\partial}{\partial z_{1 \mu}} \gamma_{\mu} \gamma^+ \right) \delta (z_1^+) \right] e^{-i k \cdot z_{10}} 
\end{gather}
and then integrate by parts to get
\begin{gather}
     \left[\bar{\psi}_{\text{eff }} \left(z_0\right)\right]^{\text{n.i}}  \nonumber \\ =  \int d^D z_1 \int \frac{d^D k}{(2 \pi)^D} \frac{1}{k^2 + i0} \left[ - \big( \bar{\psi} \overleftarrow{\Box} \big) (z_1) \theta (-z_1^+) +  \big(\bar{\psi} \overleftarrow{\slashed{\partial}} \big) (z_1) \gamma^{+} \delta (z_1^+) \right] e^{-i k \cdot z_{10}} \; .
\end{gather}
Writing $\bar{\psi} \overleftarrow{\Box} = \bar{\psi} \overleftarrow{\slashed{\partial}} \overleftarrow{\slashed{\partial}}$ and integrating by parts again, we finally find
\begin{equation}
    \left[\bar{\psi}_{\text{eff }}\left(z_0\right)\right]^{\text{n.i}} = - \int d^D z_1 \, \theta (-z_1^+) \big( \bar{\psi} \overleftarrow{\slashed{\partial}} \big) (z_1)  G_{0} (z_{10}) \; .
    \label{Final_Monopoles_AntiQuark}
\end{equation}
It is easy to verify that for the $\psi$ field, one has
\begin{equation}
    \left[ \psi_{\text{eff }}\left(z_0\right)\right]^{\text{n.i}} = \int d^D z_2 \, \theta (-z_2^+) G_{0} (z_{20}) \big( \slashed{\partial} \psi \big) (z_2)  \; . 
\label{Final_Monopoles_Quark}
\end{equation}
Eqs.~(\ref{Final_Monopoles_AntiQuark}) and~(\ref{Final_Monopoles_Quark}) contain exactly the kinetic part of the QCD equations of motion 
\begin{equation}
    \bar{\psi} \overleftarrow{\slashed{D}} = 0 \hspace{0.5 cm} \text{and} \hspace{0.5 cm} \slashed{D} \psi = 0 \; ,
\end{equation}
where
\begin{equation}
    D_{\mu} = \partial_{\mu} - i g A_{\mu} \,. 
\end{equation}
\begin{figure}
\includegraphics[scale=0.40]{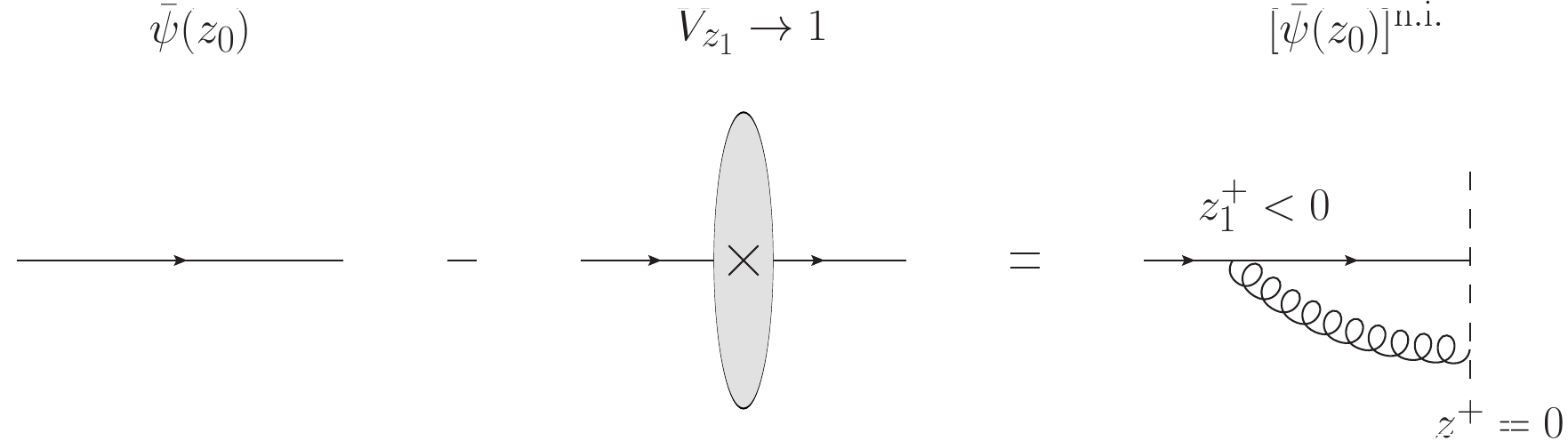}
\caption{Pictorial representation of the quark monopoles.}
  \label{fig:MonopoleSplitting}
\end{figure}
This means that eqs.~(\ref{Final_Monopoles_AntiQuark}) and~(\ref{Final_Monopoles_Quark}) can be ultimately rewritten as
\begin{equation}
    \left[\bar{\psi}_{\text{eff }}\left(z_0\right)\right]^{\text{n.i}}  = - \int d^D z_1 \, \theta (-z_1^+)  i g  \bar{\psi} (z_1) \slashed{A} (z_1)    G_{0} (z_{10}) 
    \label{Final_Monopoles_AntiQuark2}
\end{equation}
and
\begin{equation}
    \left[ \psi_{\text{eff }}\left(z_0\right)\right]^{\text{n.i}} = - \int d^D z_2 \,\theta (-z_2^+) G_{0} (z_{20}) i g \slashed{A} (z_2) \psi (z_2)  \; . 
\label{Final_Monopoles_Quark2}
\end{equation}
It is now clear that the combination of the two monopoles terms in eqs.~(\ref{Eq:EffecQBarOpe}, \ref{Eq:EffecQOpe}) corresponds to a final state emission of an additional gluon occurring at \textit{negative} light-cone time, as pictorially depicted in fig.~\ref{fig:MonopoleSplitting} for the $\bar{\psi}$ field. In the exclusive factorization, these final state emissions do not contribute to the coefficient functions, but they are part of the evolution kernel of the collinear DAs. \\

\begin{figure}
\hspace{0.05 cm} \includegraphics[scale=0.40]{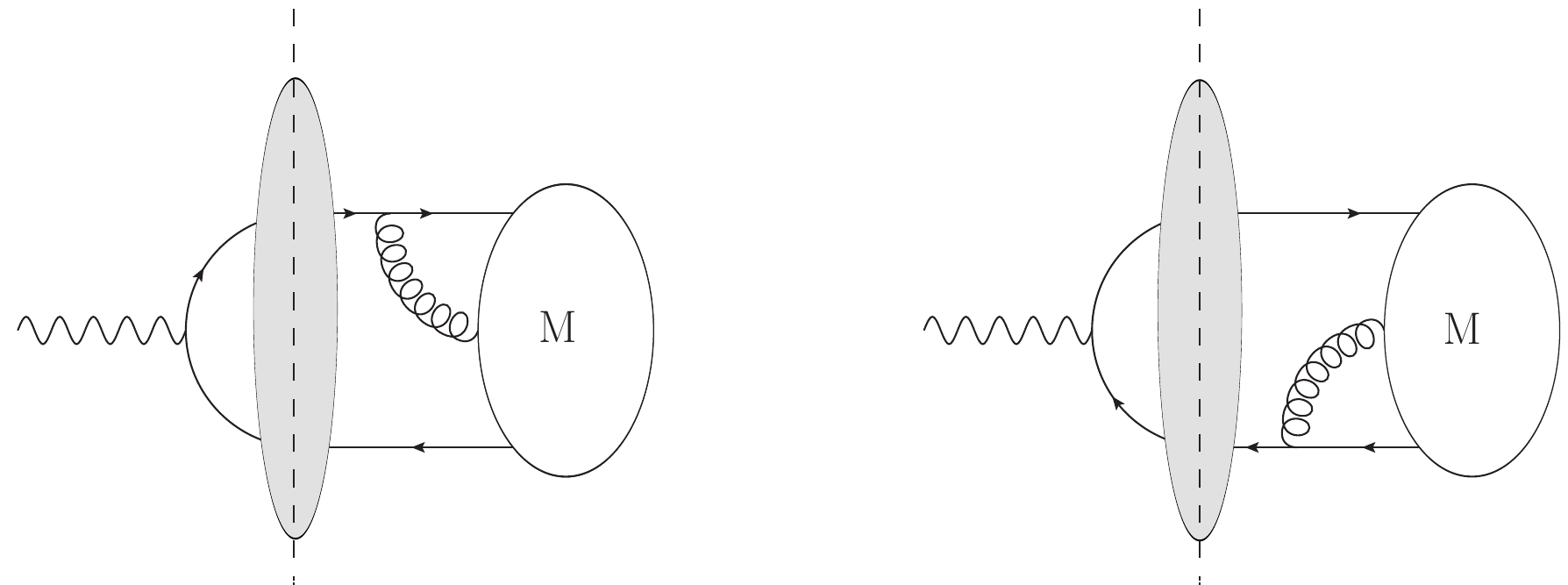} \vspace{0.5 cm} \\
\includegraphics[scale=0.40]{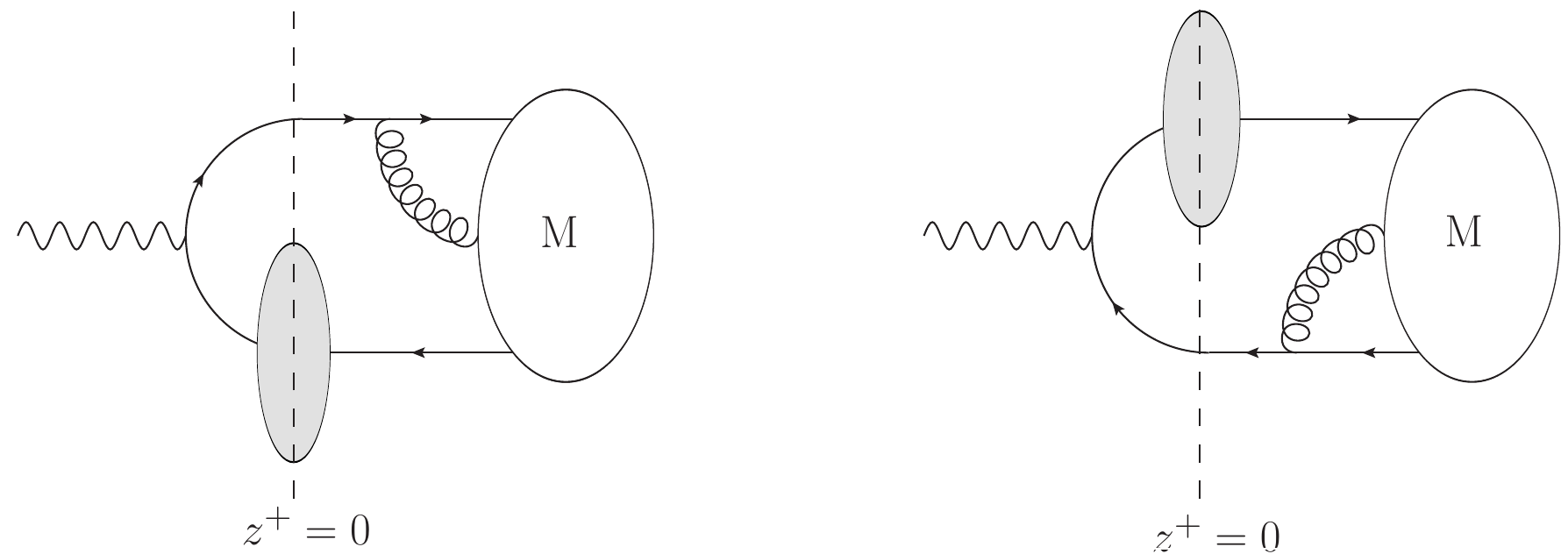}
\caption{On the top of the figure, the diagrams in which the gluon is emitted after the shockwave. In the bottom, their contributions when the Wilson line preceding the gluon emission is set to one.}
  \label{fig:MonopoleDiagrams}
\end{figure}

For illustrative purposes, let us consider the example of the impact factor for the production of a light vector meson within twist-3 accuracy, which is the subject of the present work. As anticipated in the introduction, the twist-3 calculation requires taking into account both the kinematic twist effects, i.e. with a slightly off-collinear $q\bar{q}$ pair in the final state, and the genuine twist ones, i.e. with a collinear $q\bar{q}g$ system in the final state. When the $q \bar{q}$ contribution is calculated, the product of the effective operators eqs.~(\ref{Eq:EffecQBarOpe}, \ref{Eq:EffecQOpe}), produces three contributions. The product of the interacting terms, i.e. containing the two Wilson lines, is the 2-body contribution to the impact factor, which ultimately leads to the standard dipole form of the amplitude. 
\begin{figure}
\includegraphics[scale=0.40]{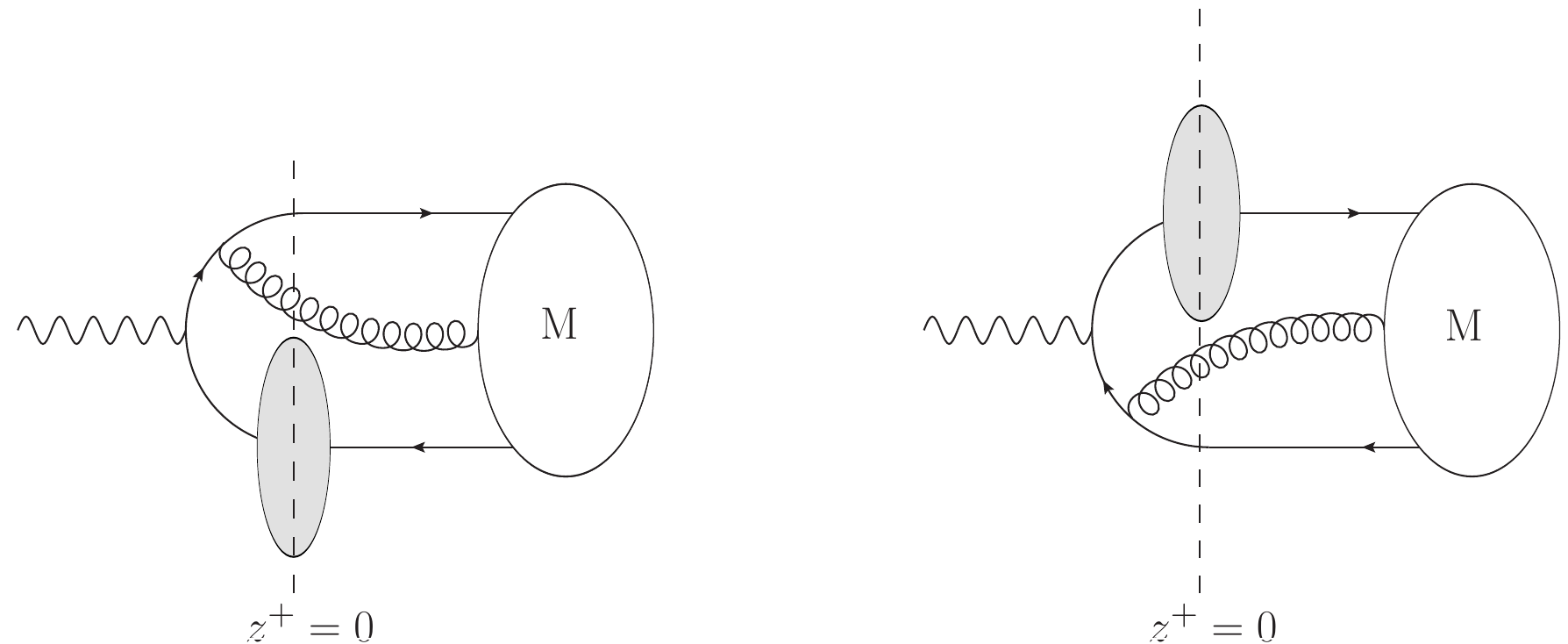}
\caption{Monopole associated diagrams in the $\gamma \rightarrow q \bar{q}$ contribution.}
  \label{fig:MonopoleDiagramsInTheImpact}
\end{figure}
The other two contributions (see fig.~\ref{fig:MonopoleDiagramsInTheImpact}) are instead the diagrams produced by the aforementioned monopole terms. They are just a small subset of the terms contributing to the evolution. In the contribution in which the $q \bar{q} g$ system is produced, there are four diagrams, in two of which the gluon emission occurs after the shockwave, i.e at \textit{positive} light-cone time $z^+ > 0$ (see top of fig.~\ref{fig:MonopoleDiagrams}). In these diagrams the single monopole term contribution correspond to a $q \bar{q} g g$ final state and can be completely neglected within twist-3 accuracy (in the impact factor). Since quark, antiquark and gluon are collinear within twist-3, any emission after the shockwave leads to a purely singular contribution that is associated with the ERBL evolution of the twist-2 DA and must be excluded from the coefficient function. It is important to stress that the gluon emission occurs strictly at $z^+>0$. 
This is compatible with the evolution since if the emission occurred before the interaction with the shockwave then the collinearity of the gluon-quark-antiquark system would not force any propagator to be on its mass shell, due to the transverse kick from the shockwave gluon(s). Note, however, that there is a single exception to this reasoning. Indeed, the Wilson lines preceding the gluon emission in diagrams in the top of fig.~\ref{fig:MonopoleDiagrams} contain an infinite resummation of interactions with the target, including the case where there is no interaction, and thus obviously no transverse kick. These terms, shown in the bottom of fig.~\ref{fig:MonopoleDiagrams}, do not contain the entire contribution associated with the ERBL evolution, since, in this configuration, also the negative light-cone time emissions should be included. These "missing" contributions are exactly the ones generated by the monopole contributions in the $q \bar{q}$ term in fig.~\ref{fig:MonopoleDiagramsInTheImpact}. Thus, the two contributions together generate the correct result. \\

We now consider the gluon monopoles and show that they have an identical interpretation. Since the gluon effective operator enters only in a genuine twist 3 contribution, the effect of the gluon monopoles starts with the twist-3 evolution. \\

\noindent \textbf{Gluon monopoles} \\

\begin{figure}
\begin{picture}(430,80)
\put(-5,20){ \includegraphics[scale=0.315]{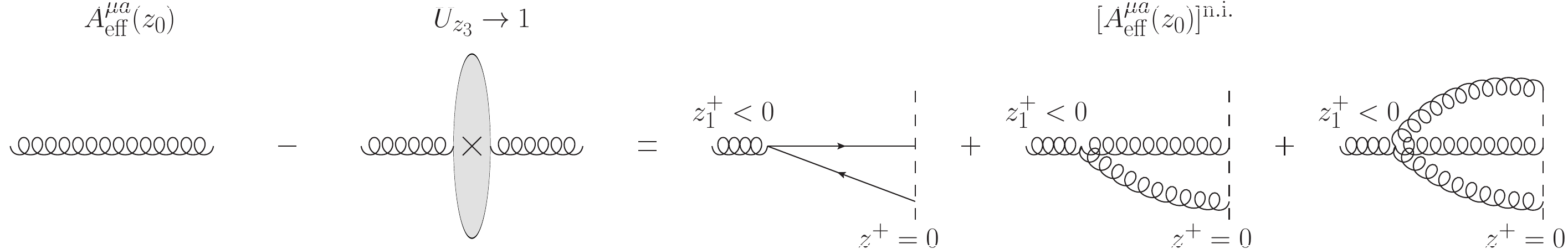}}
\end{picture}
\caption{Pictorial representation of the gluon monopoles.}
  \label{fig:GluonMonopoleSplitting}
\end{figure}

\noindent In the gluon effective operators, in eq.~(\ref{Eq:EffecGOpe}), the non-interacting term is
\begin{gather} 
{\left[A_{\text {eff }}^{a \mu}\left(z_0\right)\right]_{z_0^{+}<0}^{\mathrm{n. i.}} }  =A^{a \mu}\left(z_0\right)-2 i \int \mathrm{d}^D z_3 \, \delta\left(-z_3^{+}\right) F_{ \; \; \; \; \; \sigma}^{a+}\left(z_3\right) G^{\mu \sigma_{\perp}}\left(z_{30}\right) \nonumber \\ 
 = A^{a \mu}\left(z_0\right)-2 i \int \mathrm{d}^D z_3\left(-\partial_{+} \theta\left(-z_3^{+}\right)\right)\left(\partial^{+} A_\sigma^a\right)\left(z_3\right) G^{\mu \sigma_{\perp}}\left(z_{30}\right) \nonumber \\ 
=A^{a \mu}\left(z_0\right)-i \int \mathrm{d}^D z_3 \, \theta\left(-z_3^{+}\right)\left[\left(\square A_\sigma^a\right)\left(z_3\right) G^{\mu \sigma_{\perp}}\left(z_{30}\right)-A_\sigma^a\left(z_3\right) \square G^{\mu \sigma_{\perp}}\left(z_{30}\right)\right] \; , 
\end{gather}
where we made extensive use of integration by parts. Using the explicit form of the $\mu, \sigma_{\perp}$ component of the gluon propagator in the light-cone gauge\footnote{We specify that we choose the $n_2$ light-cone gauge with gauge fixing parameter $\xi = 0$.},
\begin{equation}
    G^{\mu \sigma_{\perp}} (z_{30}) = \int \frac{d^D l}{(2 \pi)^D} e^{-i l \cdot z_{30}} \frac{-i}{l^2+i0} \left( g^{\mu \sigma}_{\perp} - \frac{n_2^{\mu} l_{\perp}^{\sigma}}{l^+} \right)  \; ,
\end{equation}
we can prove that 
\begin{gather}
 A^{a \mu}\left(z_0\right) \; + \; i \int \mathrm{d}^D z_3 \; \theta\left(-z_3^{+}\right) A_\sigma^a\left(z_3\right)\left(\square G^{\mu \sigma}\right)\left(z_{30}\right) \nonumber \\ =  i \int \mathrm{d}^D z_3 \theta\left(-z_3^{+}\right) \partial_\sigma\left(\partial_\nu A^{a \nu}\right)\left(z_3\right) G^{\mu \sigma}\left(z_{30}\right) \; .
\end{gather}
and, therefore, finally obtain
\begin{gather}
{\left[A_{\text {eff }}^{a \mu}\left(z_0\right)\right]_{z_0^{+}<0}^{\mathrm{ni.}} } = -i \int \mathrm{d}^D z_3 \; \theta\left(-z_3^{+}\right)\left[\left(\square A_\sigma^a - \partial_{\sigma} \left( \partial_{\nu} A^{a \nu} \right) \right)\left(z_3\right) \right]  G^{\mu \sigma_{\perp}}\left(z_{30}\right) \; . 
\label{Final_Monopoles_Gluon}
\end{gather}
In eq.~(\ref{Final_Monopoles_Gluon}) the term in the square bracket is the kinetic part of the Yang-Mills equation
\begin{equation}
     \square A_\sigma - \partial_\sigma\left(\partial_\nu A^\nu\right) =-g \bar{\psi} \gamma_\sigma \psi+i g \partial_\nu\left[A^\nu, A_\sigma\right]-i g A_\nu\left(\partial_\sigma A^\nu-\partial^\nu A_\sigma\right)+g^2 A_\nu\left[A^\nu, A_\sigma\right] .
\end{equation}
Thus, the monopole terms associated with the gluon effective operator, as in the fermionic case, can be interpreted in terms of final state emissions associated with the gluon dynamics, as pictorially depicted in fig.~\ref{fig:GluonMonopoleSplitting}. \\

We have shown that the monopole terms never contribute to the impact factor, therefore, in the next section, we remove the non-interacting terms in eqs.~(\ref{Eq:EffecQBarOpe}, \ref{Eq:EffecQOpe}, \ref{Eq:EffecGOpe})  and directly use
\begin{equation}
    \left[\psi_{\text {eff }}\left(z_0\right)\right]_{z_0^{+}<0} = - \int \mathrm{d}^D z_2 G_0\left(z_{02}\right) V_{\boldsymbol{z}_2}^{\dagger} \gamma^{+} \psi\left(z_2\right) \delta\left(z_2^{+}\right) \; ,
    \label{Eq:PsiEffecNoMon}
\end{equation}
\begin{equation}
    \left[\bar{\psi}_{\text{eff }}\left(z_0\right)\right]_{z_0^{+}<0} = \int \mathrm{d}^D z_1 \bar{\psi}\left(z_1\right) \gamma^{+} V_{\boldsymbol{z}_1} G_0\left(z_{10}\right) \delta\left(z_1^{+}\right) \; ,
    \label{Eq:PsiBarEffecNoMon}
\end{equation}
\begin{equation}
   \left[ A_{\text {eff }}^{\mu a}\left(z_0\right) \right]_{z_0^{+}<0} = 2 i \int \mathrm{d}^D z_3 \delta\left(z_3^{+}\right) F_{-\sigma}^b\left(z_3\right) G^{\mu \sigma_{\perp}}\left(z_{30}\right) U_{\boldsymbol{z}_3}^{a b}\; . 
   \label{Eq:AEffecNoMon}
\end{equation}

\section{Exclusive light vector meson production in a saturation framework}
\label{Sec:exclusive-meson}

In this section, we are setting up the calculation of the amplitudes for the production of a light vector meson at the twist-3 within the shockwave approach. The computation is split into two subsections: in the first one, we compute the 2-body contribution where the meson is produced from a quark-antiquark, then we evaluate the 3-body contribution where it is procuced with an additional (non-perturbative) gluon. At this point, both contributions will contain infinite kinematic twists (i.e. the partons later forming the hadron are not expanded around the collinear direction). When expanding in twists in a later section, we will see that the 2-body contribution contains twist-2 and twist-3 terms, while the 3-body contribution is a pure twist-3 one. In the third and last subsection, we provide a summary of these intermediate results.

\subsection{2-body contribution}
\label{Eq:2-body_contribution}

\begin{figure} \includegraphics[scale=0.40]{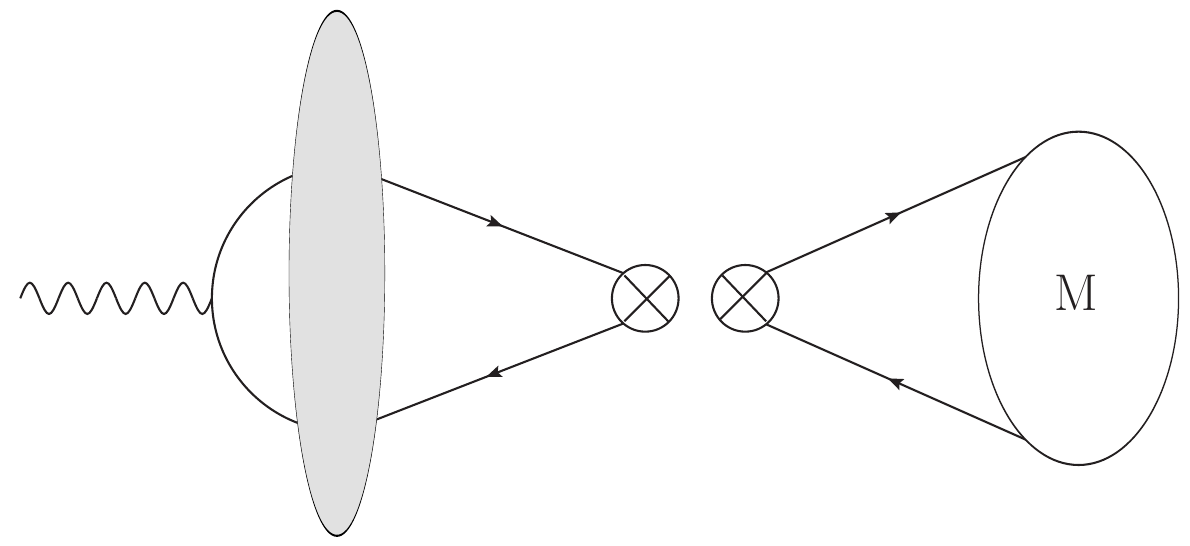}
\caption{Pictorial representation of the 2-body contribution factorization.}
  \label{fig:Shock2BodyContribution}
\end{figure}

The general form of the 2-body contribution to the EDMP amplitude reads
\begin{equation}
{\cal A}_2 =-ie_{q}\int{\rm d}^{D}z_{0}\theta\left(-z_{0}^{+}\right)\left\langle P\left(p^{\prime}\right)M\left(p_{M}\right)\left|\overline{\psi}_{{\rm eff}}\left(z_{0}\right)\slashed{\varepsilon}_{q}{\rm e}^{-i\left(q\cdot z_{0}\right)}\psi_{{\rm eff}}\left(z_{0}\right)\right|P\left(p\right)\right \rangle \; , 
\end{equation}
with the effective operators defined in eqs.~(\ref{Eq:PsiEffecNoMon}) and (\ref{Eq:PsiBarEffecNoMon}), i.e. excluding the contribution of monopole terms. Relying on rapidity factorization, also named high-energy operator product expansion (OPE)~\cite{Balitsky:1995ub}, and using Fierz decomposition in both color and Dirac space, we can factorize the scattering amplitude as pictorially depicted in Fig.~\ref{fig:Shock2BodyContribution} in order to get
\begin{gather}
{\cal A}_{2} = - ie_{q}\int{\rm d}^{D}z_{0}\int{\rm d}^{D}z_{1}\int{\rm d}^{D}z_{2}\theta\left(-z_{0}^{+}\right)\delta\left(z_{1}^{+}\right)\delta\left(z_{2}^{+}\right) \left\langle P\left(p^{\prime}\right)\big| \mathcal{U}_{12} \big| P\left(p\right)\right\rangle \nonumber \\ \times\left\langle M\left(p_{M}\right)\left|\overline{\psi}\left(z_{1}\right)\Gamma^{\lambda}\psi\left(z_{2}\right)\right|0\right\rangle \frac{1}{4}{\rm tr}\left[\gamma^{+}G_{0}\left(z_{10}\right)\slashed{\varepsilon}_{q}{\rm e}^{-i\left(q\cdot z_{0}\right)}G_{0}\left(z_{02}\right)\gamma^{+}\Gamma_{\lambda}\right] \; ,
\label{Eq:BeginTwoBodyConSub}
\end{gather}
where
\begin{equation}
\mathcal{U}_{\boldsymbol{z}_1 \boldsymbol{z}_2} \equiv 1-\frac{1}{N_{c}}{\rm tr}\left(V_{\boldsymbol{z}_{1}}V_{\boldsymbol{z}_{2}}^{\dagger}\right) 
\end{equation}
is the dipole operator, which is constructed in such a way that for $\boldsymbol{z}_1 = \boldsymbol{z}_2$ it vanishes, in agreement with color transparency when the dipole size goes to zero. Please note that eq.~(\ref{Eq:BeginTwoBodyConSub}) has been obtained after subtracting the non-interacting part of the amplitude (i.e. all Wilson lines set to 1). Fourier transforming the propagators, integrating their $-$ components
by using Cauchy theorem and then integrating w.r.t. $z_{0}$ yields
\begin{gather}
{\cal A}_{2} = - e_{q}\!\int{\rm d}z_{1}^{-}\!\int{\rm d}z_{2}^{-}\!\int{\rm d}^{d}\boldsymbol{z}_{1}{\rm d}^{d}\boldsymbol{z}_{2} \left\langle P\left(p^{\prime}\right)\big| \mathcal{U}_{\boldsymbol{z}_1 \boldsymbol{z}_2} \big| P\left(p\right)\right\rangle \!\left\langle M\left(p_{M}\right)\left|\overline{\psi}\left(z_{1}\right)\Gamma^{\lambda}\psi\left(z_{2}\right)\right|0\right\rangle _{z_{1,2}^{+}=0} \nonumber \\
 \times \int_{0}^{q^{+}}\frac{{\rm d}k^{+}}{2\pi}\int\frac{{\rm d}^{d}\boldsymbol{k}}{\left(2\pi\right)^{d}} \frac{{\rm e}^{-ik^{+}z_{12}^{-}-iq^{+}z_{2}^{-}+i\left(\boldsymbol{k}\cdot\boldsymbol{z}_{12}\right)+i\left(\boldsymbol{q}\cdot\boldsymbol{z}_{2}\right)}}{16 k^{+} \left(q^{+}-k^{+}\right)\left(q^{-}-\frac{\boldsymbol{k}^{2}}{2k^{+}}-\frac{\left(\boldsymbol{q}-\boldsymbol{k}\right)^{2}}{2\left(q^{+}-k^{+}\right)}+i0\right)}  {\rm tr} \left[ \gamma^{+}\slashed{k}\slashed{\varepsilon}_{q}\left(\slashed{q}-\slashed{k}\right)\gamma^{+}\Gamma_{\lambda} \right] .  
\end{gather}
Introducing the longitudinal fraction of the photon momenta carried by the outgoing quark, $x = k^{+}/q^{+}$ and performing the shift $\boldsymbol{k}\rightarrow\boldsymbol{k}+x\boldsymbol{q}$, we obtain 
\begin{gather}
{\cal A}_{2} =e_{q}\int \hspace{-0.1 cm} {\rm d}z_{1}^{-} \hspace{-0.1 cm} \int{\rm d}z_{2}^{-} \hspace{-0.2 cm} \int{\rm d}^{d} \boldsymbol{z}_{1}{\rm d}^{d}\boldsymbol{z}_{2} \left\langle P\left(p^{\prime}\right)\big| \mathcal{U}_{\boldsymbol{z}_1 \boldsymbol{z}_2} \big| P\left(p\right)\right\rangle \nonumber \left\langle M\left(p_{M}\right)\left|\overline{\psi}\left(z_{1}\right)\Gamma^{\lambda}\psi\left(z_{2}\right)\right|0\right\rangle _{z_{1,2}^{+}=0} \\
  \times \hspace{-0.1 cm} \int_{0}^{1} \hspace{-0.1 cm} \frac{{\rm d}x}{2\pi} \hspace{-0.1 cm} \int \hspace{-0.1 cm} \frac{{\rm d}^{d}\boldsymbol{k}}{\left(2\pi\right)^{d}} \frac{ {\rm e}^{-ixq^{+}z_{1}^{-}-i\overline{x}q^{+}z_{2}^{-}+i\left(\boldsymbol{k}\cdot\boldsymbol{z}_{12}\right)+i\boldsymbol{q}\cdot\left(x\boldsymbol{z}_{1}+\overline{x}\boldsymbol{z}_{2}\right)}}{8 \left( \boldsymbol{k}^{2}+x\overline{x}Q^{2}-i0 \right)}
{\rm tr} \left[\gamma^{+}\left(x\slashed{q}+\slashed{k}_{\perp}\right)\slashed{\varepsilon}_{q}\left(\overline{x}\slashed{q}-\slashed{k}_{\perp}\right)\gamma^{+}\Gamma_{\lambda} \right] \; .
\end{gather}
The Dirac structure inside square brackets takes the form
\begin{gather}
  \gamma^{+} \left(x\slashed{q}+\slashed{k}_{\perp}\right)\slashed{\varepsilon}_{q}\left(\overline{x}\slashed{q}-\slashed{k}_{\perp}\right)\gamma^{+}
  =4x\overline{x}q^{+}q\cdot\left(\varepsilon_{q}-\frac{\varepsilon_{q}^{+}}{q^{+}}q\right)\gamma^{+} 
  +2x\left(\varepsilon_{q}^{+}\slashed{q}_{\perp}-q^{+}\slashed{\varepsilon}_{q\perp}\right)\slashed{k}_{\perp}\gamma^{+}\nonumber \\
  -2\overline{x}\slashed{k}_{\perp}\left(\varepsilon_{q}^{+}\slashed{q}_{\perp}-q^{+}\slashed{\varepsilon}_{q\perp}\right)\gamma^{+}
  -2\left(\boldsymbol{k}^{2}+x\overline{x}Q^{2}\right)\varepsilon_{q}^{+}\gamma^{+}  \; . 
  \label{Eq:FirstDiracTrDipole}
\end{gather}
We observe that, in eq.~(\ref{Eq:FirstDiracTrDipole}), the first three terms are separately QED gauge invariant, while the fourth one is a gauge invariance breaking contribution. We can easily see that it does not contribute to the amplitude, indeed
\begin{eqnarray}
&&{\cal A}_{2}^{{\rm QED}-{\rm break}.}  \nonumber \\
&&= - e_{q}  \int \hspace{-0.1 cm} {\rm d}z_{1}^{-} \hspace{-0.1 cm} \int \hspace{-0.1 cm} {\rm d}z_{2}^{-} \hspace{-0.1 cm} \int \hspace{-0.1 cm} {\rm d}^{d}\boldsymbol{z}_{1}{\rm d}^{d}\boldsymbol{z}_{2} \left\langle P\left(p^{\prime}\right)\big| \mathcal{U}_{\boldsymbol{z}_1 \boldsymbol{z}_2} \big| P\left(p\right)\right\rangle \left\langle M\left(p_{M}\right)\left|\overline{\psi}\left(z_{1}\right)\Gamma^{\lambda}\psi\left(z_{2}\right)\right|0\right\rangle _{z_{1,2}^{+}=0} \nonumber \\
&&  \times\int_{0}^{1}\frac{{\rm d}x}{2\pi}\int\frac{{\rm d}^{d}\boldsymbol{k}}{\left(2\pi\right)^{d}}{\rm e}^{-ixq^{+}z_{1}^{-}-i\overline{x}q^{+}z_{2}^{-}+i\left(\boldsymbol{k}\cdot\boldsymbol{z}_{12}\right)+i\boldsymbol{q}\cdot\left(x\boldsymbol{z}_{1}+\overline{x}\boldsymbol{z}_{2}\right)} \frac{\left(\boldsymbol{k}^{2}+x\overline{x}Q^{2}\right)\varepsilon_{q}^{+} {\rm tr} \left[ \gamma^{+}\Gamma_{\lambda} \right]}{ 4 \left( \boldsymbol{k}^{2}+x\overline{x}Q^{2}-i0 \right) } \; . 
\end{eqnarray}
We see in this expression that the factor $\left(\boldsymbol{k}^{2}+x\overline{x}Q^{2}\right)$ in the numerator cancels the analogous in the denominator and then the only dependence on $\boldsymbol{k}$ is in the exponent. The $\boldsymbol{k}$-integral thus yields a $\delta^{d}\left(\boldsymbol{z}_{12}\right)$ function that makes the dipole operator, $\mathcal{U}_{\boldsymbol{z}_1 \boldsymbol{z}_2}$, vanish. We can therefore effectively make the replacement 
\begin{align}
 \gamma^{+}\left(x\slashed{q}+\slashed{k}_{\perp}\right)\slashed{\varepsilon}_{q}\left(\overline{x}\slashed{q}-\slashed{k}_{\perp}\right)\gamma^{+}
 \longrightarrow 2q^{+}\left(\varepsilon_{q\mu}-\varepsilon_{q}^{+}\frac{q_{\mu}}{q^{+}}\right)\left(2x\overline{x}q^{\mu}-x\gamma_{\perp}^{\mu}\slashed{k}_{\perp}+\overline{x}\slashed{k}_{\perp}\gamma_{\perp}^{\mu}\right)\gamma^{+} \nonumber \; ,
\end{align}
in eq.~(\ref{Eq:FirstDiracTrDipole}) and, thus, write the 2-body dipole amplitude in the explicitly QED gauge-invariant form:
\begin{gather}
{\cal A}_{2} =e_{q}q^{+}\left(\varepsilon_{q\mu}-\varepsilon_{q}^{+}\frac{q_{\mu}}{q^{+}}\right)\int_{0}^{1}\frac{{\rm d}x}{2\pi}\int{\rm d}z_{1}^{-}\int{\rm d}z_{2}^{-}{\rm e}^{-ixq^{+}z_{1}^{-}-i\overline{x}q^{+}z_{2}^{-}} \int{\rm d}^{d}\boldsymbol{z}_{1}{\rm d}^{d}\boldsymbol{z}_{2}{\rm e}^{i\boldsymbol{q}\cdot\left(x\boldsymbol{z}_{1}+\overline{x}\boldsymbol{z}_{2}\right)}  \nonumber \\ \times \left\langle P\left(p^{\prime}\right)\big| \mathcal{U}_{\boldsymbol{z}_1 \boldsymbol{z}_2} \big| P\left(p\right)\right\rangle  
  \left\langle M\left(p_{M}\right)\left|\overline{\psi}\left(z_{1}\right)\Gamma^{\lambda}\psi\left(z_{2}\right)\right|0\right\rangle _{z_{1,2}^{+}=0} \int\frac{{\rm d}^{d}\boldsymbol{k}}{\left(2\pi\right)^{d}}{\rm e}^{i\left(\boldsymbol{k}\cdot\boldsymbol{z}_{12}\right)} \nonumber \\
  \times \frac{1}{4}{\rm tr}\left[\frac{\left(2x\overline{x}q^{\mu}-x\gamma_{\perp}^{\mu}\slashed{k}_{\perp}+\overline{x}\slashed{k}_{\perp}\gamma_{\perp}^{\mu}\right)\gamma^{+}\Gamma_{\lambda}}{\boldsymbol{k}^{2}+x\overline{x}Q^{2}-i0}\right] \; .
\end{gather}
By using the $N=1$ case of the standard integrals
\begin{align}
\int\frac{{\rm d}^{d}\boldsymbol{k}}{\left(2\pi\right)^{d}}\frac{e^{i\left(\boldsymbol{k}\cdot\boldsymbol{r}\right)}}{\left(\boldsymbol{k}^{2}+\mu^{2}\right)^{N}} & =\left(\sqrt{\frac{\mu^{2}}{\boldsymbol{r}^{2}}}\right)^{\left(\frac{d}{2}-N\right)}\frac{K_{\frac{d}{2}-N}\left(\sqrt{\mu^{2}\boldsymbol{r}^{2}}\right)}{2^{N-1}\Gamma\left(N\right)\left(2\pi\right)^{\frac{d}{2}}} \; , \\
\int\frac{{\rm d}^{d}\boldsymbol{k}}{\left(2\pi\right)^{d}}\frac{e^{i\left(\boldsymbol{k}\cdot\boldsymbol{r}\right)}k_{\perp}^{\mu}}{\left(\boldsymbol{k}^{2}+\mu^{2}\right)^{N}} & =\left(\sqrt{\frac{\mu^{2}}{\boldsymbol{r}^{2}}}\right)^{\left(\frac{d}{2}-N+1\right)}\frac{ir_{\perp}^{\mu}K_{\frac{d}{2}-N+1}\left(\sqrt{\mu^{2}\boldsymbol{r}^{2}}\right)}{2^{N-1}\Gamma\left(N\right)\left(2\pi\right)^{\frac{d}{2}}} \; ,
\end{align}
we can integrate over $\boldsymbol{k}$ to express the result in terms of MacDonald $K_{\nu}$ functions and obtain
\begin{gather}
{\cal A}_{2} =\frac{e_{q}q^{+}}{\left(2\pi\right)^{\frac{d}{2}}}\left(\varepsilon_{q\mu}-\varepsilon_{q}^{+}\frac{q_{\mu}}{q^{+}}\right)\int_{0}^{1}\frac{{\rm d}x}{2\pi}\int{\rm d}z_{1}^{-}\int{\rm d}z_{2}^{-}{\rm e}^{-ixq^{+}z_{1}^{-}-i\overline{x}q^{+}z_{2}^{-}}
  \int{\rm d}^{d}\boldsymbol{z}_{1}{\rm d}^{d}\boldsymbol{z}_{2}{\rm e}^{i\boldsymbol{q}\cdot\left(x\boldsymbol{z}_{1}+\overline{x}\boldsymbol{z}_{2}\right)} \nonumber \\ \times \left\langle P\left(p^{\prime}\right)\big| \mathcal{U}_{\boldsymbol{z}_1 \boldsymbol{z}_2} \big| P\left(p\right)\right\rangle \left\langle M\left(p_{M}\right)\left|\overline{\psi}\left(z_{1}\right)\Gamma^{\lambda}\psi\left(z_{2}\right)\right|0\right\rangle _{z_{1,2}^{+}=0}\nonumber \\
  \times\left\{ \frac{1}{4}2x\overline{x}q^{\mu}{\rm tr}\left[\gamma^{+}\Gamma_{\lambda}\right]\left(\sqrt{\frac{x\overline{x}Q^{2}}{\boldsymbol{r}^{2}}}\right)^{\frac{d}{2}-1}K_{\frac{d}{2}-1}\left(\sqrt{x\overline{x}Q^{2}\boldsymbol{z}_{12}^{2}}\right)\right. \nonumber \\
  \left.+\frac{iz_{12\perp\nu}}{4}{\rm tr}\left[\left(-x\gamma_{\perp}^{\mu}\gamma_{\perp}^{\nu}+\overline{x}\gamma_{\perp}^{\nu}\gamma_{\perp}^{\mu}\right)\gamma^{+}\Gamma_{\lambda}\right]\left(\sqrt{\frac{x\overline{x}Q^{2}}{\boldsymbol{z}_{12}^{2}}}\right)^{\frac{d}{2}}K_{\frac{d}{2}}\left(\sqrt{x\overline{x}Q^{2}\boldsymbol{z}_{12}^{2}}\right)\right\} \; . 
\end{gather}
Until now we kept the generic dimension $d=D-2$, but since there are no divergences we can move to four dimensions and thus decompose our Fierz matrices in the usual 16 dimensional basis. Since only chiral-even generators contribute to the traces, i.e. $\left(\Gamma^{\lambda},\Gamma_{\lambda}\right)=\left(\gamma^{\lambda},\gamma_{\lambda}\right)$
and $\left(\Gamma^{\lambda},\Gamma_{\lambda}\right)=\left(\gamma^{\lambda}\gamma^{5},-\gamma_{\lambda}\gamma^{5}\right)$, we find
\begin{align}
{\cal A}_{2} & =\frac{e_{q}q^{+}}{\left(2\pi\right)^{\frac{d}{2}}}\left(\varepsilon_{q\mu}-\varepsilon_{q}^{+}\frac{q_{\mu}}{q^{+}}\right)\int_{0}^{1}\frac{{\rm d}x}{2\pi}\int{\rm d}z_{1}^{-}\int{\rm d}z_{2}^{-}{\rm e}^{-ixq^{+}z_{1}^{-}-i\overline{x}q^{+}z_{2}^{-}}\nonumber \\
 & \times\int{\rm d}^{d}\boldsymbol{z}_{1}{\rm d}^{d}\boldsymbol{z}_{2}{\rm e}^{i\boldsymbol{q}\cdot\left(x\boldsymbol{z}_{1}+\overline{x}\boldsymbol{z}_{2}\right)}\left\langle P\left(p^{\prime}\right)\left|1-\frac{1}{N_{c}}{\rm tr}\left(V_{\boldsymbol{z}_{1}}V_{\boldsymbol{z}_{2}}^{\dagger}\right)\right|P\left(p\right)\right\rangle \nonumber \\
 & \times\left\{ 2x\overline{x}q^{\mu}\left\langle M\left(p_{M}\right)\left|\overline{\psi}\left(z_{1}\right)\gamma^{+}\psi\left(z_{2}\right)\right|0\right\rangle _{z_{1,2}^{+}=0}K_{0}\left(\sqrt{x\overline{x}Q^{2}\boldsymbol{z}_{12}^{2}}\right)\right. \nonumber \\
 & -\frac{iz_{12\perp}^{\mu}}{\boldsymbol{z}_{12}^{2}}\left(x-\overline{x}\right)\left\langle M\left(p_{M}\right)\left| \overline{\psi}\left(z_{1}\right)\gamma^{+}\psi\left(z_{2}\right)\right|0\right\rangle _{z_{1,2}^{+}=0}\sqrt{x\overline{x}Q^{2}\boldsymbol{z}_{12}^{2}}K_{1}\left(\sqrt{x\overline{x}Q^{2}\boldsymbol{z}_{12}^{2}}\right)\nonumber \\
 & \left.+\frac{\epsilon^{\mu\nu+-}z_{12\perp\nu}}{\boldsymbol{z}_{12}^{2}}\left\langle M\left(p_{M}\right)\left|\overline{\psi}\left(z_{1}\right)\gamma^{+}\gamma^{5}\psi\left(z_{2}\right)\right|0\right\rangle _{z_{1,2}^{+}=0}\sqrt{x\overline{x}Q^{2}\boldsymbol{z}_{12}^{2}}K_{1}\left(\sqrt{x\overline{x}Q^{2}\boldsymbol{z}_{12}^{2}}\right)\right\} \; ,
\end{align}
where we defined the antisymmetric tensor $\epsilon^{\mu\nu\rho\sigma}$ such that\footnote{We use the convention $\epsilon^{0123}=1$.} ${\rm tr} \left[\gamma^{\mu}\gamma^{\nu}\gamma^{\rho}\gamma^{\sigma} \gamma^{5} \right]=-4i\epsilon^{\mu\nu\rho\sigma}$ and we write the dipole operator explicitly. We observe that the term involving the $K_{0}$ ($K_1$) MacDonald function describes the contribution of longitudinally (transversely) polarized incoming photon, as expected from the known expression of the impact parameter representation of the photon wavefunction~\cite{Bjorken:1970ah}. Moving to the impact parameter and dipole size variables
\begin{equation}
    \boldsymbol{b} = x \boldsymbol{z}_1 + \bar{x} \boldsymbol{z}_2 \hspace{0.5 cm} \text{and} \hspace{0.5 cm} \boldsymbol{r} = \boldsymbol{z}_1 - \boldsymbol{z}_2 \; ,
    \label{Eq:ChangeziTorb}
\end{equation}
and using translational invariance, we ultimately get 
\begin{align}
{\cal A}_{2} & =\frac{e_{q}}{\left(2\pi\right)^{\frac{d}{2}-1}} \delta\left( 1 - \frac{p_{M}^{+}}{q^{+}} \right)\left(\varepsilon_{q\mu}-\varepsilon_{q}^{+}\frac{q_{\mu}}{q^{+}}\right)\int_{0}^{1}\frac{{\rm d}x}{2\pi}\int{\rm d}^{4}r\delta\left(r^{+}\right){\rm e}^{-i\left(xp_{M}\cdot r\right)}\nonumber \\
 & \times\int{\rm d}^{d}\boldsymbol{b}{\rm e}^{i\left(\boldsymbol{q}-\boldsymbol{p}_{M}\right)\cdot\boldsymbol{b}}\left\langle P\left(p^{\prime}\right)\left|1-\frac{1}{N_{c}}{\rm tr}\left(V_{\boldsymbol{b}+\overline{x}\boldsymbol{r}}V_{\boldsymbol{b}-x\boldsymbol{r}}^{\dagger}\right)\right|P\left(p\right)\right\rangle \nonumber \\
 & \times\left\{ 2x\overline{x}q^{\mu}\left\langle M\left(p_{M}\right)\left|\overline{\psi}\left(r\right)\gamma^{+}\psi\left(0\right)\right|0\right\rangle K_{0}\left(\sqrt{x\overline{x}Q^{2}\boldsymbol{r}^{2}}\right)\right.\nonumber \\
 & -\frac{ir_{\perp}^{\mu}}{\boldsymbol{r}^{2}}\left(x-\overline{x}\right)\left\langle M\left(p_{M}\right)\left|\overline{\psi}\left(r\right)\gamma^{+}\psi\left(0\right)\right|0\right\rangle \sqrt{x\overline{x}Q^{2}\boldsymbol{r}^{2}}K_{1}\left(\sqrt{x\overline{x}Q^{2}\boldsymbol{r}^{2}}\right)\nonumber \\
 & \left.+\frac{\epsilon^{\mu\nu+-}r_{\perp\nu}}{\boldsymbol{r}^{2}}\left\langle M\left(p_{M}\right)\left|\overline{\psi}\left(r\right)\gamma^{+}\gamma^{5}\psi\left(0\right)\right|0\right\rangle \sqrt{x\overline{x}Q^{2}\boldsymbol{r}^{2}}K_{1}\left(\sqrt{x\overline{x}Q^{2}\boldsymbol{r}^{2}}\right)\right\} .  
\end{align}
We can further use the relation
\begin{equation}
    \frac{\partial}{\partial r_{\perp}^{\nu}} K_0 \left( \sqrt{x \bar{x} Q^2 \boldsymbol{r}^{2}} \right) = \frac{r_{\perp  \nu}}{\boldsymbol{r}^{2}} \sqrt{x \bar{x} Q^2 \boldsymbol{r}^{2}} K_1 \left( \sqrt{x \bar{x} Q^2 \boldsymbol{r}^{2}} \right) \; ,
\end{equation}
to express the final result in the compact form
\begin{align}
{\cal A}_{2} & =\frac{e_{q}}{\left(2\pi\right)^{\frac{d}{2}-1}} \delta\left( 1 - \frac{p_{M}^{+}}{q^{+}} \right) \left(\varepsilon_{q\mu}-\varepsilon_{q}^{+}\frac{q_{\mu}}{q^{+}}\right)\int_{0}^{1}\frac{{\rm d}x}{2\pi}\int{\rm d}^{4}r\delta\left(r^{+}\right){\rm e}^{-i\left(xp_{M}\cdot r\right)}\nonumber \\
 & \times\int{\rm d}^{d}\boldsymbol{b} \, {\rm e}^{i\left(\boldsymbol{q}-\boldsymbol{p}_{M}\right)\cdot\boldsymbol{b}}\left\langle P\left(p^{\prime}\right)\left|1-\frac{1}{N_{c}}{\rm tr}\left(V_{\boldsymbol{b}+\overline{x}\boldsymbol{r}}V_{\boldsymbol{b}-x\boldsymbol{r}}^{\dagger}\right)\right|P\left(p\right)\right\rangle \nonumber \\
 & \times\left\{ 2x\overline{x}q^{\mu}\left\langle M\left(p_{M}\right)\left|\overline{\psi}\left(r\right)\gamma^{+}\psi\left(0\right)\right|0\right\rangle K_{0}\left(\sqrt{x\overline{x}Q^{2}\boldsymbol{r}^{2}}\right)\right.\nonumber \\
 & - i \left(x-\overline{x}\right)\left\langle M\left(p_{M}\right)\left|\overline{\psi}\left(r\right)\gamma^{+}\psi\left(0\right)\right|0\right\rangle \frac{\partial}{\partial r_{\perp \mu}} K_0 \left( \sqrt{x \bar{x} Q^2 \boldsymbol{r}^{2}} \right) \nonumber \\
 & \left. + \epsilon^{\mu\nu+-}\left\langle M\left(p_{M}\right)\left|\overline{\psi}\left(r\right)\gamma^{+}\gamma^{5}\psi\left(0\right)\right|0\right\rangle \frac{\partial}{\partial r_{\perp}^{\nu}} K_0 \left( \sqrt{x \bar{x} Q^2 \boldsymbol{r}^{2}} \right) \right\} . 
 \label{Eq.GeneralFinal2bodyCon}
\end{align}
This expression has completely general vacuum-to-meson matrix elements, i.e. it is valid before any kind of twist expansion is taken. We observe that for transverse $\rho^{0}$ production, because the outgoing matrix is sandwiched between two $\gamma^{+}$ matrices from the effective Feynman rules, only longitudinal gamma matrices $\gamma^{-}$ and $\gamma^{-} \gamma^5$ contribute from the hard part side of the Fierz decomposition, which implies having only contributions from the longitudinal gamma matrices $\gamma^{+}$ and $\gamma^{+} \gamma^5$ in the non-perturbative $\rho$-meson matrix element side.

\subsection{3-body contribution}
\begin{figure}
\includegraphics[scale=0.35]{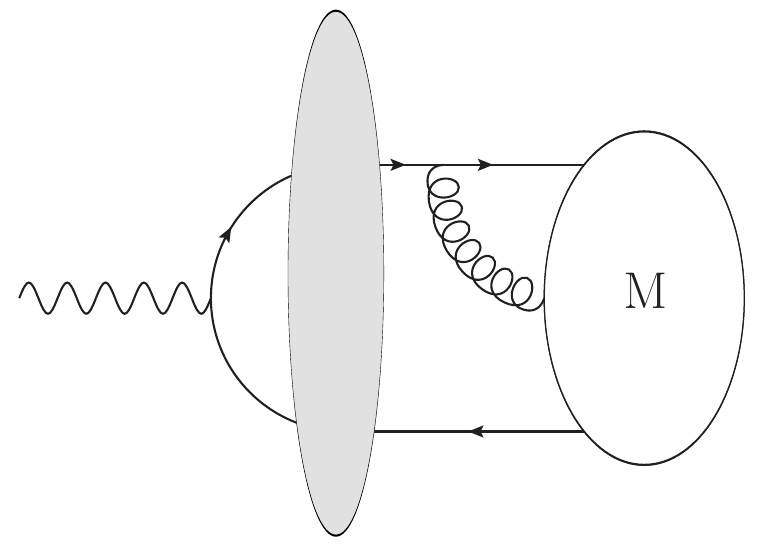} \hspace{0.4 cm} \includegraphics[scale=0.35]{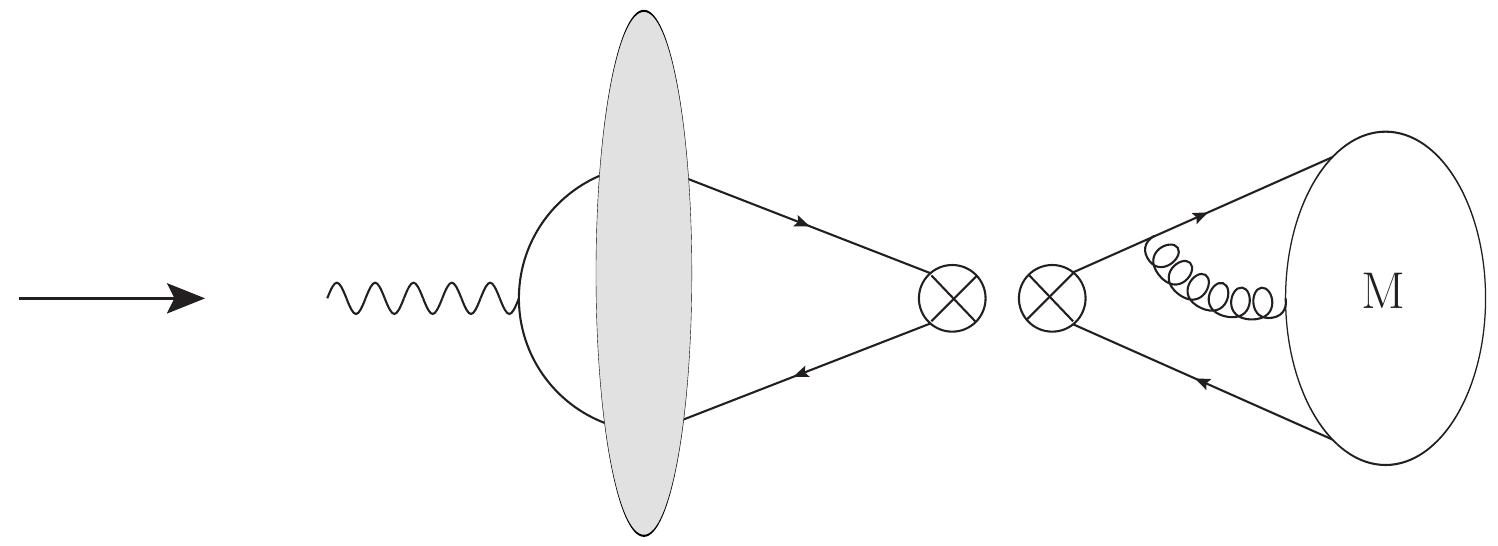}
\caption{Pictorial illustration of the factorization of external lines emissions. In the right diagram, the collinear gluon is included inside the collinear "blob" as part of the evolution kernel.}
  \label{fig:Evolution_diagrams}
\end{figure}

As one might expect, the 3-body contribution is much more involved. For this reason, it is convenient to consider separately the emission from the quark (subsection \ref{Emission_from_the_quark}) and the emission from the antiquark (subsection \ref{Emission_from_the_antiquark}). These two contributions are not separately QED gauge invariant and we will thus carefully isolate the two gauge-invariant breaking terms in both contributions to show that they actually cancel when combining them, restoring the full QED gauge invariance of the result. Since we will eventually restrict ourselves to twist-3 accuracy, diagrams in which the additional gluon is emitted from an external line (i.e. after the shockwave in the present case) are excluded from the coefficient function (i.e. the impact factor in the present case). Indeed, in the genuine contribution to the twist-3 amplitude, the three partons fly collinearly and therefore any emission after shockwave lead to a purely singular contribution that is therefore part of the ERBL evolution of the twist-2 $q$$\bar{q}$ contribution, as illustrated in fig.~\ref{fig:Evolution_diagrams}. Nonetheless, it is important to stress that the contribution of this diagram to the impact factor is only absent in the context of a twist expansion. 

\subsubsection{Emission from the quark}
\label{Emission_from_the_quark}

\begin{figure}
\includegraphics[scale=0.40]{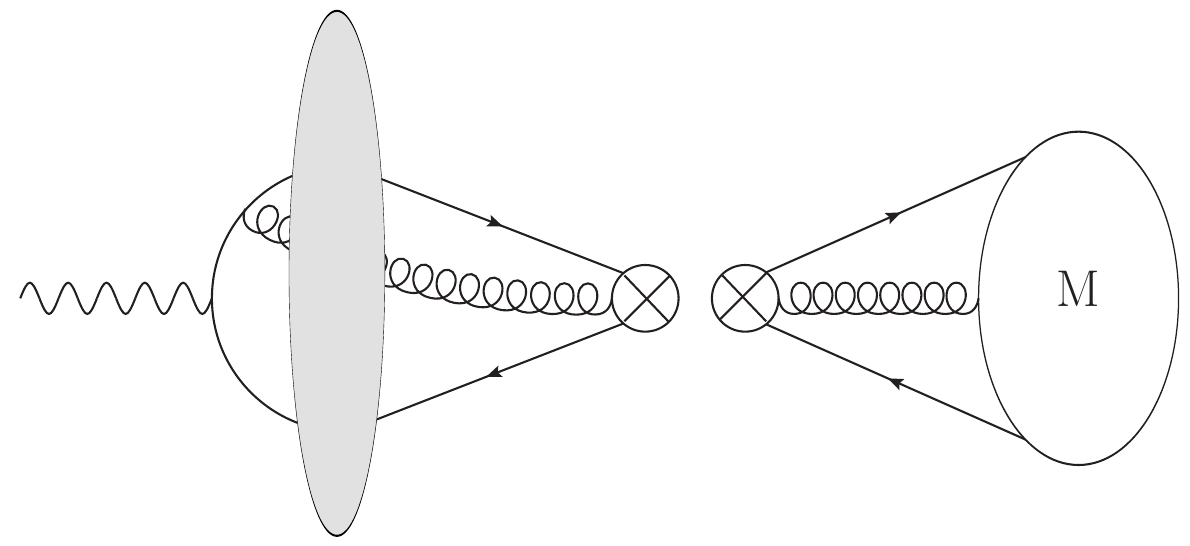}
\caption{Pictorial representation of the 3-body contribution factorization, in the case of emission from a quark.}
  \label{fig:Shock3BodyContributionQuark}
\end{figure}

The 3-body EDMP amplitude contribution where the gluon is emitted from the quark
before the shockwave reads
\begin{align}
{\cal A}_{q} & =\left(-ie_{q}\right)\left(ig\right)\int{\rm d}^{D}z_{4}{\rm d}^{D}z_{0}\theta\left(-z_{4}^{+}\right)\theta\left(-z_{0}^{+}\right) \nonumber \\
 & \times \left\langle P\left(p^{\prime}\right)M\left(p_{M}\right)\left|\overline{\psi}_{{\rm eff}}\left(z_{4}\right)\gamma_{\mu}A_{{\rm eff}}^{\mu a}\left(z_{4}\right)t^{a}G\left(z_{40}\right)\slashed{\varepsilon}_{q}{\rm e}^{-i\left(q\cdot z_{0}\right)}\psi_{{\rm eff}}\left(z_{0}\right)\right|P\left(p\right)\right\rangle \; ,
\end{align}
where, as for the 2-body contribution, the effective operators should be taken according to eqs.~(\ref{Eq:PsiEffecNoMon}), (\ref{Eq:PsiBarEffecNoMon}) and (\ref{Eq:AEffecNoMon}). In order to achieve the factorization pictorially depicted in Fig.~\ref{fig:Shock3BodyContributionQuark}, we first perform the Fierz decomposition in Dirac space to get
\begin{align}
{\cal A}_{q3} & =-\frac{2ie_{q}}{4}g\int{\rm d}^{D}z_{4}{\rm d}^{D}z_{3}{\rm d}^{D}z_{2}{\rm d}^{D}z_{1}{\rm d}^{D}z_{0}\nonumber \\
 & \times\theta\left(-z_{4}^{+}\right)\delta\left(z_{3}^{+}\right)\delta\left(z_{2}^{+}\right)\delta\left(z_{1}^{+}\right)\theta\left(-z_{0}^{+}\right){\rm e}^{-i\left(q\cdot z_{0}\right)}\left\langle P\left(p^{\prime}\right)\left|\left(V_{\boldsymbol{z}_{1}}t^{a}U_{\boldsymbol{z}_{3}}^{ab}V_{\boldsymbol{z}_{2}}^{\dagger}\right)_{ij}\right|P\left(p\right)\right\rangle \nonumber \\
 & \times{\rm tr}_{D}\left[\gamma^{+}G_{0}\left(z_{14}\right)\gamma_{\mu}G^{\mu\sigma_{\perp}}\left(z_{34}\right)G_{0}\left(z_{40}\right)\slashed{\varepsilon}_{q}G_{0}\left(z_{02}\right)\gamma^{+}\Gamma_{\lambda}\right] \nonumber \\
 & \times\left\langle M\left(p_{M}\right)\left|\overline{\psi}_{i}\left(z_{1}\right)\Gamma^{\lambda}F_{-\sigma}^{b}\left(z_{3}\right)\psi_{j}\left(z_{2}\right)\right|0\right\rangle \;.
\end{align}
Now, we separate the target and the final state matrix elements by applying Fierz decomposition in color space, i.e.
\begin{gather}
  \overline{\psi}_{i}\left(z_{1}\right)\Gamma^{\lambda}F_{-\sigma}^{b}\left(z_{3}\right)\psi_{j}\left(z_{2}\right)\left(V_{\boldsymbol{z}_{1}}t^{a}U_{\boldsymbol{z}_{3}}^{ab}V_{\boldsymbol{z}_{2}}^{\dagger}\right)_{ij} \hspace{-0.3 cm} =\overline{\psi}_{i}\left(z_{1}\right)\Gamma^{\lambda}F_{-\sigma}^{b}\left(z_{3}\right)\psi_{j}\left(z_{2}\right)\left(V_{\boldsymbol{z}_{1}}t^{a}U_{\boldsymbol{z}_{3}}^{ab}V_{\boldsymbol{z}_{2}}^{\dagger}\right)_{kl} \hspace{-0.15 cm} \delta_{ik} \delta_{jl} \nonumber \\
  =\overline{\psi}_{i}\left(z_{1}\right)\Gamma^{\lambda}F_{-\sigma}^{b}\left(z_{3}\right)\psi_{j}\left(z_{2}\right)\left(V_{\boldsymbol{z}_{1}}t^{a}U_{\boldsymbol{z}_{3}}^{ab}V_{\boldsymbol{z}_{2}}^{\dagger}\right)_{kl}\left(2t_{ij}^{c}t_{lk}^{c}+\frac{1}{N_{c}}\delta_{ij}\delta_{kl}\right) \nonumber \\ = 2{\rm tr}\overline{\psi}\left(z_{1}\right)t^{c}\Gamma^{\lambda}F_{-\sigma}^{b}\left(z_{3}\right)\psi\left(z_{2}\right){\rm tr}\left(V_{\boldsymbol{z}_{1}}t^{a}U_{\boldsymbol{z}_{3}}^{ab}V_{\boldsymbol{z}_{2}}^{\dagger}t^{c}\right)  \; ,
\end{gather}
where in the second line we discarded the second term in the round brackets because it yields ${\rm tr}\overline{\psi}\left(z_{1}\right)\Gamma^{\lambda}F_{-\sigma}^{b}\left(z_{3}\right)\psi\left(z_{2}\right)$
which is octet and therefore will cancel in the singlet matrix element. To disentangle completely the color structure, we project the target matrix element onto the color singlet exchange, which means
\begin{align}
 {\rm tr} \left(V_{\boldsymbol{z}_{1}}t^{a}U_{\boldsymbol{z}_{3}}^{ab}V_{\boldsymbol{z}_{2}}^{\dagger}t^{c}\right) = \frac{\delta^{bc}}{N_c^2-1} {\rm tr} \left(V_{\boldsymbol{z}_{1}}t^{a}U_{\boldsymbol{z}_{3}}^{ad}V_{\boldsymbol{z}_{2}}^{\dagger}t^{d}\right) 
\end{align}
and finally get
\begin{align}
 \frac{2}{N_c^2 - 1} \overline{\psi}\left(z_{1}\right)\Gamma^{\lambda}F_{-\sigma}\left(z_{3}\right)\psi\left(z_{2}\right){\rm tr}\left(V_{\boldsymbol{z}_{1}}t^{a}U_{\boldsymbol{z}_{3}}^{ab}V_{\boldsymbol{z}_{2}}^{\dagger}t^{b}\right), \nonumber 
\end{align}
where $F_{-\sigma} = t^b F_{-\sigma}^b$. Now, we recognize the operator from the BK evolution equation which, going
from the adjoint to the fundamental Wilson line representation, can be written as
\begin{align}
{\rm tr}\left(V_{\boldsymbol{z}_{1}}t^{a}U_{\boldsymbol{z}_{3}}^{ab}V_{\boldsymbol{z}_{2}}^{\dagger}t^{b}\right)  =\frac{1}{2}\left[{\rm tr}\left(V_{\boldsymbol{z}_{1}}V_{\boldsymbol{z}_{3}}^{\dagger}\right){\rm tr}\left(V_{\boldsymbol{z}_{3}}V_{\boldsymbol{z}_{2}}^{\dagger}\right)-\frac{1}{N_{c}}{\rm tr}\left(V_{\boldsymbol{z}_{1}}V_{\boldsymbol{z}_{2}}^{\dagger}\right)\right].\nonumber
\end{align}
In the following, we will keep it in its more compact left-hand side form and we will not subtract the non-interacting part of the amplitude (i.e. all Wilson lines set to 1) immediately, but only at the end. Now, the 3-body contribution from quark emission can be written as   
\begin{gather}
{\cal A}_{q3} = \frac{-ie_{q}}{N_c^2 - 1} g \int{\rm d}^{D}z_{4}{\rm d}^{D}z_{3}{\rm d}^{D}z_{2}{\rm d}^{D}z_{1}{\rm d}^{D}z_{0}\nonumber \theta\left(-z_{4}^{+}\right)\delta\left(z_{3}^{+}\right)\delta\left(z_{2}^{+}\right)\delta\left(z_{1}^{+}\right)\theta\left(-z_{0}^{+}\right){\rm e}^{-i\left(q\cdot z_{0}\right)}\nonumber \\
  \times\left\langle P\left(p^{\prime}\right)\left|{\rm tr}\left(V_{\boldsymbol{z}_{1}}t^{a}V_{\boldsymbol{z}_{2}}^{\dagger}t^{b}U_{\boldsymbol{z}_{3}}^{ab}\right)\right|P\left(p\right)\right\rangle \left\langle M \left(p_{M}\right)\left|\overline{\psi}\left(z_{1}\right)\Gamma^{\lambda}F_{-\sigma}\left(z_{3}\right)\psi\left(z_{2}\right)\right|0\right\rangle \nonumber \\
  \times{\rm tr}_{D}\left[\gamma^{+}G_{0}\left(z_{14}\right)\gamma_{\mu}G^{\mu\sigma_{\perp}}\left(z_{34}\right)G_{0}\left(z_{40}\right)\slashed{\varepsilon}_{q}G_{0}\left(z_{02}\right)\gamma^{+}\Gamma_{\lambda}\right] \; .
\end{gather}
Let us Fourier transform the propagators, integrate w.r.t. their $-$
components and integrate w.r.t. $z_{4}$ and $z_{0}$. This leads to
\begin{gather}
{\cal A}_{q3} =\frac{ie_{q}}{N_c^2-1} g\int{\rm d}z_{3}^{-}{\rm d}^{d}\boldsymbol{z}_{3}{\rm d}z_{2}^{-}{\rm d}^{d}\boldsymbol{z}_{2}{\rm d}z_{1}^{-}{\rm d}^{d}\boldsymbol{z}_{1} \left\langle M\left(p_{M}\right)\left|\overline{\psi}\left(z_{1}\right)\Gamma^{\lambda}F_{-\sigma}\left(z_{3}\right)\psi\left(z_{2}\right)\right|0\right\rangle _{z_{1,2,3}^{+}=0}\nonumber \\
  \times\left\langle P\left(p^{\prime}\right)\left|{\rm tr}\left(V_{\boldsymbol{z}_{1}}t^{a}V_{\boldsymbol{z}_{2}}^{\dagger}t^{b}U_{\boldsymbol{z}_{3}}^{ab}\right)\right|P\left(p\right)\right\rangle \int\frac{{\rm d}k_{14}^{+}{\rm d}^{d}\boldsymbol{k}_{14}}{\left(2\pi\right)^{d+1}}\frac{{\rm d}k_{40}^{+}{\rm d}^{d}\boldsymbol{k}_{40}}{\left(2\pi\right)^{d+1}}\frac{{\rm d}k_{20}^{+}{\rm d}^{d}\boldsymbol{k}_{20}}{\left(2\pi\right)^{d+1}}\frac{{\rm d}\ell^{+}{\rm d}^{d}\boldsymbol{\ell}}{\left(2\pi\right)^{d+1}}  \nonumber \\
  \times \left(2\pi\right)^{2+2d} \delta\left(q^{+}-k_{40}^{+}-k_{20}^{+}\right) \delta^{d}\left(\boldsymbol{q}-\boldsymbol{k}_{40}-\boldsymbol{k}_{20}\right)
  \delta\left(k_{40}^{+}-\ell^{+}-k_{14}^{+}\right) \delta^{d}\left(\boldsymbol{\ell}+\boldsymbol{k}_{14}-\boldsymbol{k}_{40}\right) \nonumber \\
  \times \frac{{\rm e}^{-ik_{14}^{+}z_{1}^{-}-ik_{20}^{+}z_{2}^{-}-i\ell^{+}z_{3}^{-}+i\left(\boldsymbol{k}_{14}\cdot\boldsymbol{z}_{1}\right)+i\left(\boldsymbol{k}_{20}\cdot\boldsymbol{z}_{2}\right)+i\left(\boldsymbol{\ell}\cdot\boldsymbol{z}_{3}\right)} \theta\left(k_{14}^{+}\right)\theta\left(k_{20}^{+}\right)\theta\left(\ell^{+}\right) }{\frac{\boldsymbol{\ell}^{2}-i0}{2\ell^{+}}+\frac{\boldsymbol{k}_{14}^{2}-i0}{2k_{14}^{+}}+\frac{\boldsymbol{k}_{20}^{2}-i0}{2k_{20}^{+}}-q^{-}}\frac{-1}{2\ell^{+}}\left(g_{\perp}^{\mu\sigma}-\frac{\ell_{\perp}^{\sigma}}{\ell^{+}}n_{2}^{\mu}\right)\nonumber \\
  \times{\rm tr}_{D}\left[ \gamma^{+}\frac{\slashed{k}_{14}}{2k_{14}^{+}}\gamma_{\mu} \left( \frac{\gamma^{+}}{2k_{40}^{+}} -\frac{\left(\frac{\boldsymbol{k}_{40}^{2}}{2k_{40}^{+}}\gamma^{+}+k_{40}^{+}\gamma^{-}+\slashed{k}_{40\perp}\right)}{2k_{40}^{+} \left(\frac{\boldsymbol{k}_{40}^{2}-i0}{2k_{40}^{+}}+\frac{\boldsymbol{k}_{20}^{2}-i0}{2k_{20}^{+}}-q^{-} \right)} \right) \slashed{\varepsilon}_{q}\frac{\slashed{k}_{20}}{2k_{20}^{+}}\gamma^{+}\Gamma_{\lambda} \right] \; . 
  \label{Eq:Aq3BefTrace}
\end{gather}
The energy denominators can be manipulated to get
\begin{align}
 & \frac{\boldsymbol{\ell}^{2}-i0}{2\ell^{+}}+\frac{\boldsymbol{k}_{14}^{2}-i0}{2k_{14}^{+}}+\frac{\boldsymbol{k}_{20}^{2}-i0}{2k_{20}^{+}}-q^{-}\nonumber \\
 & =\frac{k_{40}^{+}}{2\ell^{+}\left(k_{40}^{+}-\ell^{+}\right)}\left(\boldsymbol{\ell}-\frac{\ell^{+}}{k_{40}^{+}}\boldsymbol{k}_{40}\right)^{2}+\frac{q^{+}}{2k_{40}^{+}\left(q^{+}-k_{40}^{+}\right)}\left(\boldsymbol{k}_{40}-\frac{k_{40}^{+}}{q^{+}}\boldsymbol{q}\right)^{2}+\frac{Q^{2}}{2q^{+}}-i0\nonumber \; ,
\end{align}
\begin{align}
 & \frac{\boldsymbol{k}_{40}^{2}-i0}{2k_{40}^{+}}+\frac{\left(\boldsymbol{q}-\boldsymbol{k}_{40}\right)^{2}-i0}{2\left(q^{+}-k_{40}^{+}\right)}-q^{-} =\frac{q^{+}}{2k_{40}^{+}\left(q^{+}-k_{40}^{+}\right)}\left(\boldsymbol{k}_{40}-\frac{k_{40}^{+}}{q^{+}}\boldsymbol{q}\right)^{2}+\frac{Q^{2}}{2q^{+}}-i0\nonumber \; .
\end{align}
The simplification of the Dirac trace is conceptually straightforward but algebraically tedious; for the sake of compactness we report here the result only and relegate technical details on the derivation to Appendix~\ref{Sec:AppendixB1}. We have
\begin{align}
 & {\rm tr}_{D}\left[\gamma^{+}\frac{\slashed{k}_{14}}{2k_{14}^{+}}\gamma_{\mu}\left(g_{\perp}^{\mu\sigma}-\frac{\ell_{\perp}^{\sigma}}{\ell^{+}}n_{2}^{\mu}\right)\left( \frac{\gamma^{+}}{ 2k_{40}^{+} } -\frac{\left(\frac{\boldsymbol{k}_{40}^{2}}{2k_{40}^{+}}\gamma^{+}+k_{40}^{+}\gamma^{-}+\slashed{k}_{40\perp}\right)}{ 2k_{40}^{+} \left( \frac{\boldsymbol{k}_{40}^{2}-i0}{2k_{40}^{+}}+\frac{\boldsymbol{k}_{20}^{2}-i0}{2k_{20}^{+}}-q^{-} \right) } \right) \slashed{\varepsilon}_{q}\frac{\slashed{k}_{20}}{2k_{20}^{+}}\gamma^{+}\Gamma_{\lambda}\right]\nonumber \\
 & = \frac{-1}{2k_{40}^{+}}{\rm tr}_{D}\left[\gamma_{\perp}^{\sigma}\left(\slashed{\varepsilon}_{q\perp}-\frac{\varepsilon_{q}^{+}}{q^{+}}\slashed{q}_{\perp}\right) \gamma^{+}\Gamma_{\lambda}\right] -\frac{\varepsilon_{q}^{+}}{2k_{40}^{+}\left(q^{+}-k_{40}^{+}\right)}{\rm tr}_{D}\left[\gamma_{\perp}^{\sigma}\left(\slashed{k}_{40\perp}-\frac{k_{40}^{+}}{q^{+}}\slashed{q}_{\perp}\right)\gamma^{+}\Gamma_{\lambda}\right]\nonumber \\
\nonumber \\
 & \!\!+\frac{k_{40}^{+}\left(q^{+}-k_{40}^{+}\right) \left(\varepsilon_{q\rho}-\frac{\varepsilon_{q}^{+}}{q^{+}}q_{\rho}\right) q^{\rho}\left(\ell_{\perp\beta}-\frac{\ell^{+}}{k_{40}^{+}}k_{40\perp\beta}\right) }{\left(q^{+}\right)^{2}\left(k_{40}^{+}-\ell^{+}\right)\left(\left(\boldsymbol{k}_{40}-\frac{k_{40}^{+}}{q^{+}}\boldsymbol{q}\right)^{2}+\frac{k_{40}^{+}\left(q^{+}-k_{40}^{+}\right)}{\left(q^{+}\right)^{2}}Q^{2}\right)} {\rm tr}_{D} \!\left[ \left( \gamma_{\perp}^{\beta}\gamma_{\perp}^{\sigma}+2\frac{k_{40}^{+}-\ell^{+}}{\ell^{+}}g_{\perp}^{\beta\sigma}\right)\gamma^{+}\Gamma_{\lambda} \right] \nonumber \\
\nonumber \\
 & \!\!-\frac{\varepsilon_{q}^{+} \left(\ell_{\perp\beta}-\frac{\ell^{+}}{k_{40}^{+}}k_{40\perp\beta}\right)}{2\left(k_{40}^{+}-\ell^{+}\right)q^{+}} {\rm tr}_{D} \left[\! \left(2\frac{k_{40}^{+}}{\ell^{+}}g_{\perp}^{\beta\sigma}-\gamma_{\perp}^{\sigma}\gamma_{\perp}^{\beta}\right)\gamma^{+}\Gamma_{\lambda} \right]\! +\frac{\left(\varepsilon_{q\rho}-\frac{\varepsilon_{q}^{+}}{q^{+}}q_{\rho}\right)}{2q^{+}\left(k_{40}^{+}-\ell^{+}\right) }\!\! \left(\!k_{40\perp\alpha}-\frac{k_{40}^{+}}{q^{+}}q_{\perp\alpha}\!\right) \nonumber \\
 & \!\!\times \hspace{-0.1 cm} \frac{\left(\ell_{\perp\beta}-\frac{\ell^{+}}{k_{40}^{+}}k_{40\perp\beta}\right)}{ \left(\boldsymbol{k}_{40}-\frac{k_{40}^{+}}{q^{+}}\boldsymbol{q}\right)^{2} + \frac{k_{40}^{+}\left(q^{+}-k_{40}^{+}\right)Q^{2}}{q^{+}}}  {\rm tr}_{D} \hspace{-0.1 cm} \left[ \!\left( \hspace{-0.1 cm} 2\frac{k_{40}^{+}}{\ell^{+}}g_{\perp}^{\beta\sigma} \hspace{-0.2 cm} -\gamma_{\perp}^{\sigma}\gamma_{\perp}^{\beta} \hspace{-0.05 cm} \right) \hspace{-0.15 cm} \left[\!\left(q^{+}-k_{40}^{+}\right)\gamma_{\perp}^{\alpha}\gamma_{\perp}^{\rho}-k_{40}^{+}\gamma_{\perp}^{\rho}\gamma_{\perp}^{\alpha}\right]\!\gamma^{+}\Gamma_{\lambda} \right] \!.
 \label{Eq:DiracTraceQuarkFinalResult}
\end{align}
After using the $\delta$ functions in eq.~(\ref{Eq:Aq3BefTrace}), we can simplify the integrand through the two sequential shifts 
\begin{equation}
  \boldsymbol{\ell}\rightarrow\boldsymbol{\ell}+\frac{\ell^{+}}{k_{40}^{+}}\boldsymbol{k}_{40} \; , \hspace{1 cm}  \boldsymbol{k}_{40}\rightarrow\boldsymbol{k}_{40}+\frac{k_{40}^{+}}{q^{+}}\boldsymbol{q} \; ,
  \label{Eq:TransverseShiftQuark}
\end{equation}
and finally get
\begin{gather}
{\cal A}_{q3} =-\frac{ie_{q}g}{N_c^2-1}\!\int\!{\rm d}z_{3}^{-}{\rm d}z_{2}^{-}{\rm d}z_{1}^{-}\!\!\int\!{\rm d}^{d}\boldsymbol{z}_{3}{\rm d}^{d}\boldsymbol{z}_{2}{\rm d}^{d}\boldsymbol{z}_{1} \left\langle M\left(p_{M}\right)\!\left|\overline{\psi}\left(z_{1}\right)\Gamma^{\lambda}F_{-\sigma} \left(z_{3}\right)\psi\left(z_{2}\right)\right|0\right\rangle _{z_{1,2,3}^{+}=0} \nonumber \\
  \times\left\langle P\left(p^{\prime}\right)\left|{\rm tr}\left(V_{\boldsymbol{z}_{1}}t^{a}V_{\boldsymbol{z}_{2}}^{\dagger}t^{b}U_{\boldsymbol{z}_{3}}^{ab}\right)\right|P\left(p\right)\right\rangle  \int\frac{{\rm d}k_{40}^{+}{\rm d}^{d}\boldsymbol{k}_{40}}{\left(2\pi\right)^{d+1}}\frac{{\rm d}\ell^{+}{\rm d}^{d}\boldsymbol{\ell}}{\left(2\pi\right)^{d+1}}\theta\left(k_{40}^{+}-\ell^{+}\right)\theta\left(q^{+}-k_{40}^{+}\right)\theta\left(\ell^{+}\right)\nonumber \\
  \times{\rm e}^{-i\left(k_{40}^{+}-\ell^{+}\right)z_{1}^{-}-i\left(q^{+}-k_{40}^{+}\right)z_{2}^{-}-i\ell^{+}z_{3}^{-}+ i \boldsymbol{q}\cdot\left(\frac{k_{40}^{+}-\ell^{+}}{q^{+}}\boldsymbol{z}_{1}+\frac{q^{+}-k_{40}^{+}}{q^{+}}\boldsymbol{z}_{2}+\frac{\ell^{+}}{q^{+}}\boldsymbol{z}_{3}\right)-i\boldsymbol{k}_{40}\cdot\left(\boldsymbol{z}_{21}+\frac{\ell^{+}}{k_{40}^{+}}\boldsymbol{z}_{13}\right)-i\left(\boldsymbol{\ell}\cdot\boldsymbol{z}_{13}\right)}\nonumber \\
  \times\frac{1}{2\ell^{+}}\frac{1}{ \left( \frac{k_{40}^{+}}{2\ell^{+}\left(k_{40}^{+}-\ell^{+}\right)}\boldsymbol{\ell}^{2}+\frac{q^{+}}{2k_{40}^{+}\left(q^{+}-k_{40}^{+}\right)}\boldsymbol{k}_{40}^{2}+\frac{Q^{2}}{2q^{+}} \right) } 
  \left\{ -\frac{1}{2k_{40}^{+}}\left(\varepsilon_{q\rho}-\frac{\varepsilon_{q}^{+}}{q^{+}}q_{\rho}\right){\rm tr}_{D}\left[\gamma_{\perp}^{\sigma}\gamma_{\perp}^{\rho}\gamma^{+}\Gamma_{\lambda}\right]\right.\nonumber \\
  +\frac{k_{40}^{+}\left(q^{+}-k_{40}^{+}\right)q^{\rho}\ell_{\perp\beta}\left(\varepsilon_{q\rho}-\frac{\varepsilon_{q}^{+}}{q^{+}}q_{\rho}\right)}{\left(q^{+}\right)^{2}\!\left(k_{40}^{+}-\ell^{+}\right)\!\!\left(\!\boldsymbol{k}_{40}^{2}+\frac{k_{40}^{+}\left(q^{+}-k_{40}^{+}\right)}{\left(q^{+}\right)^{2}}Q^{2}\!\right)} 
  {\rm tr}_{D} \!\left[ \left( \gamma_{\perp}^{\beta}\gamma_{\perp}^{\sigma}+2\frac{k_{40}^{+}-\ell^{+}}{\ell^{+}}g_{\perp}^{\beta\sigma} \right) \gamma^{+}\Gamma_{\lambda} \right]\! +\frac{k_{40\perp\alpha}\ell_{\perp\beta}}{\left(k_{40}^{+}-\ell^{+}\right)} \nonumber \\
  \times \frac{\left(\varepsilon_{q\rho}-\frac{\varepsilon_{q}^{+}}{q^{+}}q_{\rho}\right)}{2q^{+}\! \left(\boldsymbol{k}_{40}^{2}+\frac{k_{40}^{+}\left(q^{+}-k_{40}^{+}\right)Q^{2}}{\left(q^{+}\right)^{2}}\right)}
  {\rm tr}_{D} \left[  \left(2\frac{k_{40}^{+}}{\ell^{+}}g_{\perp}^{\beta\sigma}-\gamma_{\perp}^{\sigma}\gamma_{\perp}^{\beta}\right) \left( \left(q^{+}-k_{40}^{+}\right)\gamma_{\perp}^{\alpha}\gamma_{\perp}^{\rho}-k_{40}^{+}\gamma_{\perp}^{\rho}\gamma_{\perp}^{\alpha}\right) \gamma^{+}\Gamma_{\lambda} \right] \nonumber \\
  -\frac{\varepsilon_{q}^{+} \ell_{\perp\beta} }{2q^{+}\left(k_{40}^{+}-\ell^{+}\right)} {\rm tr}_{D} \left[ \left(2\frac{k_{40}^{+}}{\ell^{+}}g_{\perp}^{\beta\sigma}-\gamma_{\perp}^{\sigma}\gamma_{\perp}^{\beta}\right)\gamma^{+}\Gamma_{\lambda} \right]
  \left. -\frac{\varepsilon_{q}^{+} k_{40\perp\beta} }{2k_{40}^{+}\left(q^{+}-k_{40}^{+}\right)}{\rm tr}_{D}\left[\gamma_{\perp}^{\sigma}\gamma_{\perp}^{\beta}\gamma^{+}\Gamma_{\lambda}\right]\right\}  \; .
\label{Eq:QuarkContrBeforeFraction}
\end{gather}
Now, we perform the following change of variables 
\begin{equation}
x_{1}=\frac{k_{40}^{+}-\ell^{+}}{q^{+}} \; , \hspace{1 cm} x_{2}=\frac{q^{+}-k_{40}^{+}}{q^{+}} \; , \hspace{1 cm}
x_{3}=\frac{\ell^{+}}{q^{+}} \; ,
\end{equation}
to introduce the fractions of photon momenta carried by the quark ($x_1$), antiquark ($x_2$) and gluon ($x_3$), in accordance with the appearance of the light-cone positions $z_1^-, z_2^-, z_3^-$ in the exponential factor of eq.~(\ref{Eq:QuarkContrBeforeFraction}). Then, we obtain 
\begin{gather}
{\cal A}_{q3} = \frac{-ie_{q}g}{N_c^2-1}\int\frac{{\rm d}x_{1}}{2\pi}\frac{{\rm d}x_{2}}{2\pi}\frac{{\rm d}x_{3}}{2\pi}2\pi\delta\left(1-x_{1}-x_{2}-x_{3}\right)\theta\left(x_{1}\right)\theta\left(x_{2}\right)\theta\left(x_{3}\right) \int{\rm d} z_{3}^{-}{\rm d}z_{2}^{-}{\rm d}z_{1}^{-}  \nonumber \\ \times  \int{\rm d}^{d}\boldsymbol{z}_{3}{\rm d}^{d}\boldsymbol{z}_{2}{\rm d}^{d} \boldsymbol{z}_{1} {\rm e}^{-ix_{1}q^{+}z_{1}^{-}-ix_{2}q^{+}z_{2}^{-}-ix_{3}q^{+}z_{3}^{-}} \left\langle P\left(p^{\prime}\right)\left|{\rm tr}\left(V_{\boldsymbol{z}_{1}}t^{a}V_{\boldsymbol{z}_{2}}^{\dagger}t^{b}U_{\boldsymbol{z}_{3}}^{ab}\right)\right|P\left(p\right)\right\rangle \nonumber \\
  \times {\rm e}^{i\boldsymbol{q}\cdot\left(x_{1}\boldsymbol{z}_{1}+x_{2}\boldsymbol{z}_{2}+x_{3}\boldsymbol{z}_{3}\right)} \!\left\langle M\left(p_{M}\right)\!\left|\overline{\psi}\left(z_{1}\right)\Gamma^{\lambda}F_{-\sigma}\left(z_{3}\right)\psi\left(z_{2}\right)\right|0\right\rangle _{z_{1,2,3}^{+}=0} \int\!\frac{{\rm d}^{d}\boldsymbol{k}_{40}}{\left(2\pi\right)^{d}}{\rm e}^{i\boldsymbol{k}_{40}\cdot\left(\frac{x_{1}\boldsymbol{z}_{12}+x_{3}\boldsymbol{z}_{32}}{x_{1}+x_{3}}\right)} \nonumber \\
  \times\int\frac{{\rm d}^{d}\boldsymbol{\ell}}{\left(2\pi\right)^{d}}\frac{{\rm e}^{-i\left(\boldsymbol{\ell}\cdot\boldsymbol{z}_{13}\right)}}{\boldsymbol{\ell}^{2}+\frac{x_{1}x_{3}}{x_{1}+x_{3}}\left(\frac{\boldsymbol{k}_{40}^{2}}{x_{2}\left(1-x_{2}\right)}+Q^{2}\right)} \frac{1}{x_{1}+x_{3}} \left\{ \frac{-x_{1}q^{+}}{2\left(1-x_{2}\right)}\left(\varepsilon_{q\rho}-\frac{\varepsilon_{q}^{+}}{q^{+}}q_{\rho}\right) \right. \nonumber \\
  \times \left. {\rm tr}_{D}\left[\gamma_{\perp}^{\sigma}\gamma_{\perp}^{\rho}\gamma^{+}\Gamma_{\lambda}\right]\right. 
  +q^{+}\frac{x_{2}\left(1-x_{2}\right)q^{\rho}\ell_{\perp\beta}\left(\varepsilon_{q\rho}-\frac{\varepsilon_{q}^{+}}{q^{+}}q_{\rho}\right)}{\boldsymbol{k}_{40}^{2}+x_{2}\left(1-x_{2}\right)Q^{2}}{\rm tr}_{D} \left[ \left(2\frac{x_{1}}{x_{3}}g_{\perp}^{\beta\sigma}+\gamma_{\perp}^{\beta}\gamma_{\perp}^{\sigma}\right)\gamma^{+}\Gamma_{\lambda} \right] \nonumber \\
  +q^{+}\frac{k_{40\perp\alpha}\ell_{\perp\beta}\left(\varepsilon_{q\rho}-\frac{\varepsilon_{q}^{+}}{q^{+}}q_{\rho}\right){\rm tr}_{D} \left[ \left(2\frac{x_{1}}{x_{3}}g_{\perp}^{\beta\sigma}+\gamma_{\perp}^{\beta}\gamma_{\perp}^{\sigma}\right)\left(x_{2}\gamma_{\perp}^{\alpha}\gamma_{\perp}^{\rho}-\overline{x}_{2}\gamma_{\perp}^{\rho}\gamma_{\perp}^{\alpha}\right)\gamma^{+}\Gamma_{\lambda} \right]}{2\left(\boldsymbol{k}_{40}^{2}+x_{2}\left(1-x_{2}\right)Q^{2}\right)}\nonumber \\
  -\frac{1}{2}\varepsilon_{q}^{+}\ell_{\perp\beta}{\rm tr}_{D} \left[ \left(2\frac{x_{1}}{x_{3}}g_{\perp}^{\beta\sigma}+\gamma_{\perp}^{\beta}\gamma_{\perp}^{\sigma}\right)\gamma^{+}\Gamma_{\lambda} \right]
  \left.-\frac{x_{1}\varepsilon_{q}^{+}}{2x_{2}\left(1-x_{2}\right)}k_{40\perp\beta}{\rm tr}_{D}\left[\gamma_{\perp}^{\sigma}\gamma_{\perp}^{\beta}\gamma^{+}\Gamma_{\lambda}\right]\right\}  \; .
  \label{Eq:3-bodyQuarkContBeforAnyTransverseMomInt}
\end{gather}
We calculate the integrals over transverse momentum in Appendix~\ref{Sec:AppedixC1} and we use them below to obtain the result. We first integrate over $\boldsymbol{\ell}$, using eq.~(\ref{Eq:TrasvMomIntQuarkElle}), in order to cast the contribution in the form of an integral over a Schwinger parameter $t$ and, in $d=2$ dimensions, we get 
\begin{gather}
{\cal A}_{q3} =-\frac{ie_{q}g}{N_c^2-1}\int\frac{{\rm d}x_{1}}{2\pi}\frac{{\rm d}x_{2}}{2\pi}\frac{{\rm d}x_{3}}{2\pi}2\pi\delta\left(1-x_{1}-x_{2}-x_{3}\right)\theta\left(x_{1}\right)\theta\left(x_{2}\right)\theta\left(x_{3}\right)  \int{\rm d}z_{3}^{-}{\rm d}z_{2}^{-}{\rm d}z_{1}^{-} \nonumber \\
  \times \int{\rm d}^{2}\boldsymbol{z}_{3}{\rm d}^{2}\boldsymbol{z}_{2}{\rm d}^{2}\boldsymbol{z}_{1} {\rm e}^{-ix_{1}q^{+}z_{1}^{-}-ix_{2}q^{+}z_{2}^{-}-ix_{3}q^{+}z_{3}^{-}} \left\langle P\left(p^{\prime}\right)\left|{\rm tr}\left(V_{\boldsymbol{z}_{1}}t^{a}V_{\boldsymbol{z}_{2}}^{\dagger}t^{b}U_{\boldsymbol{z}_{3}}^{ab}\right)\right|P\left(p\right)\right\rangle \nonumber \\
  \times {\rm e}^{i\boldsymbol{q}\cdot\left(x_{1}\boldsymbol{z}_{1}+x_{2}\boldsymbol{z}_{2}+x_{3}\boldsymbol{z}_{3}\right)} \!\left\langle M\left(p_{M}\right)\!\left|\overline{\psi}\left(z_{1}\right)\Gamma^{\lambda}F_{-\sigma}\left(z_{3}\right)\psi\left(z_{2}\right)\right|0\right\rangle _{z_{1,2,3}^{+}=0} \int\!\frac{{\rm d}^{2}\boldsymbol{k}_{40}}{\left(2\pi\right)^{2}}{\rm e}^{i\boldsymbol{k}_{40}\cdot\left(\frac{x_{1}\boldsymbol{z}_{12}+x_{3}\boldsymbol{z}_{32}}{x_{1}+x_{3}}\right)} \nonumber \\
  \times \frac{1}{4\pi}\int_{0}^{+\infty}\frac{{\rm d}t}{t}{\rm e}^{i\frac{\boldsymbol{z}_{13}^{2}}{4t}-it\frac{x_{1}x_{3}}{x_{2}\left(1-x_{2}\right)^{2}}\left(\boldsymbol{k}_{40}^{2}+x_{2}\left(1-x_{2}\right)Q^{2}\right)-i0} \frac{1}{x_{1}+x_{3}}\left\{ -\frac{x_{1}q^{+}}{2\left(1-x_{2}\right)}\left(\varepsilon_{q\rho}-\frac{\varepsilon_{q}^{+}}{q^{+}}q_{\rho}\right) \right. \nonumber \\ \times {\rm tr}_{D}\left[\gamma_{\perp}^{\sigma}\gamma_{\perp}^{\rho}\gamma^{+}\Gamma_{\lambda}\right] 
  -\frac{q^{+}}{2t}\frac{x_{2}\left(1-x_{2}\right)q^{\rho}z_{13\perp\beta}\left(\varepsilon_{q\rho}-\frac{\varepsilon_{q}^{+}}{q^{+}}q_{\rho}\right)}{\boldsymbol{k}_{40}^{2}+x_{2}\left(1-x_{2}\right)Q^{2}}{\rm tr}_{D} \left[ \left(2\frac{x_{1}}{x_{3}}g_{\perp}^{\beta\sigma}+\gamma_{\perp}^{\beta}\gamma_{\perp}^{\sigma}\right)\gamma^{+}\Gamma_{\lambda} \right] \nonumber \\
  -\frac{q^{+}}{4t}\frac{k_{40\perp\alpha}z_{13\perp\beta}\left(\varepsilon_{q\rho}-\frac{\varepsilon_{q}^{+}}{q^{+}}q_{\rho}\right){\rm tr}_{D} \left[ \left(2\frac{x_{1}}{x_{3}}g_{\perp}^{\beta\sigma}+\gamma_{\perp}^{\beta}\gamma_{\perp}^{\sigma}\right)\left(x_{2}\gamma_{\perp}^{\alpha}\gamma_{\perp}^{\rho}-\overline{x}_{2}\gamma_{\perp}^{\rho}\gamma_{\perp}^{\alpha}\right)\gamma^{+}\Gamma_{\lambda} \right] }{\boldsymbol{k}_{40}^{2}+x_{2}\left(1-x_{2}\right)Q^{2}} \nonumber \\
  +\frac{1}{4t}\varepsilon_{q}^{+}z_{13\perp\beta}{\rm tr}_{D} \left[ \left(2\frac{x_{1}}{x_{3}}g_{\perp}^{\beta\sigma}+\gamma_{\perp}^{\beta}\gamma_{\perp}^{\sigma}\right)\gamma^{+}\Gamma_{\lambda} \right] \left.-\frac{x_{1}\varepsilon_{q}^{+}}{2x_{2}\left(1-x_{2}\right)}k_{40\perp\beta}{\rm tr}_{D}\left[\gamma_{\perp}^{\sigma}\gamma_{\perp}^{\beta}\gamma^{+}\Gamma_{\lambda}\right]\right\} \; .
\label{Eq:3BodyQuarkBothQEDGaugeInAndNot}
\end{gather}
At this point, we separate the QED gauge invariant and QED gauge breaking part of the amplitude ${\cal A}_{q3}$ in eq.~(\ref{Eq:3BodyQuarkBothQEDGaugeInAndNot}). Then, using eqs.~(\ref{Eq:QuarkTransvMomIntegralK40WithoutDen}) and (\ref{Eq:QuarkTransvMomIntegralK40WDen}), we finally get
\begin{gather}
{\cal A}_{q3}^{{\rm QED}-{\rm inv.}}  =-\frac{e_{q}g \; q^{+} }{4\pi} \frac{1}{N_c^2-1}\! \left(\varepsilon_{q\rho}-\frac{\varepsilon_{q}^{+}}{q^{+}}q_{\rho}\right)\!\!\int\!{\rm d}z_{3}^{-}{\rm d}z_{2}^{-}{\rm d}z_{1}^{-}\int{\rm d}^{2}\boldsymbol{z}_{3}{\rm d}^{2}\boldsymbol{z}_{2}{\rm d}^{2}\boldsymbol{z}_{1} \!
 \int\frac{{\rm d}x_{1}}{2\pi}\frac{{\rm d}x_{2}}{2\pi}\frac{{\rm d}x_{3}}{2\pi} \nonumber \\ \times \delta\left(1-x_{1}-x_{2}-x_{3}\right)\theta\left(x_{1}\right)\theta\left(x_{2}\right)\theta\left(x_{3}\right) {\rm e}^{-ix_{1}q^{+}z_{1}^{-}-ix_{2}q^{+}z_{2}^{-}-ix_{3}q^{+}z_{3}^{-}} {\rm e}^{i\boldsymbol{q}\cdot\left(x_{1}\boldsymbol{z}_{1}+x_{2}\boldsymbol{z}_{2}+x_{3}\boldsymbol{z}_{3}\right)} \nonumber \\
  \times \left\langle P\left(p^{\prime}\right)\left|{\rm tr}\left(V_{\boldsymbol{z}_{1}}t^{a}V_{\boldsymbol{z}_{2}}^{\dagger}t^{b}U_{\boldsymbol{z}_{3}}^{ab}\right)\right|P\left(p\right)\right\rangle \left\langle M\left(p_{M}\right)\left|\overline{\psi}\left(z_{1}\right)\Gamma^{\lambda}F_{-\sigma}\left(z_{3}\right)\psi\left(z_{2}\right)\right|0\right\rangle _{z_{1,2,3}^{+}=0}\nonumber \\
  \times\!\left\{ \!\frac{ 2x_{2} z_{13\perp\beta}}{\boldsymbol{z}_{13}^{2}}q^{\rho}K_{0}\left(QZ\right){\rm tr}_{D} \hspace{-0.15 cm} \left[ \left(2\frac{x_{1}}{x_{3}}g_{\perp}^{\beta\sigma}+\gamma_{\perp}^{\beta}\gamma_{\perp}^{\sigma}\right)\gamma^{+}\Gamma_{\lambda} \right] \right. \hspace{-0.15 cm} -\frac{i x_{1}x_{2}}{1-x_{2}}{\rm tr}_{D} \hspace{-0.10 cm}\left[\gamma_{\perp}^{\sigma}\gamma_{\perp}^{\rho}\gamma^{+}\Gamma_{\lambda}\right]\frac{Q}{Z}K_{1}\left(QZ\right) \nonumber \\
  +i\frac{z_{13\perp\beta}}{\boldsymbol{z}_{13}^{2}}{\rm tr}_{D} \left[ \left(2\frac{x_{1}}{x_{3}}g_{\perp}^{\beta\sigma}+\gamma_{\perp}^{\beta}\gamma_{\perp}^{\sigma}\right)\left(x_{2}\gamma_{\perp}^{\alpha}\gamma_{\perp}^{\rho}-\overline{x}_{2}\gamma_{\perp}^{\rho}\gamma_{\perp}^{\alpha}\right)\gamma^{+}\Gamma_{\lambda} \right] \nonumber \\ 
  \left. \times \left(\frac{x_{1}x_{2}z_{12\perp\alpha}+x_{3}x_{2}z_{32\perp\alpha}}{x_{1}+x_{3}}\right)\frac{Q}{Z}K_{1}\left(QZ\right) \right\}  \; ,
  \label{Eq:Quark3BodyConFinalExpr}
\end{gather}
for the QED gauge invariant part of the amplitude and 
\begin{gather}
{\cal A}_{q3}^{{\rm QED}-{\rm break}.}  = \frac{e_{q}g\varepsilon_{q}^{+}}{\left(4\pi\right)} \frac{1}{N_c^2-1} \int{\rm d}z_{3}^{-}{\rm d}z_{2}^{-}{\rm d}z_{1}^{-} \int{\rm d}^{2}\boldsymbol{z}_{3}{\rm d}^{2}\boldsymbol{z}_{2}{\rm d}^{2}\boldsymbol{z}_{1} \int\frac{{\rm d}x_{1}}{2\pi}\frac{{\rm d}x_{2}}{2\pi} \frac{{\rm d}x_{3}}{2\pi}  \nonumber \\
  \times  \delta\left(1-x_{1}-x_{2}-x_{3}\right)\theta\left(x_{1}\right)\theta\left(x_{2}\right)\theta\left(x_{3}\right) {\rm e}^{-ix_{1}q^{+}z_{1}^{-}-ix_{2}q^{+}z_{2}^{-}-ix_{3}q^{+}z_{3}^{-}} {\rm e}^{i\boldsymbol{q}\cdot\left(x_{1}\boldsymbol{z}_{1}+x_{2}\boldsymbol{z}_{2}+x_{3}\boldsymbol{z}_{3}\right)} \nonumber \\
  \times \left\langle P\left(p^{\prime}\right)\left|{\rm tr}\left(V_{\boldsymbol{z}_{1}}t^{a}V_{\boldsymbol{z}_{2}}^{\dagger}t^{b}U_{\boldsymbol{z}_{3}}^{ab}\right)\right|P\left(p\right)\right\rangle \left\langle M\left(p_{M}\right)\left|\overline{\psi}\left(z_{1}\right)\Gamma^{\lambda}F_{-\sigma}\left(z_{3}\right)\psi\left(z_{2}\right)\right|0\right\rangle _{z_{1,2,3}^{+}=0} \nonumber \\
  \times x_{1}x_{2}\frac{Q^{2}}{\boldsymbol{Z}^{2}}K_{2} \left( QZ \right) {\rm tr}_{D} \left[ \left(\slashed{z}_{1\perp}\gamma_{\perp}^{\sigma}+\gamma_{\perp}^{\sigma}\slashed{z}_{2\perp}-2z_{3\perp}^{\sigma}\right)\gamma^{+}\Gamma_{\lambda} \right] 
  \label{Eq:QEDBreaking3BodyQuark}
\end{gather}
for the QED gauge invariance violating term.

\subsubsection{Emission from the antiquark}
\label{Emission_from_the_antiquark}
\begin{figure} \includegraphics[scale=0.40]{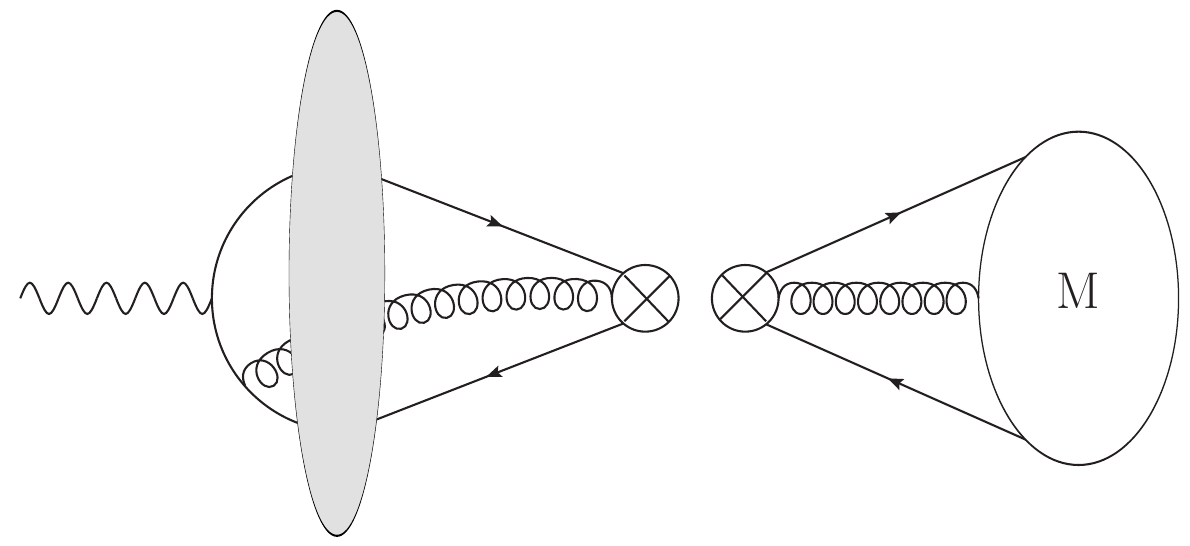}
\caption{Pictorial representation of the 3-body contribution factorization, in the case of emission from a antiquark.}
  \label{fig:Shock3BodyContributionAntiQuark}
\end{figure}
The 3-body EDMP amplitude contribution where the gluon is emitted from the antiquark
before the shockwave reads
\begin{align}
{\cal A}_{\overline{q}} & =\left(-ie_{q}\right)\left(ig\right)\int{\rm d}^{D}z_{4}{\rm d}^{D}z_{0}\theta\left(-z_{4}^{+}\right)\theta\left(-z_{0}^{+}\right) \nonumber \\
 & \times\left\langle P\left(p^{\prime}\right)M\left(p_{M}\right)\left|\overline{\psi}_{{\rm eff}}\left(z_{0}\right)\slashed{\varepsilon}_{q}{\rm e}^{-i\left(q\cdot z_{0}\right)}G_{0}\left(z_{04}\right)t^{a}\gamma_{\mu}A_{{\rm eff}}^{\mu a}\left(z_{4}\right)\psi_{{\rm eff}}\left(z_{4}\right)\right|P\left(p\right)\right\rangle \; ,
\end{align}
where, again, the effective operators should be taken as in eqs.~(\ref{Eq:PsiEffecNoMon}), (\ref{Eq:PsiBarEffecNoMon}) and (\ref{Eq:AEffecNoMon}). In order to achieve the factorization pictorially depicted in Fig.~\ref{fig:Shock3BodyContributionAntiQuark}, we proceed analogously to the quark case and get
\begin{gather}
{\cal A}_{\overline{q}3} = \frac{- i e_{q}}{N_c^2-1} g \int{\rm d}^{D}z_{4}{\rm d}^{D}z_{3}{\rm d}^{D}z_{2}{\rm d}^{D}z_{1}{\rm d}^{D}z_{0} \theta\left(-z_{4}^{+}\right)\delta\left(z_{3}^{+}\right)\delta\left(z_{2}^{+}\right)\delta\left(z_{1}^{+}\right)\theta\left(-z_{0}^{+}\right){\rm e}^{-i\left(q\cdot z_{0}\right)} \nonumber \\
  \times\left\langle P\left(p^{\prime}\right)\left|{\rm tr}\left(V_{\boldsymbol{z}_{1}}t^{a}U_{\boldsymbol{z}_{3}}^{ab}V_{\boldsymbol{z}_{2}}^{\dagger}t^{b}\right)\right|P\left(p\right)\right\rangle \left\langle M\left(p_{M}\right)\left|\overline{\psi}\left(z_{1}\right)\Gamma^{\lambda}F_{-\sigma}\left(z_{3}\right)\psi\left(z_{2}\right)\right|0\right\rangle \nonumber \\
  \times {\rm tr}_{D}\left[\gamma^{+}G_{0}\left(z_{10}\right)\slashed{\varepsilon}_{q}G_{0}\left(z_{04}\right)\gamma_{\mu}G^{\mu\sigma_{\perp}}\left(z_{34}\right)G_{0}\left(z_{42}\right)\gamma^{+}\Gamma_{\lambda}\right] \; .
\end{gather}
Let us Fourier transform the propagators, integrate w.r.t. their $-$
components and integrate with respect to $z_{4}$ and $z_{0}$:
\begin{gather}
{\cal A}_{\overline{q}3} =\frac{ie_{q}g}{N_c^2-1}\int{\rm d}^{D}z_{3}{\rm d}^{D}z_{2}{\rm d}^{D}z_{1}\delta\left(z_{3}^{+}\right)\delta\left(z_{2}^{+}\right)\delta\left(z_{1}^{+}\right) \left\langle P\left(p^{\prime}\right)\left|{\rm tr}\left(V_{\boldsymbol{z}_{1}}t^{a}U_{\boldsymbol{z}_{3}}^{ab}V_{\boldsymbol{z}_{2}}^{\dagger}t^{b}\right)\right|P\left(p\right)\right\rangle \nonumber \\
  \times \left\langle M\left(p_{M}\right)\left|\overline{\psi}\left(z_{1}\right)\Gamma^{\lambda}F_{-\sigma}\left(z_{3}\right)\psi\left(z_{2}\right)\right|0\right\rangle  \int {\rm d} k_{10}^{+}{\rm d}^{d}\boldsymbol{k}_{10} \; {\rm d}k_{40}^{+}{\rm d}^{d}\boldsymbol{k}_{40} \frac{{\rm d}k_{24}^{+}{\rm d}^{d}\boldsymbol{k}_{24}}{\left(2\pi\right)^{d+1}}\frac{{\rm d}\ell^{+}{\rm d}^{d}\boldsymbol{\ell}}{\left(2\pi\right)^{d+1}}\theta\left(k_{10}^{+}\right) \nonumber \\
  \times \theta\left(k_{24}^{+}\right)\theta\left(\ell^{+}\right) \delta^{d}\left(\boldsymbol{k}_{40}-\boldsymbol{\ell}-\boldsymbol{k}_{24}\right) \delta\left(q^{+}-k_{40}^{+}-k_{10}^{+}\right) \delta^{d}\left(\boldsymbol{q}-\boldsymbol{k}_{40}-\boldsymbol{k}_{10}\right) \delta\left(k_{40}^{+}-\ell^{+}-k_{24}^{+}\right) \nonumber \\ \times \frac{{\rm e}^{-ik_{10}^{+}z_{1}^{-}-ik_{24}^{+}z_{2}^{-}-i\ell^{+}z_{3}^{-}+i\left(\boldsymbol{k}_{10}\cdot\boldsymbol{z}_{1}\right)+i\left(\boldsymbol{k}_{24}\cdot\boldsymbol{z}_{2}\right)+i\left(\boldsymbol{\ell}\cdot\boldsymbol{z}_{3}\right)}}{\frac{\boldsymbol{k}_{24}^{2}-i0}{2k_{24}^{+}} + \frac{\boldsymbol{k}_{10}^{2}-i0}{2k_{10}^{+}}+\frac{\boldsymbol{\ell}^{2}-i0}{2\ell^{+}}-q^{-}} 
  \frac{1}{2k_{10}^{+}2k_{40}^{+}2k_{24}^{+}2\ell^{+}}\left(g_{\perp}^{\mu\sigma}-\frac{\ell_{\perp}^{\sigma}}{\ell^{+}}n_{2}^{\mu}\right)\nonumber \\
  \times{\rm tr}_{D}\left[\gamma^{+}\slashed{k}_{10}\slashed{\varepsilon}_{q}\left(\gamma^{+}-\frac{\slashed{k}_{40}\gamma^{+}\slashed{k}_{40}}{2k_{40}^{+}\left(\frac{\boldsymbol{k}_{10}^{2}-i0}{2k_{10}^{+}}+\frac{\boldsymbol{k}_{40}^{2}-i0}{2k_{40}^{+}}-q^{-}\right)}\right)\gamma_{\mu}\slashed{k}_{24}\gamma^{+}\Gamma_{\lambda}\right] \; .
\end{gather}
Writing the energy denominators as
\begin{gather}
 \frac{\boldsymbol{k}_{24}^{2}-i0}{2k_{24}^{+}}+\frac{\boldsymbol{k}_{10}^{2}-i0}{2k_{10}^{+}}+\frac{\boldsymbol{\ell}^{2}-i0}{2\ell^{+}}-q^{-} =\frac{k_{40}^{+}}{2\ell^{+}\left(k_{40}^{+}-\ell^{+}\right)}\left(\boldsymbol{\ell}-\frac{\ell^{+}}{k_{40}^{+}}\boldsymbol{k}_{40}\right)^{2} \nonumber \\ + \frac{q^{+}}{2k_{40}^{+}\left(q^{+}-k_{40}^{+}\right)}\left(\boldsymbol{k}_{40}-\frac{k_{40}^{+}}{q^{+}}\boldsymbol{q}\right)^{2}+\frac{Q^{2}}{2q^{+}}-i0 \; ,
\end{gather}
\begin{align}
 & \frac{\boldsymbol{k}_{10}^{2}-i0}{2k_{10}^{+}}+\frac{\boldsymbol{k}_{40}^{2}-i0}{2k_{40}^{+}}-q^{-} =\frac{q^{+}}{2k_{40}^{+}\left(q^{+}-k_{40}^{+}\right)}\left(\boldsymbol{k}_{40}-\frac{k_{40}^{+}}{q^{+}}\boldsymbol{q}\right)^{2}+\frac{Q^{2}}{2q^{+}}-i0 
\end{align}
and using the $\delta$ functions to perform some integrations, we end up with
\begin{gather}
{\cal A}_{\overline{q}3} =\frac{ie_{q}g}{N_c^2-1} \!\int{\rm d}z_{3}^{-}{\rm d}z_{2}^{-}{\rm d}z_{1}^{-}\int{\rm d}^{d}\boldsymbol{z}_{3}{\rm d}^{d}\boldsymbol{z}_{2}{\rm d}^{d}\boldsymbol{z}_{1} \!\left\langle M\left(p_{M}\right)\left|\overline{\psi}\left(z_{1}\right)\Gamma^{\lambda}F_{-\sigma}\left(z_{3}\right)\psi\left(z_{2}\right)\right|0\right\rangle _{z_{1,2,3}^{+}=0} \nonumber \\
  \times \left\langle P\left(p^{\prime}\right)\left|{\rm tr}\left(V_{\boldsymbol{z}_{1}}t^{a}U_{\boldsymbol{z}_{3}}^{ab}V_{\boldsymbol{z}_{2}}^{\dagger}t^{b}\right)\right|P\left(p\right)\right\rangle
 \int\frac{{\rm d}k_{40}^{+}{\rm d}^{d}\boldsymbol{k}_{40}}{\left(2\pi\right)^{d+1}}\frac{{\rm d}\ell^{+}{\rm d}^{d}\boldsymbol{\ell}}{\left(2\pi\right)^{d+1}}\theta\left(q^{+}-k_{40}^{+}\right)\theta\left(k_{40}^{+}-\ell^{+}\right)\theta\left(\ell^{+}\right)\nonumber \\
  \times \frac{{\rm e}^{-i\left(q^{+}-k_{40}^{+}\right)z_{1}^{-}-i\left(k_{40}^{+}-\ell^{+}\right)z_{2}^{-}-i\ell^{+}z_{3}^{-}}{\rm e}^{i\left(\boldsymbol{q}-\boldsymbol{k}_{40}\right)\cdot\boldsymbol{z}_{1}+i\left(\boldsymbol{k}_{40}-\boldsymbol{\ell}\right)\cdot\boldsymbol{z}_{2}+i\left(\boldsymbol{\ell}\cdot\boldsymbol{z}_{3}\right)}}{\frac{k_{40}^{+}}{2\ell^{+}\left(k_{40}^{+}-\ell^{+}\right)}\left(\boldsymbol{\ell}-\frac{\ell^{+}}{k_{40}^{+}}\boldsymbol{k}_{40}\right)^{2}+\frac{q^{+}}{2k_{40}^{+}\left(q^{+}-k_{40}^{+}\right)}\left(\boldsymbol{k}_{40}-\frac{k_{40}^{+}}{q^{+}}\boldsymbol{q}\right)^{2}+\frac{Q^{2}}{2q^{+}}-i0}\nonumber \\
  \times\frac{1}{2\left(q^{+}-k_{40}^{+}\right)2k_{40}^{+}2\left(k_{40}^{+}-\ell^{+}\right)2\ell^{+}}\left(g_{\perp}^{\mu\sigma}-\frac{\ell_{\perp}^{\sigma}}{\ell^{+}}n_{2}^{\mu}\right)\nonumber \\
  \times{\rm tr}_{D}\left[\gamma^{+}\left(\slashed{q}-\slashed{k}_{40}\right)\slashed{\varepsilon}_{q}\left(\gamma^{+}-\frac{\left(\frac{q^{+}-k_{40}^{+}}{q^{+}}\right)\slashed{k}_{40}\gamma^{+}\slashed{k}_{40}}{\left(\boldsymbol{k}_{40}-\frac{k_{40}^{+}}{q^{+}}\boldsymbol{q}\right)^{2}+\frac{k_{40}^{+}\left(q^{+}-k_{40}^{+}\right)}{\left(q^{+}\right)^{2}}Q^{2}-i0}\right)\gamma_{\mu}\left(\slashed{k}_{40}-\slashed{\ell}\right)\gamma^{+}\Gamma_{\lambda}\right]\; . 
\end{gather}
Now, we again perform the shifts in (\ref{Eq:TransverseShiftQuark}) and then we move to the variables
\begin{equation}
   x_{1}=\frac{q^{+}-k_{40}^{+}}{q^{+}}, \hspace{1 cm} x_{2}=\frac{k_{40}^{+}-\ell^{+}}{q^{+}} , \hspace{1 cm} x_{3}=\frac{\ell^{+}}{q^{+}} \; ,
\end{equation}
which again correspond to the fractions of photon momenta carried by the quark ($x_1$), antiquark ($x_2$) and gluon ($x_3$). Then, we obtain
\begin{gather}
{\cal A}_{\overline{q}3} =\frac{ie_{q}g}{N_c^2-1} \int{\rm d}z_{3}^{-}{\rm d}z_{2}^{-}{\rm d}z_{1}^{-}\int{\rm d}^{d}\boldsymbol{z}_{3}{\rm d}^{d}\boldsymbol{z}_{2}{\rm d}^{d}\boldsymbol{z}_{1} \int\frac{{\rm d}x_{1}}{2\pi}\frac{{\rm d}x_{2}}{2\pi}\frac{{\rm d}x_{3}}{2\pi}\left(2\pi\right)\delta\left(1-x_{1}-x_{2}-x_{3}\right) \nonumber \\
  \times \theta\left(x_{1}\right) \theta\left(x_{2}\right)\theta\left(x_{3}\right) {\rm e}^{-ix_{1}q^{+}z_{1}^{-}-ix_{2}q^{+}z_{2}^{-}-ix_{3}q^{+}z_{3}^{-}}{\rm e}^{i\boldsymbol{q}\cdot\left(x_{1}\boldsymbol{z}_{1}+x_{2}\boldsymbol{z}_{2}+x_{3}\boldsymbol{z}_{3}\right)} \frac{1}{8x_{1}x_{2}x_{3}\left(x_{2}+x_{3}\right)q^{+}}  \nonumber \\
  \times\left\langle P\left(p^{\prime}\right)\left|{\rm tr}\left(V_{\boldsymbol{z}_{1}}t^{a}U_{\boldsymbol{z}_{3}}^{ab}V_{\boldsymbol{z}_{2}}^{\dagger}t^{b}\right)\right|P\left(p\right)\right\rangle \left\langle M\left(p_{M}\right)\left|\overline{\psi}\left(z_{1}\right)\Gamma^{\lambda}F_{-\sigma}\left(z_{3}\right)\psi\left(z_{2}\right)\right|0\right\rangle _{z_{1,2,3}^{+}=0}  \nonumber \\
  \times \int\frac{{\rm d}^{d}\boldsymbol{k}_{40}}{\left(2\pi\right)^{d}}\frac{{\rm d}^{d}\boldsymbol{\ell}}{\left(2\pi\right)^{d}}\frac{e^{-i\boldsymbol{k}_{40}\cdot\left(\boldsymbol{z}_{1}-\frac{x_{2}\boldsymbol{z}_{2}+x_{3}\boldsymbol{z}_{3}}{x_{2}+x_{3}}\right)-i\left(\boldsymbol{\ell}\cdot\boldsymbol{z}_{23}\right)}}{\frac{x_{2}+x_{3}}{x_{2}x_{3}}\boldsymbol{\ell}^{2}+\frac{\boldsymbol{k}_{40}^{2}+x_{1}\left(1-x_{1}\right)Q^{2}}{x_{1}\left(1-x_{1}\right)}} \left(g_{\perp}^{\mu\sigma}-\left(\frac{\ell_{\perp}^{\sigma}}{\ell^{+}}+\frac{k_{40\perp}^{\sigma}}{k_{40}^{+}}+\frac{q_{\perp}^{\sigma}}{q^{+}}\right)n_{2}^{\mu}\right)\nonumber \\
  \times{\rm tr}_{D} \left[ \gamma^{+}\left(x_{1}\slashed{q}-\slashed{k}_{40\perp}\right)\slashed{\varepsilon}_{q}
  \left(\gamma^{+}-\frac{x_{1}\left[\left(x_{2}+x_{3}\right)\slashed{q}+\slashed{k}_{40\perp}\right]\gamma^{+}\left[\left(x_{2}+x_{3}\right)\slashed{q}+\slashed{k}_{40\perp}\right]}{\boldsymbol{k}_{40}^{2}+x_{1}\left(1-x_{1}\right)Q^{2}-i0}\right) \right. \nonumber \\
   \times\gamma_{\mu}\left(x_{2}\slashed{q}+\frac{x_{2}}{x_{2}+x_{3}}\slashed{k}_{40\perp}-\slashed{\ell}_{\perp}\right)\gamma^{+}\Gamma_{\lambda} \Bigg] \; .
 \label{Eq:DiracStepInTheAntiQuark}
\end{gather}
We provide in Appendix~\ref{Sec:AppendixB2} useful details on the calculation of Dirac traces. After this step, as for the quark case, we can isolate the gauge invariance breaking term, which reads
\begin{gather}
{\cal A}_{\overline{q}3}^{{\rm QED}-{\rm break}.} \!= \frac{ie_{q}g}{N_c^2-1} \!\int{\rm d}z_{3}^{-}{\rm d}z_{2}^{-}{\rm d}z_{1}^{-}\!\!\int{\rm d}^{d}\boldsymbol{z}_{3}{\rm d}^{d}\boldsymbol{z}_{2}{\rm d}^{d}\boldsymbol{z}_{1} \!\!\int\frac{{\rm d}x_{1}}{2\pi}\frac{{\rm d}x_{2}}{2\pi}\frac{{\rm d}x_{3}}{2\pi}2\pi\delta\left(1-x_{1}-x_{2}-x_{3}\right) \nonumber \\
  \times \theta\left(x_{1}\right)\theta\left(x_{2}\right)\theta\left(x_{3}\right) {\rm e}^{-ix_{1}q^{+}z_{1}^{-}-ix_{2}q^{+}z_{2}^{-}-ix_{3}q^{+}z_{3}^{-}}{\rm e}^{i\boldsymbol{q}\cdot\left(x_{1}\boldsymbol{z}_{1}+x_{2}\boldsymbol{z}_{2}+x_{3}\boldsymbol{z}_{3}\right)} \nonumber \\
  \times \left\langle M\left(p_{M}\right)\left|\overline{\psi}\left(z_{1}\right)\Gamma^{\lambda}F_{-\sigma}\left(z_{3}\right)\psi\left(z_{2}\right)\right|0\right\rangle _{z_{1,2,3}^{+}=0} \left\langle P\left(p^{\prime}\right)\left|{\rm tr}\left(V_{\boldsymbol{z}_{1}}t^{a}U_{\boldsymbol{z}_{3}}^{ab}V_{\boldsymbol{z}_{2}}^{\dagger}t^{b}\right)\right|P\left(p\right)\right\rangle \nonumber \\
  \times\int\frac{{\rm d}^{d}\boldsymbol{k}_{40}}{\left(2\pi\right)^{d}}\frac{{\rm d}^{d}\boldsymbol{\ell}}{\left(2\pi\right)^{d}}\frac{e^{-i\boldsymbol{k}_{40}\cdot\left(\boldsymbol{z}_{1}-\frac{x_{2}\boldsymbol{z}_{2}+x_{3}\boldsymbol{z}_{3}}{x_{2}+x_{3}}\right)-i\left(\boldsymbol{\ell}\cdot\boldsymbol{z}_{23}\right)}}{\frac{x_{2}+x_{3}}{x_{2}x_{3}}\boldsymbol{\ell}^{2}+\frac{\boldsymbol{k}_{40}^{2}+x_{1}\left(1-x_{1}\right)Q^{2}}{x_{1}\left(1-x_{1}\right)}} \frac{-\varepsilon_{q}^{+}}{2x_{1}x_{2}x_{3}\left(x_{2}+x_{3}\right)} \nonumber \\ 
  \times {\rm tr}_{D} \left[ \left( x_{2}\slashed{k}_{40\perp}\gamma_{\perp}^{\sigma}+x_{1}\left(1-x_{1}\right)\left(\gamma_{\perp}^{\sigma}\slashed{\ell}_{\perp}+2\frac{x_{2}}{x_{3}}\ell_{\perp}^{\sigma}\right) \right) \gamma^{+}\Gamma_{\lambda} \right]  \; .
  \label{Eq:QEDBreaking3BodyAntiQuark}
\end{gather}
The transverse momentum integration can be done by using the result in eq.~(\ref{Eq:TrasvMomIntAntiQuarkQEDBreak}), then in 4 dimensions the QED gauge invariance breaking part reads
\begin{gather}
{\cal A}_{\overline{q}3}^{{\rm QED}-{\rm break}.}\!\! = \!\frac{e_{q}g \varepsilon_{q}^{+} }{4\pi} \frac{-1}{N_c^2-1} \!\!\int \!{\rm d} z_{3}^{-}{\rm d}z_{2}^{-}{\rm d}z_{1}^{-}\!\!\!\int\!\!{\rm d}^{2}\boldsymbol{z}_{3}{\rm d}^{2}\boldsymbol{z}_{2}{\rm d}^{2}\boldsymbol{z}_{1}\! \!\int\!\!\frac{{\rm d}x_{1}}{2\pi}\frac{{\rm d}x_{2}}{2\pi}\frac{{\rm d}x_{3}}{2\pi} \delta\left(1-x_{1}-x_{2}-x_{3}\right) \nonumber \\
  \times \theta\left(x_{1}\right)\theta\left(x_{2}\right)\theta\left(x_{3}\right) {\rm e}^{-ix_{1}q^{+}z_{1}^{-}-ix_{2}q^{+}z_{2}^{-}-ix_{3}q^{+}z_{3}^{-}}{\rm e}^{i\boldsymbol{q}\cdot\left(x_{1}\boldsymbol{z}_{1}+x_{2}\boldsymbol{z}_{2}+x_{3}\boldsymbol{z}_{3}\right)}\frac{Q^{2}}{Z^{2}}K_{2}\left(QZ\right)\nonumber \\
  \times \left\langle P\left(p^{\prime}\right)\left|{\rm tr}\left(V_{\boldsymbol{z}_{1}}t^{a}U_{\boldsymbol{z}_{3}}^{ab}V_{\boldsymbol{z}_{2}}^{\dagger}t^{b}\right)\right|P\left(p\right)\right\rangle \left\langle M\left(p_{M}\right)\left|\overline{\psi}\left(z_{1}\right)\Gamma^{\lambda}F_{-\sigma}\left(z_{3}\right)\psi\left(z_{2}\right)\right|0\right\rangle _{z_{1,2,3}^{+}=0}  \nonumber \\
  \times x_{1}x_{2}{\rm tr}_{D} \left[ \left( z_{1\perp\mu}\gamma_{\perp}^{\mu}\gamma_{\perp}^{\sigma}+z_{2\perp\mu}\gamma_{\perp}^{\sigma}\gamma_{\perp}^{\mu}-2z_{3\perp}^{\sigma}\right) \gamma^{+}\Gamma_{\lambda} \right] \; . 
\end{gather}
Let us now move on to the discussion of the gauge invariant part, which reads
\begin{gather}
{\cal A}_{\overline{q}3}^{{\rm QED}-{\rm inv}.} \!=\!\frac{ie_{q} \; g \; q^{+} }{2 (N_c^2 -1)} \!\int\!{\rm d}z_{3}^{-}{\rm d}z_{2}^{-}{\rm d}z_{1}^{-}\!\int\!{\rm d}^{d}\boldsymbol{z}_{3}{\rm d}^{d}\boldsymbol{z}_{2}{\rm d}^{d}\boldsymbol{z}_{1} \!\int\!\frac{{\rm d}x_{1}}{2\pi}\frac{{\rm d}x_{2}}{2\pi}\frac{{\rm d}x_{3}}{2\pi}2\pi\delta\left(1-x_{1}-x_{2}-x_{3}\right) \nonumber \\
  \times \theta\left(x_{1}\right)\theta\left(x_{2}\right)\theta\left(x_{3}\right) {\rm e}^{-ix_{1}q^{+}z_{1}^{-}-ix_{2}q^{+}z_{2}^{-}-ix_{3}q^{+}z_{3}^{-}}{\rm e}^{i\boldsymbol{q}\cdot\left(x_{1}\boldsymbol{z}_{1}+x_{2}\boldsymbol{z}_{2}+x_{3}\boldsymbol{z}_{3}\right)}  \nonumber \\
  \times \left\langle P\left(p^{\prime}\right)\left|{\rm tr}\left(V_{\boldsymbol{z}_{1}}t^{a}U_{\boldsymbol{z}_{3}}^{ab}V_{\boldsymbol{z}_{2}}^{\dagger}t^{b}\right)\right|P\left(p\right)\right\rangle \left\langle M\left(p_{M}\right)\left|\overline{\psi}\left(z_{1}\right)\Gamma^{\lambda}F_{-\sigma}\left(z_{3}\right)\psi\left(z_{2}\right)\right|0\right\rangle _{z_{1,2,3}^{+}=0} \nonumber \\
  \times \!\left(\varepsilon_{q\rho}-\frac{\varepsilon_{q}^{+}}{q^{+}}q_{\rho}\right)\!\! \int\!\frac{{\rm d}^{d}\boldsymbol{k}_{40}}{\left(2\pi\right)^{d}}\frac{{\rm d}^{d}\boldsymbol{\ell}}{\left(2\pi\right)^{d}}\frac{e^{-i\boldsymbol{k}_{40}\cdot\left(\boldsymbol{z}_{1}-\frac{x_{2}\boldsymbol{z}_{2}+x_{3}\boldsymbol{z}_{3}}{x_{2}+x_{3}}\right)-i\left(\boldsymbol{\ell}\cdot\boldsymbol{z}_{23}\right)}}{\left(x_{2}+x_{3}\right)\left[\boldsymbol{\ell}^{2}+\frac{x_{2}x_{3}\left(\boldsymbol{k}_{40}^{2}+x_{1}\left(1-x_{1}\right)Q^{2}\right)}{x_{1}\left(1-x_{1}\right)^{2}}\right]} 
  \left\{ \frac{1}{\boldsymbol{k}_{40}^{2}+x_{1}\left(1-x_{1}\right)Q^{2}}\right. \nonumber \\
  \times{\rm tr}_{D} \left[ \left( 2x_{1}\left(1-x_{1}\right)q^{\rho}+x_{1}\gamma_{\perp}^{\rho}\slashed{k}_{40\perp}-\left(1-x_{1}\right)\slashed{k}_{40\perp}\gamma_{\perp}^{\rho} \right) \left(\gamma_{\perp}^{\sigma}\slashed{\ell}_{\perp}+2\frac{x_{2}}{x_{3}}\ell_{\perp}^{\sigma}\right)\gamma^{+}\Gamma_{\lambda} \right] \nonumber \\
  \left.-\frac{x_{2}}{x_{2}+x_{3}}{\rm tr}_{D} \left[ \gamma_{\perp}^{\rho}\gamma_{\perp}^{\sigma}\gamma^{+}\Gamma_{\lambda} \right] \right\} \; .
  \label{Eq:3BodyContAntiQuarkBeforeTransvMomInt}
\end{gather}
We can use the integral in (\ref{Eq:TrasvMomIntAntiQuarkElle}) to integrate over $\boldsymbol{\ell}$, in order to get
\begin{gather}
{\cal A}_{\overline{q}3}^{{\rm QED}-{\rm inv}.} \!\!=\!\frac{ie_{q}g}{2 (N_c^2-1)}q^{+}\!\!\!\int\!\!{\rm d}z_{3}^{-}{\rm d}z_{2}^{-}{\rm d}z_{1}^{-}\!\!\!\int\!{\rm d}^{d}\boldsymbol{z}_{3}{\rm d}^{d}\boldsymbol{z}_{2}{\rm d}^{d}\boldsymbol{z}_{1} \!\!\int\!\frac{{\rm d}x_{1}}{2\pi}\frac{{\rm d}x_{2}}{2\pi}\frac{{\rm d}x_{3}}{2\pi}2\pi\delta\left(1-x_{1}-x_{2}-x_{3}\right) \nonumber \\
  \times \theta\left(x_{1}\right)\theta\left(x_{2}\right)\theta\left(x_{3}\right) {\rm e}^{-ix_{1}q^{+}z_{1}^{-}-ix_{2}q^{+}z_{2}^{-}-ix_{3}q^{+}z_{3}^{-}}{\rm e}^{i\boldsymbol{q}\cdot\left(x_{1}\boldsymbol{z}_{1}+x_{2}\boldsymbol{z}_{2}+x_{3}\boldsymbol{z}_{3}\right)} \nonumber \\
  \times\left\langle P\left(p^{\prime}\right)\left|{\rm tr}\left(V_{\boldsymbol{z}_{1}}t^{a}U_{\boldsymbol{z}_{3}}^{ab}V_{\boldsymbol{z}_{2}}^{\dagger}t^{b}\right)\right|P\left(p\right)\right\rangle \left\langle M\left(p_{M}\right)\left|\overline{\psi}\left(z_{1}\right)\Gamma^{\lambda}F_{-\sigma}\left(z_{3}\right)\psi\left(z_{2}\right)\right|0\right\rangle _{z_{1,2,3}^{+}=0} \nonumber \\
  \times \frac{\left(\varepsilon_{q\rho}-\frac{\varepsilon_{q}^{+}}{q^{+}}q_{\rho}\right)}{4\pi} \int\frac{{\rm d}^{d}\boldsymbol{k}_{40}}{\left(2\pi\right)^{d}}\frac{e^{-i\boldsymbol{k}_{40} \cdot\left(\frac{x_{2}\boldsymbol{z}_{12}+x_{3}\boldsymbol{z}_{13}}{x_{2}+x_{3}}\right)}}{\left(x_{2}+x_{3}\right)}  \int_{0}^{\infty}{\rm d}t{\rm e}^{i\frac{\boldsymbol{z}_{23}^{2}}{4t}-it\left(\frac{x_{2}x_{3}}{x_{1}\left(1-x_{1}\right)^{2}}\left(\boldsymbol{k}_{40}^{2}+x_{1}\left(1-x_{1}\right)Q^{2}\right)-i0 \right)} \nonumber \\
  \left\{ -\frac{1}{t^{2}}\frac{z_{23\perp\nu}}{2}\frac{1}{\boldsymbol{k}_{40}^{2}+x_{1}\left(1-x_{1}\right)Q^{2}}\right. 
  {\rm tr}_{D} \left[ \left(2x_{1}\left(1-x_{1}\right)q^{\rho}+x_{1}\gamma_{\perp}^{\rho}\slashed{k}_{40\perp}-\overline{x}_{1}\slashed{k}_{40\perp}\gamma_{\perp}^{\rho}\right) \right.
\nonumber \\
  \left. \left. \times \left(\gamma_{\perp}^{\sigma}\gamma_{\perp}^{\nu}+2\frac{x_{2}}{x_{3}}g_{\perp}^{\sigma\nu}\right)\gamma^{+}\Gamma_{\lambda} \right] -\frac{1}{t}\frac{x_{2}}{x_{2}+x_{3}}{\rm tr}_{D} \left[ \gamma_{\perp}^{\rho}\gamma_{\perp}^{\sigma}\gamma^{+}\Gamma_{\lambda} \right] \right\} \; . 
\end{gather}
After this first integration, we have again two classes of integral, calculated in (\ref{Eq:AntiQuarkTransvMomIntegralK40WithoutDen}), (\ref{Eq:AntiQuarkTransvMomIntegralK40WDen}). These integrals allow us to cast the QED gauge invariant part in 4 dimensions as
\begin{gather}
{\cal A}_{\overline{q}3}^{{\rm QED}-{\rm inv}.} \!\!=\!\frac{e_{q} \; g \; q^{+} }{4 \pi} \frac{1}{N_c^2-1} \!\int\!{\rm d}z_{3}^{-}{\rm d}z_{2}^{-}{\rm d}z_{1}^{-}\!\!\!\int\!\!{\rm d}^{2}\boldsymbol{z}_{3}{\rm d}^{2}\boldsymbol{z}_{2}{\rm d}^{2}\boldsymbol{z}_{1} \!\!\int\!\frac{{\rm d}x_{1}}{2\pi}\frac{{\rm d}x_{2}}{2\pi}\frac{{\rm d}x_{3}}{2\pi} \delta\left(1-x_{1}-x_{2}-x_{3}\right) \nonumber \\
  \times \theta\left(x_{1}\right)\theta\left(x_{2}\right)\theta\left(x_{3}\right) 
  \left(\varepsilon_{q\rho}-\frac{\varepsilon_{q}^{+}}{q^{+}}q_{\rho}\right){\rm e}^{-ix_{1}q^{+}z_{1}^{-}-ix_{2}q^{+}z_{2}^{-}-ix_{3}q^{+}z_{3}^{-}}{\rm e}^{i\boldsymbol{q}\cdot\left(x_{1}\boldsymbol{z}_{1}+x_{2}\boldsymbol{z}_{2}+x_{3}\boldsymbol{z}_{3}\right)}\nonumber \\
  \times \left\langle P\left(p^{\prime}\right)\left|{\rm tr}\left(V_{\boldsymbol{z}_{1}}t^{a}U_{\boldsymbol{z}_{3}}^{ab}V_{\boldsymbol{z}_{2}}^{\dagger}t^{b}\right)\right|P\left(p\right)\right\rangle \left\langle M\left(p_{M}\right)\left|\overline{\psi}\left(z_{1}\right)\Gamma^{\lambda}F_{-\sigma}\left(z_{3}\right)\psi\left(z_{2}\right)\right|0\right\rangle _{z_{1,2,3}^{+}=0} \nonumber \\
  \times\!\left\{\! 2x_{1}\frac{z_{23\perp\nu}}{\boldsymbol{z}_{23}^{2}}q^{\rho}K_{0}\left(QZ\right){\rm tr}_{D} \!\left[\! \left(\gamma_{\perp}^{\sigma}\gamma_{\perp}^{\nu}+2\frac{x_{2}}{x_{3}}g_{\perp}^{\sigma\nu}\right)\!\gamma^{+}\Gamma_{\lambda} \right] \right.\!\!
  -i\frac{x_{1}x_{2}}{1-x_{1}}\frac{Q}{Z}K_{1}\!\left(QZ\right){\rm tr}_{D} \left[ \gamma_{\perp}^{\rho}\gamma_{\perp}^{\sigma}\gamma^{+}\Gamma_{\lambda} \right] \nonumber \\
  -i\frac{z_{23\perp\nu}}{\boldsymbol{z}_{23}^{2}}{\rm tr}_{D} \left[ \left(x_{1}\gamma_{\perp}^{\rho}\gamma_{\perp}^{\alpha}-\overline{x}_{1}\gamma_{\perp}^{\alpha}\gamma_{\perp}^{\rho}\right)\left(\gamma_{\perp}^{\sigma}\gamma_{\perp}^{\nu}+2\frac{x_{2}}{x_{3}}g_{\perp}^{\sigma\nu}\right)\gamma^{+}\Gamma_{\lambda} \right] \nonumber \\
  \left.\times\left(\frac{x_{1}x_{2}z_{12\perp\alpha}+x_{1}x_{3}z_{13\perp\alpha}}{x_{2}+x_{3}}\right)\frac{Q}{Z}K_{1}\left(QZ\right)\right\} \; .
  \label{Eq:AntiQuark3BodyConFinalExpr}
\end{gather}

\subsection{Summary of the results at the amplitude level}
As seen in section~\ref{Eq:2-body_contribution}
\begin{equation}
   \mathcal{A}_2^{{\rm QED}-{\rm break}.}  = 0 \; , \nonumber \\
\end{equation}
while, we observe here that combining eq.~(\ref{Eq:QEDBreaking3BodyQuark}) and eq.~(\ref{Eq:QEDBreaking3BodyAntiQuark}), we get
\begin{equation}
    {\cal A}_{q3}^{{\rm QED}-{\rm break}.}+{\cal A}_{\overline{q}3}^{{\rm QED}-{\rm break}.} = 0 \nonumber \; , 
\end{equation}
which shows that our result is explicitly QED gauge invariant. \\

\noindent The full \textbf{2-body contribution} (see eq.~(\ref{Eq.GeneralFinal2bodyCon})) to the amplitude reads
\begin{gather}
  {\cal A}_{2} = e_{q} \delta\left( 1 - \frac{p_{M}^{+}}{q^{+}} \right)\left(\varepsilon_{q\mu}-\varepsilon_{q}^{+}\frac{q_{\mu}}{q^{+}}\right)\int_{0}^{1}\frac{{\rm d}x}{2\pi}\int{\rm d}^{4}r\delta\left(r^{+}\right){\rm e}^{-i\left(xp_{M}\cdot r\right)}\nonumber \\
  \times\int{\rm d}^{2}\boldsymbol{b}\,{\rm e}^{i\left(\boldsymbol{q}-\boldsymbol{p}_{M}\right)\cdot\boldsymbol{b}}\left\langle P\left(p^{\prime}\right)\left|1-\frac{1}{N_{c}}{\rm tr}\left(V_{\boldsymbol{b}+\overline{x}\boldsymbol{r}}V_{\boldsymbol{b}-x\boldsymbol{r}}^{\dagger}\right)\right|P\left(p\right)\right\rangle \nonumber \\
  \times\left\{ 2x\overline{x}q^{\mu}\left\langle M\left(p_{M}\right)\left|\overline{\psi}\left(r\right)\gamma^{+}\psi\left(0\right)\right|0\right\rangle K_{0}\left(\sqrt{x\overline{x}Q^{2}\boldsymbol{r}^{2}}\right)\right.\nonumber \\
  - i \left(x-\overline{x}\right)\left\langle M\left(p_{M}\right)\left|\overline{\psi}\left(r\right)\gamma^{+}\psi\left(0\right)\right|0\right\rangle \frac{\partial}{\partial r_{\perp, \mu}} K_0 \left( \sqrt{x \bar{x} Q^2 \boldsymbol{r}^{2}} \right) \nonumber \\
  \left. + \epsilon^{\mu\nu+-}\left\langle M\left(p_{M}\right)\left|\overline{\psi}\left(r\right)\gamma^{+}\gamma^{5}\psi\left(0\right)\right|0\right\rangle \frac{\partial}{\partial r_{\perp}^{\nu}} K_0 \left( \sqrt{x \bar{x} Q^2 \boldsymbol{r}^{2}} \right) \right\} .  \vspace{0.15 cm}
  \label{Eq.GeneralFinal2bodyConSummary}
\end{gather}
To obtain the full 3-body contribution, in accordance with the fact that the two gauge invariance breaking terms cancel, we have to sum eqs.~(\ref{Eq:Quark3BodyConFinalExpr}) and (\ref{Eq:AntiQuark3BodyConFinalExpr}), i.e. 
\begin{gather}
  {\cal A}_{3} \hspace{-0.1 cm} = \hspace{-0.1 cm}  {\cal A}_{3q}^{\rm inv.} \hspace{-0.1 cm} + \hspace{-0.1 cm} {\cal A}_{3\overline{q}}^{\rm inv.} \hspace{-0.1 cm}  = \hspace{-0.1 cm} \frac{e_{q}g q^{+} }{4 \pi (N_c^2-1)} \left(\varepsilon_{q\rho}-\frac{\varepsilon_{q}^{+}}{q^{+}}q_{\rho}\right) \hspace{-0.1 cm} \int{\rm d} \hspace{-0.1 cm}  z_{3}^{-}  {\rm d}z_{2}^{-}  {\rm d}z_{1}^{-} \hspace{-0.1 cm} \int \hspace{-0.1 cm} {\rm d}^{2} \boldsymbol{z}_{3}{\rm d}^{2}\boldsymbol{z}_{2}{\rm d}^{2}\boldsymbol{z}_{1} \hspace{-0.1 cm} \int\frac{{\rm d}x_{1}}{2\pi}\frac{{\rm d}x_{2}}{2\pi}\frac{{\rm d}x_{3}}{2\pi} \nonumber \\
  \times \delta\left(1-x_{1}-x_{2}-x_{3}\right)\theta\left(x_{1}\right)\theta\left(x_{2}\right)\theta\left(x_{3}\right) {\rm e}^{-ix_{1}q^{+}z_{1}^{-}-ix_{2}q^{+}z_{2}^{-}-ix_{3}q^{+}z_{3}^{-}} {\rm e}^{i\boldsymbol{q}\cdot\left(x_{1}\boldsymbol{z}_{1}+x_{2}\boldsymbol{z}_{2}+x_{3}\boldsymbol{z}_{3}\right)} \nonumber \\
\times \left\langle P\left(p^{\prime}\right)\left|{\rm tr}\left(V_{\boldsymbol{z}_{1}}t^{a}V_{\boldsymbol{z}_{2}}^{\dagger}t^{b}U_{\boldsymbol{z}_{3}}^{ab}\right)\right|P\left(p\right)\right\rangle  \left\langle M\left(p_{M}\right)\left|\overline{\psi}\left(z_{1}\right)\Gamma^{\lambda}F_{-\sigma}\left(z_{3}\right)\psi\left(z_{2}\right)\right|0\right\rangle _{z_{1,2,3}^{+}=0} \nonumber \\ \times \left\{ ix_{1}x_{2}{\rm tr}_{D}\left[\left(\frac{\gamma_{\perp}^{\sigma}\gamma_{\perp}^{\rho}}{1-x_{2}}-\frac{\gamma_{\perp}^{\rho}\gamma_{\perp}^{\sigma}}{1-x_{1}}\right)\gamma^{+}\Gamma_{\lambda}\right]\frac{Q}{Z}K_{1}\left(QZ\right) \right. + 2q^{\rho}K_{0}\left(QZ\right) \nonumber \\ \times {\rm tr}_{D} \left[ \left(2\left(\frac{x_{1}x_{2}z_{23\perp\nu}}{x_{3}\boldsymbol{z}_{23}^{2}}-\frac{x_{1}x_{2}z_{13\perp\nu}}{x_{3}\boldsymbol{z}_{13}^{2}}\right)g_{\perp}^{\sigma\nu}+\frac{x_{1}z_{23\perp\nu}}{\boldsymbol{z}_{23}^{2}}\gamma_{\perp}^{\sigma}\gamma_{\perp}^{\nu}-\frac{x_{2}z_{13\perp\nu}}{\boldsymbol{z}_{13}^{2}}\gamma_{\perp}^{\nu}\gamma_{\perp}^{\sigma}\right)\gamma^{+}\Gamma_{\lambda} \right] \nonumber \\
  -i\frac{z_{23\perp\nu}}{\boldsymbol{z}_{23}^{2}}  \left(\frac{x_{1}x_{2}z_{12\perp\alpha}+x_{1}x_{3}z_{13\perp\alpha}}{x_{2}+x_{3}}\right)\frac{Q}{Z}K_{1}\left(QZ\right)  \nonumber \\ \times  {\rm tr}_{D} \left[ \left(x_{1}\gamma_{\perp}^{\rho}\gamma_{\perp}^{\alpha}-\overline{x}_{1}\gamma_{\perp}^{\alpha}\gamma_{\perp}^{\rho}\right)\left(\gamma_{\perp}^{\sigma}\gamma_{\perp}^{\nu}+2\frac{x_{2}}{x_{3}}g_{\perp}^{\sigma\nu}\right)\gamma^{+}\Gamma_{\lambda} \right] 
  \nonumber \\
  -i\frac{z_{13\perp\nu}}{\boldsymbol{z}_{13}^{2}} \left(\frac{x_{1}x_{2}z_{12\perp\alpha}+x_{3}x_{2}z_{32\perp\alpha}}{x_{1}+x_{3}}\right)\frac{Q}{Z}K_{1}\left(QZ\right) \nonumber \\ \times {\rm tr}_{D} \left[ \left(2\frac{x_{1}}{x_{3}}g_{\perp}^{\nu\sigma}+\gamma_{\perp}^{\nu}\gamma_{\perp}^{\sigma}\right)\left(x_{2}\gamma_{\perp}^{\alpha}\gamma_{\perp}^{\rho}-\overline{x}_{2}\gamma_{\perp}^{\rho}\gamma_{\perp}^{\alpha}\right)\gamma^{+}\Gamma_{\lambda} \right]
  \left.  \right. \bigg \}  \; .
\end{gather}
By using
\begin{align}
\left(x_{1}x_{2}z_{12\perp\alpha}+x_{1}x_{3}z_{13\perp\alpha}\right)\frac{Q}{Z}K_{1}\left(QZ\right) & =\frac{\partial}{\partial z_{1\perp}^{\alpha}}K_{0}\left(QZ\right) 
\end{align}
and
\begin{align}
\left(x_{1}x_{2}z_{21\perp\alpha}+x_{2}x_{3}z_{23\perp\alpha}\right)\frac{Q}{Z}K_{1}\left(QZ\right) & =\frac{\partial}{\partial z_{2\perp}^{\alpha}}K_{0}\left(QZ\right) \; ,
\end{align}
we can alternatively write the full \textbf{3-body contribution} as
\begin{gather}
  {\cal A}_{3} = \frac{e_{q} g q^{+}}{16\pi^{2}} \frac{2}{N_c^2-1} \left(\varepsilon_{q\rho}-\frac{\varepsilon_{q}^{+}}{q^{+}}q_{\rho}\right)\int{\rm d}z_{3}^{-}{\rm d}z_{2}^{-}{\rm d}z_{1}^{-}\int{\rm d}^{2}\boldsymbol{z}_{3}{\rm d}^{2}\boldsymbol{z}_{2}{\rm d}^{2}\boldsymbol{z}_{1} \int\frac{{\rm d}x_{1}}{2\pi}\frac{{\rm d}x_{2}}{2\pi}\frac{{\rm d}x_{3}}{2\pi} \nonumber \\
  \times 2\pi\delta\left(1-x_{1}-x_{2}-x_{3}\right) \theta\left(x_{1}\right)\theta\left(x_{2}\right)\theta\left(x_{3}\right) {\rm e}^{-ix_{1}q^{+}z_{1}^{-}-ix_{2}q^{+}z_{2}^{-}-ix_{3}q^{+}z_{3}^{-}+i\boldsymbol{q}\cdot\left(x_{1}\boldsymbol{z}_{1}+x_{2}\boldsymbol{z}_{2}+x_{3}\boldsymbol{z}_{3}\right)} \nonumber \\
  \times \left\langle P\left(p^{\prime}\right)\left|{\rm tr}\left(V_{\boldsymbol{z}_{1}}t^{a}V_{\boldsymbol{z}_{2}}^{\dagger}t^{b}U_{\boldsymbol{z}_{3}}^{ab}\right)\right|P\left(p\right)\right\rangle \left\langle M\left(p_{M}\right)\left|\overline{\psi}\left(z_{1}\right)\Gamma^{\lambda}F_{-\sigma}\left(z_{3}\right)\psi\left(z_{2}\right)\right|0\right\rangle _{z_{1,2,3}^{+}=0} \nonumber \\
  \times \left\{ ix_{1}x_{2}{\rm tr}_{D}\left[\left(\frac{\gamma_{\perp}^{\sigma}\gamma_{\perp}^{\rho}}{1-x_{2}}-\frac{\gamma_{\perp}^{\rho}\gamma_{\perp}^{\sigma}}{1-x_{1}}\right)\gamma^{+}\Gamma_{\lambda}\right]\frac{Q}{Z}K_{1}\left(QZ\right)\right. \nonumber \\
  +{\rm tr}_{D} \left[ \left(2q^{\rho}-i\frac{x_{1}\gamma_{\perp}^{\rho}\gamma_{\perp}^{\alpha}-\overline{x}_{1}\gamma_{\perp}^{\alpha}\gamma_{\perp}^{\rho}}{x_{1}\overline{x}_{1}}\frac{\partial}{\partial z_{1\perp}^{\alpha}}\right)\left(2\frac{x_{1}x_{2}}{x_{3}}g_{\perp}^{\sigma\nu}+x_{1}\gamma_{\perp}^{\sigma}\gamma_{\perp}^{\nu}\right)\gamma^{+}\Gamma_{\lambda} \right] \frac{z_{23\perp\nu}}{\boldsymbol{z}_{23}^{2}}K_{0}\left(QZ\right) \nonumber \\ 
  -{\rm tr}_{D} \left[ \left(2\frac{x_{1}x_{2}}{x_{3}}g_{\perp}^{\sigma\nu}+x_{2}\gamma_{\perp}^{\nu}\gamma_{\perp}^{\sigma}\right)\left(2q^{\rho}-i\frac{x_{2}\gamma_{\perp}^{\alpha}\gamma_{\perp}^{\rho}-\overline{x}_{2}\gamma_{\perp}^{\rho}\gamma_{\perp}^{\alpha}}{x_{2}\overline{x}_{2}}\frac{\partial}{\partial z_{2\perp}^{\alpha}}\right)\gamma^{+}\Gamma_{\lambda} \right] \left. \frac{z_{13\perp\nu}}{\boldsymbol{z}_{13}^{2}}K_{0}\left(QZ\right)\right\} .
  \label{Eq:AntiQuark3BodyConFinalExprSummary}
\end{gather}

\subsection{Extraction of the impact factors: coordinate space}
We will now extract the impact factors for the $\gamma^{(*)} \rightarrow M (\rho, \varphi, \omega)$ transition in coordinate space.
\subsubsection{2-body contribution}
Let us introduce the standard form of the dipole amplitude
\begin{equation}
 \mathcal{A}_2 = \int_0^1 \hspace{-0.35 cm} \mathrm{~d} x \hspace{-0.1 cm} \int \hspace{-0.15 cm} \mathrm{d}^d \boldsymbol{z}_1 \hspace{-0.1 cm} \mathrm{~d}^d \boldsymbol{z}_2 \mathrm{e}^{i (\boldsymbol{q}-\boldsymbol{p}_M) \cdot\left(x \boldsymbol{z}_1+\bar{x} \boldsymbol{z}_2\right)} \Psi_2 \left(x, \boldsymbol{z}_{12}\right)\left\langle P\left(p^{\prime}\right)\left|1-\frac{1}{N_c} \operatorname{tr}\left(V_{\boldsymbol{z}_1} V_{\boldsymbol{z}_2}^{\dagger}\right)\right| P(p)\right\rangle ,
\end{equation}
where $\Psi_2 \left(x, \boldsymbol{z}_{12}\right)$, in the present case, is the overlap of the photon and meson wave functions. Performing the change of variables in eq.~(\ref{Eq:ChangeziTorb}), the amplitude can be written as
\begin{equation}
   \mathcal{A}_2 = \int_{0}^{1} {\rm d} x \int{\rm d}^{2} \boldsymbol{r} \Psi_2 \left(x, \boldsymbol{r} \right) \int{\rm d}^{d} \boldsymbol{b} \; {\rm e}^{i (\boldsymbol{q}-\boldsymbol{p}_M) \cdot\boldsymbol{b}}\left\langle P\left(p^{\prime}\right)\left|1-\frac{1}{N_{c}}{\rm tr}\left(V_{\boldsymbol{b}+\overline{x}\boldsymbol{r}}V_{\boldsymbol{b}-x\boldsymbol{r}}^{\dagger}\right)\right|P\left(p\right)\right\rangle \; .
\label{Eq:StandardDipoleAmp-rb}
\end{equation}
Comparing the last expression with eq.~(\ref{Eq.GeneralFinal2bodyConSummary}), we can immediately extract the  photon/meson wave function overlap of the 2-body contribution
\begin{gather}
    \Psi_2 \left(x, \boldsymbol{r} \right) = e_q \delta \left( 1 - \frac{p_M^+}{q^+} \right) \left(\varepsilon_{q\mu}-\frac{\varepsilon_{q}^{+}}{q^{+}}q_{\mu} \right)  \nonumber \\
  \times  \left[ \phi_{\gamma^+} (x, \boldsymbol{r}) \left( 2 x \bar{x} q^{\mu} - i (x-\bar{x}) \frac{\partial}{\partial r_{\perp \mu}} \right) + \epsilon^{\mu\nu+-} \phi_{\gamma^+ \gamma^5} (x, \boldsymbol{r}) \frac{\partial}{\partial r_{\perp}^{\nu}} \right] K_0 \left( \sqrt{x \bar{x} Q^2 \boldsymbol{r}^{2}} \right) \; ,
  \label{Eq:Psi_2_Full_KinTwist}
\end{gather}
where we have introduced the distributions 
\begin{equation}
    \phi_{\gamma^+} (x, \boldsymbol{r}) = \frac{1}{2 \pi} \int_{-\infty}^{\infty} d r^- e^{-i x p_M^+ r^- + i x \boldsymbol{p}_M \cdot\boldsymbol{r}} \left\langle M\left(p_{M}\right)\left|\overline{\psi}\left(r\right)\gamma^{+}\psi\left(0\right)\right|0\right\rangle_{r^+=0} \; ,
    \label{Phi_+_Distrib_Coordinate}
\end{equation}
\begin{equation}
    \phi_{\gamma^+ \gamma^5} (x, \boldsymbol{r}) = \frac{1}{2 \pi} \int_{-\infty}^{\infty} d r^- e^{-i x p_M^+ r^- + i x \boldsymbol{p}_M \cdot\boldsymbol{r}} \left\langle M\left(p_{M}\right)\left|\overline{\psi}\left(r\right)\gamma^{+} \gamma^5 \psi\left(0\right)\right|0\right\rangle_{r^+=0} \; .
    \label{Phi_+5_Distrib_Coordinate}
\end{equation}

\subsubsection{3-body contribution}
To extract the 3-body photon/meson wave functions overlap, one should first express the target correlator in terms of dipole and double dipole operators, i.e.
\begin{gather}
    \left\langle P\left(p^{\prime}\right)\left|{\rm tr}\left(V_{\boldsymbol{z}_{1}}t^{a}V_{\boldsymbol{z}_{2}}^{\dagger}t^{b}U_{\boldsymbol{z}_{3}}^{ab}\right)\right|P\left(p\right)\right\rangle \nonumber \\ = \frac{1}{2} \left\langle P\left(p^{\prime}\right)\left| \left[{\rm tr}\left(V_{\boldsymbol{z}_{1}}V_{\boldsymbol{z}_{3}}^{\dagger}\right){\rm tr}\left(V_{\boldsymbol{z}_{3}}V_{\boldsymbol{z}_{2}}^{\dagger}\right)-\frac{1}{N_{c}}{\rm tr}\left(V_{\boldsymbol{z}_{1}}V_{\boldsymbol{z}_{2}}^{\dagger}\right)\right] \right|P\left(p\right)\right\rangle \nonumber \\ 
    = \frac{N_c^2}{2} \left\langle P\left(p^{\prime}\right)\left| \mathcal{U}_{\boldsymbol{z}_1 \boldsymbol{z}_3} \mathcal{U}_{\boldsymbol{z}_3 \boldsymbol{z}_2} - \mathcal{U}_{\boldsymbol{z}_1 \boldsymbol{z}_3} -\mathcal{U}_{\boldsymbol{z}_3 \boldsymbol{z}_2} + \mathcal{U}_{\boldsymbol{z}_1 \boldsymbol{z}_2}  + \frac{N_c^2 - 1}{N_c^2} (1-\mathcal{U}_{\boldsymbol{z}_1 \boldsymbol{z}_2}) \right|P\left(p\right)\right\rangle \; ,
\end{gather}
and then subtract the non-interacting part to obtain
\begin{gather}
    \left\langle P\left(p^{\prime}\right)\left|{\rm tr}\left(V_{\boldsymbol{z}_{1}}t^{a}V_{\boldsymbol{z}_{2}}^{\dagger}t^{b}U_{\boldsymbol{z}_{3}}^{ab}\right)\right|P\left(p\right)\right\rangle \; - \; {\rm n.i.} \nonumber \\ = \frac{N_c^2}{2} \left\langle P\left(p^{\prime}\right)\left| \mathcal{U}_{\boldsymbol{z}_1 \boldsymbol{z}_3} \mathcal{U}_{\boldsymbol{z}_3 \boldsymbol{z}_2} - \mathcal{U}_{\boldsymbol{z}_1 \boldsymbol{z}_3} -\mathcal{U}_{\boldsymbol{z}_3 \boldsymbol{z}_2} + \frac{1}{N_c^2} \mathcal{U}_{\boldsymbol{z}_1 \boldsymbol{z}_2}  \right|P\left(p\right)\right\rangle \; .
\end{gather}
We can now write the general three-body small-$x$ amplitude as
\begin{gather}
    \mathcal{A}_3 = \left( \prod_{i=1}^3 \int d x_i \theta (x_i) \right) \delta (1 - x_1 - x_2 - x_3) \int d^2 \boldsymbol{z}_1 d^2 \boldsymbol{z}_2 d^2 \boldsymbol{z}_3 e^{i \boldsymbol{q} (x_1 \boldsymbol{z}_1 + x_2 \boldsymbol{z}_2 + x_3 \boldsymbol{z}_3)} \nonumber \\
   \times \Psi_3 \left( x_1, x_2, x_3 , \boldsymbol{z}_1, \boldsymbol{z}_2, \boldsymbol{z}_3\right) \left\langle P\left(p^{\prime}\right)\left| \mathcal{U}_{\boldsymbol{z}_1 \boldsymbol{z}_3} \mathcal{U}_{\boldsymbol{z}_3 \boldsymbol{z}_2} - \mathcal{U}_{\boldsymbol{z}_1 \boldsymbol{z}_3} -\mathcal{U}_{\boldsymbol{z}_3 \boldsymbol{z}_2} + \frac{1}{N_c^2} \mathcal{U}_{\boldsymbol{z}_1 \boldsymbol{z}_2}  \right|P\left(p\right)\right\rangle \; .
   \label{Eq:GenealStructure3bodyWaveFunOver}
\end{gather}
Comparing eqs.~(\ref{Eq:AntiQuark3BodyConFinalExprSummary}) and~(\ref{Eq:GenealStructure3bodyWaveFunOver}), we can extract the wave functions overlap
\begin{gather}
    \Psi_3 \left( x_1, x_2, x_3 , \boldsymbol{z}_1, \boldsymbol{z}_2, \boldsymbol{z}_3\right) = \frac{e_{q} q^{+}}{2 (4 \pi)} \frac{N_c^2}{N_c^2-1} \left(\varepsilon_{q\rho}-\frac{\varepsilon_{q}^{+}}{q^{+}}q_{\rho}\right)   \nonumber \\
    \times \bigg \{ \chi_{\gamma^+ \sigma } \left[ \left( 4 i g_{\perp}^{ \rho \sigma } \frac{x_1 x_2}{1-x_2} \frac{Q}{Z} K_1 (QZ) +  T_1^{\sigma \rho \nu} (x_1, x_2, x_3) \frac{z_{23 \perp \nu}}{\boldsymbol{z}_{23}^{2}} K_0 (QZ) \right) - \left( 1 \leftrightarrow 2 \right) \right] \nonumber \\ 
    - \chi_{\gamma^+ \gamma^5 \sigma} \left[ \left( 4 \epsilon^{ \sigma \rho + - } \frac{x_1 x_2}{1-x_2} \frac{Q}{Z} K_1 (QZ) +  T_2^{\sigma \rho \nu} (x_1, x_2, x_3) \frac{z_{23 \perp \nu}}{\boldsymbol{z}_{23}^{2}} K_0 (QZ) \right) + \left( 1 \leftrightarrow 2 \right) \right] \bigg \} \; .
\label{Eq:GeneralPhotonMesonWaveFunOv}
\end{gather}
In eq.~(\ref{Eq:GeneralPhotonMesonWaveFunOv}) the $\chi$ functions are Fourier transform of the 3-body vacuum-to-meson matrix elements, i.e.
\begin{gather}
    \chi_{\gamma^+ , \sigma } \equiv \chi_{\gamma^+ , \sigma}  (x_1, x_2, x_3, \boldsymbol{z}_1, \boldsymbol{z}_2, \boldsymbol{z}_3 ) \nonumber \\ 
   = \hspace{-0.1 cm} \int \hspace{-0.1 cm} \frac{ {\rm d} z_{1}^{-}}{2 \pi} \frac{ {\rm d} z_{2}^{-}}{2 \pi} \frac{ {\rm d} z_{3}^{-}}{2 \pi} {\rm e}^{-ix_{1}q^{+}z_{1}^{-}-ix_{2}q^{+}z_{2}^{-}-ix_{3}q^{+}z_{3}^{-}}  \hspace{-0.15 cm} \left\langle M\left(p_{M}\right)\left|\overline{\psi}\left(z_{1}\right) g \gamma^+ F_{-\sigma}\left(z_{3}\right)\psi\left(z_{2}\right)\right|0\right\rangle _{z_{1,2,3}^{+}=0} ,
   \label{Eq:Chi+_Vector_function}
\end{gather}
\begin{gather}
    \chi_{\gamma^+ \gamma^5 , \sigma} \equiv \chi_{\gamma^+ \gamma^5 , \sigma}  (x_1, x_2, x_3, \boldsymbol{z}_1, \boldsymbol{z}_2, \boldsymbol{z}_3 ) \nonumber \\ \hspace{-0.1 cm}
   = \hspace{-0.1 cm} \int \hspace{-0.1 cm} \frac{ {\rm d} z_{1}^{-}}{2 \pi} \frac{ {\rm d} z_{2}^{-}}{2 \pi} \frac{ {\rm d} z_{3}^{-}}{2 \pi} {\rm e}^{-ix_{1}q^{+}z_{1}^{-}-ix_{2}q^{+}z_{2}^{-}-ix_{3}q^{+}z_{3}^{-}} \hspace{-0.1 cm} \left\langle M\left(p_{M}\right)\left|\overline{\psi}\left(z_{1}\right) g \gamma^+ \gamma^5 F_{-\sigma}\left(z_{3}\right)\psi\left(z_{2}\right)\right|0\right\rangle _{z_{1,2,3}^{+}=0} ,
   \label{Eq:Chi+5_Axial_function}
\end{gather}
and the tensor structures, $T_i^{\sigma \rho \nu}$, are given by
\begin{gather}
   T_1^{\sigma \rho \nu} (x_1, x_2, x_3) = {\rm tr}_{D} \left[ \left(2q^{\rho}-i\frac{x_{1}\gamma_{\perp}^{\rho}\gamma_{\perp}^{\alpha}-\overline{x}_{1}\gamma_{\perp}^{\alpha}\gamma_{\perp}^{\rho}}{x_{1}\overline{x}_{1}}\frac{\partial}{\partial z_{1\perp}^{\alpha}}\right)\left(2\frac{x_{1}x_{2}}{x_{3}}g_{\perp }^{\sigma\nu}+x_{1}\gamma_{\perp}^{\sigma}\gamma_{\perp}^{\nu}\right)\gamma^{+} \gamma^{-} \right] \nonumber \\
   \hspace{-0.15 cm} = 4 \left[ 2 \frac{x_1 (\bar{x}_1 + x_2)}{x_3} g_{\perp }^{\sigma \nu} q^{\rho} + \left( \frac{(\bar{x}_1 + x_2) (\bar{x}_1 - x_1)}{\bar{x}_1 x_3} g_{\perp}^{\sigma \nu} g_{\perp}^{\rho \mu} - \frac{1}{\bar{x}_1} \left(  g_{\perp}^{\nu \rho} g_{\perp}^{\sigma \mu} - g_{\perp}^{\rho \sigma} g_{\perp}^{\nu \mu} \right) \right) i \partial_{z_{1\perp} \mu}  \right] \nonumber \\ = 4\left[2x_{1}q^{\rho}\left(2\frac{x_{2}}{x_{3}}+1\right)g_{\perp}^{\sigma\nu}-\left(2\frac{x_{1}-x_{2}}{x_{3}}g_{\perp}^{\sigma\nu}g_{\perp}^{\rho\mu}-\frac{g_{\perp}^{\mu\nu}g_{\perp}^{\rho\sigma}+g_{\perp}^{\sigma\nu}g_{\perp}^{\rho\mu}-g_{\perp}^{\mu\sigma}g_{\perp}^{\rho\nu}}{x_{2}+x_{3}}\right)i \partial_{z_{1\perp} \mu} \right] \; ,
\end{gather} 
\begin{gather}
    T_2^{\sigma \rho \nu} (x_1, x_2, x_3) = {\rm tr}_{D} \hspace{-0.1 cm} \left[ \left(2q^{\rho}-i\frac{x_{1}\gamma_{\perp}^{\rho}\gamma_{\perp}^{\alpha}-\overline{x}_{1}\gamma_{\perp}^{\alpha}\gamma_{\perp}^{\rho}}{x_{1}\overline{x}_{1}}\frac{\partial}{\partial z_{1\perp}^{\alpha}}\right) \hspace{-0.1 cm} \left(2\frac{x_{1}x_{2}}{x_{3}}g_{\perp }^{\sigma\nu}+x_{1}\gamma_{\perp}^{\sigma}\gamma_{\perp}^{\nu}\right)\gamma^{+} \gamma^{-} \gamma^{5} \right] \nonumber \\
   \hspace{-0.2 cm} = \frac{ 4 i }{ \bar{x}_1 } \left[ 2 x_1 \bar{x}_1 q^{\rho} \epsilon^{ \nu \sigma + -} - \left( \left( 1 + \frac{2 x_2}{x_3} \right) \left( g_{\perp}^{ \sigma \mu } \epsilon^{\nu \rho + -} -  g_{\perp}^{ \rho \sigma } \epsilon^{\nu \mu + -}  \right)  + (x_1 - \bar{x}_1) g_{\perp}^{ \rho \mu }  \epsilon^{\nu \sigma + -} \right ) i \partial_{z_{1\perp} \mu}  \right ] \; 
\end{gather}
and the $(1 \leftrightarrow 2)$ denotes the following exchange
\begin{equation}
    \left( x_1 , \boldsymbol{z}_1 \right) \leftrightarrow \left( x_2 , \boldsymbol{z}_2 \right) \; .
\end{equation}

\subsection{Extraction of the impact factors: momentum space}
We will now extract the impact factors for the $\gamma^{(*)} \rightarrow M (\rho, \varphi, \omega)$ transition in momentum space.
\subsubsection{2-body contribution}
The wave function overlap is related to the momentum space impact factor by Fourier transform, i.e.
\begin{equation}
   \Psi_2 \left(x, \boldsymbol{r} \right) = \int \frac{d^2 \boldsymbol{p}}{(2 \pi)^2}  {\rm e}^{i \boldsymbol{p} \cdot\boldsymbol{r}} \Phi_2 \left(x, \boldsymbol{p} \right) \hspace{0.3 cm} \implies \hspace{0.3 cm} \Phi_2 \left(x, \boldsymbol{p} \right) = \int d^2 \boldsymbol{r} \; {\rm e}^{-i \boldsymbol{p} \cdot\boldsymbol{r}} \Psi_2 \left(x, \boldsymbol{r} \right) \; ,
\end{equation}
Introducing the Fourier transforms of $\phi_{\Gamma^{\lambda}} (x, \boldsymbol{r})$,
\begin{equation}
    \tilde{\phi}_{\gamma^+} (x, \boldsymbol{l}) = \int d^2 \boldsymbol{r} \; {\rm e}^{- i \boldsymbol{l} \cdot\boldsymbol{r}} \phi_{\gamma^+} (x, \boldsymbol{r})  \; ,
\end{equation}
\begin{equation}
   \tilde{\phi}_{\gamma^+ \gamma^5} (x, \boldsymbol{l}) = \int d^2 \boldsymbol{r} \; {\rm e}^{- i \boldsymbol{l} \cdot\boldsymbol{r}}  \phi_{\gamma^+ \gamma^5} (x, \boldsymbol{r}) \; ,
\end{equation}
At this point, the vacuum-to-meson matrix elements with full dependence on both longitudinal and transverse momenta are not explicitly QCD gauge invariant. Indeed, the Wilson line structure which should connect the $\bar{\psi}$ and $\psi$ operator is then 3-dimensional so in general it cannot be gauged away. In a lot of studies of vector meson production, it is assumed that light cone gauge with the right subgauge condition allows to neglect this gauge link structure, which then means we can directly relate these matrix elements to the Fock state expansion of the light front wave function of the meson. However a more rigorous study of such objects, such as the one for so-called Transverse Momentum Distribution Amplitudes (TMDA) in ref.~\cite{Radyushkin:2015gpa}, may be required. Note that our purpose is to proceed to a twist expansion, so an approach for these tridimensional matrix elements is not required. \\

\noindent Using the integral~(\ref{Eq:FourTrasfBessel1}), we get the momentum space impact factor
\begin{gather}
    \Phi_2 \left(x, \boldsymbol{p} \right) = 2 \pi e_q \delta \left( 1 - \frac{p_M^+}{q^+} \right) \left(\varepsilon_{q\mu}-\frac{\varepsilon_{q}^{+}}{q^{+}}q_{\mu} \right) \int \frac{d^2 \boldsymbol{l}}{(2 \pi)^2} \frac{1}{ (\boldsymbol{p} - \boldsymbol{l} \; )^2 + x \bar{x} Q^2 } \nonumber \\
    \times \left[ \left( 2 x \bar{x} q^{\mu} - (x-\bar{x}) (p-l)_{\perp}^{\mu} \right)  \tilde{\phi}_{\gamma^+} (x, \boldsymbol{l}) - i \epsilon^{\mu\nu+-} (p-l)_{\perp \nu} \; \tilde{\phi}_{\gamma^+ \gamma^5} (x, \boldsymbol{l}) \right] \; .
    \label{Eq:2BodyMomSpaceImpactFact}
\end{gather}
Introducing the Fourier transform of the dipole operator
\begin{equation}
   \mathcal{U}_{\boldsymbol{z}_1 \boldsymbol{z}_2} = \int \frac{d^2 \boldsymbol{k}_1}{(2 \pi)^2} \int \frac{d^2 \boldsymbol{k}_2}{(2 \pi)^2} {\rm e}^{i \boldsymbol{k}_1 \cdot\boldsymbol{z}_1} {\rm e}^{i \boldsymbol{k}_2 \cdot\boldsymbol{z}_2} \tilde{\mathcal{U}}_{\boldsymbol{k}_1 \boldsymbol{k}_2} \; ,
\end{equation}
it is easy to see that $\Phi_2$ enters the $\boldsymbol{k}_T$-factorization formula
\begin{equation}
   \mathcal{A}_2 = \int_0^1 dx \int \frac{d^2 \boldsymbol{k}_1}{(2 \pi)^2} \langle P (p') | \tilde{\mathcal{U}}_{\boldsymbol{k}_1 - x (\boldsymbol{q}-\boldsymbol{p}_M) , -\boldsymbol{k}_1 - \bar{x} (\boldsymbol{q}-\boldsymbol{p}_M) } | P (p) \rangle \Phi_2 \left(x, -\boldsymbol{k}_1 \right) \; .
    \label{Eq:TwoBodyKTFactForm}
\end{equation}

\subsubsection{3-body contribution}
The momentum space impact factor is related to the three body wave functions overlap by Fourier transform, i.e.
\begin{equation}
    \Phi_3 \left( \{ x \} , \{ \boldsymbol{p} \} \right) = \left( \prod_{j=1}^{3} \int d^2 \boldsymbol{z}_j  e^{- i \boldsymbol{z}_j \boldsymbol{p}_j } \right) \Psi_3 \left( \{ x \} , \{ \boldsymbol{z} \} \right) \; ,
\end{equation}
where we have introduced the shorthand notations $ \{ x \} = x_1, x_2, x_3$ and $ \{ \boldsymbol{a} \} = \boldsymbol{a}_1, \boldsymbol{a}_2, \boldsymbol{a}_3$ for the longitudinal and transverse vector variables. 
Introducing the three-body vacuum-to-meson matrix elements through the relation
\begin{equation}
    \chi_{\Gamma^{\lambda}, \sigma} \left( \{ x \} , \{ \boldsymbol{z} \} \right) = \left( \prod_{j=1}^{3} \int \frac{ d^2 \boldsymbol{k}_j}{(2 \pi)^2}  e^{i \boldsymbol{z}_j \boldsymbol{k}_j } \right) \tilde{\chi}_{\Gamma^{\lambda}, \sigma} \left( \{ x \} , \{ \boldsymbol{k} \} \right) \; \equiv \left( \prod_{j=1}^{3} \int \frac{ d^2 \boldsymbol{k}_j}{(2 \pi)^2}  e^{i \boldsymbol{z}_j \boldsymbol{k}_j } \right) \tilde{\chi}_{\Gamma^{\lambda}, \sigma} 
\end{equation}
and armed with the integrals in the appendix~\ref{Sec:AppedixD}, we find
\begin{gather}
    \Phi_3 \left( \{ x \} , \{ \boldsymbol{p} \} \right) = \frac{e_{q} q^{+}}{2 (4\pi)} c_f   \left(\varepsilon_{q\rho}-\frac{\varepsilon_{q}^{+}}{q^{+}}q_{\rho}\right) \left( \prod_{j=1}^{3} \int \frac{ d^2 \boldsymbol{k}_j}{(2 \pi)^2 x_j} \right) \frac{ \displaystyle (2 \pi)^4 \delta^{(2)} \left( \sum_{i=1}^{3} \boldsymbol{k}_i - \boldsymbol{p}_i \right)}{ \left[  Q^2 + \sum_{i=1}^3 (\boldsymbol{k}_i - \boldsymbol{p}_i)^2 / x_i \right]} \nonumber \\ 
    \times \Bigg \{ \tilde{\chi}_{\gamma^{+}, \sigma}  \left[ \left( 4 i g^{\rho \sigma}_{\perp} \frac{x_1 x_2}{1-x_2} + i \tilde{T}_1^{\sigma \rho \nu} ( \{ x \} ) x_1 \frac{ x_3 ( k_2 - p_2)_{\perp \nu} - x_2 ( k_3 - p_3)_{\perp \nu} }{(\boldsymbol{k}_1 - \boldsymbol{p}_1)^2 + x_1 (1-x_1) Q^2 } \right)  - \left( 1 \leftrightarrow 2 \right) \right] \nonumber \\
    -\tilde{\chi}_{\gamma^{+} \gamma^5, \sigma}  \left[ \left( 4 \epsilon^{\sigma \rho + -} \frac{x_1 x_2}{1-x_2} + i \tilde{T}_2^{\sigma \rho \nu} ( \{ x \} ) x_1 \frac{x_3 ( k_2 - p_2)_{\perp \nu} - x_2 ( k_3 - p_3)_{\perp \nu} }{(\boldsymbol{k}_1 - \boldsymbol{p}_1)^2 + x_1 (1-x_1) Q^2} \right)  + \left( 1 \leftrightarrow 2 \right) \right] \; ,
\label{Eq:ImpactFactorThreeBodyMom}
\end{gather}
where
\begin{equation}
    c_f = \frac{N_c^2}{N_c^2 - 1} \; ,
\end{equation}
\begin{gather}
    \tilde{T}_1^{\sigma \rho \nu} ( \{ x \} ) = 4 \left[ 2 \frac{x_1 (\bar{x}_1 + x_2)}{x_3} g_{\perp}^{\sigma \nu} q^{\rho}  - ( k_1 - p_1 )_{ \perp \mu}  \right. \nonumber \\ \left. \times \left( \frac{(\bar{x}_1 + x_2) (\bar{x}_1 - x_1)}{\bar{x}_1 x_3} g_{\perp}^{\sigma \nu} g_{\perp}^{\rho \mu} - \frac{1}{\bar{x}_1} \left(  g_{\perp}^{\nu \rho} g_{\perp}^{\sigma \mu} - g_{\perp}^{\rho \sigma} g_{\perp}^{\nu \mu} \right) \right)   \right] \; ,
    \label{Eq:T1Mom}
\end{gather}
\begin{gather}
    \tilde{T}_2^{\sigma \rho \nu} ( \{ x \} ) = \frac{ 4 i }{ \bar{x}_1 } \bigg[ 2 x_1 \bar{x}_1 q^{\rho} \epsilon^{ \nu \sigma + -} + ( k_1 - p_1 )_{ \perp \mu} \nonumber \\ \left. \times \left( \left( 1 + \frac{2 x_2}{x_3} \right) \left( g_{\perp}^{ \sigma \mu } \epsilon^{\nu \rho + -} -  g_{\perp}^{ \rho \sigma } \epsilon^{\nu \mu + -}  \right)  + (x_1 - \bar{x}_1) g_{\perp}^{ \rho \mu }  \epsilon^{\nu \sigma + -} \right )  \right ] \; ,
    \label{Eq:T2Mom}
\end{gather}
and, in eq.~(\ref{Eq:ImpactFactorThreeBodyMom}), $(1 \leftrightarrow 2)$ should be understood as
\begin{equation}
    \left( x_1 , \boldsymbol{p}_1 \right) \leftrightarrow \left( x_2 , \boldsymbol{p}_2 \right) \; .
\end{equation}
The impact factor $\Phi_3 \left( \{ x \} , \{ \boldsymbol{p} \} \right)$ enters the $\boldsymbol{k}_T$-factorization formula
\begin{gather}
  \mathcal{A}_3 = \left( \prod_{j=1}^{3} \int d x_j \theta (x_j) \int \frac{ d^2 \boldsymbol{k}_j}{(2 \pi)^2} \right) (2 \pi)^2 \delta (1-x_1-x_2-x_3) \Phi_3 \left( \{ x \} , \{ - \boldsymbol{k} - x \boldsymbol{q} \} \right) \nonumber \\
  \times \left\langle P\left(p^{\prime}\right) \left| \frac{ \widetilde{\mathcal{U}_{\boldsymbol{k}_1 \boldsymbol{k}_3} \mathcal{U}_{\boldsymbol{k}_3 \boldsymbol{k}_2}} }{(2 \pi)^2} - \delta^{(2)} (\boldsymbol{k}_2) \tilde{\mathcal{U}}_{\boldsymbol{k}_1 \boldsymbol{k}_3} - \delta^{(2)} (\boldsymbol{k}_1)\tilde{\mathcal{U}}_{\boldsymbol{k}_3 \boldsymbol{k}_2} + \frac{\delta^{(2)} (\boldsymbol{k}_3) }{N_c^2} \tilde{\mathcal{U}}_{\boldsymbol{k}_1 \boldsymbol{k}_2}  \right|P\left(p\right)\right\rangle \; ,
  \label{Eq:ThreeBodyKTFactForm}
\end{gather}
where the Fourier transform of the double dipole operator is introduced through the relation
\begin{equation}
   \mathcal{U}_{\boldsymbol{z}_1 \boldsymbol{z}_3} \mathcal{U}_{\boldsymbol{z}_3 \boldsymbol{z}_2} = \int \frac{d^2 \boldsymbol{k}_1 }{(2 \pi)^2} \frac{d^2 \boldsymbol{k}_2 }{(2 \pi)^2} 
    \frac{d^2 \boldsymbol{k}_3 }{(2 \pi)^2}  e^{i \boldsymbol{z}_1 \boldsymbol{k}_1 } e^{i \boldsymbol{z}_2 \boldsymbol{k}_2 } e^{i \boldsymbol{z}_3 \boldsymbol{k}_3 } \widetilde{\mathcal{U}_{\boldsymbol{k}_1 \boldsymbol{k}_3} \mathcal{U}_{\boldsymbol{k}_3 \boldsymbol{k}_2}} \; .
\end{equation}

\section{Collinear factorization for $\gamma^{*} \;P \rightarrow M (\rho, \varphi, \omega) \; P$ beyond the leading twist}
\label{Sec:CovColl}

\begin{figure} \includegraphics[scale=0.50]{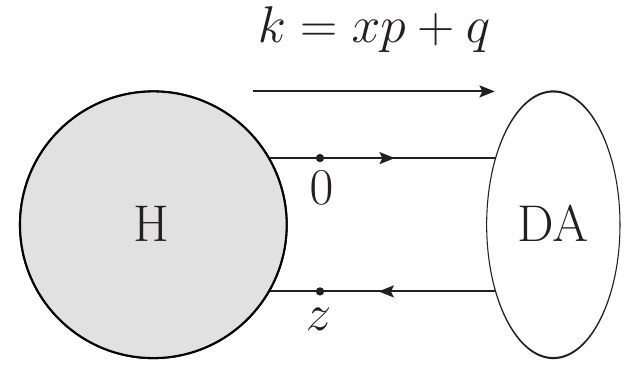} \hspace{2 cm}
\includegraphics[scale=0.50]{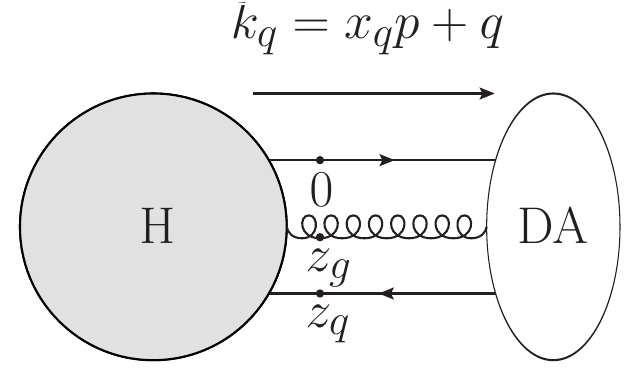}
\caption{On the left the 2-body contribution to the EDMP. On the right the 3-body contribution to the EDMP.}
  \label{fig:GenericAmp}
\end{figure}
So far, no kinematic twist expansion has been taken. The purpose of this section is to detail how to perform it for any process involving the production of a light vector meson. Since in this paper we deal with EDMP up to twist 3, it is sufficient to consider 2- and 3-body contributions as illustrated in fig.~\ref{fig:GenericAmp}. The 2-body contribution to the amplitude for the exclusive production of a light meson in collinear factorization is given by a convolution of the hard part for the production of a quark-antiquark pair $H_{\alpha \beta}^{ij}$ and the non-perturbative vacuum-to-meson matrix element $\langle M (p_M) | \overline{\psi}^{\; i}_{\alpha} \left(z\right) \psi_{\beta}^{\; j} \left(0\right) | 0 \rangle$ (see fig.~\ref{fig:GenericAmp} (a)), i.e.
\begin{equation}
    {\cal A}_{2} = \int \frac{{\rm d}^4 k}{(2 \pi)^4} \int {\rm d}^4 z \, {\rm e}^{- i (k \cdot z)} \langle M (p_M) |\, \overline{\psi}^{\; i}_{\alpha} (z) \psi_{\beta}^{\; j} (0) | 0 \rangle H_{2, \alpha \beta}^{ij}(k) \; .
\end{equation}
By using Fierz identity in both Dirac and color space, (\ref{Eq:FierzIdentityDirac}) and (\ref{Eq:FierzIdentityColor}) respectively, one can immediately factorize the amplitude as   
\begin{equation}
{\cal A}_{2}=\frac{1}{4N_{c}}\int\frac{{\rm d}^{4}k}{(2\pi)^{4}}\int{\rm d}^{4}z\,{\rm e}^{-i(k\cdot z)}\left\langle M(p_M)\left|\, \overline{\psi}(z)\Gamma_{\lambda}\psi(0)\right|0\right\rangle {\rm tr}\left[H_{2}(k)\Gamma^{\lambda}\right] \; ,
\label{Eq:A2AfterFierz}
\end{equation}
where we denote with the label tr the trace over Dirac and color indices. When needed we will distinguish the two types of trace, using ${\rm tr}_D$ for the trace over Dirac indices and ${\rm tr }_c$ for the one over color indices. \\

When going at higher twists, new contributions appear with increasing number of legs (both quarks or gluons carrying physical polarization) connecting the hard and the non-perturbative part of the amplitude, the so called genuine twist contributions. Here, we deal with the next-to-leading twist (twist-3); in this case, it is therefore necessary to consider also the 3-body contribution, represented in (b) of fig.~\ref{fig:GenericAmp}, i.e.
\begin{align}
{\cal A}_{3} & = \int\frac{{\rm d}^{4}k_{q}}{(2\pi)^{4}}\frac{{\rm d}^{4}k_{g}}{(2\pi)^{4}}\int{\rm d}^{4}z_{q}\,{\rm d}^{4}z_{g}\,{\rm e}^{-i\left(k_{q}\cdot z_{q}\right)-i(k_{g}\cdot z_{g})} \nonumber \\
 & \times\left\langle M(p_M)\left|\,\overline{\psi}_{\alpha}^{\; i} (z_{q}) \Gamma_{\lambda} \; g A_{\mu}^{a} (z_{g}) \psi_{\beta}^{\; j} (0)\right|0\right\rangle {\rm tr}\left[H_{3, \alpha \beta \mu}^{ij a} (k_{q},k_{g})\Gamma^{\lambda}\right] \; .
\end{align}
Using Fierz decompositions, the amplitude can be expressed as 
\begin{align}
{\cal A}_{3} & =\frac{1}{2 (N_c^2-1)}\int\frac{{\rm d}^{4}k_{q}}{\left(2\pi\right)^{4}}\frac{{\rm d}^{4}k_{g}}{\left(2\pi\right)^{4}}\int{\rm d}^{4}z_{q}\,{\rm d}^{4}z_{g}\,{\rm e}^{-i(k_{q}\cdot z_{q})-i(k_{g}\cdot z_{g})} \nonumber \\
 & \times\left\langle M(p_M)\left|\,\overline{\psi}(z_{q})\Gamma_{\lambda}gA_{\mu}(z_{g})\psi(0)\right|0\right\rangle {\rm tr}\left[t^b H_{3}^{\mu , b} (k_{q},k_{g})\Gamma^{\lambda}\right] \; .
 \label{Eq:A3AfterFierz}
\end{align}
Within twist-3 accuracy, any meson mass corrections to the amplitude are neglected and therefore we take $p_M^{2}=0$. In this section, we present the factorization of the EDMP up to twist-3 amplitudes in the Covariant Collinear Factorization (CCF) scheme~\cite{Ball:1998sk}. In appendix \ref{LCCF_and_comparison_with_CCF}, we present an alternative approach, Light Cone Collinear Factorization (LCCF)~\cite{Anikin:2002wg,Anikin:2009hk,Anikin:2009bf}. 

\subsection{Covariant collinear factorization and distribution amplitudes}
Let us consider eqs.~(\ref{Eq:A2AfterFierz}) and (\ref{Eq:A3AfterFierz}), i.e. 
\begin{equation*}
{\cal A}_{2}=\frac{1}{4N_{c}}\int\frac{{\rm d}^{4}k}{\left(2\pi\right)^{4}}\int{\rm d}^{4}z\,{\rm e}^{-i(k\cdot z)}\left\langle M(p)\left| \overline{\psi}(z)\Gamma_{\lambda}\psi(0)\right|0\right\rangle {\rm tr}\left[H_{2}(k)\Gamma^{\lambda}\right] \; ,
\end{equation*}
\begin{align*}
{\cal A}_{3} & =\frac{1}{2 (N_c^2-1)}\int\frac{{\rm d}^{4}k_{q}}{\left(2\pi\right)^{4}}\frac{{\rm d}^{4}k_{g}}{\left(2\pi\right)^{4}}\int{\rm d}^{4}z_{q}\,{\rm d}^{4}z_{g}\,{\rm e}^{-i(k_{q}\cdot z_{q})-i(k_{g}\cdot z_{g})}  \\
 & \times\left\langle M(p_M)\left|\overline{\psi}(z_{q})\Gamma_{\lambda}gA_{\mu}(z_{g})\psi(0)\right|0\right\rangle {\rm tr}\left[t^b H_{3}^{\mu , b} (k_{q},k_{g})\Gamma^{\lambda}\right] \; .
\end{align*}
In the CCF approach, a twist-expansion at operator level is performed. Within the twist-3 accuracy, we need 2- and 3-body operators in the following gauge invariant form:
\begin{gather}
    \overline{\psi} (z) \Gamma_{\lambda} [ z, 0 ] \psi (0) \; , \label{2BodyGaugeCCF} \\
    \overline{\psi} (z) \gamma_{\lambda} [ z, tz  ] F^{\mu \nu} (tz) [ tz, 0  ] \psi (0) \; ,
    \label{3BodyGaugeCCF1} \\
     \overline{\psi} (z) \gamma_{\lambda} \gamma^5 [ z, tz  ] \tilde{F}^{\mu \nu} (tz) [ tz, 0  ] \psi (0) \; ,
    \label{3BodyGaugeCCF2}
\end{gather}
where 
\begin{equation}
    F_{\mu \nu}^a (z) = \partial_{\mu} A_{\nu}^a (z) - \partial_{\nu} A_{\mu}^a (z) - g f^{abc} A_{\mu}^b (z) A_{\nu}^{b} (z) 
\end{equation}
is the field strenght tensor,
\begin{equation}
    F^{a \mu \nu} (z) = \frac{1}{2} \varepsilon^{\mu \nu \alpha \beta} \tilde{F}^a_{\alpha \beta} (z) 
\end{equation}
its dual and the gauge link reads
\begin{equation}
    [z,0] = \mathcal{P} \, {\rm exp} \left[ \, i g \int_0^1 dt A^{\mu} \left( tz \right) z_{\mu} \right] \; .
\end{equation}
The twist expansion is performed by taking the $z^2 \rightarrow 0$ limit of the above correlators, which amounts to perform the operator product expansion (OPE). This latter is done in the spirit of Balitsky and Braun~\cite{Balitsky:1987bk}, i.e. the off light-cone correlators are expanded in powers of the deviation from the light-cone $z^2  = 0$ and each coefficient of this expansion is a finite sum of on light-cone non-local correlators\footnote{This is different from the more standard Wilson OPE, in which each coefficient of the $r^2$ expansion is a series of local operators.}. In our case, the leading term in this expansion is represented by the initial correlator evaluated for light-like separation. The subleading contributions to the operator product expansion (OPE) are at least of twist-4~\cite{Ball:1998sk} and can be ignored for our purposes. Although the non-local correlators are considered on the light-cone, within twist-3 accuracy, the parametrization of the vacuum-to-meson non-perturbative matrix element introduces contributions of different kinematic twists. Indeed, the matrix element is a linear combination of the available four-vectors: $p_{M \mu}, z_{\mu}$ and $\varepsilon_{M \mu}^{*}$ (the latter is the meson polarization vector), with some coefficients depending on the available Lorentz invariants, $p_M \cdot z, \varepsilon_M^{*} \cdot z$ and $m_M^2$\footnote{We stress that at this point $z^2=0$.}. These quantities have different scaling in the $ Q \rightarrow \infty$ limit and this leads to contributions of different  kinematic twist. \\

In the parametrization of our amplitudes, the correlators do not appear in explicitly gauge invariant form (i.e. they appear without the gauge link). It will be necessary to estimate the contribution of the link to subtract it from the gauge invariant 2-body contribution to get the right parametrization for our 2-body correlators. Our goal will therefore be to start from the parametrization in terms of the explicitly gauge invariant matrix elements and deduce a parametrization for the matrix elements appearing in eqs.~(\ref{Eq:A2AfterFierz}) and (\ref{Eq:A3AfterFierz}). \\

Let us first consider the parametrization of the 3-body contributions, since, as we will see, their parametrizations will be necessary to connect the matrix elements with and without the gauge link in the 2-body contributions. We stress that the formulas given below, are obtained from~\cite{Ball:1998sk}, where the kinematic twist expansion of the correlator is performed to obtain the explicit twist-3 parametrization. 

\subsection{Explicit parametrizations up to twist-3}

\subsubsection{3-body vector matrix element} 
The parametrization of the 3-body vector matrix element reads
\begin{gather}
 \left\langle M(p_M)\left| \overline{\psi}(z_{q})[z_{q},z_{g}]\gamma^{\lambda}gF^{\mu\nu}(z_{g})[z_{g},z_{\overline{q}}]\psi(z_{\overline{q}})\right|0\right\rangle
  = \hspace{-0.1 cm}i \; m_{M} f_{3M}^{V} p_M^{\lambda}(p_M^{\nu}\varepsilon_M^{\ast\mu}-p_M^{\mu}\varepsilon_M^{\ast\nu}) \nonumber \\ \times \int \hspace{-0.1 cm} {\rm d} x_{q}\,{\rm d}x_{\overline{q}}\,{\rm d}x_{g} \, \delta(1-x_{q}-x_{\overline{q}}-x_{g}){\rm e}^{ip_M \cdot(x_{q}z_{q}+x_{\overline{q}}z_{\overline{q}}+x_{g}z_{g})}V(x_{q},x_{\overline{q}}).  
\end{gather}
Within our twist 3 accuracy the gauge links can be expanded at the first order, i.e. 
\begin{gather}
 \left\langle M(p_M)\left|\, \overline{\psi}\left(z_{q}\right)\gamma^{\lambda}gF^{\mu\nu}\left(z_{g}\right)\psi\left(z_{\overline{q}}\right)\right|0\right\rangle
 \nonumber \\ \simeq\left\langle M(p_M)\left| \overline{\psi}(z_{q})[z_{q},z_{g}]\gamma^{\lambda}gF^{\mu\nu}(z_{g})[z_{g},z_{\overline{q}}]\psi(z_{\overline{q}})\right|0\right\rangle 
  = \hspace{-0.1 cm}i \; m_{M} f_{3M}^{V} p_M^{\lambda}(p_M^{\nu}\varepsilon_M^{\ast\mu}-p_M^{\mu}\varepsilon_M^{\ast\nu}) \nonumber \\ \times \int \hspace{-0.1 cm} {\rm d} x_{q}\,{\rm d}x_{\overline{q}}\,{\rm d}x_{g} \, \delta(1-x_{q}-x_{\overline{q}}-x_{g}){\rm e}^{ip_M\cdot(x_{q}z_{q}+x_{\overline{q}}z_{\overline{q}}+x_{g}z_{g})}V(x_{q},x_{\overline{q}}).  
  \label{Eq:3_Body_Vector_Matrix_Twist_Exp}
\end{gather}
Commonly, the amplitude requires the parametrization of the matrix
element with $A^{\mu}$ rather than the field strength tensor. We can relate the two matrix elements by exploiting the fact that, in a given $n$ light-cone gauge,
\begin{equation}
A^{\mu}(z)=\int_{0}^{\infty}{\rm d} \sigma \; {\rm e}^{-\epsilon\sigma}n_{\nu}F^{\mu\nu}(z+\sigma n).
\end{equation}
Then
\begin{gather}
  \left\langle M(p_M)\left|\, \overline{\psi}\left(z_{q}\right)\gamma^{\lambda}gA^{\mu}\left(z_{g}\right)\psi\left(z_{\overline{q}}\right)\right|0\right\rangle\nonumber \\ = \int_{0}^{\infty} {\rm d} \sigma \, {\rm e}^{-\epsilon\sigma} n_{\nu}\left\langle M(p_M)\left| \overline{\psi}\left(z_{q}\right)\gamma^{\lambda}gF^{\mu\nu}\left(z_{g}+\sigma n\right)\psi\left(z_{\overline{q}}\right)\right|0\right\rangle  =m_{M}f_{3M}^{V}p_M^{\lambda} \nonumber \\ \times \frac{(p_M^{\mu}\varepsilon_M^{\ast} -\varepsilon_M^{\ast\mu}p_M) \cdot n}{(p_M\cdot n)} \int\frac{{\rm d}x_{q}\,{\rm d}x_{\overline{q}}\,{\rm d}x_{g}}{x_{g}+i\epsilon} \delta(1-x_{q}-x_{\overline{q}}-x_{g}){\rm e}^{ip_M \cdot(x_{q}z_{q}+x_{\overline{q}}z_{\overline{q}}+x_{g}z_{g})}V(x_{q},x_{\overline{q}}).  
\end{gather}

\subsubsection{3-body axial-vector matrix element} 
Similarly, the 3-body axial-vector correlator can be parametrized as
\begin{gather}
  \left\langle M(p_M)\left|\,\overline{\psi} \left( z_{q} \right)\gamma^{\lambda} \gamma^5 [z_{q}, z_{g}] g \tilde{F}^{\mu\nu} \left( z_{g} \right)[z_{g}, z_{\bar{q}}] \psi\left( z_{\bar{q}} \right) \right|0\right\rangle = -p_M^{\lambda}(p_M^{\mu}\varepsilon_M^{\ast\nu}-p_M^{\nu}\varepsilon_M^{\ast\mu})m_{M}f_{3M}^{A} \nonumber \\ \times \int{\rm d}x_{q}\,{\rm d}x_{\overline{q}}\,{\rm d}x_{g}\,\delta(1-x_{q}-x_{\overline{q}}-x_{g}) A(x_{q},x_{\overline{q}})\,{\rm e}^{ip_M\cdot(x_{q}z_{q}+x_{\bar{q}} z_{\bar{q}}+x_{g}z_{g})}. 
   \label{Eq:3_Body_Axial_Matrix_Twist_Exp}
\end{gather}
From this, it is easy to see that the matrix element containing the field $A^{\mu} (z)$ reads
\begin{gather}
 \left\langle M(p_M)\left|\,\overline{\psi}\left(z_{q}\right)\gamma^{\lambda}\gamma_{5}gA^{\mu}\left(z_{g}\right)\psi\left(z_{\overline{q}}\right)\right|0\right\rangle 
  =-ip_M^{\lambda}\epsilon^{\mu\nu\alpha\beta}\frac{p_{M \alpha}n_{\nu}}{p_M \cdot n}\varepsilon_{M\beta}^{\ast}m_{M}f_{3M}^{A} \nonumber \\ \times \int\frac{{\rm d}x_{q}\,{\rm d}x_{\overline{q}}\,{\rm d}x_{g}}{x_{g}+i\epsilon} \delta(1-x_{q}-x_{\overline{q}}-x_{g})A(x_{q},x_{\overline{q}}){\rm e}^{ip_M \cdot(x_{q}z_{q}+x_{\overline{q}}z_{\overline{q}}+x_{g}z_{g})} . 
\end{gather}

\subsubsection{2-body vector matrix element} 
The parametrization of the gauge invariant 2-body vector correlator reads

\begin{align}
 & \left\langle M(p_M)\left|\,\overline{\psi}\left(z\right)\gamma^{\lambda}[z,0]\psi\left(0\right)\right|0\right\rangle \nonumber \\
 & =f_{M}m_{M}\int_{0}^{1}\!{\rm d}x\,{\rm e}^{ix(p_M \cdot z)}\left[\,p_M^{\lambda}\frac{(\varepsilon_M^{\ast}\cdot z)}{(p_M \cdot z)}\phi(x)+\varepsilon_{M, T}^{\ast\lambda}g_{\perp}^{(v)}(x)\right] \nonumber \\
 & =f_{M}m_{M}\int_{0}^{1}\!{\rm d}x\,{\rm e}^{ix(p_M \cdot z)}\left[\,p_M^{\lambda}\frac{(\varepsilon_M^{\ast}\cdot z)}{(p_M \cdot z)}\phi(x)+\left(\varepsilon_M^{\ast\lambda}-\frac{(\varepsilon_M^{\ast}\cdot z)}{(p_M\cdot z)}p_M^{\lambda}\right)g_{\perp}^{(v)}(x)\right]\nonumber \\
 & =f_{M}m_{M}\int_{0}^{1}\!{\rm d}x\,{\rm e}^{ix(p_M\cdot z)}\left[\,p_M^{\lambda}\frac{(\varepsilon_M^{\ast}\cdot z)}{(p_M\cdot z)}\left(\phi(x)-g_{\perp}^{(v)}(x)\right)+\varepsilon_M^{\ast\lambda}g_{\perp}^{(v)}(x)\right] \; ,
\end{align}
where by $\varepsilon_{M,T}$ we denote the components of the polarization vector transverse in the $(p, z)$ basis, which should not be confused with $\varepsilon_{M \perp}$ which is transverse in the $(p, n_2)$ basis. We now define the quantity
\begin{equation}
h(x)=\int_{0}^{x}{\rm d}u\left(\phi(u)-g_{\perp}^{(v)}(u)\right) \; ,
\end{equation}
which satisfies $h(0)=0$ and $h(1) = 0$ due to the normalization of the DAs. Performing an integration by parts, we get
\begin{gather}
  \left\langle M(p_M)\left|\overline{\psi}(z)\gamma^{\lambda}\left[z,0\right]\psi\left(0\right)\right|0\right\rangle 
  =f_{M}m_{M}\int_{0}^{1}\!{\rm d}x\,{\rm e}^{ix\left(p_M\cdot z\right)}\left[\,p_M^{\lambda}\frac{\left(\varepsilon_M^{\ast}\cdot z\right)}{\left(p_M\cdot z\right)}\frac{{\rm d}h}{{\rm d}x}\left(x\right)+\varepsilon_M^{\ast\lambda}g_{\perp}^{\left(v\right)}\left(x\right)\right] \nonumber \\
  =f_{M}m_{M} \left[ p_{M}^{\lambda}\frac{\left(\varepsilon_M^{\ast}\cdot z\right)}{\left(p_M\cdot z\right)} \left({\rm e}^{i\left(p_M\cdot z\right)}h\left(1\right)-h\left(0\right)\right) \right. \nonumber \\
 \left. -   \int_{0}^{1}\!{\rm d}x\,p_M^{\lambda}\frac{(\varepsilon_M^{\ast}\cdot z)}{(p_M\cdot z)}h(x)\frac{\partial}{\partial x}{\rm e}^{ix\left(p_M\cdot z\right)} 
  + \int_{0}^{1}\!{\rm d}x\,{\rm e}^{ix\left(p_M\cdot z\right)}\varepsilon_M^{\ast\lambda}g_{\perp}^{(v)}(x) \right]
\end{gather}
and thus
\begin{gather}
  \left\langle M(p_M)\left|\,\overline{\psi}\left(z\right)\gamma^{\lambda}\left[z,0\right]\psi\left(0\right)\right|0\right\rangle \nonumber \\
  =f_{M}m_{M} \int_{0}^{1}\!{\rm d}x\,{\rm e}^{ix\left(p_M\cdot z\right)}\left[\,\varepsilon_M^{\ast\lambda}g_{\perp}^{(v)}(x)-ip_M^{\lambda}(\varepsilon_M^{\ast}\cdot z)h(x)\right]. 
\end{gather}
We want to obtain the matrix element without the gauge link $[z,0]$. We can expand the gauge link as
\begin{equation}
    [z,0] = \mathcal{P} \, {\rm exp} \left[ i g \int_0^1 dt A^{\mu} ( tz ) z_{\mu} \right] = 1 + ig \int_0^1 dt A^{\mu} \left( tz \right) z_{\mu} + {\rm higher \; twist }
\end{equation}
and find
\begin{gather}
  \left\langle M\left(p\right)\left|\,\overline{\psi}\left(z\right)\gamma^{\lambda}\psi\left(0\right)\right|0\right\rangle = f_{M} m_{M} \int_{0}^{1}\!{\rm d}x\,{\rm e}^{ix\left(p_M \cdot z\right)} \nonumber \\
 \times \left[\, \varepsilon_M^{\ast\lambda}g_{\perp}^{\left(v\right)}\left(x\right)-ip_M^{\lambda}\left(\varepsilon_M^{\ast}\cdot z\right)\left[\, h\left(x\right)-\widetilde{h}\left(x\right)\right] -ip_M^{\lambda}\frac{p_M \cdot z}{p_M \cdot n}\left(\varepsilon_M^{\ast}\cdot n\right)\widetilde{h}\left(x\right)\right] \; ,
\end{gather}
where 
\begin{equation}
  \widetilde{h}\left(x\right)=\frac{f_{3M}^{V}}{f_{M}}\int_{0}^{x}{\rm d}x_{q}\int_{0}^{1-x}\!{\rm d}x_{\overline{q}}\,\frac{V\left(x_{q},x_{\overline{q}}\right)}{\left(1-x_{q}-x_{\overline{q}}\right)^{2}} \; .  
\end{equation}
Through another integration by parts, we obtain
\begin{gather}
  \left\langle M\left(p\right)\left|\,\overline{\psi}\left(z\right)\gamma^{\lambda}\psi\left(0\right)\right|0\right\rangle \\
  =f_{M}m_{M}\!\!\int_{0}^{1}\!\!{\rm d}x\,{\rm e}^{ix\left(p_M\cdot z\right)}\!\left[\,\varepsilon_M^{\ast\lambda}g_{\perp}^{(v)}(x)-ip_M^{\lambda}\left(\varepsilon_M^{\ast}\cdot z\right)\left[ \, h\left(x\right)-\widetilde{h}\left(x\right)\right] +p_M^{\lambda}\left(\frac{\varepsilon_M^{\ast}\cdot n}{p_M \cdot n}\right)\frac{{\rm d}\widetilde{h}}{{\rm d}x}\left(x\right)\right] \! . \nonumber
\end{gather}
For the scope of our calculation, we will choose the $n_2$ light-cone gauge and have $z^+=0$ (see eqs.~(\ref{Phi_+_Distrib_Coordinate}, \ref{Phi_+5_Distrib_Coordinate})), because of the shockwave approximation. Then, the expression can be rewritten as
\begin{gather}
    \left\langle M\left(p\right)\left|\,\overline{\psi}\left(z\right)\gamma^{\lambda}\psi\left(0\right)\right|0\right\rangle \nonumber \\ = m_{M} f_{M} \int_{0}^{1}{\rm d}x\,{\rm e}^{ix\left(p_M \cdot z\right)} \left[ \varepsilon_{M}^{* \lambda} \phi (x) - i p_M^{\lambda} \left( \varepsilon_{M \mu}^{*} - \frac{\varepsilon_M^{*+}}{p_M^+} p_{M \mu} \right) z_{\perp}^{\mu} \left[ \, h\left(x\right)-\widetilde{h}\left(x\right)\right] \right] \; ,
    \label{Eq:VectorDAsNoGaugeLink}
\end{gather}
where we used the fact that, with $z^+=0$,
\begin{equation}
    \left( \varepsilon_{M \mu}^{*} - \frac{\varepsilon_M^{*+}}{p_M^+} p_{M \mu} \right) z^{\mu} = \left( \varepsilon_{M \mu}^{*} - \frac{\varepsilon_M^{*+}}{p_M^+} p_{M \mu} \right) z_{\perp}^{\mu} \; .
\end{equation}
In eq.~(\ref{Eq:VectorDAsNoGaugeLink}), the term proportional to $\phi (x)$ is the twist-2 contribution for the production of a longitudinally polarized meson, while the remaining part is the twist-3 contribution for the production of a transversely polarized meson. 

\subsubsection{2-body axial-vector matrix element} 
The parametrization of the gauge invariant 2-body axial-vector correlator reads
\begin{gather}
  \left\langle M(p_M)\left|\,\overline{\psi}\left(z\right)\left[z,0\right]\gamma^{\lambda}\gamma_{5}\psi\left(0\right)\right|0\right\rangle 
  =\frac{1}{4}f_{M}m_{M}\epsilon^{\lambda\alpha\beta\delta}\varepsilon_{M \alpha}^{\ast}p_{M \beta}z_{\delta}\int_{0}^{1}\!{\rm d}x\,{\rm e}^{ix\left(p_M \cdot z\right)}g_{\perp}^{(a)}(x) \; . 
\end{gather}
Proceeding the same way as for the vector case, one can get rid of the gauge link up to twist-3 accuracy, and obtain 
\begin{gather}
  \left\langle M(p_M)\left|\,\overline{\psi}\left(z\right)\gamma^{\lambda}\gamma_{5}\psi\left(0\right)\right|0\right\rangle \nonumber \\
  =f_{M}m_{M}\varepsilon_{M \alpha}^{\ast}p_{M \beta}z_{\delta}\int_{0}^{1}\!{\rm d}x\,{\rm e}^{ix\left(p_{M}\cdot z\right)}\frac{1}{4}\left[\epsilon^{\lambda\alpha\beta\delta}g_{\perp}^{(a)}(x)-\epsilon^{\nu\alpha\beta\delta}\frac{p_M^{\lambda}n_{\nu}}{p_M \cdot n}\widetilde{g}_{\perp}^{(a)}(x)\right] \; ,
  \label{Eq:AxialDAsNoGaugeLink}
\end{gather}
with
\begin{equation}
\widetilde{g}_{\perp}^{\left(a\right)}\left(x\right)=4\frac{f_{3M}^{A}}{f_{M}}\int_{0}^{x}{\rm d}x_{q}\int_{0}^{1-x}\!{\rm d}x_{\overline{q}}\,\frac{A\left(x_{q},x_{\overline{q}}\right)}{(1-x_{q}-x_{\overline{q}}+i\epsilon)^{2}} \; .
\end{equation}

\section{Impact factors at twist 3}
\label{Sec:impact-factor-twist3}

Equipped with the results of the two previous sections, we can perform the kinematic twist expansion of the wave function overlaps and of the momentum space impact factors. Our result also contains the LO contribution for the production of a longitudinally polarized meson, which is already known at NLO accuracy~\cite{Boussarie:2016bkq, Ivanov:2004pp, Mantysaari:2022bsp}. We will omit it in the following discussion.

\subsection{Kinematic twist expansion: impact factors with DAs in coordinate space}

\subsubsection{2-body contribution}
We perform the twist expansion of the wave function overlaps as described in section~\ref{Sec:CovColl}. Therefore, we start from eqs.~(\ref{Phi_+_Distrib_Coordinate},~\ref{Phi_+5_Distrib_Coordinate}), we substitute the vacuum-to-meson matrix elements by their explicit parameterization in eqs.~(\ref{Eq:VectorDAsNoGaugeLink},~\ref{Eq:AxialDAsNoGaugeLink}) and finally, we integrate over $r^-$. This amounts to performing the following replacements:
\begin{gather}
    \phi_{\gamma^+} (x, \boldsymbol{r}) \xrightarrow{\rm twist-expansion} - i \; m_M f_M \left( \varepsilon_{M \mu}^{*} - \frac{\varepsilon^{*+}_M}{p_M^+} p_{M \mu} \right) r_{\perp}^{\mu} (h(x)-\tilde{h}(x)) 
    \label{Eq:Twist_Expan_Corr_Vector_2_Body}
\end{gather}
and
\begin{gather}
    \phi_{\gamma^+ \gamma^5} (x, \boldsymbol{r}) \xrightarrow{\rm twist-expansion} m_M f_M \epsilon^{+ \alpha - \delta} \left( \varepsilon_{M \alpha}^{*} - \frac{\varepsilon^{*+}_M}{p_M^+} p_{M \alpha} \right) r_{\perp \delta} \frac{g^{(a)}_{\perp} (x) - \tilde{g}_{\perp}^{(a)} (x)}{4} \; .
    \label{Eq:Twist_Expan_Corr_Axial_2_Body}
\end{gather}
After performing this replacement in eq.~(\ref{Eq:Psi_2_Full_KinTwist}), we get the twist-3 2-body wave function overlaps\footnote{For simplicity of notation, we still name this object $\Psi_2$, although this is twist expanded. We will do the same for the twist-expanded versions of $\Psi_3$, $\Phi_2$ and $\Phi_3$.}
\begin{gather}
     \Psi_2 \left(x, \boldsymbol{r} \right) = e_q m_M f_M \delta \left( 1 - \frac{p_M^+}{q^+} \right) \left(\varepsilon_{q\mu}-\frac{\varepsilon_{q}^{+}}{q^{+}}q_{\mu} \right) \left(\varepsilon_{M \alpha}^{*}-\frac{\varepsilon_{M}^{*+}}{p_M^{+}} p_{M \alpha} \right) \nonumber \\ \times \left[ i r_{\perp}^{\alpha} (h(x) - \tilde{h}(x)) \left(  (x-\bar{x}) \frac{i \partial}{ \partial r_{\perp \mu}} - 2 x \bar{x} q^{\mu} \right) + \epsilon^{\mu \nu + -} \epsilon^{+ \alpha - \delta} r_{\perp \delta} \left( \frac{g^{(a)}_{\perp} (x) - \tilde{g}_{\perp}^{(a)} (x)}{4} \right) \frac{\partial}{\partial r_{\perp}^{\nu}} \right] \nonumber \\ \times K_0 \left( \sqrt{x \bar{x} Q^2 \boldsymbol{r}^{2}} \right) .
\end{gather}
\subsubsection{3-body contribution}
We proceed similarly as described before. Thus, we start from eqs.~(\ref{Eq:Chi+_Vector_function},~\ref{Eq:Chi+5_Axial_function}), we substitute the vacuum-to-meson matrix elements by their explicit paramentrizations in eqs.~(\ref{Eq:3_Body_Vector_Matrix_Twist_Exp},~\ref{Eq:3_Body_Axial_Matrix_Twist_Exp}) and, finally, we integrate over $r^-$. This amounts to perform the following replacements:
\begin{align}
    \chi_{\gamma^+ , \sigma } ( \{ x \}, \{ \boldsymbol{z} \} ) \xrightarrow{\rm twist-expansion} & - i \; m_M f_{3M}^{V}  \left( \varepsilon^{*}_{M \sigma} - \frac{p_{M \sigma}}{p_M^+} \varepsilon_{M}^{*+} \right) \delta (q^+ - p_M^+) V (x_1, x_2) \nonumber \\ & \times \left( \prod_{j=1}^3 \theta (x_j) \theta (1-x_j) e^{- i x_j \boldsymbol{p}_M \boldsymbol{z}_j } \right)
    \label{Eq:Geometric_Twist_Exp3Body1}
\end{align}
and
\begin{align}
    \chi_{\gamma^+ \gamma^5, \sigma } ( \{ x \}, \{ \boldsymbol{z} \} ) \xrightarrow{\rm twist-expansion} \; & \epsilon_{- + \sigma \beta} m_M f_{3M}^{A} \left( \varepsilon^{* \beta }_{M} - \frac{p_{M}^{\beta}}{p_M^+} \varepsilon_{M}^{*+} \right)  \delta (q^+ - p_M^+) A (x_1, x_2) \nonumber \\ & \times \left( \prod_{j=1}^3 \theta (x_j) \theta (1-x_j) e^{- i x_j \boldsymbol{p}_M \boldsymbol{z}_j } \right)   \; ,
    \label{Eq:Geometric_Twist_Exp3Body2}
\end{align} 
The 3-body twist-3 wave functions overlap then becomes
\begin{gather}
    \Psi_3 \hspace{-0.1 cm} \left( \{ x \} , \{ \boldsymbol{z} \} \right) \hspace{-0.1 cm} = \hspace{-0.1 cm} \frac{e_{q} m_M }{2 (4\pi)}  \hspace{-0.1 cm} \left( \hspace{-0.1 cm} \varepsilon_{q\rho}-\frac{\varepsilon_{q}^{+}}{q^{+}}q_{\rho} \hspace{-0.1 cm} \right)  \hspace{-0.1 cm} \left( \hspace{-0.1 cm} \varepsilon^{* \mu }_{M} - \frac{p_{M}^{\mu}}{p_M^+} \varepsilon_{M}^{*+} \hspace{-0.1 cm} \right) \hspace{-0.1 cm} \delta \hspace{-0.1 cm} \left( \hspace{-0.1 cm} 1 - \frac{p_M^+}{q^+} \hspace{-0.1 cm} \right) \hspace{-0.15 cm} \left( \prod_{j=1}^{3} \theta (x_j) \theta (1-x_j) e^{-i x_j \boldsymbol{p}_M \boldsymbol{z}_j } \hspace{-0.1 cm} \right)   \nonumber \\
    \times c_f \bigg \{ f_{3M}^{V} g_{\sigma \mu} V(x_1, x_2) \left[ \left( 4 g_{\perp}^{ \rho \sigma } \frac{x_1 x_2}{1-x_2} \frac{Q}{Z} K_1 (QZ)  - i T_1^{\sigma \rho \nu} (\{ x \}) \frac{z_{23 \perp \nu}}{\boldsymbol{z}_{23}^{2}} K_0 (QZ) \right) - \left( 1 \leftrightarrow 2 \right) \right] \nonumber \\ 
    - \epsilon_{- + \sigma \beta} f_{3M}^{A} g^{\beta}_{\perp \mu} A (x_1, x_2) \hspace{-0.1 cm} \left[ \left( \hspace{-0.1 cm} 4 \epsilon^{ \sigma \rho + - } \hspace{-0.05 cm} \frac{x_1 x_2}{1-x_2} \frac{Q}{Z} K_1 (QZ) \hspace{-0.1 cm} + \hspace{-0.1 cm} T_2^{\sigma \rho \nu} (\{ x \}) \frac{z_{23 \perp \nu}}{\boldsymbol{z}_{23}^{2}} K_0 (QZ) \hspace{-0.1 cm} \right) \hspace{-0.1 cm} + \hspace{-0.1 cm} \left( 1 \leftrightarrow 2 \right) \hspace{-0.05 cm} \right] \hspace{-0.1 cm} \bigg \} .
\end{gather}
We observe that, due to the properties
\begin{equation}
    A(x_2, x_1) = A (x_1, x_2) \; , \hspace{1 cm}  V (x_2, x_1) = - V (x_1, x_2) \; ,
\end{equation}
the expression is fully symmetric under the exchange $(1 \leftrightarrow 2)$. Thus, we can alternatively write
\begin{gather}
    \Psi_3 \left( \{ x \} , \{ \boldsymbol{z} \} \right) = \frac{e_{q} m_M }{2 (4\pi)} c_f \left(\varepsilon_{q\rho}-\frac{\varepsilon_{q}^{+}}{q^{+}}q_{\rho}\right)  \; \left( \varepsilon^{* \mu }_{M} - \frac{p_{M}^{\mu}}{p_M^+} \varepsilon_{M}^{*+} \right) \left( \prod_{j=1}^{3} \theta (x_j) \theta (1-x_j) e^{-i x_j \boldsymbol{p}_M \boldsymbol{z}_j } \right)   \nonumber \\
    \times \delta \left( 1 - \frac{p_M^+}{q^+} \right)  \bigg \{  f_{3M}^{V} g_{\sigma \mu} V(x_1, x_2) \left[ \left( 4 g_{\perp}^{ \rho \sigma } \frac{x_1 x_2}{1-x_2} \frac{Q}{Z} K_1 (QZ) - i T_1^{\sigma \rho \nu} (\{ x \}) \frac{z_{23 \perp \nu}}{\boldsymbol{z}_{23}^{2}} K_0 (QZ) \right) \right] \nonumber \\ 
    - \epsilon_{- + \sigma \beta} f_{3M}^{A} g^{\beta}_{\perp \mu} A (x_1, x_2) \left[ \left( 4 \epsilon^{ \sigma \rho + - } \frac{x_1 x_2}{1-x_2} \frac{Q}{Z} K_1 (QZ) +  T_2^{\sigma \rho \nu} (\{ x \}) \frac{z_{23 \perp \nu}}{\boldsymbol{z}_{23}^{2}} K_0 (QZ) \right) \right] \bigg \} \nonumber \\
    + (1 \leftrightarrow 2) \; . 
\end{gather}

\subsection{Kinematic twist expansion: impact factors with DAs in momentum space}

\subsubsection{2-body contribution}
We can express the momentum space impact factor in eq.~(\ref{Eq:2BodyMomSpaceImpactFact}) as
\begin{gather}
    \Phi_2 \left(x, \boldsymbol{p} \right) = 2 \pi e_q \delta \left( 1 - \frac{p_M^+}{q^+} \right) \left(\varepsilon_{q\mu}-\frac{\varepsilon_{q}^{+}}{q^{+}}q_{\mu} \right) \int \frac{d^2 \boldsymbol{l}}{(2 \pi)^2} \int d^2 \boldsymbol{r} \frac{e^{-i \boldsymbol{l} \boldsymbol{r}} }{ (\boldsymbol{p} - \boldsymbol{l} \; )^2 + x \bar{x} Q^2 } \nonumber \\
    \times \left[ \left( 2 x \bar{x} q^{\mu} - (x-\bar{x}) (p-l)_{\perp}^{\mu} \right)  \phi_{\gamma^+} (x, \boldsymbol{r}) - i \epsilon^{\mu\nu+-} (p-l)_{\perp \nu} \; \phi_{\gamma^+ \gamma^5} (x, \boldsymbol{r}) \right] \; 
\end{gather}
and then use the twist expansions in~(\ref{Eq:Twist_Expan_Corr_Vector_2_Body}, \ref{Eq:Twist_Expan_Corr_Axial_2_Body}) to get
\begin{gather}
    \Phi_2 \left(x, \boldsymbol{p} \right) = 2 \pi e_q m_M f_M \delta \left( 1 - \frac{p_M^+}{q^+} \right) \left(\varepsilon_{q\mu}-\frac{\varepsilon_{q}^{+}}{q^{+}} q_{\mu} \right) \left(\varepsilon_{M \alpha}^{*}-\frac{\varepsilon_{M}^{*+}}{p_M^{+}} p_{M \alpha} \right) \nonumber \\ \int d^2 \boldsymbol{l} \; \delta^{(2)} ( \boldsymbol{l})   \bigg \{ (h(x) - \tilde{h} (x)) \frac{\partial}{\partial l_{\perp \alpha}} \left[ \frac{ 2 x \bar{x} q^{\mu} - (x-\bar{x}) (p-l)_{\perp}^{\mu} }{ (\boldsymbol{p} - \boldsymbol{l} \; )^2 + x \bar{x} Q^2 } \right] \nonumber \\ + \left( \frac{g^{(a)}_{\perp} (x) - \tilde{g}_{\perp}^{(a)} (x)}{4} \right) \epsilon^{\mu\nu+-} \epsilon^{+ \alpha - \delta} \frac{\partial}{\partial l_{\perp}^{\delta}} \left[ \frac{(p-l)_{\perp \nu}}{(\boldsymbol{p} - \boldsymbol{l} )^2 + x \bar{x} Q^2}  \right]  \bigg \} \; ,
\end{gather}
where we have performed an integration by parts. Performing also the integration over $\boldsymbol{l}$, we finally get
\begin{gather}
    \Phi_2 \left(x, \boldsymbol{p} \right) = \frac{2 \pi e_q m_M f_M \delta \left( 1 - \frac{p_M^+}{q^+} \right) \left(\varepsilon_{q\mu} - \frac{\varepsilon_{q}^{+}}{q^{+}}q_{\mu} \right) \left(\varepsilon_{M \alpha}^{*} - \frac{\varepsilon_{M}^{*+}}{p_M^{+}} p_{M \alpha} \right)}{ \left[ \boldsymbol{p}^{2} + x \bar{x} Q^2 \right]^2} \nonumber \\ \times \bigg \{ (h(x) - \tilde{h} (x))  \bigg( g_{\perp}^{\mu \alpha} (x - \bar{x}) \left[ \boldsymbol{p}^{2} + x \bar{x} Q^2 \right] - 2 p_{\perp}^{\alpha} \left( 2 x \bar{x} q^{\mu} - (x-\bar{x}) p_{\perp}^{\mu} \right) \bigg) \nonumber \\ - \left( \frac{g^{(a)}_{\perp} (x) - \tilde{g}_{\perp}^{(a)} (x)}{4} \right) \epsilon^{\mu \nu + - } \epsilon^{+ \alpha - \delta} \left( 2 p_{\perp \nu} p_{\perp \delta} + g_{\delta \nu}^{\perp} [\boldsymbol{p}^{2} + x \bar{x} Q^2] \right) \bigg \} \; .
    \label{Eq:ImpactFactorDAMomentumNonForGeneralFrame}
\end{gather}
For later scope, it is useful to present the expression for $\Phi_2$ in the frame in which the outgoing meson has no transverse momenta, $\boldsymbol{p}_M = 0$, i.e.
\begin{gather}
    \Phi_2 \left(x, \boldsymbol{p} \right) = \frac{4 \pi m_M f_M e_q \delta \left( 1 - \frac{p_M^+}{q^+} \right) \left(\frac{\varepsilon_{q}^{+}}{q^{+}}q_{\mu} - \varepsilon_{q\mu} \right)}{ \left[ \boldsymbol{p}^{\; 2} + x \bar{x} Q^2 \right]^2} \nonumber \\ \times \bigg \{ (h(x) - \tilde{h} (x))  \bigg( \left( 2 x \bar{x} q^{\mu} - (x-\bar{x}) p_{\perp}^{\mu} \right) (\varepsilon_{M \perp}^{*} \cdot p_{\perp}) - \frac{(x - \bar{x}) \left[ \boldsymbol{p}^{\; 2} + x \bar{x} Q^2 \right] \varepsilon_{M \perp}^{* \mu } }{2} \bigg) \nonumber \\ + \frac{g^{(a)}_{\perp} (x) - \tilde{g}_{\perp}^{(a)} (x)}{4} \left( \varepsilon_{M \perp}^{* \nu } g_{\perp}^{\delta \mu} - \varepsilon_{M \perp}^{* \mu } g_{\perp}^{\delta \nu} \right) \left( p_{\perp \nu} p_{\perp \delta}  + \frac{ g_{\delta \nu}^{\perp} (\boldsymbol{p}^{\; 2} + x \bar{x} Q^2) }{2} \right) \bigg \} \; ,
    \label{Eq:ImpactFactorDAMomentumNonFor}
\end{gather}
where we have used
\begin{equation}
    \epsilon^{\mu\nu+-} \epsilon^{+ \alpha - \delta} = g^{\mu \delta}_{\perp} g^{\nu \alpha}_{\perp} - g^{\mu \alpha}_{\perp} g^{\nu \delta}_{\perp} \; .
\end{equation}

\subsubsection{3-body contribution}
To perform the kinematic twist expansion in momentum space, we write the momentum space versions of expressions (\ref{Eq:Geometric_Twist_Exp3Body1}, \ref{Eq:Geometric_Twist_Exp3Body2}), i.e. 
\begin{align}
    \chi_{\gamma^+ , \sigma } ( \{ x \}, \{ \boldsymbol{k} \} ) \xrightarrow{\rm twist-expansion} & - i \; m_M f_{3M}^{V}  \left( \varepsilon^{*}_{M \sigma} - \frac{p_{M \sigma}}{p_M^+} \varepsilon_{M}^{*+} \right) \delta (q^+ - p_M^+) V (x_1, x_2) \nonumber \\ & \times \left( \prod_{j=1}^{3} (2 \pi)^2 \delta^{(2)} \left( \boldsymbol{k}_j + x_j \boldsymbol{p}_M \right) \theta (x_j) \theta (1-x_j) \right)
    \label{Eq:Geometric_Twist_Exp3Body1Mom}
\end{align}
and 
\begin{align}
    \chi_{\gamma^+ \gamma^5, \sigma } ( \{ x \}, \{ \boldsymbol{k} \} ) \xrightarrow{\rm twist-expansion} & \,\,\epsilon_{- + \sigma \beta} m_M f_{3M}^{A} \left( \varepsilon^{* \beta }_{M} - \frac{p_{M}^{\beta}}{p_M^+} \varepsilon_{M}^{*+} \right)  \delta (q^+ - p_M^+) A (x_1, x_2) \nonumber \\ & \times \left( \prod_{j=1}^{3} (2 \pi)^2 \delta^{(2)} \left( \boldsymbol{k}_j + x_j \boldsymbol{p}_M \right) \theta (x_j) \theta (1-x_j) \right)   \; ,
    \label{Eq:Geometric_Twist_Exp3Body2Mom}
\end{align}
obtained by Fourier transformation. Then, starting from eq.~(\ref{Eq:ImpactFactorThreeBodyMom}) and making the replacements~(\ref{Eq:Geometric_Twist_Exp3Body1Mom},~\ref{Eq:Geometric_Twist_Exp3Body2Mom}), we get
\begin{gather}
    \Phi_3 \left( \{ x \} , \{ \boldsymbol{p} \} \right) = \frac{e_{q} q^+ m_M}{4} c_f   \left(\varepsilon_{q\rho}-\frac{\varepsilon_{q}^{+}}{q^{+}}q_{\rho}\right) \left( \varepsilon^{* \beta }_{M} - \frac{p_{M}^{\beta}}{p_M^+} \varepsilon_{M}^{*+} \right) \delta (q^+ - p_M^+) \nonumber \\ \times  \left( \prod_{j=1}^{3} \frac{ \theta (1 - x_j) \theta (x_j) }{x_j} \right) \frac{ (2 \pi)^3 \delta^{(2)} \left( \sum_{i=1}^3 (\boldsymbol{p}_i + x_i \boldsymbol{p}_M) \right)}{ \left[  Q^2 + \sum_{i=1}^3 (\boldsymbol{p}_i+x_i \boldsymbol{p}_M)^2 / x_i \right]}  
    \Bigg \{ g_{\beta \sigma} f_{3 M}^{V} V (x_1, x_2) \left( 4  g^{\rho \sigma}_{\perp} \frac{x_1 x_2}{1-x_2} \right. \nonumber \\ \left.  + \tilde{T}_1^{\sigma \rho \nu} ( \{ x \} ) \big |_{\boldsymbol{k}_i = - x_i \boldsymbol{p}_M } \frac{x_1 x_2 (p_{3} + x_3 p_{M })_{\perp \nu} - x_1 x_3 (p_{2} + x_2 p_{M })_{\perp \nu} }{(\boldsymbol{p}_1+x_1 \boldsymbol{p}_M)^2 + x_1 (1-x_1) Q^2 } \right)  - \epsilon_{-+ \sigma \beta} f_{3 M}^{A} A (x_1, x_2) \nonumber \\
   \times \left( 4 \frac{x_1 x_2}{1-x_2} \epsilon^{\sigma \rho + -} + i \tilde{T}_2^{\sigma \rho \nu} ( \{ x \} ) \big |_{\boldsymbol{k}_i = - x_i \boldsymbol{p}_M } \frac{x_1 x_2 (p_{3} + x_3 p_{M })_{\perp \nu} - x_1 x_3 (p_{2} + x_2 p_{M })_{\perp \nu} }{(\boldsymbol{p}_1+x_1 \boldsymbol{p}_M)^2 + x_1 (1-x_1) Q^2 } \right)  \Bigg \} \nonumber \\ + \left( 1 \leftrightarrow 2 \right) \; .
   \label{Eq:MomentumSpaceImpTwistExp}
\end{gather}
Using eq.~(\ref{Eq:ThreeBodyKTFactForm}), we observe that the Dirac delta function in the second line of eq.~(\ref{Eq:MomentumSpaceImpTwistExp}) imposes the transverse momentum conservation, i.e.
\begin{equation}
      \boldsymbol{q} - \boldsymbol{p}_M = -(\boldsymbol{k}_1 + \boldsymbol{k}_2 + \boldsymbol{k}_3) = \boldsymbol{p}'-\boldsymbol{p} \equiv \boldsymbol{\Delta} \; .
      \label{Eq:3BodyConservMom}
\end{equation}

\section{Dilute limit and forward case}
\label{Sec:dilute-limit}

It is interesting to investigate the so-called dilute limit of our result, i.e. the limit in which the target is not extremely dense and saturation effects are absent. In this limit, the small-$x$ evolution of the target is purely driven by BFKL dynamics. Considering this limit has a double utility. First of all, it allows us to provide explicit results for the impact factors in the BFKL scheme~\cite{Fadin:1998fv,Fadin:1998sh}, which can be useful for the investigation of the preasymptotic high-energy regime of QCD. Indeed, it is important to stress that, although the evolution (linear or non-linear) concerns just the target and not the projectile, there is still a freedom in defining the impact factors. In general, as we show below, it is a combination of the LO impact factors obtained in the saturation scheme (which we will call the BK scheme) that gives the impact factor in the linear scheme (which we will call the BFKL scheme)\footnote{At NLO the correspondence is more complicated (see~\cite{Fadin:2007de,Fadin:2009za,Fadin:2009gh,Fadin:2011jg}).}. Moreover, the transversely-polarized-photon-to-transversely-polarized-meson part of our impact factor, in the forward and dilute limit, must coincide with that found by some of us in refs.~\cite{Anikin:2009hk,Anikin:2009bf}\footnote{We stress that since we have the non-forward result, we can also describe the longitudinal-photon-to-transverse-meson transition.}. In those references, the result was obtained by combining the BFKL technique, based on momentum space calculations with Reggeized gluons in the $t$-channel, and the higher-twist formalisms. Remarkably, the result was achieved by using both the LCCF and the CCF approaches and showing their equivalence for the considered case (summarized by us in more generality in Appendix~\ref{LCCF_and_comparison_with_CCF}). This emphasizes the non-triviality of this agreement, since we compare results obtained through two completely independent techniques: from one side BFKL supplemented by LCCF~\cite{Anikin:2009hk,Anikin:2009bf} and from the other the semi-classical approach supplemented by CCF. \\

In order to find the BFKL results, we will carry out expansion of the Wilson lines in Reggeized gluon fields, following the procedure introduced by S. Caron-Huot in ref.~\cite{Caron-Huot:2013fea}.

\subsection{Reggeized gluon expansion and BFKL impact factors}
Following ref.~\cite{Caron-Huot:2013fea}, we define the Reggeized gluon field as the logarithm of the Wilson line in the adjoint representation,
\begin{equation}
    R^a(\boldsymbol{z}) \equiv \frac{f^{a b c}}{g C_A} \ln \left(U_{\boldsymbol{z}}^{b c}\right) \; .
    \label{Eq:Reggeized_Gluon}
\end{equation}
We stress here that the Wilson line depends on the rapidity cut-off $e^{\eta} p_{\gamma}^+$. Similarly to what happens in the case of the dipole operator, whose evolution with respect to the rapidity cut-off is governed by the B-JIMWLK equations, the evolution of the operator defined in (\ref{Eq:Reggeized_Gluon}) is governed by the Reggeized gluon trajectory within next-to-leading logarithmic accuracy~\cite{Caron-Huot:2013fea}. \\

Using the definition~(\ref{Eq:Reggeized_Gluon}), we can perform a systematic expansion of the Wilson lines in terms of Reggeized gluons $R^a (\boldsymbol{z})$:
\begin{equation}
    V_{\boldsymbol{z}_1}=1+i g \boldsymbol{t}^a R^a\left(\boldsymbol{z}_1\right)-\frac{1}{2} g^2 \boldsymbol{t}^a \boldsymbol{t}^b R^a\left(\boldsymbol{z}_1\right) R^b\left(\boldsymbol{z}_1\right)+O\left(g^3\right) \; ,
\end{equation}
\begin{equation}
V_{\boldsymbol{z}_2}^{\dagger}=1-i g \boldsymbol{t}^b R^b\left(\boldsymbol{z}_2\right)-\frac{1}{2} g^2 \boldsymbol{t}^a \boldsymbol{t}^b R^a\left(\boldsymbol{z}_2\right) R^b\left(\boldsymbol{z}_2\right)+O\left(g^3\right) \; .
\end{equation}
We truncate the expansion to order $g^2$ because we are interested in the BFKL Pomeron (two-Reggeized-gluons-to-two-Reggeized-gluons Green function). 
Introducing the momentum space Reggeized gluon field according to the relation
\begin{equation}
    R^a(\boldsymbol{z}) = \int \frac{\mathrm{d}^d \boldsymbol{q}}{(2 \pi)^d} \mathrm{e}^{i(\boldsymbol{q} \cdot \boldsymbol{z})} R^a(\boldsymbol{q}) \; ,
\end{equation}
the target correlator can then be expressed as 
\begin{gather} \left\langle P\left(p^{\prime}\right)\left|1-\frac{1}{N_c} \operatorname{tr}\left(V_{\boldsymbol{z}_1} V_{\boldsymbol{z}_2}^{\dagger}\right)\right| P(p)\right\rangle \nonumber \\ =\frac{g^2}{4 N_c} \int \frac{\mathrm{d}^d \boldsymbol{\ell}}{(2 \pi)^d}\left(\mathrm{e}^{i\left(\boldsymbol{\ell}-\frac{\Delta}{2}\right) \cdot \boldsymbol{z}_1}-\mathrm{e}^{i\left(\boldsymbol{\ell}-\frac{\Delta}{2}\right) \cdot \boldsymbol{z}_2}\right)\left(\mathrm{e}^{-i\left(\boldsymbol{\ell}+\frac{\Delta}{2}\right) \cdot \boldsymbol{z}_1}-\mathrm{e}^{-i\left(\boldsymbol{\ell}+\frac{\Delta}{2}\right) \cdot \boldsymbol{z}_2}\right) \nonumber \\ \times \int \mathrm{d}^d \boldsymbol{v} \, \mathrm{e}^{-i(\boldsymbol{\ell} \cdot \boldsymbol{v})}\left\langle P\left(p^{\prime}\right)\left|R^a\left(\frac{\boldsymbol{v}}{2}\right) R^a\left(-\frac{\boldsymbol{v}}{2}\right)\right| P(p)\right\rangle+O\left(g^3\right) \; ,
\end{gather}
where the last line defines the unintegrated gluon density in the BFKL sense, i.e.
\begin{equation}
    \mathcal{U}(\boldsymbol{\ell}) \equiv \int \mathrm{d}^d \boldsymbol{v} \, \mathrm{e}^{-i(\boldsymbol{\ell} \cdot \boldsymbol{v})}\left\langle P\left(p^{\prime}\right)\left|R^a\left(\frac{\boldsymbol{v}}{2}\right) R^a\left(-\frac{\boldsymbol{v}}{2}\right)\right| P(p)\right\rangle \; .
\end{equation}
Then, the target matrix element of the dipole operator becomes
\begin{gather} 
\left\langle P\left(p^{\prime}\right)\left|1-\frac{1}{N_c} \operatorname{tr}\left(V_{\boldsymbol{z}_1} V_{\boldsymbol{z}_2}^{\dagger}\right)\right| P(p)\right\rangle \nonumber \\ = \frac{g^2}{4 N_c} \int \frac{\mathrm{d}^d \boldsymbol{\ell}}{(2 \pi)^d}\left(\mathrm{e}^{i\left(\boldsymbol{\ell}-\frac{\Delta}{2}\right) \cdot \boldsymbol{z}_1}-\mathrm{e}^{i\left(\boldsymbol{\ell}-\frac{\Delta}{2}\right) \cdot \boldsymbol{z}_2}\right)\left(\mathrm{e}^{-i\left(\boldsymbol{\ell}+\frac{\Delta}{2}\right) \cdot \boldsymbol{z}_1}-\mathrm{e}^{-i\left(\boldsymbol{\ell}+\frac{\Delta}{2}\right) \cdot \boldsymbol{z}_2}\right) \mathcal{U}(\ell)+O\left(g^3\right) \; .
\label{Eq:DipoleTargetForm1}
\end{gather}
Using this result, the general small-$x$ amplitude in eq.~(\ref{Eq:StandardDipoleAmp-rb}) can be written as
\begin{gather}
\mathcal{A}_2^{\rm dilute} =\frac{g^2}{4 N_c}(2 \pi)^d \delta^d(\boldsymbol{q}-\boldsymbol{p}_M-\boldsymbol{\Delta}) \int \frac{\mathrm{d}^d \boldsymbol{\ell}}{(2 \pi)^d} \mathcal{U}(\boldsymbol{\ell}) \int_0^1 \mathrm{~d} x \int \mathrm{d}^d \boldsymbol{r} \Psi_2 (x, \boldsymbol{r}) \nonumber \\ \times \left(\mathrm{e}^{-i(\bar{x} \boldsymbol{\Delta} \cdot \boldsymbol{r})}-\mathrm{e}^{i\left(\boldsymbol{\ell}+\frac{x-\bar{x}}{2} \boldsymbol{\Delta}\right) \cdot \boldsymbol{r}}-\mathrm{e}^{-i\left(\boldsymbol{\ell}-\frac{x-\bar{x}}{2} \Delta\right) \cdot \boldsymbol{r}}+\mathrm{e}^{i x(\boldsymbol{\Delta} \cdot \boldsymbol{r})}\right)+O\left(g^3\right) \; ,
\end{gather}
or, in terms of impact factors, as
\begin{gather} 
\mathcal{A}_2^{\rm dilute} =-\frac{g^2}{4 N_c}(2 \pi)^d \delta^d(\boldsymbol{q}-\boldsymbol{p}_M-\boldsymbol{\Delta}) \int \frac{\mathrm{d}^d \ell}{(2 \pi)^d} \mathcal{U}(\ell) \int_0^1 \mathrm{~d} x \nonumber \\  \times\left[\Phi_2 \left(x, \boldsymbol{\ell}-\frac{x-\bar{x}}{2} \boldsymbol{\Delta}\right) + \Phi_2 \left(x,-\boldsymbol{\ell}-\frac{x-\bar{x}}{2} \boldsymbol{\Delta}\right) - \Phi_2 (x, \bar{x} \boldsymbol{\Delta})- \Phi_2 (x,-x \boldsymbol{\Delta})\right]+O\left(g^3\right) \; .
\label{Eq:BKBFKLImpactFactorsRelation1}
\end{gather}
The last line in eq~(\ref{Eq:BKBFKLImpactFactorsRelation1}), which is a combination of four BK impact factors, defines the BFKL impact factor
\begin{gather} 
\Phi_{2, {\rm BFKL}} \left(x, \boldsymbol{\ell}, \boldsymbol{\Delta} \right) \nonumber \\ \equiv \left[\Phi_2 \left(x, \boldsymbol{\ell}-\frac{x-\bar{x}}{2} \boldsymbol{\Delta}\right) + \Phi_2 \left(x,-\boldsymbol{\ell}-\frac{x-\bar{x}}{2} \boldsymbol{\Delta}\right) - \Phi_2 (x, \bar{x} \boldsymbol{\Delta})- \Phi_2 (x,-x \boldsymbol{\Delta})\right]  \; ,
\label{Eq:BKBFKLImpactFactorsRelation2}
\end{gather}
which vanishes explicitly for $\boldsymbol{\ell}= \pm \boldsymbol{\Delta}/2$. \\

A similar procedure can be applied to the three-body amplitude of eq.~(\ref{Eq:GenealStructure3bodyWaveFunOver}). Neglecting the double-dipole operator in the target correlator, we have
\begin{gather}
    \mathcal{A}_3 = - \left( \prod_{i=1}^3 \int d x_i \theta (x_i) \right) \delta (1 - x_1 - x_2 - x_3) \int {\rm d}^2 \boldsymbol{z}_1 {\rm d}^2 \boldsymbol{z}_2 {\rm d}^2 \boldsymbol{z}_3 \, {\rm e}^{i \boldsymbol{q} (x_1 \boldsymbol{z}_1 + x_2 \boldsymbol{z}_2 + x_3 \boldsymbol{z}_3)} \nonumber \\
   \times \Psi_3 \left( x_1, x_2, x_3 , \boldsymbol{z}_1, \boldsymbol{z}_2, \boldsymbol{z}_3\right) \left\langle P\left(p^{\prime}\right)\left| \mathcal{U}_{\boldsymbol{z}_1 \boldsymbol{z}_3} + \mathcal{U}_{\boldsymbol{z}_3 \boldsymbol{z}_2} - \frac{1}{N_c^2} \mathcal{U}_{\boldsymbol{z}_1 \boldsymbol{z}_2}  \right|P\left(p\right)\right\rangle \; .
\label{Eq:Three_body_linearized}
\end{gather}
The three dipole terms in eq.~(\ref{Eq:Three_body_linearized}) can be treated separately, following the procedure described above for the two-body amplitude; e.g. for the first term we get
\begin{gather}
    \mathcal{A}_{3, \rm first}^{\rm dilute} = \left( \prod_{i=1}^3 \int d x_i \theta (x_i) \right) \delta (1 - x_1 - x_2 - x_3) \frac{-g^2}{4 N_c} \int \frac{ d^d \boldsymbol{\ell}}{(2 \pi)^d} \mathcal{U}(\ell) \nonumber \\ \bigg \{   \Phi_3 \left( \{ x \}, - x_1 \boldsymbol{q} + \boldsymbol{\Delta}, -x_2 \boldsymbol{q}, -x_3 \boldsymbol{q} \right) - \Phi_3 \left( \{ x \} , - x_1 \boldsymbol{q}  - \boldsymbol{\ell} + \frac{\boldsymbol{\Delta}}{2}, -x_2 \boldsymbol{q}, -x_3 \boldsymbol{q} + \boldsymbol{\ell} + \frac{\boldsymbol{\Delta}}{2} \right) \nonumber \\ - \Phi_3 \left( \{ x \}, - x_1 \boldsymbol{q} + \boldsymbol{\ell} + \boldsymbol{\Delta}, -x_2 \boldsymbol{q}, -x_3 \boldsymbol{q} - \boldsymbol{\ell} + \frac{\boldsymbol{\Delta}}{2} \right) + \Phi_3 \left( \{ x \} , - x_1 \boldsymbol{q}, -x_2 \boldsymbol{q}, -x_3 \boldsymbol{q} + \boldsymbol{\Delta} \right) \bigg \} \; .
     \label{Eq:ExpansionOfOneOfTheThreeBodyTerms}
\end{gather}
eq.~(\ref{Eq:ExpansionOfOneOfTheThreeBodyTerms}) can be expressed in a more elegant form, in which the fact that impact factor depends only on $\boldsymbol{\Delta} = \boldsymbol{q}-\boldsymbol{p}_M$ is evident. To this aim, we observe that in $\Phi_3 ( \{ x \}, \{ \boldsymbol{p} \} )$, any momentum $\boldsymbol{p}_i$ and $x_i \boldsymbol{p}_M$ appears only through the sum $\boldsymbol{p}_i + x_i \boldsymbol{p}_M$. Therefore we can define a new impact factor as
\begin{equation}
    \Phi_3' ( \{ x \}, \{ \boldsymbol{p} + x \boldsymbol{p}_M \} ) \equiv  \Phi_3 ( \{ x \}, \{ \boldsymbol{p} \} ) \; ,
\end{equation}
where $\{ \boldsymbol{p} + x \boldsymbol{p}_M \} = \{ \boldsymbol{p}_1 + x_1 \boldsymbol{p}_M, \boldsymbol{p}_2 + x_2 \boldsymbol{p}_M  , \boldsymbol{p}_3 + x_3 \boldsymbol{p}_M \}$.
Using this new definition and the condition (\ref{Eq:3BodyConservMom}), we get that
\begin{gather}
    \mathcal{A}_{3, \rm first}^{\rm dilute} = \left( \prod_{i=1}^3 \int d x_i \theta (x_i) \right) \delta (1 - x_1 - x_2 - x_3) \frac{-g^2}{4 N_c} \int \frac{ d^d \boldsymbol{\ell}}{(2 \pi)^d} \mathcal{U}(\ell) \nonumber \\ \bigg \{   \Phi_3' \left( \{ x \}, \bar{x}_1 \boldsymbol{\Delta}, -x_2 \boldsymbol{\Delta}, -x_3 \boldsymbol{\Delta} \right) - \Phi_3' \left( \{ x \} ,  \frac{1 -2 x_1}{2}  \boldsymbol{\Delta}  - \boldsymbol{\ell} , -x_2 \boldsymbol{\Delta},  \frac{1-2x_3}{2}  \boldsymbol{\Delta} + \boldsymbol{\ell} \right) \nonumber \\ - \Phi_3' \left( \{ x \},  \frac{1 - 2 x_1}{2}  \boldsymbol{\Delta} + \boldsymbol{\ell}, -x_2 \boldsymbol{\Delta},     \frac{1-2 x_3}{2}  \boldsymbol{\Delta} - \boldsymbol{\ell} \right) + \Phi_3' \left( \{ x \} , - x_1 \boldsymbol{\Delta}, -x_2 \boldsymbol{\Delta}, \bar{x}_3 \boldsymbol{\Delta} \right) \bigg \} \; .
\end{gather} 
Proceeding in analogous way for the last two terms in eq.~(\ref{Eq:Three_body_linearized}), the complete $\mathcal{A}_3^{\rm dilute}$ amplitude reads
\begin{gather}
   \mathcal{A}_{3}^{\rm dilute} = \left( \prod_{i=1}^3 \int d x_i \theta (x_i) \right) \delta (1 - x_1 - x_2 - x_3) \frac{-g^2}{4 N_c} \int \frac{ d^d \boldsymbol{\ell}}{(2 \pi)^d} \mathcal{U}(\ell) \nonumber \\ \bigg \{   \Phi_3' \left( \{ x \}, \bar{x}_1 \boldsymbol{\Delta}, -x_2 \boldsymbol{\Delta}, -x_3 \boldsymbol{\Delta} \right) - \Phi_3' \left( \{ x \} ,  \frac{1 -2 x_1}{2}  \boldsymbol{\Delta}  - \boldsymbol{\ell} , -x_2 \boldsymbol{\Delta},  \frac{1-2x_3}{2}  \boldsymbol{\Delta} + \boldsymbol{\ell} \right) \nonumber \\ - \Phi_3' \left( \{ x \}, \frac{1 - 2 x_1}{2} \boldsymbol{\Delta} + \boldsymbol{\ell}, -x_2 \boldsymbol{\Delta},  \frac{1-2 x_3}{2}  \boldsymbol{\Delta} - \boldsymbol{\ell} \right) + \Phi_3' \left( \{ x \} , - x_1 \boldsymbol{\Delta}, -x_2 \boldsymbol{\Delta}, \bar{x}_3 \boldsymbol{\Delta} \right) \nonumber \\
   + \Phi_3' \left( \{ x \}, -x_1 \boldsymbol{\Delta}, -x_2 \boldsymbol{\Delta}, \bar{x}_3 \boldsymbol{\Delta} \right) - \Phi_3' \left( \{ x \} , -x_1 \boldsymbol{\Delta} ,   \frac{1-2x_2}{2}  \boldsymbol{\Delta} + \boldsymbol{\ell},  \frac{1-2x_3}{2}  \boldsymbol{\Delta} - \boldsymbol{\ell} \right) \nonumber \\ - \Phi_3' \left( \{ x \}, -x_1 \boldsymbol{\Delta} ,   \frac{1-2x_2}{2}  \boldsymbol{\Delta} - \boldsymbol{\ell},   \frac{1-2 x_3}{2}  \boldsymbol{\Delta} + \boldsymbol{\ell} \right) + \Phi_3' \left( \{ x \} , - x_1 \boldsymbol{\Delta}, \bar{x}_2 \boldsymbol{\Delta}, -x_3 \boldsymbol{\Delta} \right) \nonumber \\
   - \frac{1}{N_c^2} \bigg[ \Phi_3' \left( \{ x \}, \bar{x}_1 \boldsymbol{\Delta}, -x_2 \boldsymbol{\Delta}, -x_3 \boldsymbol{\Delta} \right) - \Phi_3' \left( \{ x \} ,  \frac{1 -2 x_1}{2}  \boldsymbol{\Delta}  - \boldsymbol{\ell} ,  \frac{1-2 x_2}{2}  \boldsymbol{\Delta} + \boldsymbol{\ell} , -x_3 \boldsymbol{\Delta} \right) \nonumber \\ - \Phi_3' \left( \{ x \},  \frac{1 - 2 x_1}{2}  \boldsymbol{\Delta} + \boldsymbol{\ell},  \frac{1 - 2 x_2}{2}  \boldsymbol{\Delta} - \boldsymbol{\ell}, - x_3 \boldsymbol{\Delta} \right) + \Phi_3' \left( \{ x \} , - x_1 \boldsymbol{\Delta}, \bar{x}_2 \boldsymbol{\Delta}, -x_3 \boldsymbol{\Delta} \right) \bigg] \bigg \} \; .
   \label{Eq:FullThreeBodyLinearizedNicely}
\end{gather}
Again, the combination of impact factors vanishes for $\boldsymbol{\ell} = \pm \boldsymbol{\Delta}/2$.
The expression~(\ref{Eq:FullThreeBodyLinearizedNicely}) greatly simplifies in the forward limit as
\begin{gather}
   \mathcal{A}_{3T, \boldsymbol{\Delta}=0}^{\rm dilute}  = \left( \prod_{i=1}^3 \int d x_i \theta (x_i) \right) \delta (1 - x_1 - x_2 - x_3) \frac{g^2}{4 N_c} \int \frac{ d^d \boldsymbol{\ell}}{(2 \pi)^d} \mathcal{U}(\ell) \nonumber \\ \bigg \{ \Phi_3' \left( \{ x \} , - \boldsymbol{\ell}, \boldsymbol{0}, \boldsymbol{\ell} \right) + \Phi_3' \left( \{ x \}, \boldsymbol{\ell}, \boldsymbol{0}, - \boldsymbol{\ell} \right) + \Phi_3' \left( \{ x \}, \boldsymbol{0}, -\boldsymbol{\ell}, \boldsymbol{\ell} \right) + \Phi_3' \left( \{ x \}, \boldsymbol{0}, \boldsymbol{\ell}, - \boldsymbol{\ell} \right)   \; \nonumber \\
  - 4 \Phi_3' \left( \{ x \}, \boldsymbol{0}, \boldsymbol{0}, \boldsymbol{0} \right) - \frac{1}{N_c^2} \left( \Phi_3' \left( \{ x \}, -\boldsymbol{\ell}, \boldsymbol{\ell}, \boldsymbol{0} \right) + \Phi_3' \left( \{ x \}, \boldsymbol{\ell}, -\boldsymbol{\ell}, \boldsymbol{0} \right) - 2 \Phi_3' \left( \{ x \}, \boldsymbol{0}, \boldsymbol{0}, \boldsymbol{0} \right) \right) \bigg \} \; .
\end{gather}
It is easy to observe that the sign of $\boldsymbol{\ell}$ is inessential\footnote{This property is true for the transverse-photon-to-transverse-meson part of the impact factor, which is the only one surviving in the $\boldsymbol{\Delta} = 0$ limit.}, then
\begin{gather}
   \mathcal{A}_{3T, \boldsymbol{\Delta}=0}^{\rm dilute}  = \left( \prod_{i=1}^3 \int d x_i \theta (x_i) \right) \delta (1 - x_1 - x_2 - x_3) \frac{g^2}{4 N_c} \int \frac{ d^d \boldsymbol{\ell}}{(2 \pi)^d} \mathcal{U}(\ell) \nonumber \\ \bigg \{ 2 \Phi_3' \left( \{ x \} , - \boldsymbol{\ell}, \boldsymbol{0}, \boldsymbol{\ell} \right) + 2 \Phi_3' \left( \{ x \}, \boldsymbol{0}, -\boldsymbol{\ell}, \boldsymbol{\ell} \right)
  - 4 \Phi_3' \left( \{ x \}, \boldsymbol{0}, \boldsymbol{0}, \boldsymbol{0} \right) \nonumber \\ - \frac{1}{N_c^2} \left( 2 \Phi_3' \left( \{ x \}, \boldsymbol{\ell}, -\boldsymbol{\ell}, \boldsymbol{0} \right) - 2 \Phi_3' \left( \{ x \}, \boldsymbol{0}, \boldsymbol{0}, \boldsymbol{0} \right) \right) \bigg \} \; .
\end{gather}
The last expression can be further simplified by nothing that both $\Phi_3 \left( \{ x \}, \boldsymbol{0}, \boldsymbol{0}, \boldsymbol{0} \right)$ and $\Phi_3 \left( \{ x \}, \boldsymbol{\ell}, -\boldsymbol{\ell}, \boldsymbol{0} \right)$ are symmetric quantities under the exchange $x_1 \leftrightarrow x_2$ while
\begin{equation}
    \Phi_3' \left( x_1, x_2, x_3, \boldsymbol{0}, -\boldsymbol{\ell}, \boldsymbol{\ell} \right) = \Phi_3' \left( x_2, x_1, x_3, - \boldsymbol{\ell}, \boldsymbol{0}, \boldsymbol{\ell} \right) .
\end{equation}
The final form of the amplitude is thus
\begin{gather}
   \mathcal{A}_{3T, \boldsymbol{\Delta}=0}^{\rm dilute}  = \left( \prod_{i=1}^3 \int d x_i \theta (x_i) \right) \delta (1 - x_1 - x_2 - x_3) \frac{g^2}{4 N_c} \int \frac{ d^d \boldsymbol{\ell}}{(2 \pi)^d} \mathcal{U}(\ell) \nonumber \\ \bigg \{ 2 \left( \Phi_3' \left( x_1, x_2, x_3 , - \boldsymbol{\ell}, \boldsymbol{0}, \boldsymbol{\ell} \right)
  -  \Phi_3' \left( x_1, x_2, x_3 , \boldsymbol{0}, \boldsymbol{0}, \boldsymbol{0} \right) \right) \nonumber \\ - \frac{1}{N_c^2} \left(  \Phi_3' \left( x_1, x_2, x_3, \boldsymbol{\ell}, -\boldsymbol{\ell}, \boldsymbol{0} \right) -  \Phi_3' \left( x_1, x_2, x_3 , \boldsymbol{0}, \boldsymbol{0}, \boldsymbol{0} \right) \right) \bigg \} 
 + (x_1 \leftrightarrow x_2) \; , 
\end{gather}
We observe that, as expected, the combination in the curly bracket vanishes when $\boldsymbol{\ell}=0$.

\subsubsection{Explicit 2-body contribution in the dilute limit and forward case}
The BFKL impact factor for the forward case was computed in ref.~\cite{Anikin:2009bf}. 
By taking the forward limit of the result presented in eq.~(\ref{Eq:ImpactFactorDAMomentumNonFor}) and the relation~(\ref{Eq:BKBFKLImpactFactorsRelation2}), we can obtain the BFKL impact factor, which serves as an independent check of both results. The result in Ref.~\cite{Anikin:2009bf} is presented in a reference frame in which the meson has no transverse momenta, $\boldsymbol{p}_M=0$, therefore, for simplicity of comparison, the result of the next two subsections will be obtained in the same frame. For consistency of notations in the comparison, here we will write $\boldsymbol{p}$ instead of $\boldsymbol{\ell}$ for the transverse momentum in the unintegrated gluon density. Starting with the 2-body contribution, we take the forward limit of~(\ref{Eq:ImpactFactorDAMomentumNonFor}) and, in complete analogy to ref.~\cite{Anikin:2009bf}, we split our impact factor in the helicity flip contribution and helicity non-flip contribution. The first one reads
\begin{equation}
    \Phi_{2,\boldsymbol{\Delta}=0}^{\rm flip} \left(x, \boldsymbol{p} \right) = 2 \pi m_M f_M e_q \delta \left( 1 - \frac{p_M^+}{q^+} \right) \frac{2 \boldsymbol{p}^2}{\left[ \boldsymbol{p}^2 + x \bar{x} Q^2 \right]^2} T_{\rm f.} \; \phi_{2, \rm f.} (x) \; ,
\end{equation}
where
\begin{equation}
    T_{\rm f.} = \frac{ ( \boldsymbol{\varepsilon}_{q} \cdot \boldsymbol{p}) ( \boldsymbol{\varepsilon}_{M}^{*} \cdot \boldsymbol{p}) }{\boldsymbol{p}^2} - \frac{\boldsymbol{\varepsilon}_{q} \cdot \boldsymbol{\varepsilon}_{M}^{*}}{2} 
\end{equation}
and
\begin{equation}
    \phi_{2, \rm f.} (x) = (2x-1) (h(x)-\tilde{h}(x)) - \frac{g^{(a)}_{\perp} (x) - \tilde{g}_{\perp}^{(a)} (x)}{4} \; .
\end{equation}
The second takes the form
\begin{equation}    \Phi_{2,\boldsymbol{\Delta}=0}^{\rm no-flip} \left(x, \boldsymbol{p} \right) = - 2 \pi m_M f_M e_q \delta \left( 1 - \frac{p_M^+}{q^+} \right) \frac{x \bar{x} Q^2}{\left[ \boldsymbol{p}^2 + x \bar{x} Q^2 \right]^2} T_{\rm n. f.} \; \phi_{2, \rm n. f.} (x) \; , 
\end{equation}
where
\begin{equation}
    T_{\rm n. f.} = \boldsymbol{\varepsilon}_{q} \cdot \boldsymbol{\varepsilon}_{M}^{*} \; 
\end{equation}
and
\begin{equation}
    \phi_{2,\rm n. f.} (x) = (2x-1) (h(x)-\tilde{h}(x)) + \frac{g^{(a)}_{\perp} (x) - \tilde{g}_{\perp}^{(a)} (x)}{4} \; .
\end{equation}
Now, in the forward case, eq.~(\ref{Eq:BKBFKLImpactFactorsRelation2}) implies that
\begin{equation}
    \Phi_{2,\boldsymbol{\Delta}=0, \rm BFKL} \left(x, \boldsymbol{p} \right) = 2 \left( \Phi_{2,\boldsymbol{\Delta}=0} \left(x, \boldsymbol{p} \right) - \Phi_{2,\boldsymbol{\Delta}=0} \left(x, \boldsymbol{0} \right) \right) \; ,
\end{equation}
from which we immediately obtain
\begin{equation}
    \Phi_{2,\boldsymbol{\Delta}=0, \rm BFKL}^{\rm flip} \left(x, \boldsymbol{p} \right) = 4 \pi m_M f_M e_q \delta \left( 1 - \frac{p_M^+}{q^+} \right) \frac{2 \boldsymbol{p}^2}{\left[ \boldsymbol{p}^2 + x \bar{x} Q^2 \right]^2} T_{\rm f.} \; \phi_{\rm f.} (x) 
    \label{Eq:TwoBodyFlipBFKLScheme}
\end{equation}
and
\begin{equation}
     \Phi_{2,\boldsymbol{\Delta}=0, \rm BFKL}^{\rm no-flip} \left(x, \boldsymbol{p} \right) = 4 \pi m_M f_M e_q \delta \left( 1 - \frac{p_M^+}{q^+} \right) \frac{ \boldsymbol{p}^2 (\boldsymbol{p}^2 + 2 x \bar{x} Q^2)}{x \bar{x} Q^2 \left[ \boldsymbol{p}^2 + x \bar{x} Q^2 \right]^2} T_{\rm n. f.} \; \phi_{\rm n.f. } (x) \; .
     \label{Eq:TwoBodyNoFlipBFKLScheme}
\end{equation}
Results in eqs.~(\ref{Eq:TwoBodyFlipBFKLScheme},~\ref{Eq:TwoBodyNoFlipBFKLScheme}) are in agreement with eq.~(169) of ref.~\cite{Anikin:2009bf}. An important feature of our computation can be emphasized here. Because of QCD gauge invariance, the full impact factor, i.e. the sum of the 2 and 3-body contributions, should vanish for $\boldsymbol{p} = 0$. In the calculation in ref.~\cite{Anikin:2009bf}, QCD gauge invariance breaking terms appears separately in the 2- and 3-body contributions and cancel each other in a non-trivial way in the full result. In the present case, since we are dealing with the dipole form~(\ref{Eq:DipoleTargetForm1}) of the amplitudes, these terms are absent from the beginning for both the 2- and the 3-body terms.

\subsubsection{Explicit 3-body contribution in the dilute limit and forward case}
Comparison of the 3-body contribution of the present paper with the one of ref.~\cite{Anikin:2009bf} is quite involved. We start by defining an auxiliary impact factor $\varphi_3 (\{ x \}, \{ p \} )$, through the relation 
\begin{gather}
    \Phi_3' ( \{ x \}, \{ p \} ) \equiv \frac{e_q m_N}{4 \pi^2} c_f  (2 \pi) \delta \left(1 - \frac{p_M^+}{q^+} \right) \nonumber \\ \times (2 \pi)^2 \delta^2 (\boldsymbol{q} - \boldsymbol{p}_M - \boldsymbol{\Delta}) \left( \prod_{j=1}^{3} \theta (1 - x_j) \theta (x_j)  \right) \varphi_3 (\{ x \}, \{ p \} )  \; .
\end{gather}
Since in ref.~\cite{Anikin:2009bf}, only the amplitude for the transition from a transverse photon to a transverse meson is considered, in order to compare with our calculation we need to remove from our result the term related to the longitudinal-to-transverse transition. This amounts to removing the terms that are proportional to $q^{\rho}$ in the structures $\tilde{T}_1^{\sigma \rho \nu}$ and $\tilde{T}_2^{\sigma \rho \nu}$, given respectively by eqs.~(\ref{Eq:T1Mom}) and~(\ref{Eq:T2Mom}). In terms of the auxiliary impact factor $\varphi_3$, the 3-body transverse amplitude $\mathcal{A}_{3 T, \boldsymbol{\Delta}=0 }^{\text {dilute }}$ thus reads 
\begin{gather}
    \mathcal{A}_{3 T, \boldsymbol{\Delta}=0 }^{\text {dilute }} = e_q m_M \frac{g^2}{4 N_c} (2 \pi) \delta \left( 1 -\frac{p_M^{+}}{q^+} \right) \delta^2\left(\boldsymbol{q}-\boldsymbol{p}_M\right) \left( \prod_{i=1}^3 \int_0^1 d x_i \right)  \delta (1 - x_1 - x_2 - x_3) \nonumber \\ \times \int \frac{\mathrm{d}^d \boldsymbol{\ell}}{(2 \pi)^d} \mathcal{U} \left( \boldsymbol{\ell} \right)  \bigg \{ 2 c_f \bigg[\varphi_{3T} \left(\left\{x \right\} ; \boldsymbol{\ell}, \mathbf{0},-\boldsymbol{\ell}\right)-\varphi_{3T}\left(\left\{x \right\} ; \mathbf{0}, \mathbf{0}, \mathbf{0}\right)\bigg] + 2 c_f \bigg[\varphi_{3T} \left(\left\{x \right\} ; \mathbf{0},\boldsymbol{\ell}, -\boldsymbol{\ell} \right) \nonumber  \\ -\varphi_{3T} \left(\left\{x \right\} ; \mathbf{0}, \mathbf{0}, \mathbf{0}\right) \bigg] + 2 \left(1-c_f\right) \bigg[\varphi_{3T} \left(\left\{x \right\} ; \boldsymbol{\ell} ,-\boldsymbol{\ell}, \mathbf{0}\right)-\varphi_{3T} \left(\left\{x \right\} ; \mathbf{0}, \mathbf{0}, \mathbf{0}\right)\bigg] \bigg\} \; ,
\end{gather}
where we have used again the fact that the sign of $\boldsymbol{\ell}$ is not important. Now, 
we split $\varphi_{3T}$ in three quantities:
\begin{gather} 
\varphi_T^{(1)}\left(\left\{x \right\} , \left\{ \boldsymbol{\ell} \right\}\right) =x_1 x_2 g_{\perp}^{\rho \sigma} \frac{(2 \pi)^2}{x_1 x_2 x_3\left(\frac{\boldsymbol{\ell}_1^2}{x_1}+\frac{\boldsymbol{\ell}_2^2}{x_2}+\frac{\boldsymbol{\ell}_3^2}{x_3}+Q^2\right)}\left(\varepsilon_{q \perp}-\frac{\varepsilon_q^{+}}{q^{+}} q_{\perp}\right)\left(\varepsilon_{M \sigma}^*-\frac{p_{M \sigma}}{p_M^{+}} \varepsilon_{M}^{*+}\right) \nonumber \\ \times\left[-\left(\frac{1}{x_1+x_3}+\frac{1}{x_2+x_3}\right) f_{3 M}^A A\left(x_1, x_2\right)+\left(\frac{1}{x_1+x_3}-\frac{1}{x_2+x_3}\right) f_{3 M}^V V\left(x_1, x_2\right)\right] \; ,
\end{gather}
\begin{gather}
    \varphi_T^{(2)}\left(\left\{x \right\} , \left\{ \boldsymbol{\ell} \right\}\right) =\frac{(2 \pi)^2 \ell_{1 \perp \alpha}\left(x_3 \ell_{2 \perp \nu}-x_2 \ell_{3 \perp \nu}\right)}{x_2 x_3\left(x_2+x_3\right)\left(\boldsymbol{\ell}_1^2+x_1 \bar{x}_1 Q^2\right)\left(\frac{\boldsymbol{\ell}_1^2}{x_1}+\frac{\boldsymbol{\ell}_2^2}{x_2}+\frac{\boldsymbol{\ell}_3^2}{x_3}+Q^2\right)}\left(\varepsilon_{q \rho}-\frac{\varepsilon_q^{+}}{q^{+}} q_\rho\right) \nonumber \\ \times \left(\varepsilon_{M\sigma}^*-\frac{p_{M \sigma}}{p_M^{+}} \varepsilon_M^{*+}\right) \left\{-\left[\left(2 x_1-1\right) g_{\perp}^{\alpha \rho} g_{\perp}^{\nu \sigma}-\left(1+2 \frac{x_2}{x_3}\right)\left(g_{\perp}^{\alpha \nu} g_{\perp}^{\rho \sigma}-g_{\perp}^{\alpha \sigma} g_{\perp}^{\rho \nu}\right)\right] f_{3 M}^A A\left(x_1, x_2\right)\right. \nonumber \\ \left.+\left[\left(2 x_1-1\right)\left(1+2 \frac{x_2}{x_3}\right) g_{\perp}^{\alpha \rho} g_{\perp}^{\sigma \nu}+g_{\perp}^{\alpha \sigma} g_{\perp}^{\rho \nu}-g_{\perp}^{\alpha \nu} g_{\perp}^{\rho \sigma}\right] f_{3 M}^V V\left(x_1, x_2\right)\right\} \; ,
\end{gather}
\begin{gather} 
\varphi_T^{(3)}\left(\left\{x \right\} , \left \{ \boldsymbol{\ell} \right \} \right) =\frac{(2 \pi)^2 \ell_{2 \perp \alpha}\left(x_3 \ell_{1 \perp \nu}-x_1 \ell_{3 \perp \nu}\right)}{x_1 x_3\left(x_1+x_3\right)\left(\boldsymbol{\ell}_2^2+x_2 \bar{x}_2 Q^2\right)\left(\frac{\boldsymbol{\ell}_1^2}{x_1}+\frac{\boldsymbol{\ell}_2^2}{x_2}+\frac{\boldsymbol{\ell}_3^2}{x_3}+Q^2\right)}\left(\varepsilon_{q \rho}-\frac{\varepsilon_q^{+}}{q^{+}} q_\rho\right) \nonumber \\ \times \left(\varepsilon_{M \sigma}^*-\frac{p_{M \sigma}}{p_M^{+}} \varepsilon_{M}^{*+}\right) \left\{-\left[\left(1+2 \frac{x_1}{x_3}\right)\left(g_{\perp}^{\rho \nu} g_{\perp}^{\alpha \sigma}-g_{\perp}^{\rho \sigma} g_{\perp}^{\alpha \nu}\right)+\left(2 x_2-1\right) g_{\perp}^{\alpha \rho} g_{\perp}^{\nu \sigma}\right] f_{3 M}^A A\left(x_1, x_2\right)\right. \nonumber \\ \left.-\left[\left(2 x_2-1\right)\left(1+2 \frac{x_1}{x_3}\right) g_{\perp}^{\alpha \rho} g_{\perp}^{\nu \sigma}+g_{\perp}^{\alpha \sigma} g_{\perp}^{\nu \rho}-g_{\perp}^{\alpha \nu} g_{\perp}^{\rho \sigma}\right] f_{3 M}^V V\left(x_1, x_2\right)\right\}
\end{gather}
and then consider the combinations
\begin{gather}
    \phi_{3T}^{(i)} \left( \{ x \} , \boldsymbol{\ell}  \right) \equiv 2 c_f \bigg[\varphi_{3T}^{(i)} \left(\left\{x \right\} ; \boldsymbol{\ell}, \mathbf{0},-\boldsymbol{\ell}\right)-\varphi_{3T}^{(i)} \left(\left\{x \right\} ; \mathbf{0}, \mathbf{0}, \mathbf{0}\right)\bigg] + 2 c_f \bigg[\varphi_{3T}^{(i)}  \left(\left\{x \right\} ; \mathbf{0},\boldsymbol{\ell}, -\boldsymbol{\ell} \right) \nonumber  \\ -\varphi_{3T}^{(i)}  \left(\left\{ x \right\} ; \mathbf{0}, \mathbf{0}, \mathbf{0}\right) \bigg] + 2 \left(1-c_f\right) \bigg[\varphi_{3T}^{(i)}  \left(\left\{x \right\} ; \boldsymbol{\ell} ,-\boldsymbol{\ell}, \mathbf{0}\right)-\varphi_{3T}^{(i)}  \left(\left\{x \right\} ; \mathbf{0}, \mathbf{0}, \mathbf{0}\right)\bigg] \; .
    \label{Eq:CombThreeBodyOrg}
\end{gather}
With that, our amplitude reads
\begin{gather}
    \mathcal{A}_{3 T, \boldsymbol{\Delta}=0 }^{\text {dilute }} = e_q m_M \frac{g^2}{4 N_c} (2 \pi) \delta \left( 1 -\frac{p_M^{+}}{q^+} \right) \delta^2\left(\boldsymbol{q}-\boldsymbol{p}_M\right)  \nonumber \\ \times \left( \prod_{i=1}^3 \int_0^1 d x_i \right)  \delta (1 - x_1 - x_2 - x_3) \int \frac{\mathrm{d}^d \boldsymbol{\ell}}{(2 \pi)^d} \mathcal{U} \left( \boldsymbol{\ell} \right) \sum_{j} \phi_{3T}^{(j)} \left( \{ x \} , \boldsymbol{\ell} \right) \; .
    \label{Eq:DiluteAmpInTermsOfPhi}
\end{gather}
In the forward limit and in the frame in which the meson has no transverse momentum, only the transverse part (in the sense of light-cone variables) of $\left(\varepsilon_{q \rho}-\frac{\varepsilon_q^{+}}{q^{+}} q_{\rho}\right) \left(\varepsilon_{M\sigma}^*-\frac{p_{M \sigma}}{p_M^{+}} \varepsilon_M^{*+}\right)$ survives. Then, from eqs.~(\ref{Eq:CombThreeBodyOrg}), we obtain
\begin{gather} 
\phi_T^{(1)}\left(\left\{x \right\} , \boldsymbol{\ell} \right) =-2 \frac{(2 \pi)^2}{x_3}\left(\varepsilon_{q \perp} \cdot \varepsilon_{M \perp}^*\right) \frac{\ell^2}{Q^2} \nonumber \\ \times\left[-\left(\frac{1}{x_1+x_3}+\frac{1}{x_2+x_3}\right) f_{3 M}^A A\left(x_1, x_2\right)+\left(\frac{1}{x_1+x_3}-\frac{1}{x_2+x_3}\right) f_{3 M}^V V\left(x_1, x_2\right)\right] \nonumber \\ \times\left(\frac{\left(1-c_f\right)}{\boldsymbol{\ell}^2+\frac{x_1 x_2}{x_1+x_2} Q^2}+\frac{c_f}{\boldsymbol{\ell}^2+\frac{x_1 x_3}{x_1+x_3} Q^2}+\frac{c_f}{\boldsymbol{\ell}^2+\frac{x_2 x_3}{x_2+x_3} Q^2}\right) \; ,
\end{gather}
\begin{gather} \phi_T^{(2)}\left(\left\{x \right\} ,\boldsymbol{\ell} \right) =\frac{2(2 \pi)^2}{x_2 x_3\left(x_2+x_3\right)} \frac{\varepsilon_{q \rho} \varepsilon_{M \sigma}^* \boldsymbol{\ell}^2}{\boldsymbol{\ell}^2+x_1 \bar{x}_1 Q^2}\left(\frac{x_2 c_f}{\frac{x_1+x_3}{x_1 x_3} \boldsymbol{\ell}^2+Q^2}-\frac{x_3\left(1-c_f\right)}{\frac{x_1+x_2}{x_1 x_2} \boldsymbol{\ell}^2+Q^2}\right) \nonumber \\  \times\left\{-\left[2 \frac{\left(x_2+x_3\right)\left(x_1+x_2\right)}{x_3} \frac{\ell_{\perp}^\rho \ell_{\perp}^\sigma}{\boldsymbol{\ell}^2}+\left(1+2 \frac{x_2}{x_3}\right) g_{\perp}^{\rho \sigma}\right] f_{3 M}^A A\left(x_1, x_2\right)\right. \nonumber \\  \left.+\left[2\left(x_1+2 \frac{x_1 x_2}{x_3}-\frac{x_2}{x_3}\right) \frac{\ell_{\perp}^\rho \ell_{\perp}^\sigma}{\boldsymbol{\ell}^2}+g_{\perp}^{\rho \sigma}\right] f_{3 M}^V V\left(x_1, x_2\right)\right\} \; ,
\end{gather}
\begin{gather} 
\phi_T^{(3)}\left(\left\{x \right\} ;\left\{ \boldsymbol{\ell} \right\}\right) =\frac{2(2 \pi)^2}{x_1 x_3\left(x_1+x_3\right)} \frac{\varepsilon_{q \rho} \varepsilon_{M \sigma}^* \boldsymbol{\ell}^2}{\boldsymbol{\ell}^2+x_2 \bar{x}_2 Q^2}\left(\frac{x_1 c_f}{\frac{x_2+x_3}{x_2 x_3} \boldsymbol{\ell}^2+Q^2}-\frac{x_3\left(1-c_f\right)}{\frac{x_1+x_2}{x_1 x_2} \boldsymbol{\ell}^2+Q^2}\right) \nonumber \\ \times\left\{-\left[2 \frac{\left(x_1+x_2\right)\left(x_1+x_3\right)}{x_3} \frac{\ell_{\perp}^\rho \ell_{\perp}^\sigma}{\boldsymbol{\ell}^2}+\left(1+2 \frac{x_1}{x_3}\right) g_{\perp}^{\rho \sigma}\right] f_{3 M}^A A\left(x_1, x_2\right)\right. \nonumber \\ \left.-\left[2\left(x_2+2 \frac{x_1 x_2}{x_3}-\frac{x_1}{x_3}\right) \frac{\ell_{\perp}^\rho \ell_{\perp}^\sigma}{\boldsymbol{\ell}^2}+g_{\perp}^{\rho \sigma}\right] f_{3 M}^V V\left(x_1, x_2\right)\right\} \; .
\end{gather}
We can now isolate the term that appears only in the helicity flip part of the amplitude, i.e. the one proportional to $\displaystyle \frac{\ell_{\perp}^{\rho} \ell_{\perp}^{\sigma} }{\boldsymbol{\ell}^2}$. We get
\begin{gather}
    \sum_{j} \phi_{3T}^{(j)} \left( \{ x \} , \boldsymbol{\ell} \right) \big |_{ \displaystyle \frac{\ell_{\perp}^{\rho} \ell_{\perp}^{\sigma} }{\boldsymbol{\ell}^2}} = -\frac{4(2 \pi)^2}{x_1 x_2 x_3^2} \frac{\left(\boldsymbol{\ell} \cdot \boldsymbol{\varepsilon}_q\right)\left(\boldsymbol{\ell} \cdot \boldsymbol{\varepsilon}_M^*\right)}{\boldsymbol{l}^2} \nonumber \\ \times  \frac{\boldsymbol{l^2}}{Q^2} \left[\left(x_1+x_2\right) f_{3 M}^A A\left(x_1, x_2\right) + \left(x_2-x_1\right) f_{3 M}^V V\left(x_1, x_2\right)\right] \nonumber \\  \times\left(\frac{x_3 c_f}{\boldsymbol{\ell}^2+\frac{x_2 x_3}{x_2+x_3} Q^2}+\frac{x_3 c_f}{\boldsymbol{\ell}^2+\frac{x_1 x_3}{x_1+x_3} Q^2}-\frac{\bar{x}_3\left(1-c_f\right)}{\boldsymbol{\ell}^2+\frac{x_1 x_2}{x_1+x_2} Q^2}+\frac{x_2-\bar{x}_1 c_f}{\boldsymbol{\ell}^2+x_1 \bar{x}_1 Q^2}+\frac{x_1-\bar{x}_2 c_f}{\boldsymbol{\ell}^2+x_2 \bar{x}_2 Q^2}\right) \; .
    \label{Eq:Dulutel_perpl_perp}
\end{gather}
The remaining part is proportional to $g^{\rho \sigma}_{\perp}$ and reads
\begin{gather}
   \sum_j \phi_T^{(j)}\left(\left\{x_i\right\} ;\left\{\ell_i\right\}\right) \big|_{ \displaystyle g_{\perp}^{\rho \sigma}} \nonumber \\ =  - \frac{2(2 \pi)^2\left(\varepsilon_{q \perp} \cdot \varepsilon_{M \perp}^*\right) \boldsymbol{\ell}^2}{x_1 x_2 x_3\left(x_1+x_3\right) Q^2} \nonumber \left[\left(1+2 \frac{x_1}{x_3} \right) f_{3 M}^A A\left(x_1, x_2\right) + f_{3 M}^V V\left(x_1, x_2\right)\right] \nonumber \\ \times \left( \frac{x_3 c_f}{\boldsymbol{\ell}^2+\frac{x_2 x_3}{x_2+x_3} Q^2}-\frac{x_1\left(1-c_f\right)}{\boldsymbol{\ell}^2+\frac{x_1 x_2}{x_1+x_2} Q^2}+\frac{x_1-\bar{x}_2 c_f}{\boldsymbol{\ell}^2+x_2 \bar{x}_2 Q^2} \right) \nonumber \\ +\frac{2 (2 \pi)^2\left(\varepsilon_{q \perp} \cdot \varepsilon_{M \perp}^*\right) \boldsymbol{\ell}^2}{x_1 x_2 x_3\left(x_1+x_3\right) Q^2}\left[f_{3 M}^A A\left(x_1, x_2\right) - f_{3 M}^V V\left(x_1, x_2\right)\right] \nonumber \\ \times\left(\frac{x_1 x_2\left(1-c_f\right)}{\boldsymbol{\ell}^2+\frac{x_1 x_2}{x_1+x_2} Q^2}+\frac{x_1 x_2 c_f}{\boldsymbol{\ell}^2+\frac{x_1 x_3}{x_1+x_3} Q^2}+\frac{x_1 x_2 c_f}{\boldsymbol{\ell}^2+\frac{x_2 x_3}{x_2+x_3} Q^2}\right) \nonumber \\ -\frac{2(2 \pi)^2\left(\varepsilon_{q \perp} \cdot \varepsilon_{M \perp}^*\right) \boldsymbol{\ell}^2}{x_1 x_2 x_3\left(x_2+x_3\right) Q^2}\left[\left(1+2 \frac{x_2}{x_3}\right) f_{3 M}^A A\left(x_1, x_2\right) - f_{3 M}^V V\left(x_1, x_2\right)\right] \nonumber \\ \times\left(\frac{x_3 c_f}{\boldsymbol{\ell}^2+\frac{x_1 x_3}{x_1+x_3} Q^2}-\frac{\left(1-c_f\right) x_2}{\boldsymbol{\ell}^2+\frac{x_1 x_2}{x_1+x_2} Q^2}+\frac{x_2-\bar{x}_1 c_f x_2}{\boldsymbol{\ell}^2+x_1 \bar{x}_1 Q^2}\right) \nonumber \\ +\frac{2(2 \pi)^2\left(\varepsilon_{q \perp} \cdot \varepsilon_{M \perp}^*\right) \boldsymbol{\ell}^2}{x_1 x_2 x_3\left(x_2+x_3\right) Q^2}\left[f_{3 M}^A A\left(x_1, x_2\right) + f_{3 M}^V V\left(x_1, x_2\right)\right] \nonumber \\ \times\left(\frac{x_1 x_2\left(1-c_f\right)}{\boldsymbol{\ell}^2+\frac{x_1 x_2}{x_1+x_2} Q^2}+\frac{x_1 x_2 c_f}{\boldsymbol{\ell}^2+\frac{x_1 x_3}{x_1+x_3} Q^2}+\frac{x_1 x_2 c_f}{\boldsymbol{\ell}^2+\frac{x_2 x_3}{x_2+x_3} Q^2}\right) \; .
   \label{Eq:g_perpl_perp} 
\end{gather}
We perform the manipulation,
\begin{gather}
 \frac{ ( \boldsymbol{\varepsilon}_{q} \cdot \boldsymbol{\ell}) ( \boldsymbol{\varepsilon}_{M}^{*} \cdot \boldsymbol{\ell}) }{\boldsymbol{\ell}^2} =  \left( \frac{ ( \boldsymbol{\varepsilon}_{q} \cdot \boldsymbol{\ell}) ( \boldsymbol{\varepsilon}_{M}^{*} \cdot \boldsymbol{\ell}) }{\boldsymbol{\ell}^2} + \frac{\varepsilon_{q \perp} \cdot \varepsilon_{M \perp}^{*}}{2} \right) - \frac{\varepsilon_{q \perp} \cdot \varepsilon_{M \perp}^{*}}{2} = T_{\rm f.} + \frac{T_{\rm n. f.}}{2}  \; ,
   \label{Eq:TfandTnfManip}
\end{gather}
in eq.~(\ref{Eq:Dulutel_perpl_perp}) so that, keeping just the $T_{\rm f.}$, we get the full flip contribution
\begin{gather}
\sum_j \phi_T^{(j)}\left(\left\{x_i\right\} ;\left\{\ell_i\right\}\right)\bigg|_{\mathrm{flip}} =\frac{4(2 \pi)^2 \boldsymbol{\ell}^2}{x_1 x_2 x_3^2 Q^2} T_{\rm f.} \left[f_{3 M}^V V\left(x_1, x_2\right)-f_{3 M}^A A\left(x_1, x_2\right)\right] \nonumber \\ \times 2 x_1\left(\frac{x_3 c_f}{\boldsymbol{\ell}^2+\frac{x_2 x_3}{x_2+x_3} Q^2}+\frac{x_3 c_f}{\boldsymbol{\ell}^2+\frac{x_1 x_3}{x_1+x_3} Q^2}-\frac{\bar{x}_3\left(1-c_f\right)}{\boldsymbol{\ell}^2+\frac{x_1 x_2}{x_1+x_2} Q^2}+\frac{x_2-\bar{x}_1 c_f}{\boldsymbol{\ell}^2+x_1 \bar{x}_1 Q^2}+\frac{x_1-\bar{x}_2 c_f}{\boldsymbol{\ell}^2+x_2 \bar{x}_2 Q^2}\right) .
\end{gather}
Now, we can add the second term in the last equality of eq.~(\ref{Eq:TfandTnfManip}) to eq.~(\ref{Eq:g_perpl_perp}) to get the full flip contribution. Using the fact that the function is integrated over a symmetric domain with respect to the $x_1 \leftrightarrow x_2$ exchange, after some tedious algebraic manipulations, we get  
\begin{gather}
    \sum_j \phi_T^{(j)}\left(\left\{x_i\right\} ;\left\{\ell_i\right\}\right) \bigg|_{\text {n.f. }} = - \frac{4(2 \pi)^2 \boldsymbol{\ell}^2}{x_1 x_2 x_3^2 Q^2} T_{\rm n.f. } \left[f_{3 M}^V V\left(x_1, x_2\right)+f_{3 M}^A A\left(x_1, x_2\right)\right] \nonumber \\  \times\left(\frac{\left(1-c_f\right) x_1 \bar{x}_3}{\bar{x}_3 \boldsymbol{\ell}^2+x_1 x_2 Q^2}-\frac{c_f x_3^2}{\bar{x}_1 \boldsymbol{\ell}^2+x_2 x_3 Q^2}-\frac{\left(x_2-\bar{x}_1 c_f\right) x_1 x_2}{\bar{x}_1\left(\boldsymbol{\ell}^2+x_1 \bar{x}_1 Q^2\right)}-\frac{\left(x_1-\bar{x}_2 c_f\right) \bar{x}_2}{\left(\boldsymbol{\ell}^2+x_2 \bar{x}_2 Q^2\right)} \right) \; .
\end{gather}
From eq.~(\ref{Eq:DiluteAmpInTermsOfPhi}), the final form of the amplitude is
\begin{gather}
     \mathcal{A}_{3 T, \boldsymbol{\Delta}=\mathbf{0}}^{\text {dilute }} = e_q m_M \frac{g^2}{N_c}(2 \pi) q^{+} \delta\left(q^{+}-p_M^{+}\right)(2 \pi)^2 \delta^2\left(\boldsymbol{q}-\boldsymbol{p}_M\right) \int \frac{\mathrm{d}^d \boldsymbol{\ell}}{(2 \pi)^d} \mathcal{U} (\boldsymbol{\ell}) \nonumber \\   \times \left( \prod_{i=1}^3 \int_0^1 \frac{d x_i}{x_i} \right)  \frac{\delta (1 - x_1 - x_2 - x_3)}{x_3} \frac{\boldsymbol{\ell}^2}{Q^2} \bigg \{ T_{\rm f.} \left[f_{3 M}^V V\left(x_1, x_2\right)-f_{3 M}^A A\left(x_1, x_2\right)\right] \nonumber \\  \times 2 x_1\left(\frac{x_3 c_f}{\boldsymbol{\ell}^2+\frac{x_2 x_3}{x_2+x_3} Q^2}+\frac{x_3 c_f}{\boldsymbol{\ell}^2+\frac{x_1 x_3}{x_1+x_3} Q^2}-\frac{\bar{x}_3\left(1-c_f\right)}{\boldsymbol{\ell}^2+\frac{x_1 x_2}{x_1+x_2} Q^2}+\frac{x_2-\bar{x}_1 c_f}{\boldsymbol{\ell}^2+x_1 \bar{x}_1 Q^2}+\frac{x_1-\bar{x}_2 c_f}{\boldsymbol{\ell}^2+x_2 \bar{x}_2 Q^2}\right) \nonumber \\  - T_{\rm n. f.} \left[f_{3 M}^V V\left(x_1, x_2\right)+f_{3 M}^A A\left(x_1, x_2\right)\right] \nonumber \\  \left.\times\left(\frac{\left(1-c_f\right) x_1 \bar{x}_3}{\bar{x}_3 \boldsymbol{\ell}^2+x_1 x_2 Q^2}-\frac{c_f x_3^2}{\bar{x}_1 \boldsymbol{\ell}^2+x_2 x_3 Q^2}-\frac{\left(x_2-\bar{x}_1 c_f\right) x_1 x_2}{\bar{x}_1\left(\boldsymbol{\ell}^2+x_1 \bar{x}_1 Q^2\right)}-\frac{\left(x_1-\bar{x}_2 c_f\right) \bar{x}_2}{\left(\boldsymbol{\ell}^2+x_2 \bar{x}_2 Q^2\right)}\right)\right\} \; ,
\end{gather}
which is again in agreement with eq.~(167) of ref.~\cite{Anikin:2009bf}. \\

In summary, the agreement of final results is very remarkable, especially considering how different the methodologies developed in ref.~\cite{Anikin:2009bf} and in the present paper are.




\section{Conclusion and outlook}
\label{Sec:Conc}

Diffractive productions are important channels to perform gluon tomography in the nucleon. Achieving an appropriate level of precision calls for both full NLO description and beyond leading power corrections within a saturation framework. Our present result is the first step towards this second need. We obtain, for the very first time, a complete description of the whole $\gamma^{(*)} \to M(\rho, \phi, \omega)$ transition amplitude with arbitrary polarizations of both incoming photon and outgoing light vector meson, including the ones which appear at next-to-leading power. The non-forward kinematics permits to probe the Wigner function of the target and to capture information on the perturbative $t$-dependence of the cross-section. This is of particular interest in large-$t$ exclusive diffraction meson photoproduction. \\  

The $\gamma^{(*)} p \to M(\rho, \phi, \omega) p$ processes have already been experimentally investigated at HERA and are of special interest at the LHC and future EIC experiments. The EDMPs in the case of light vector mesons are particularly sensitive to saturation effects due to the large size of light meson wavefunctions, which indeed probe the region in which gluonic saturation changes the shape of the cross-sections. It is already known experimentally that, for exclusive light pseudo-scalar meson production, higher-twist corrections are essential to describe medium energy JLab data~\cite{JeffersonLabHallA:2016wye}. We want to stress here that, in order to be sensitive to gluonic saturation, the hard scale of the process (e.g. $Q^2$ in the electroproduction case), should be rather small (below the saturation scale). Because of that, power suppressed contributions are expected to be considerably sizable also at high energy. Thus, the newly developed method, both systematic and universal, can be directly employed to determine power-suppressed contributions for twist-2 dominated processes, which would presumably require accounting for higher power contributions.

\acknowledgments
We thank Valerio Bertone, Guillame Beuf, Giovanni A. Chirilli, Andrey V. Grabovsky, Edmond Iancu, Saad Nabeebaccus, Alessandro Papa, Simone Rodini and Jakob Schoenleber for useful discussions. This  project  has  received  funding  from  the  European  Union’s  Horizon  2020  research  and  innovation program under grant agreement STRONG–2020 (WP 13 "NA-Small-x"). The work by M.~F. is supported by Agence Nationale de la Recherche under the contract ANR-17-CE31-0019. M. F. acknowledges support from the Italian Foundation “Angelo della Riccia”. The  work of  L.~S. is  supported  by  the  grant  2019/33/B/ST2/02588  of  the  National  Science Center in  Poland. L.~S. thanks the P2IO Laboratory of Excellence (Programme Investissements d'Avenir ANR-10-LABEX-0038) and the P2I - Graduate School of Physics of Paris-Saclay University for support. This work was also partly supported by the French CNRS via the GDR QCD.

\appendix

\section{Light-cone components and Fierz decomposition}
\label{sec:Light-conecomponentsPlusFierz}
For convenience of the reader we remind some conventions which we use in the main text.
\subsection{Basic definition and light-cone components}
\label{sec:Light-conecomponents}
We introduce two light-cone four-vectors $n_1$ and $n_2$ such that $n_1 \cdot n_2 = 1$ and decompose any arbitrary four-momenta as
\begin{equation}
    a = a^+ n_1^{\mu} +a^- n_2^{\mu} + a_{\perp} \; ,
\end{equation}
where, by definition, $a_{\perp}$ is a four-vector orthogonal to both $n_1$ and $n_2$ and
\begin{equation}
    a^+ = a \cdot n_2 \; , \hspace{1 cm} a^{-} = a \cdot n_1 \; . 
\end{equation}
A scalar product is written as
\begin{equation}
    a \cdot b = a^{+} b^{-} + a^{-} b^{+} + a_{\perp} \cdot b_{\perp} \; .
\end{equation}
In general, transverse vectors in Minkowski space will be denoted with $\perp$ subscripts, and in
Euclidean space they will be denoted as bold characters. \\

\noindent We extensively use the free quark propagator in momentum space,
\begin{equation}
    G_0 (k) = \frac{i \slashed{k}}{k^2+i0} \; ,
\end{equation}
and the gluon propagator in coordinate space, in the $n_2$ light-cone gauge,
\begin{equation}
    G^{\mu \sigma_{\perp}} (z) = \int \frac{d^D l}{(2 \pi)^D} e^{-i l \cdot z} \frac{-i}{l^2+i0} \left( g^{\mu \sigma}_{\perp} - \frac{n_2^{\mu} l_{\perp}^{\sigma}}{l^+} \right)  \; .
\end{equation}
\subsection{Fierz decomposition in Dirac space}
\begin{figure}
    \centering
    \includegraphics[scale=0.50]{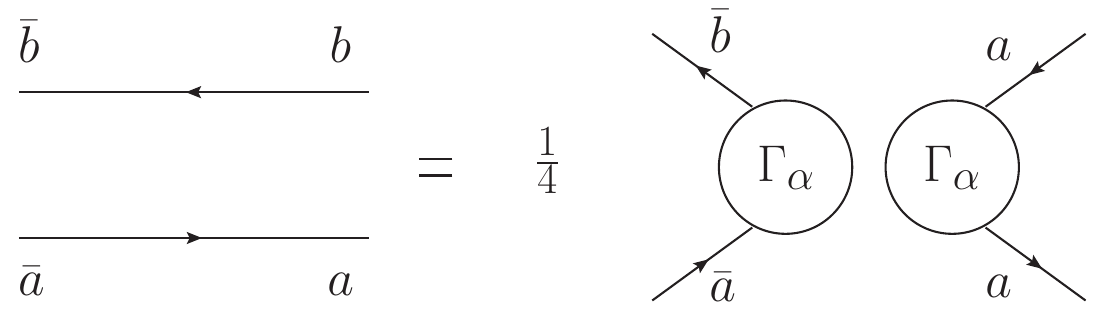}
\caption{Fierz identity in Dirac space.}
  \label{fig:FierzDirac}
\end{figure}
In 4-dimensions, there are 16-independent matrices
\begin{center}
\begin{tabular}{ c c c c c}
 $\Gamma_s$ & $\Gamma_V^{\mu}$ & $\Gamma_T^{\mu \nu}$ & $\Gamma_A^{\mu}$ & $\Gamma_P$ \vspace{0.4 cm} \\ 
 $I$ & $\gamma^{\mu}$ & $\sigma^{\mu \nu} = \frac{i}{2} \left[ \gamma^{\mu} , \gamma^{\nu} \right]$ & $ \gamma^5 \gamma^{\mu}$ & $i \gamma^5$   
\end{tabular}
\end{center}
If we denote one element of this basis as $\Gamma^{\alpha}$ with $\alpha \in [1,16]$, then, we can define the set of inverse matrices  $\Gamma_{\alpha} = \left( \Gamma^{\alpha} \right)^{-1}$, as
\begin{center}
\begin{tabular}{ c c c c c}
 $\Gamma_s$ & $\Gamma_{V, \mu}$ & $\Gamma_{T, \mu \nu}$ & $\Gamma_{A, \mu}$ & $\Gamma_P^{-1}$ \vspace{0.4 cm} \\ 
 $I$ & $\gamma_{\mu}$ & $\sigma_{\mu \nu} = \frac{i}{2} \left[ \gamma_{\mu} , \gamma_{\nu} \right]$ & $ -\gamma^5 \gamma_{\mu}$ & $-i \gamma^5$   
\end{tabular}
\end{center}
It is easy to observe that
\begin{equation}
   \left[ \bar{\psi} \Gamma^{\alpha} \psi \right]^{\dagger} = \bar{\psi} \Gamma^{\alpha} \psi \; , 
\end{equation}
i.e. any $\bar{\psi} \Gamma^{\alpha} \psi$ quantity is hermitian. The 16 Fierz matrices also satisfy the relation
\begin{equation}
    {\rm tr}_D \left[ \Gamma^{\alpha} \Gamma_{\beta} \right] = 4 \delta^{\alpha}_{\beta} \; ,
\end{equation}
from which it is easy to understand that any $4 \times 4$ matrix can expanded as
\begin{equation}
    X = x_{\alpha} \Gamma^{\alpha} \hspace{1.5 cm} {\rm with} \hspace{1.5 cm} x_{\alpha} = \frac{1}{4} {\rm tr}_D \left[ X \Gamma_{\alpha} \right] \; .
\end{equation}
The matrix element $X_{\bar{b} \bar{a}}$ thus satisfy (denoting bispinor indices with latin letters and labelling the Dirac
matrices with greek letters)
\begin{equation}
    X_{ \bar{b} \bar{a} } = \delta_{\bar{b} b} \delta_{\bar{a} a} X_{ b a } = \frac{1}{4} \Gamma_{\alpha, \bar{b} \bar{a}} \Gamma^{\alpha}_{a b} X_{ b a } 
\end{equation}
from which we obtain the Fierz identity in Dirac space
\begin{equation}
    \delta_{\bar{b} b} \delta_{\bar{a} a} = \frac{1}{4} \Gamma_{\alpha, \bar{b} \bar{a}} \Gamma^{\alpha}_{a b} \; ,
    \label{Eq:FierzIdentityDirac}
\end{equation}
pictorially depicted in fig.~(\ref{fig:FierzDirac}).

\subsection{Fierz decomposition in color space}
\begin{figure}
 \includegraphics[scale=0.50]{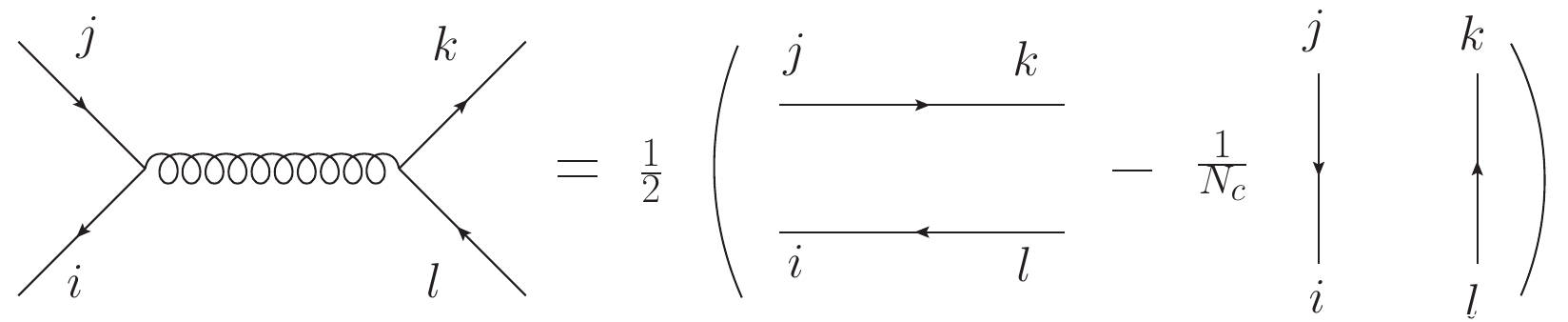}
\caption{Fierz identity in color space.}
  \label{fig:FierzColor}
\end{figure}

In color space, any $N \times N$ matrix can be expanded as
\begin{equation}
    A = c^0 I + c^a t^a \; .
\end{equation}
Then, by using the normalization 
\begin{equation}
    {\rm tr} \left( t^a t^b \right) = \frac{1}{2} \delta^{ab} \; , 
\end{equation}
a Fierz decomposition in color space can be obtained and reads
\begin{equation}
    t^{a}_{lk} t^{a}_{ij} = \frac{1}{2} \left( \delta_{il} \delta_{jk} - \frac{1}{N_c} \delta_{ij} \delta_{lk} \right) \; .
     \label{Eq:FierzIdentityColor}
\end{equation}
It is pictorially depicted in fig.~\ref{fig:FierzColor}.


\section{LCCF and comparison with CCF}
\label{LCCF_and_comparison_with_CCF}
In the present work, we rely on the CCF approach in order to calculate the higher twist corrections. An alternative approach is represented by the LCCF approach~\cite{Anikin:2002wg,Anikin:2009bf,Anikin:2009hk}. For completeness, we provide here a brief summary of the LCCF framework and we review the comparision with the CCF one. The equivalence of the two approaches has been already established in~\cite{Anikin:2009bf,Anikin:2009hk}.
\subsection{Light Cone Collinear Factorization}
The general idea of LCCF is to perform a Taylor expansion of any loop momentum $k$ in fig.~\ref{fig:GenericAmp} around the direction collinear to the produced meson momentum. LCCF is easiest to set up within a given choice of reference frame and gauge. In particular, we will choose the frame in which the meson flies in the +-direction and we will choose $n_2$ light cone gauge. We make use of the Sudakov decomposition described in appendix~\ref{sec:Light-conecomponents}, where one light-cone vector is provided by $p_M$\footnote{Within twist-3, any mass effect is neglected.}. We fix $p_M=p_M^{+}n_{1}$ and we parametrize $k\equiv k^{+}n_{1}+l \equiv xp_M+l$ and expand for small values of $l$. If $p_M^{+}$ is the dominant direction and by definition $l^{+}=0$, $l_{\perp}$ is a first order correction, while $l^{-}$ is a second order correction ($l^{-}\sim\frac{\boldsymbol{l}^{2}}{2xp_M^{+}}\ll\boldsymbol{l}$). Since $x=k^{+}/p_M^{+}$ and having $l=k-xp_M$, we get
\begin{align}
{\cal A}_{2} & =\frac{p_M^{+}}{4N_{c}}\int\frac{{\rm d} x}{2\pi} \int\frac{{\rm d}l^{-}}{2\pi}\int\frac{{\rm d}^{2}\boldsymbol{l}}{\left(2\pi\right)^{2}}\int{\rm d}^{4}z \; {\rm e}^{-ixp_M^{+}z^{-}-il^{-}z^{+}+i(\boldsymbol{l}\cdot\boldsymbol{z})}\nonumber \\
 & \times\left\langle M(p_M)\left|\,\overline{\psi}(z)\Gamma_{\lambda}\psi(0)\right|0\right\rangle {\rm tr}\left[H_{2}(xp_M+l)\Gamma^{\lambda}\right] \; .
\end{align}
Now, we perform a Taylor expansion of the hard part around $k=xp_M$, i.e. 
\begin{equation}
H_{2}(xp_M+l) = H_{2}(xp_M) + l_{\perp\mu} \left[ \frac{\partial}{\partial l_{\perp \mu}} H_{2}(x p_M + l) \right]_{k = x p_M} \hspace{-0.9 cm} + {\rm h.t.} \equiv H_{2}(xp_M) + l_{\perp\mu} \partial_{\perp}^{\mu} H_{2}(x p_M)+{\rm h.t.} ,
\end{equation}
where h.t. stays for higher-twist contributions, which will be always denoted similarly everywhere in the following.
The $l^{-}$ integral yields a simple $\delta(z^{+})$. We integrate the $l_{\perp}^{\mu}$ factor by parts into a derivative acting on the operators, then the $l_{\perp}$ integrals yields a $\delta$ function as well and we get
\begin{align}
{\cal A}_{2} & =\frac{p_M^{+}}{4N_{c}}\int{\rm d}x\int\frac{{\rm d}z^{-}}{2\pi}{\rm e}^{-ixp_M^{+}z^{-}} \nonumber \\
 & \times\left\{ \left\langle M(p_M)\left|\,\overline{\psi}(z^{-})\Gamma_{\lambda}\psi(0)\right|0\right\rangle {\rm tr}\left[H_{2}(xp_M)\Gamma^{\lambda}\right]\right.\nonumber \\
 & \left.+i\left\langle M(p_M)\left|\,\overline{\psi}(z^{-})\overleftrightarrow{\partial}_{\perp\mu}\Gamma_{\lambda}\psi(0)\right|0\right\rangle {\rm tr}\left[\partial_{\perp}^{\mu}H_{2} (xp_M)\Gamma^{\lambda}\right]\right\} \; ,
 \label{Eq:2BodyConCollFact}
\end{align}
where we used 
\begin{equation}
\frac{\partial}{\partial z_{\perp\mu}}\left\langle M(p_M)\left|\,\overline{\psi}(z)\Gamma_{\lambda}\psi(0)\right|0\right\rangle =-\left\langle M(p_M)\left|\,\overline{\psi}(z)\overleftrightarrow{\partial}_{\perp\mu}\Gamma_{\lambda}\psi(0)\right|0\right\rangle \; ,
\end{equation}
with $\overleftrightarrow{\partial}_{\perp\mu}=\frac{1}{2}(\overrightarrow{\partial}_{\perp\mu}-\overleftarrow{\partial}_{\perp\mu})$. \\

The second term in eq.~(\ref{Eq:2BodyConCollFact}) breaks QCD gauge invariance, that can be only restored including the 3-body term to complete the standard derivative into a covariant derivative. This 3-body term is easier to write since only the first term in the Taylor expansion contributes:
\begin{align}
{\cal A}_{3} & =\frac{\left(p_M^{+}\right)^{2}}{2(N_c^2-1)}\int{\rm d}x_{q}\,{\rm d}x_{g}\int\frac{{\rm d}z_{q}^{-}}{2\pi}\frac{{\rm d}z_{g}^{-}}{2\pi}{\rm e}^{-ix_{q}p_M^{+}z_{q}^{-}-ix_{g}p_M^{+}z_{g}^{-}} \nonumber \\
 & \times\left\langle M(p_M)\left|\,\overline{\psi}(z_{q}^{-})\Gamma_{\lambda}gA_{\mu}\left(z_{g}^{-}\right)\psi(0)\right|0\right\rangle {\rm tr}\left[t^b H_{3}^{ \mu , b }(x_{q}p_M,x_{g}p_M)\Gamma^{\lambda}\right] \; .
 \label{Eq:3BodyConCollFact}
\end{align}

\subsubsection{Distribution amplitudes parametrization}
Now, it is necessary to give the parametrizations of the various DAs. We have\footnote{See~\cite{Anikin:2009bf} and references therein.}: \\

\noindent \textbf{2-body vector matrix element without derivative} 
\begin{gather}
  p_M^{+}\int\frac{{\rm d}z^{-}}{2\pi}{\rm e}^{-ixp_M^{+}z^{-}}\left\langle M(p_M)\left|\,\overline{\psi}(z^{-})\gamma_{\lambda}\psi(0)\right|0\right\rangle \nonumber \\ = m_{M}f_{M}\left[\varphi_{1}(x)\frac{\varepsilon_M^{\ast+}}{p_M^{+}} p_{M \lambda} + \varphi_{3}(x)\varepsilon_{M \perp\lambda}^{\ast}\right] \; .
\end{gather}
\textbf{2-body axial-vector matrix element without derivative} 
\begin{gather}
 - p_M^{+}\int\frac{{\rm d}z^{-}}{2\pi}{\rm e}^{-ixp_M^{+}z^{-}}\left\langle M(p_M)\left|\,\overline{\psi}(z^{-})\gamma_{\lambda}\gamma_{5}\psi(0)\right|0\right\rangle \nonumber \\
  = im_{M}f_{M}\varphi_{A}(x)\epsilon_{\lambda\alpha+-}\varepsilon_{M \perp}^{\ast\alpha} \; .
\end{gather}
\textbf{2-body vector matrix element with derivative} 
\begin{gather}
 i p_M^{+}\int\frac{{\rm d}z^{-}}{2\pi}{\rm e}^{-ixp_M^{+}z^{-}}\left\langle M(p_M)\left|\,\overline{\psi}(z^{-})\overleftrightarrow{\partial}_{\perp\mu}\gamma_{\lambda}\psi(0)\right|0\right\rangle \nonumber \\
  = m_{M}f_{M}\varphi_{1}^{T}(x)\varepsilon_{M \perp\mu}^{\ast}p_{M \lambda} \; .
\end{gather}
\textbf{2-body axial-vector matrix element with derivative} 
\begin{gather}
  -i p_M^{+}\int\frac{{\rm d}z^{-}}{2\pi}{\rm e}^{-ixp_M^{+}z^{-}}\left\langle M(p_M)\left|\,\overline{\psi}(z^{-})\overleftrightarrow{\partial}_{\perp\mu}\gamma_{\lambda}\gamma_{5}\psi(0)\right|0\right\rangle \nonumber \\
  = i \; m_{M}f_{M}\varphi_{A}^{T}(x)p_{M\lambda}\epsilon_{\mu\alpha+-}\varepsilon_{M\perp}^{\ast\alpha} \; .
\end{gather}
\textbf{3-body vector matrix element} 
\begin{gather}
  (p_M^{+})^{2}\int\frac{{\rm d}z_{q}^{-}}{2\pi}\frac{{\rm d}z_{g}^{-}}{2\pi}{\rm e}^{-ix_{q}p_M^{+}z_{q}^{-}-ix_{g}p_M^{+}z_{g}^{-}} \nonumber \\ \times \left\langle M(p_M)\left|\,\overline{\psi}(z_{q}^{-})\gamma_{\lambda}gA_{\mu}(z_{g}^{-})\psi(0)\right|0\right\rangle = m_{M}f_{3M}^{V}B(x_{q},x_{q}+x_{g};x_{g})p_{M \lambda}\varepsilon_{M \perp\mu}^{\ast} \; .
\end{gather}
\textbf{3-body axial-vector matrix element} 
\begin{gather}
  - (p_M^{+})^{2}\int\frac{{\rm d}z_{q}^{-}}{2\pi}\frac{{\rm d}z_{g}^{-}}{2\pi}{\rm e}^{-ix_{q}p_M^{+}z_{q}^{-}-ix_{g}p_M^{+}z_{g}^{-}} \nonumber \\ \times \left\langle M(p_M)\left|\,\overline{\psi}(z_{q}^{-})\gamma_{\lambda}\gamma_{5}gA_{\mu}(z_{g}^{-})\psi(0)\right|0\right\rangle 
  = - im_{M}f_{3M}^{A}D(x_{q},x_{q}+x_{g};x_{g})p_{M \lambda}\epsilon_{\mu\nu-+}\varepsilon_{M\perp}^{\ast\nu} \; .
\end{gather}

\subsubsection{LCCF amplitude for the production of a transverse meson}
Using the explicit parametrizations, given in the previous subsection, in eqs.~(\ref{Eq:2BodyConCollFact}) and (\ref{Eq:3BodyConCollFact}), we can get the explicit form of the amplitudes for the production of a light vector meson up to twist-3 in the LCCF framework, we have: \\

\noindent\textbf{2-body vector amplitude}
\begin{align}
{\cal A}_{2V} & =\frac{m_{M}f_{M}}{4N_{c}}\int{\rm d}x\left\{ {\rm tr}\left[H_{2}(xp_M)\slashed{\varepsilon}_{M \perp}^{\ast}\right]\varphi_{3}(x)+{\rm tr}\left[(\varepsilon_{M \perp}^{\ast}\cdot\partial)H_{2}(xp_M)\slashed{p}_M\right]\varphi_{1}^{T}(x)\right\} \; .
\label{Eq:AppLCCFDAs1}
\end{align}
\textbf{2-body axial-vector amplitude}
\begin{align}
{\cal A}_{2A} & =\frac{im_{M}f_{M}}{4N_{c}}\epsilon^{\mu\alpha-+}\varepsilon_{M \perp\alpha}^{\ast} \nonumber \\ & \times \int{\rm d}x\left\{ \varphi_{A}(x){\rm tr}\left[H_{2}(xp_M)\gamma_{\mu}\gamma_{5}\right]+\varphi_{A}^{T}(x){\rm tr}\left[(\partial_{\mu}H_{2})(xp_M)\slashed{p}_M\gamma_{5}\right]\right\} \; . 
\label{Eq:AppLCCFDAs2}
\end{align}
\textbf{3-body vector amplitude}
\begin{align}
{\cal A}_{3V} & =\frac{m_{M}f_{3M}^{V}}{2 (N_c^2-1)}\int{\rm d}x_{q}\,{\rm d}x_{\overline{q}}\,{\rm d}x_{g}\,\delta(1-x_{q}-x_{\overline{q}}-x_{g})\nonumber \\
 & \times B(x_{q},1-x_{\overline{q}};x_{g})\,\varepsilon_{M \perp}^{\ast\mu}{\rm tr}\left[t^b H_{3\mu}^{b} (x_{q}p_M,x_{g}p_M)\slashed{p}_M\right] \; .
 \label{Eq:AppLCCFDAs3}
\end{align}
\textbf{3-body axial-vector amplitude}
\begin{align}
{\cal A}_{3A} & =-i\frac{m_{M}f_{3M}^{A}}{2 (N_c^2-1)}\int{\rm d}x_{q}\,{\rm d}x_{\overline{q}}\,{\rm d}x_{g}\,\delta(1-x_{q}-x_{\overline{q}}-x_{g}) \nonumber \\
 & \times D(x_{q},1-x_{\overline{q}};x_{g})\,\epsilon^{\mu}_{\; \; \nu-+}\varepsilon_{M \perp}^{\ast\nu}{\rm tr}\left[t^b H_{3\mu}^{b} (x_{q}p_M,x_{g}p_M)\slashed{p}_M\gamma_{5}\right] \; .
 \label{Eq:AppLCCFDAs4}
\end{align}

\subsection{CCF amplitudes for the production of a transverse meson}
With the aim of finding the matching between DAs in the LCCF and in the CCF, we write the analogous expressions for the amplitudes in eqs.~(\ref{Eq:AppLCCFDAs1},~\ref{Eq:AppLCCFDAs2},~\ref{Eq:AppLCCFDAs3},~\ref{Eq:AppLCCFDAs4}) in the CCF framework. The amplitudes for the production of a light-meson up to twist-3 are \\

\noindent \textbf{2-body vector amplitude} \vspace{0.2 cm} \\
\begin{gather}
{\cal A}_{2V} =\frac{f_{M}m_{M}}{4N_{c}}\int_{0}^{1}\!{\rm d}x\int\frac{{\rm d}^{4}k}{(2\pi)^{4}}\int{\rm d}^{4}z\,{\rm e}^{i\left(xp_M-k\right)\cdot z} \nonumber \\
 \times \left\{ g_{\perp}^{(v)}(x){\rm tr}\left[H_{2}\left(k\right)\slashed{\varepsilon}_M^{\ast}\right]-i\left(\varepsilon_M^{\ast}\cdot z\right)\left[h\left(x\right)-\widetilde{h}\left(x\right)\right]{\rm tr}\left[H_{2}\left(k\right)\slashed{p}_M\right]\right\} \; . 
\end{gather}
Integrating by parts the $\left(\varepsilon_M^{\ast}\cdot z\right)$ term, the $z$ integral yields a $\delta$ function and we
find
\begin{gather}
{\cal A}_{2V} \hspace{-0.1 cm} = \hspace{-0.1 cm} \frac{f_{M}m_{M}}{4N_{c}} \hspace{-0.2 cm} \int_{0}^{1} \hspace{-0.2 cm} {\rm d}x\left\{ {\rm tr}\left[H_{2}\left(xp_M\right)\slashed{\varepsilon}_M^{\ast}\right]g_{\perp}^{(v)}(x)-{\rm tr}\left[\left(\varepsilon_M^{\ast}\cdot\partial\right)H_{2}\left(xp_M\right)\slashed{p}_M\right]\left[h\left(x\right)-\widetilde{h}\left(x\right)\right]\right\} .
\end{gather}

\noindent \textbf{2-body axial-vector amplitude}
\begin{align}
{\cal A}_{2A} & =i\frac{f_{M}m_{M}}{4N_{c}}\frac{1}{4}\varepsilon_{M \alpha}^{\ast}p_{M \beta} \int_{0}^{1}{\rm d}x \; {\rm tr}\left[\partial_{\delta}H_{2}\left(xp_M\right)\gamma_{\lambda}\gamma_{5}\right]\left[\epsilon^{\lambda\alpha\beta\delta}g_{\perp}^{(a)}(x)-\epsilon^{\nu\alpha\beta\delta}\frac{p_M^{\lambda}n_{\nu}}{p_M \cdot n}\widetilde{g}_{\perp}^{(a)}(x)\right] .
\end{align}

\noindent \textbf{3-body vector amplitude}
\begin{gather}
{\cal A}_{3V} =\frac{m_{M}f_{3M}^{V}}{2 (N_c^2-1)} \frac{\left(p_M^{\mu}\varepsilon_M^{\ast}-\varepsilon_M^{\ast\mu}p_M\right)\cdot n}{\left(p_M \cdot n\right)} \int\frac{{\rm d}x_{q}\,{\rm d}x_{\overline{q}}\,{\rm d}x_{g}}{x_{g}+i\epsilon}V(x_{q},x_{\overline{q}}) \nonumber \\ \times\delta\left(1-x_{q}-x_{\overline{q}}-x_{g}\right){\rm tr} \left[ t^b H_{3\mu}^{b} (x_{q}p_M,x_{g}p_M)\slashed{p}_M\right] .  
\end{gather}

\noindent \textbf{3-body axial vector amplitude}
\begin{gather}
{\cal A}_{3A}  =i\epsilon^{\mu\nu\alpha\beta}\frac{m_{M}f_{3M}^{A}}{2 (N_c^2-1)}\frac{p_{M \alpha}n_{\nu}}{p_M \cdot n}\varepsilon_{M \beta}^{\ast}\int\frac{{\rm d}x_{q}\,{\rm d}x_{\overline{q}}\,{\rm d}x_{g}}{x_{g}+i\epsilon} \nonumber \\
  \times{\rm tr}\left[t^b H_{3\mu}^{b} \left(x_{q}p_M,x_{g}p_M\right)\slashed{p}_M \gamma_{5}\right]
  \delta\left(1-x_{q}-x_{\overline{q}}-x_{g}\right)A\left(x_{q},x_{\overline{q}}\right).
\end{gather}

In order to compare the LCCF and the CCF approach, we need to fix the same frame and gauge: a frame where $p_M=p_M^{+} n_{1}$ and the $n_2$ light-cone gauge. Then, the amplitudes become: \\ 

\noindent \textbf{2-body vector amplitude}
\begin{align}
{\cal A}_{2V} & =\frac{f_{M}m_{M}}{4N_{c}} \int_{0}^{1}{\rm d}x \nonumber \\ & \times \left\{ {\rm tr}\left[H_{2}\left(xp_M\right)\slashed{\varepsilon}_{M \perp}^{\ast}\right]g_{\perp}^{\left(v\right)}\left(x\right)-\varepsilon_{M \perp}^{\ast\mu}{\rm tr}\left[\partial_{\perp\mu}H_{2}\left(xp_M\right)\slashed{p}_M\right]\left[h\left(x\right)-\widetilde{h}\left(x\right)\right]\right\} .
\end{align}
\noindent \textbf{2-body axial-vector amplitude}
\begin{gather}
{\cal A}_{2A} = i\frac{f_{M}m_{M}}{4N_{c}}\frac{1}{4}\varepsilon_{M \perp\alpha}^{\ast}p_M^{+}\int_{0}^{1}{\rm d}x\left[\epsilon^{\lambda\alpha-\delta}g_{\perp}^{\left(a\right)}\left(x\right)-\epsilon^{+\alpha-\delta}n_{1}^{\lambda}\widetilde{g}_{\perp}^{\left(a\right)}\left(x\right)\right]{\rm tr}\left[\partial_{\delta}H_{2}\left(xp_M\right)\gamma_{\lambda}\gamma_{5}\right] \; .
\end{gather}
Indices $\alpha$ and $\delta$ in the Levi-Civita $\epsilon$ tensor of the second term can
only be transverse. In the first one, however, one of them is allowed
to be $+$. We have
\begin{gather}
{\cal A}_{2A} =-i\frac{f_{M}m_{M}}{4N_{c}}\frac{1}{4}\varepsilon_{M \perp\alpha}^{\ast}\epsilon^{\mu\alpha+-}\int_{0}^{1}{\rm d}xg_{\perp}^{\left(a\right)}\left(x\right){\rm tr}\left[p_M^{+}\partial^{-}H_{2}\left(xp_M\right)\gamma_{\perp\mu}\gamma_{5}\right] \nonumber \\ 
+i\frac{f_{M}m_{M}}{4N_{c}}\frac{1}{4}\varepsilon_{M \perp\alpha}^{\ast}\epsilon^{\alpha\mu-+}\int_{0}^{1}{\rm d}x\left[g_{\perp}^{\left(a\right)}\left(x\right)-\widetilde{g}_{\perp}^{\left(a\right)}\left(x\right)\right]{\rm tr}\left[\partial_{\perp\mu}H_{2}\left(xp_M\right)\slashed{p}_M \gamma_{5}\right]. 
\end{gather}
We can actually rewrite $p_M^{+}\partial^{-}H_{2}\left(xp_M\right)$ into
$\partial_{x}H_{2}\left(xp_M\right)$ and get
\begin{align}
{\cal A}_{2A} & =-i\frac{f_{M}m_{M}}{4N_{c}}\frac{1}{4}\varepsilon_{M \perp\alpha}^{\ast}\epsilon^{\mu\alpha+-}\int_{0}^{1}{\rm d}xg_{\perp}^{\left(a\right)}\left(x\right){\rm tr}\left[\frac{\partial H_{2}}{\partial x}\left(xp_M\right)\gamma_{\perp\mu}\gamma_{5}\right] \nonumber \\ &
  +i\frac{f_{M}m_{M}}{4N_{c}}\frac{1}{4}\varepsilon_{M \perp\alpha}^{\ast}\epsilon^{\alpha\mu-+}\int_{0}^{1}{\rm d}x\left[g_{\perp}^{\left(a\right)}\left(x\right)-\widetilde{g}_{\perp}^{\left(a\right)}\left(x\right)\right]{\rm tr}\left[\partial_{\perp\mu}H_{2}\left(xp_M\right)\slashed{p}_M\gamma_{5}\right] \; .
\end{align}
Integrating by parts finally yields
\begin{align}
{\cal A}_{2A} & =-i\frac{f_{M}m_{M}}{4N_{c}}\frac{1}{4}\varepsilon_{M \perp\alpha}^{\ast}\epsilon^{\mu\alpha-+}\int_{0}^{1}{\rm d}x\frac{\partial g_{\perp}^{\left(a\right)}}{\partial x}\left(x\right){\rm tr}\left[H_{2}\left(xp_M\right)\gamma_{\perp\mu}\gamma_{5}\right] \nonumber \\
 & -i\frac{f_{M}m_{M}}{4N_{c}}\frac{1}{4}\varepsilon_{M \perp\alpha}^{\ast}\epsilon^{\mu\alpha-+}\int_{0}^{1}{\rm d}x\left[g_{\perp}^{\left(a\right)}\left(x\right)-\widetilde{g}_{\perp}^{\left(a\right)}\left(x\right)\right]{\rm tr}\left[\partial_{\perp\mu}H_{2}\left(xp_M\right)\slashed{p}_M\gamma_{5}\right]. 
\end{align}
\noindent \textbf{3-body vector amplitude}
\begin{align}
{\cal A}_{3V} & =-\frac{1}{2 (N_c^2-1)}m_{M}f_{3M}^{V}\int\frac{{\rm d}x_{q}{\rm d}x_{\overline{q}}{\rm d}x_{g}}{x_{g}+i\epsilon}V\left(x_{q},x_{\overline{q}}\right) \nonumber \\
 & \times\delta\left(1-x_{q}-x_{\overline{q}}-x_{g}\right)\varepsilon_{M \perp}^{\ast\mu}{\rm tr}\left[ t^b H_{3\mu}^b \left(x_{q}p_M,x_{g}p_M\right)\slashed{p}_M\right] \; . 
\end{align}
\noindent \textbf{3-body axial vector amplitude}
\begin{align}
{\cal A}_{3A} & =i\epsilon^{\mu\nu\alpha\beta}\frac{m_{M}f_{3M}^{A}}{2(N_c^2-1)}\frac{p_{M \alpha}n_{\nu}}{p_M\cdot n}\varepsilon_{M \beta}^{\ast}\int\frac{{\rm d}x_{q}{\rm d}x_{\overline{q}}{\rm d}x_{g}}{x_{g}+i\epsilon}\nonumber \\ & \times{\rm tr}\left[t^b H_{3\mu}^{b} \left(x_{q}p_M,x_{g}p_M\right)\slashed{p}_M\gamma_{5}\right] \delta\left(1-x_{q}-x_{\overline{q}}-x_{g}\right)A\left(x_{q},x_{\overline{q}}\right).  
\end{align}

\subsection{Matching relations}

Comparing the results from last section and from the LCCF amplitude
section, we get the matching relations between the schemes, i.e. 
\begin{equation}
    \varphi_{1} (x) = \phi(x) \; ,
\end{equation}
\begin{equation}
\varphi_{1}^{T}\left(x\right)=\widetilde{h}\left(x\right)-h\left(x\right) \; ,
\end{equation}
\begin{equation}
B\left(x_{q},x_{\overline{q}};x_{g}\right)=\frac{V\left(x_{q},1-x_{\overline{q}};x_{g}\right)}{x_{q}-x_{\overline{q}}} \; ,
\end{equation}
\begin{equation}
D\left(x_{q},x_{\overline{q}};x_{g}\right)=\frac{A\left(x_{q},1-x_{\overline{q}};x_{g}\right)}{x_{q}-x_{\overline{q}}} \; ,
\end{equation}
\begin{equation}
\varphi_{A}\left(x\right)=-\frac{1}{4}\frac{\partial g_{\perp}^{\left(a\right)}\left(x\right)}{\partial x} \; ,
\end{equation}
\begin{equation}
\varphi_{A}^{T}\left(x\right)=-\frac{1}{4}\left[g_{\perp}^{\left(a\right)}\left(x\right)-\widetilde{g}_{\perp}^{\left(a\right)}\left(x\right)\right] \; ,
\end{equation}
\begin{equation}
\varphi_{3}\left(x\right) = g_{\perp}^{\left(v\right)}\left(x\right) \; .
\end{equation}

\section{Dirac trace in the 3-body contributions}
\label{Sec:AppendixB}
Since the simplifications of Dirac traces in the 3-body case are quite tedious, we give here some useful details for the reader's convenience.

\subsection{Quark emission case}
\label{Sec:AppendixB1}
Our aim is to evaluate the trace
\begin{gather}
    {\rm tr}_{D}\left[ \gamma^{+}\frac{\slashed{k}_{14}}{2k_{14}^{+}}\gamma_{\mu} \left(g_{\perp}^{\mu\sigma}-\frac{\ell_{\perp}^{\sigma}}{\ell^{+}}n_{2}^{\mu}\right) \left( \frac{\gamma^{+}}{2k_{40}^{+}} -\frac{\left(\frac{\boldsymbol{k}_{40}^{2}}{2k_{40}^{+}}\gamma^{+}+k_{40}^{+}\gamma^{-}+\slashed{k}_{40\perp}\right)}{2k_{40}^{+} \left(\frac{\boldsymbol{k}_{40}^{2}-i0}{2k_{40}^{+}}+\frac{\boldsymbol{k}_{20}^{2}-i0}{2k_{20}^{+}}-q^{-} \right)} \right) \slashed{\varepsilon}_{q}\frac{\slashed{k}_{20}}{2k_{20}^{+}}\gamma^{+}\Gamma_{\lambda} \right] \nonumber \\
    = \frac{1}{8 k_{14}^{+} k_{40}^{+} k_{20}^{+} } {\rm tr}_{D}\left[ \gamma^{+} \slashed{k}_{14} \gamma_{\mu} \left(g_{\perp}^{\mu\sigma}-\frac{\ell_{\perp}^{\sigma}}{\ell^{+}}n_{2}^{\mu}\right) \gamma^{+}  \slashed{\varepsilon}_{q} \slashed{k}_{20} \gamma^{+}\Gamma_{\lambda} \right] -  \frac{1}{16 k_{14}^{+} (k_{40}^{+})^2 k_{20}^{+} } \nonumber \\
    \times \frac{1}{\left(\frac{\boldsymbol{k}_{40}^{2}-i0}{2k_{40}^{+}}+\frac{\boldsymbol{k}_{20}^{2}-i0}{2k_{20}^{+}}-q^{-} \right)} {\rm tr}_{D}\left[ \gamma^{+} \slashed{k}_{14} \gamma_{\mu} \left(g_{\perp}^{\mu\sigma}-\frac{\ell_{\perp}^{\sigma}}{\ell^{+}}n_{2}^{\mu}\right) \slashed{k}_{40} \gamma^+ \slashed{k}_{40}  \slashed{\varepsilon}_{q} \slashed{k}_{20} \gamma^{+}\Gamma_{\lambda} \right] \; ,
    \label{Eq:TraceAppB1Start}
\end{gather}
where in the second term we first used $2 k_{40}^{+} = \left( \gamma^+ \slashed{k}_{40} + \slashed{k}_{40} \gamma^+ \right)$ and then removed the vanishing term proportional to $k_{40}^2=0$.\\
Now, we give some useful relations involving the object inside the traces: 
\begin{gather}
  \gamma^{+}\slashed{k}_{40}\slashed{\varepsilon}_{q}\left(\slashed{q}-\slashed{k}_{40}\right)\gamma^{+} 
  =  -2\varepsilon_{q}^{+}\left[\left(\boldsymbol{k}_{40}-\frac{k_{40}^{+}}{q^{+}}\boldsymbol{q}\right)^{2}+\frac{k_{40}^{+}\left(q^{+}-k_{40}^{+}\right)}{\left(q^{+}\right)^{2}}Q^{2}\right]\gamma^{+} + \left(\varepsilon_{q\rho}-\frac{\varepsilon_{q}^{+}}{q^{+}}q_{\rho}\right) \nonumber \\ \times \left\{ 4\frac{k_{40}^{+}\left(q^{+}-k_{40}^{+}\right)}{q^{+}}q^{\rho}\gamma^{+}\right. 
  \left.+2\left(k_{40\perp\alpha}-\frac{k_{40}^{+}}{q^{+}}q_{\perp\alpha}\right)\left[\left(q^{+}-k_{40}^{+}\right)\gamma_{\perp}^{\alpha}\gamma_{\perp}^{\rho}-k_{40}^{+}\gamma_{\perp}^{\rho}\gamma_{\perp}^{\alpha}\right]\gamma^{+}\right\} \; ,
\end{gather}
\begin{gather}
  \gamma^{+}\slashed{\varepsilon}_{q}\slashed{k}_{20}\gamma^{+}\Gamma_{\lambda} = -2\left[\varepsilon_{q}^{+}\left(\slashed{k}_{40\perp}-\frac{k_{40}^{+}}{q^{+}}\slashed{q}_{\perp}\right)+\left(q^{+}-k_{40}^{+}\right)\left(\slashed{\varepsilon}_{q\perp}-\frac{\varepsilon_{q}^{+}}{q^{+}}\slashed{q}_{\perp}\right)\right]\gamma^{+}\Gamma_{\lambda} \; , 
\end{gather}
\begin{gather}
  \gamma^{+}\left(\slashed{k}_{40}-\slashed{\ell}\right)\gamma_{\mu}\left(g_{\perp}^{\mu\sigma}-\frac{\ell_{\perp}^{\sigma}}{\ell^{+}}n_{2}^{\mu}\right)\slashed{k}_{40}\gamma^{+} = 2\left[\left(k_{40}^{+}-\ell^{+}\right)\gamma_{\perp}^{\sigma}\slashed{k}_{40\perp}+k_{40}^{+}\left(\slashed{k}_{40\perp}-\slashed{\ell}_{\perp}\right)\gamma_{\perp}^{\sigma}\right]\gamma^{+}\nonumber \\
  -4\frac{k_{40}^{+}\left(k_{40}^{+}-\ell^{+}\right)}{\ell^{+}}\ell_{\perp}^{\sigma}\gamma^{+} 
  =-2k_{40}^{+}\left(\ell_{\perp\beta}-\frac{\ell^{+}}{k_{40}^{+}}k_{40\perp\beta}\right)\left[\gamma_{\perp}^{\beta}\gamma_{\perp}^{\sigma}+2\left(\frac{k_{40}^{+}-\ell^{+}}{\ell^{+}}\right)g_{\perp}^{\beta\sigma}\right]\gamma^{+} 
\end{gather}
and
\begin{gather}
  \gamma^{+}\slashed{k}_{14}\gamma_{\mu}\left(g_{\perp}^{\mu\sigma}-\frac{\ell_{\perp}^{\sigma}}{\ell^{+}}n_{2}^{\mu}\right)\slashed{k}_{40}\gamma^{+} =2\left[k_{14}^{+}\gamma_{\perp}^{\sigma}\slashed{k}_{40\perp}+k_{40}^{+}\slashed{k}_{14\perp}\gamma_{\perp}^{\sigma}-2\frac{k_{14}^{+}k_{40}^{+}}{\ell^{+}}\ell_{\perp}^{\sigma}\right]\gamma^{+}  \; .
\end{gather}
Then, the two traces in eq.~(\ref{Eq:TraceAppB1Start}) read 
\begin{align}
 & {\rm tr}_{D}\left[\gamma^{+}\slashed{k}_{14}\gamma_{\mu}\left(g_{\perp}^{\mu\sigma}-\frac{\ell_{\perp}^{\sigma}}{\ell^{+}}n_{2}^{\mu}\right)\slashed{k}_{40}\gamma^{+}\slashed{k}_{40}\slashed{\varepsilon}_{q}\slashed{k}_{20}\gamma^{+}\Gamma_{\lambda}\right]\nonumber \\
 & =-8\frac{\left(k_{40}^{+}\right)^{2}\left(q^{+}-k_{40}^{+}\right)}{q^{+}}\left(\varepsilon_{q\rho}-\frac{\varepsilon_{q}^{+}}{q^{+}}q_{\rho}\right) \nonumber \\
 & \times q^{\rho}\left(\ell_{\perp\beta}-\frac{\ell^{+}}{k_{40}^{+}}k_{40\perp\beta}\right){\rm tr}_{D} \left[ \left(\gamma_{\perp}^{\beta}\gamma_{\perp}^{\sigma}+2\frac{k_{40}^{+}-\ell^{+}}{\ell^{+}}g_{\perp}^{\beta\sigma}\right)\gamma^{+}\Gamma_{\lambda} \right] \nonumber \\
 & -4k_{40}^{+}\left(\varepsilon_{q\rho}-\frac{\varepsilon_{q}^{+}}{q^{+}}q_{\rho}\right)\left(k_{40\perp\alpha}-\frac{k_{40}^{+}}{q^{+}}q_{\perp\alpha}\right)\left(\ell_{\perp\beta}-\frac{\ell^{+}}{k_{40}^{+}}k_{40\perp\beta}\right)\nonumber \\
 & \times{\rm tr}_{D} \left[ \left(2\frac{k_{40}^{+}}{\ell^{+}}g_{\perp}^{\beta\sigma}-\gamma_{\perp}^{\sigma}\gamma_{\perp}^{\beta}\right)\left[\left(q^{+}-k_{40}^{+}\right)\gamma_{\perp}^{\alpha}\gamma_{\perp}^{\rho}-k_{40}^{+}\gamma_{\perp}^{\rho}\gamma_{\perp}^{\alpha}\right]\gamma^{+}\Gamma_{\lambda} \right] \nonumber \\
 & +4k_{40}^{+}\varepsilon_{q}^{+}\left[\left(\boldsymbol{k}_{40}-\frac{k_{40}^{+}}{q^{+}}\boldsymbol{q}\right)^{2}+\frac{k_{40}^{+}\left(q^{+}-k_{40}^{+}\right)}{\left(q^{+}\right)^{2}}Q^{2}\right] \nonumber \\
 & \times\left(\ell_{\perp\beta}-\frac{\ell^{+}}{k_{40}^{+}}k_{40\perp\beta}\right){\rm tr}_{D} \left[\left(2\frac{k_{40}^{+}}{\ell^{+}}g_{\perp}^{\beta\sigma}-\gamma_{\perp}^{\sigma}\gamma_{\perp}^{\beta}\right) \gamma^{+}\Gamma_{\lambda} \right] \;  
\end{align}
and
\begin{align}
 & {\rm tr}_{D}\left[\gamma^{+}\slashed{k}_{14}\gamma_{\mu}\left(g_{\perp}^{\mu\sigma}-\frac{\ell_{\perp}^{\sigma}}{\ell^{+}}n_{2}^{\mu}\right)\gamma^{+}\slashed{\varepsilon}_{q}\slashed{k}_{20}\gamma^{+}\Gamma_{\lambda}\right] \nonumber \\
 & =-4\left(k_{40}^{+}-\ell^{+}\right)\left(q^{+}-k_{40}^{+}\right){\rm tr}_{D}\left[\gamma_{\perp}^{\sigma}\left(\slashed{\varepsilon}_{q\perp}-\frac{\varepsilon_{q}^{+}}{q^{+}}\slashed{q}_{\perp}\right)\gamma^{+}\Gamma_{\lambda}\right]\nonumber \\
 & -4\left(k_{40}^{+}-\ell^{+}\right)\varepsilon_{q}^{+}{\rm tr}_{D}\left[\gamma_{\perp}^{\sigma}\left(\slashed{k}_{40\perp}-\frac{k_{40}^{+}}{q^{+}}\slashed{q}_{\perp}\right)\gamma^{+}\Gamma_{\lambda}\right]  \; .
\end{align}
Using these last two results in eq.~(\ref{Eq:TraceAppB1Start}), we get the result in eq.~(\ref{Eq:DiracTraceQuarkFinalResult}).

\subsection{Antiquark emission case}
\label{Sec:AppendixB2}
We use the two useful relations
\begin{gather}
  \gamma^{+}\left[\left(x_{2}+x_{3}\right)\slashed{q}+\slashed{k}_{40\perp}\right] \left( \gamma_{\perp}^{\sigma}-\left(\frac{\ell_{\perp}^{\sigma}}{\ell^{+}}+\frac{k_{40\perp}^{\sigma}}{k_{40}^{+}}+\frac{q_{\perp}^{\sigma}}{q^{+}}\right) \gamma^{+} \right) \nonumber \left(x_{2}\slashed{q}+\frac{x_{2}}{x_{2}+x_{3}}\slashed{k}_{40\perp}-\slashed{\ell}_{\perp}\right)\gamma^{+}\Gamma_{\lambda}\\
  =-2\left(x_{2}+x_{3}\right)q^{+}\gamma^{+}\left(\gamma_{\perp}^{\sigma}\slashed{\ell}_{\perp}+2\frac{x_{2}}{x_{3}}\ell_{\perp}^{\sigma}\right)\Gamma_{\lambda}\nonumber 
\end{gather}
and
\begin{gather}
  \gamma^{+}\left(x_{1}\slashed{q}-\slashed{k}_{40\perp}\right)\slashed{\varepsilon}_{q}\left[\left(x_{2}+x_{3}\right)\slashed{q}+\slashed{k}_{40\perp}\right]\gamma^{+}  =-2\varepsilon_{q}^{+}\left[\boldsymbol{k}_{40}^{2}+x_{1}\left(1-x_{1}\right)Q^{2}\right]\gamma^{+} \nonumber \\ +2q^{+}\left(\varepsilon_{q\rho}-\frac{\varepsilon_{q}^{+}}{q^{+}}q_{\rho}\right)\left[2x_{1}\left(1-x_{1}\right)q^{\rho}+x_{1}\gamma_{\perp}^{\rho}\slashed{k}_{40\perp}-\left(1-x_{1}\right)\slashed{k}_{40\perp}\gamma_{\perp}^{\rho}\right]\gamma^{+} \nonumber
\end{gather}
in order to obtain
\begin{gather}
 - x_1   {\rm tr}_{D} \left[ \gamma^{+}\left(x_{1}\slashed{q}-\slashed{k}_{40\perp}\right)\slashed{\varepsilon}_{q} \left(\gamma^{+}-\frac{x_{1}\left[\left(x_{2}+x_{3}\right)\slashed{q}+\slashed{k}_{40\perp}\right]\gamma^{+}\left[\left(x_{2}+x_{3}\right)\slashed{q}+\slashed{k}_{40\perp}\right]}{\boldsymbol{k}_{40}^{2}+x_{1}\left(1-x_{1}\right)Q^{2}-i0}\right) \right. \nonumber \\ \left.
  \times\gamma_{\mu}\left(x_{2}\slashed{q}+\frac{x_{2}}{x_{2}+x_{3}}\slashed{k}_{40\perp}-\slashed{\ell}_{\perp}\right)\gamma^{+}\Gamma_{\lambda} \right] = \frac{4 x_1 (1-x_1) (q^+)^2 \left( \varepsilon_{q, \rho} - \frac{\varepsilon_q^+}{q^+} q_{\rho} \right) }{\boldsymbol{k}_{40}^{2}+x_{1}\left(1-x_{1}\right)Q^{2}} \nonumber \\  \times{\rm tr}_{D} \left[ \left( 2x_{1}\left(1-x_{1}\right)q^{\rho}+x_{1}\gamma_{\perp}^{\rho}\slashed{k}_{40\perp}-\left(1-x_{1}\right)\slashed{k}_{40\perp}\gamma_{\perp}^{\rho}\right) \left(\gamma_{\perp}^{\sigma}\slashed{\ell}_{\perp}+2\frac{x_{2}}{x_{3}}\ell_{\perp}^{\sigma}\right)\gamma^{+}\Gamma_{\lambda} \right] \nonumber \\ 
  - 4 x_{1}\left( 1 - x_{1} \right) q^+ \varepsilon_q^+ {\rm tr}_{D} \left[ \left(\gamma_{\perp}^{\sigma}\slashed{\ell}_{\perp}+2\frac{x_{2}}{x_{3}}\ell_{\perp}^{\sigma}\right) \gamma^{+}\Gamma_{\lambda} \right] \; ,
\end{gather}
which is the most complicated term appearing in eq.~(\ref{Eq:DiracStepInTheAntiQuark}). The simplest one in eq.~(\ref{Eq:DiracStepInTheAntiQuark}) instead reads
\begin{gather}
  {\rm tr}_{D} \hspace{-0.1 cm} \left[ \gamma^{+} \hspace{-0.15 cm} \left(x_{1}\slashed{q}-\slashed{k}_{40\perp}\right)\slashed{\varepsilon}_{q}\gamma^{+} \hspace{-0.15 cm} \left( \hspace{-0.1 cm} \gamma_{\perp}^{\sigma}-\left(\frac{\ell_{\perp}^{\sigma}}{\ell^{+}}+\frac{k_{40\perp}^{\sigma}}{k_{40}^{+}}+\frac{q_{\perp}^{\sigma}}{q^{+}}\right) \gamma^{+} \hspace{-0.1 cm} \right) \hspace{-0.1 cm} \left(x_{2}\slashed{q}+\frac{x_{2}}{x_{2}+x_{3}}\slashed{k}_{40\perp}-\slashed{\ell}_{\perp}\right)\gamma^{+}\Gamma_{\lambda} \right] \nonumber \\
  =4x_{2}q^{+}{\rm tr}_{D} \left[ \left(x_{1}q^{+}\left(\slashed{\varepsilon}_{q\perp}-\frac{\varepsilon_{q}^{+}}{q^{+}}\slashed{q}_{\perp}\right)+\varepsilon_{q}^{+}\slashed{k}_{40\perp}\right)\gamma^{+}\gamma_{\perp}^{\sigma}\Gamma_{\lambda} \right] \; .
\end{gather}

\section{Transverse momentum integrals in the 3-body contributions}
\label{Sec:AppedixC}
Since the calculations of transverse momentum integrals in the 3-body case are quite involved, we again give here some useful details for the reader's convenience. 

\subsection{Quark emission case}
\label{Sec:AppedixC1}
In eq.~(\ref{Eq:3-bodyQuarkContBeforAnyTransverseMomInt}), the first integration we perform is that with respect to $\boldsymbol{\ell}$, which reads 
\begin{align}
 & \int\frac{{\rm d}^{d}\boldsymbol{\ell}}{\left(2\pi\right)^{d}}\frac{{\rm e}^{-i\left(\boldsymbol{\ell}\cdot\boldsymbol{z}_{13}\right)}\left[a_{0}+a_{1\perp\nu}\ell_{\perp}^{\nu}\right]}{\boldsymbol{\ell}^{2}+\frac{x_{1}x_{3}}{x_{2}\left(1-x_{2}\right)^{2}}\left(\boldsymbol{k}_{40}^{2}+x_{2}\left(1-x_{2}\right)Q^{2}\right)}\nonumber \\
 & =\int\frac{{\rm d}^{d}\boldsymbol{\ell}}{\left(2\pi\right)^{d}}{\rm e}^{-i\left(\boldsymbol{\ell}\cdot\boldsymbol{z}_{13}\right)}i\left(a_{0}+a_{1\perp\nu}\ell_{\perp}^{\nu}\right)\int_{0}^{+\infty}{\rm d}t{\rm e}^{-it\left(\boldsymbol{\ell}^{2}+\frac{x_{1}x_{3}}{x_{2}\left(1-x_{2}\right)^{2}}\left(\boldsymbol{k}_{40}^{2}+x_{2}\left(1-x_{2}\right)Q^{2}\right)-i0\right)} \nonumber \\
 & =i\left(\frac{-i}{4\pi}\right)^{\frac{d}{2}}\int_{0}^{+\infty}{\rm d}t\left(a_{0}t^{-\frac{d}{2}}-\frac{a_{1\perp\nu}}{2}z_{13\perp}^{\nu}t^{-\frac{d}{2}-1}\right){\rm e}^{i\frac{\boldsymbol{z}_{13}^{2}}{4t}-it \left( \frac{x_{1}x_{3}}{x_{2}\left(1-x_{2}\right)^{2}}\left(\boldsymbol{k}_{40}^{2}+x_{2}\left(1-x_{2}\right)Q^{2}\right)-i0 \right)} \; .
 \label{Eq:TrasvMomIntQuarkElle}
\end{align}
After using this integral in eq.~(\ref{Eq:3-bodyQuarkContBeforAnyTransverseMomInt}), we have two families of integrals over $\boldsymbol{k}_{40}$: \\

\noindent \textbf{1) Integral without denominator} \vspace{0.2 cm} \\
This integral is almost immediate, we have 
\begin{align}
 & \int\frac{{\rm d}^{2}\boldsymbol{k}_{40}}{\left(2\pi\right)^{2}} \hspace{-0.1 cm} \int_{0}^{+\infty} \hspace{-0.05 cm} \frac{{\rm d}t}{t}{\rm e}^{i\frac{\boldsymbol{z}_{13}^{2}}{4t}-i t \left( \frac{x_{1}x_{3}}{x_{2}\left(1-x_{2}\right)^{2}}\left(\boldsymbol{k}_{40}^{2}+x_{2}\left(1-x_{2}\right)Q^{2}\right)-i0\right) +i\boldsymbol{k}_{40}\cdot\left(\frac{x_{1}\boldsymbol{z}_{12}+x_{3}\boldsymbol{z}_{32}}{x_{1}+x_{3}}\right)} \hspace{-0.1 cm} \left(a_{0}+\frac{\overline{a}_{0}}{t}+a_{\nu}k_{\perp}^{\nu}\right) \nonumber \\
 & =\frac{1}{4\pi}\frac{-ix_{2}\left(1-x_{2}\right)^{2}}{x_{1}x_{3}}\int_{0}^{+\infty}\frac{{\rm d}t}{t^{2}}{\rm e}^{i\frac{\boldsymbol{z}_{13}^{2}+x_{1}x_{2}x_{3}\left(\frac{x_{1}\boldsymbol{z}_{12}+x_{3}\boldsymbol{z}_{32}}{x_{1}x_{3}}\right)^{2}}{4t}-it \left(\frac{x_{1}x_{3}}{x_{1}+x_{3}}Q^{2}-i0 \right)} \nonumber \\
 & \times\left(a_{0}+\frac{\overline{a}_{0}}{t}+a_{\nu}\frac{x_{2}\left(1-x_{2}\right)}{2t}\left(\frac{x_{1}z_{12\perp\nu}+x_{3}z_{32\perp\nu}}{x_{1}x_{3}}\right)\right)\nonumber \\
 & =\frac{1}{\pi}x_{2}\left(1-x_{2}\right)\frac{Q}{Z}K_{1}\left(QZ\right)a_{0} \nonumber \\
 & +\frac{1}{\pi}2ix_{1}x_{2}x_{3}\frac{Q^{2}}{\boldsymbol{Z}^{2}}K_{2}\left(QZ\right)\left(\overline{a}_{0}+a_{\nu}\frac{x_{2}\left(1-x_{2}\right)}{2}\left(\frac{x_{1}z_{12\perp\nu}+x_{3}z_{32\perp\nu}}{x_{1}x_{3}}\right)\right)  \; ,
 \label{Eq:QuarkTransvMomIntegralK40WithoutDen}
\end{align}
with 
\begin{equation}
\boldsymbol{Z}^{2}=x_{1}x_{2}\boldsymbol{z}_{12}^{2}+x_{1}x_{3}\boldsymbol{z}_{13}^{2}+x_{2}x_{3}\boldsymbol{z}_{23}^{2} \; .
\end{equation}

\noindent \textbf{2) Integral with denominator} \vspace{0.2 cm} \\
This integral is slightly more tedious. We start performing some straightforward manipulations, i.e. 
\begin{align}
 & \int\frac{{\rm d}^{2}\boldsymbol{k}_{40}}{\left(2\pi\right)^{2}}\int_{0}^{+\infty}{\rm d}t \; {\rm e}^{-it \left( \frac{x_{1}x_{3}}{x_{2}\left(1-x_{2}\right)^{2}}\left(\boldsymbol{k}_{40}^{2}+x_{2}\left(1-x_{2}\right)Q^{2}\right)-i0 \right) +i\boldsymbol{k}_{40}\cdot\left(\frac{x_{1}\boldsymbol{z}_{12}+x_{3}\boldsymbol{z}_{32}}{x_{1}+x_{3}}\right)}\nonumber \\
 & \times\frac{1}{\boldsymbol{k}_{40}^{2}+x_{2}\left(1-x_{2}\right)Q^{2}}\left(a_{0}+a_{1\mu}k_{40\perp}^{\mu}\right)\frac{1}{t^{2}}{\rm e}^{i\frac{\boldsymbol{z}_{13}^{2}}{4t}}\nonumber \\
\nonumber \\
 & =\int\frac{{\rm d}^{2}\boldsymbol{k}_{40}}{\left(2\pi\right)^{2}}\int_{0}^{+\infty}{\rm d}t \; {\rm e}^{-it \left( \frac{x_{1}x_{3}}{x_{2}\left(1-x_{2}\right)^{2}}\left(\boldsymbol{k}_{40}^{2}+x_{2}\left(1-x_{2}\right)Q^{2}\right)-i0 \right)+i\boldsymbol{k}_{40}\cdot\left(\frac{x_{1}\boldsymbol{z}_{12}+x_{3}\boldsymbol{z}_{32}}{x_{1}+x_{3}}\right)}\nonumber \\
 & \times\frac{1}{\boldsymbol{k}_{40}^{2}+x_{2}\left(1-x_{2}\right)Q^{2}}\left(a_{0}+a_{1\mu}k_{40\perp}^{\mu}\right)\frac{\partial}{\partial t}\left(\frac{4i}{\boldsymbol{z}_{13}^{2}}{\rm e}^{i\frac{\boldsymbol{z}_{13}^{2}}{4t}}\right) \nonumber \\
\nonumber \\
 & =\frac{-4}{\boldsymbol{z}_{13}^{2}}\int\frac{{\rm d}^{2}\boldsymbol{k}_{40}}{\left(2\pi\right)^{2}} {\rm e}^{i\boldsymbol{k}_{40}\cdot\left(\frac{x_{1}\boldsymbol{z}_{12}+x_{3}\boldsymbol{z}_{32}}{x_{1}+x_{3}}\right)} \int_{0}^{+\infty} \hspace{-0.4 cm} {\rm d}t \; {\rm e}^{i\frac{\boldsymbol{z}_{13}^{2}}{4t}-it \left( \frac{x_{1}x_{3}}{x_{2}\left(1-x_{2}\right)^{2}}\left(\boldsymbol{k}_{40}^{2}+x_{2}\left(1-x_{2}\right)Q^{2}\right) -i 0 \right)} \nonumber \\
 & \times \frac{x_{1}x_{3}}{x_{2}\left(1-x_{2}\right)^{2}} \left(a_{0}+a_{1\mu}k_{40\perp}^{\mu}\right) \; ,
\end{align}
where we integrated by part with respect to $t$. We can now easily integrate over $\boldsymbol{k}_{40}$:
\begin{gather}
  \int\frac{{\rm d}^{2}\boldsymbol{k}_{40}}{\left(2\pi\right)^{2}}\int_{0}^{+\infty} \hspace{-0.55 cm} {\rm d}t \; \frac{ {\rm e}^{-it\left( \frac{x_{1}x_{3}}{x_{2}\left(1-x_{2}\right)^{2}}\left(\boldsymbol{k}_{40}^{2}+x_{2}\left(1-x_{2}\right)Q^{2}\right)-i0 \right) +i\boldsymbol{k}_{40}\cdot\left(\frac{x_{1}\boldsymbol{z}_{12}+x_{3}\boldsymbol{z}_{32}}{x_{1}+x_{3}}\right)+i\frac{\boldsymbol{z}_{13}^{2}}{4t}}}{t^2 (\boldsymbol{k}_{40}^{2}+x_{2}\left(1-x_{2}\right)Q^{2})} \left(a_{0}+a_{1\mu}k_{40\perp}^{\mu}\right) \nonumber \\
  =\frac{i}{\boldsymbol{z}_{13}^{2} \pi} \hspace{-0.1 cm} \int_{0}^{+\infty}\hspace{-0.45 cm} {\rm d}t \; \frac{{\rm e}^{i\frac{\boldsymbol{z}_{13}^{2}+\frac{x_{2}}{x_{1}x_{3}}\left(x_{1}\boldsymbol{z}_{12}+x_{3}\boldsymbol{z}_{32}\right)^{2}}{4t}-it\left( \frac{x_{1}x_{3}}{x_{1}+x_{3}}Q^{2} - i0 \right)}}{t} \nonumber \\ \times \left(a_{0}+a_{1\mu}\frac{x_{2}\left(1-x_{2}\right)}{2t}\left(\frac{x_{1}z_{12\perp}^{\mu}+x_{3}z_{32\perp}^{\mu}}{x_{1}x_{3}}\right)\right) \; . 
\end{gather}
Then, the exponent can written as
\begin{equation}
\boldsymbol{z}_{13}^{2}+\frac{x_{2}}{x_{1}x_{3}}\left(x_{1}\boldsymbol{z}_{12}+x_{3}\boldsymbol{z}_{32}\right)^{2}=\frac{\left(1-x_{2}\right)\boldsymbol{Z}^{2}}{x_{1}x_{3}}
\end{equation}
and, finally, the integral reads
\begin{align}
 & \int\frac{{\rm d}^{2}\boldsymbol{k}_{40}}{\left(2\pi\right)^{2}} \hspace{-0.1 cm} \int_{0}^{+\infty} \hspace{-0.4 cm} {\rm d}t \; \frac{{\rm e}^{-it \left( \frac{x_{1}x_{3}}{x_{2}\left(1-x_{2}\right)^{2}}\left(\boldsymbol{k}_{40}^{2}+x_{2}\left(1-x_{2}\right)Q^{2}\right)-i0 \right) +i\boldsymbol{k}_{40}\cdot\left(\frac{x_{1}\boldsymbol{z}_{12}+x_{3}\boldsymbol{z}_{32}}{x_{1}+x_{3}}\right)+i\frac{\boldsymbol{z}_{13}^{2}}{4t}}}{t^2 (\boldsymbol{k}_{40}^{2}+x_{2}\left(1-x_{2}\right)Q^{2})} \left(a_{0}+a_{1\mu}k_{40\perp}^{\mu}\right)\nonumber \\
 & =\frac{2i}{\pi\boldsymbol{z}_{13}^{2}}\left[a_{0}K_{0}\left(QZ\right)+ix_{2}\left(1-x_{2}\right)a_{1\mu}\left(\frac{x_{1}z_{12\perp}^{\mu}+x_{3}z_{32\perp}^{\mu}}{x_{1}+x_{3}}\right)\frac{Q}{Z}K_{1}\left(QZ\right)\right] \; .
 \label{Eq:QuarkTransvMomIntegralK40WDen}
\end{align}

\subsection{Antiquark emission case}

The transverse momentum integral of the QED gauge invariance breaking part is 
\begin{align}
 & \int\frac{{\rm d}^{d}\boldsymbol{k}_{40}}{\left(2\pi\right)^{d}}\frac{{\rm d}^{d}\boldsymbol{\ell}}{\left(2\pi\right)^{d}}\frac{{\rm e}^{-i\left(\boldsymbol{\ell}\cdot\boldsymbol{z}_{23}\right)-i\boldsymbol{k}_{40}\cdot\left(\boldsymbol{z}_{1}-\frac{x_{2}\boldsymbol{z}_{2}+x_{3}\boldsymbol{z}_{3}}{x_{2}+x_{3}}\right)}}{\frac{x_{2}+x_{3}}{x_{2}x_{3}}\boldsymbol{\ell}^{2}+\frac{\boldsymbol{k}_{40}^{2}+x_{1}\left(1-x_{1}\right)Q^{2}}{x_{1}\left(1-x_{1}\right)}}\left(a_{\mu}k_{40\perp}^{\mu}+b_{\mu}\ell_{\perp}^{\mu}\right) \nonumber \\
 & =-ix_{1}x_{2}x_{3}\frac{1}{4\pi^{2}}\frac{Q^{2}}{Z^{2}}K_{2}\left(QZ\right)\left[\left(x_{1}x_{2}z_{12\perp}^{\mu}+x_{1}x_{3}z_{13\perp}^{\mu}\right)a_{\mu}+\frac{x_{2}x_{3}z_{23\perp}^{\mu}}{x_{2}+x_{3}}b_{\mu}\right] \; .
 \label{Eq:TrasvMomIntAntiQuarkQEDBreak}
\end{align}
In the gauge invariant part, the integral with respect to $\boldsymbol{\ell}$ is 
\begin{align}
 & \int\frac{{\rm d}^{d}\boldsymbol{\ell}}{\left(2\pi\right)^{d}}\frac{{\rm e}^{-i\left(\boldsymbol{\ell}\cdot\boldsymbol{z}_{23}\right)}\left[a_{0}+a_{1\nu}\ell_{\perp}^{\nu}\right]}{\boldsymbol{\ell}^{2}+\frac{x_{2}x_{3}}{x_{1}\left(1-x_{1}\right)^{2}}\left(\boldsymbol{k}_{40}^{2}+x_{1}\left(1-x_{1}\right)Q^{2}\right)} \nonumber \\
 & =i\left(\frac{-i}{4\pi}\right)^{\frac{d}{2}}\int_{0}^{\infty}{\rm d}t{\rm e}^{i\frac{\boldsymbol{z}_{23}^{2}}{4t}-it \left( \frac{x_{2}x_{3}}{x_{1}\left(1-x_{1}\right)^{2}}\left(\boldsymbol{k}_{40}^{2}+x_{1}\left(1-x_{1}\right)Q^{2}\right)-i0 \right)}\left(t^{-\frac{d}{2}}a_{0}-a_{1\nu}\frac{z_{23\perp}^{\nu}}{2}t^{-\frac{d}{2}-1}\right) \; . 
 \label{Eq:TrasvMomIntAntiQuarkElle}
\end{align}
After using this integral in eq.~(\ref{Eq:3BodyContAntiQuarkBeforeTransvMomInt}), we have two families of integrals over $\boldsymbol{k}_{40}$: \\

\noindent \textbf{1) Integral without denominator} \vspace{0.2 cm} \\
This integral reads 
\begin{align}
 & \int\frac{{\rm d}^{2}\boldsymbol{k}_{40}}{\left(2\pi\right)^{2}}\int_{0}^{\infty}\frac{{\rm d}t}{t}{\rm e}^{i\frac{\boldsymbol{z}_{23}^{2}}{4t}-it \left( \frac{x_{2}x_{3}}{x_{1}\left(1-x_{1}\right)^{2}}\left(\boldsymbol{k}_{40}^{2}+x_{1}\left(1-x_{1}\right)Q^{2}\right)-i0 \right) -i\boldsymbol{k}_{40}\cdot\left(\frac{x_{2}\boldsymbol{z}_{12}+x_{3}\boldsymbol{z}_{13}}{x_{2}+x_{3}}\right)}\nonumber \\
 & =\frac{1}{\pi}x_{1}\left(1-x_{1}\right)\frac{Q}{Z}K_{1}\left(QZ\right) \; .
 \label{Eq:AntiQuarkTransvMomIntegralK40WithoutDen}
\end{align}

\noindent \textbf{2) Integral with denominator} \vspace{0.2 cm} \\
The second integral is
\begin{align}
 & \int\frac{{\rm d}^{2}\boldsymbol{k}_{40}}{\left(2\pi\right)^{2}} \hspace{-0.1 cm} \int_{0}^{\infty} \hspace{-0.3 cm} {\rm d}t \; \frac{{\rm e}^{-it \left( \frac{x_{2}x_{3}}{x_{1}\left(1-x_{1}\right)^{2}}\left(\boldsymbol{k}_{40}^{2}+x_{1}\left(1-x_{1}\right)Q^{2}\right)-i0 \right) -i\boldsymbol{k}_{40}\cdot\left(\frac{x_{2}\boldsymbol{z}_{12}+x_{3}\boldsymbol{z}_{13}}{x_{2}+x_{3}}\right)+i\frac{\boldsymbol{z}_{23}^{2}}{4t}}}{t^{2}(\boldsymbol{k}_{40}^{2}+x_{1}\left(1-x_{1}\right)Q^{2})} \left(a_{0}+a_{1\nu}k_{40\perp}^{\nu}\right) \nonumber \\
 & =\frac{1}{\pi}\frac{i}{\boldsymbol{z}_{23}^{2}}\int_{0}^{\infty}\frac{{\rm d}t}{t}{\rm e}^{i\frac{\left(x_{2}+x_{3}\right)\boldsymbol{Z}^{2}}{4x_{2}x_{3}t}-it\left(\frac{x_{2}x_{3}}{x_{2}+x_{3}}Q^{2}-i0 \right)} 
\left(a_{0}-\frac{x_{1}\left(1-x_{1}\right)}{2t}\left(\frac{x_{2}z_{12\perp}^{\nu}+x_{3}z_{13\perp}^{\nu}}{x_{2}x_{3}}\right)a_{1\nu}\right)\nonumber \\
 & =\frac{2i}{\pi\boldsymbol{z}_{23}^{2}}\left[a_{0}K_{0}\left(QZ\right)-i \left(x_{1}x_{2}z_{12\perp}^{\nu}+x_{1}x_{3}z_{13\perp}^{\nu}\right)a_{1\nu} \frac{Q}{Z}K_{1} \left(QZ\right)\right] \; .
  \label{Eq:AntiQuarkTransvMomIntegralK40WDen}
\end{align}

\section{Integrals necessary for the Fourier transforms}
\label{Sec:AppedixD}
In the main text we need the following integrals:
\begin{equation}
    \int d^2 \boldsymbol{r} e^{-i \boldsymbol{r} \cdot \boldsymbol{k}} K_0 \left( \sqrt{x \bar{x} Q^2 \boldsymbol{r}^2} \right) = \frac{ 2 \pi }{\boldsymbol{k}^2 + x \bar{x} Q^2} \; ,
\label{Eq:FourTrasfBessel1}
\end{equation}
\begin{equation}
\int \mathrm{d}^d \boldsymbol{z}_1 \mathrm{~d}^d \boldsymbol{z}_2 \mathrm{~d}^d \boldsymbol{z}_3 \mathrm{e}^{i\left(\boldsymbol{\ell}_1 \cdot \boldsymbol{z}_1\right)+i\left(\boldsymbol{\ell}_2 \cdot \boldsymbol{z}_2\right)+i\left(\boldsymbol{\ell}_3 \cdot \boldsymbol{z}_3\right)} \frac{Q}{Z} K_1(Q Z) =\frac{(2 \pi)^4 \delta^2\left(\boldsymbol{\ell}_1+\boldsymbol{\ell}_2+\boldsymbol{\ell}_3\right)}{x_1 x_2 x_3\left(\frac{\boldsymbol{\ell}_1^2}{x_1}+\frac{\boldsymbol{\ell}_2^2}{x_2}+\frac{\boldsymbol{\ell}_3^2}{x_3}+Q^2\right)} \; ,
\label{Eq:FourTrasfBessel2}
\end{equation}
\begin{gather}  \int \mathrm{d}^d \boldsymbol{z}_1 \mathrm{~d}^d \boldsymbol{z}_2 \mathrm{~d}^d \boldsymbol{z}_3 \mathrm{e}^{i\left(\boldsymbol{\ell}_1 \cdot \boldsymbol{z}_1\right)+i\left(\boldsymbol{\ell}_2 \cdot \boldsymbol{z}_2\right)+i\left(\boldsymbol{\ell}_3 \cdot \boldsymbol{z}_3\right)} \frac{z_{23 \perp \nu}}{\boldsymbol{z}_{23}^2} K_0(Q Z) \nonumber \\  = \frac{i(2 \pi)^4 \delta^2\left(\boldsymbol{\ell}_1+\boldsymbol{\ell}_2+\boldsymbol{\ell}_3\right)\left(x_3 \ell_{2 \perp \nu}-x_2 \ell_{3 \perp \nu}\right)}{x_2 x_3\left(\ell_1^2+x_1 \bar{x}_1 Q^2\right)\left(\frac{\ell_1^2}{x_1}+\frac{\boldsymbol{\ell}_2^2}{x_2}+\frac{\boldsymbol{\ell}_3^2}{x_3}+Q^2\right)} \; ,
\label{Eq:FourTrasfBessel3}
\end{gather}
and
\begin{gather}
    \int \mathrm{d}^d \boldsymbol{z}_1 \mathrm{~d}^d \boldsymbol{z}_2 \mathrm{~d}^d \boldsymbol{z}_3 \mathrm{e}^{i\left(\boldsymbol{\ell}_1 \cdot \boldsymbol{z}_1\right)+i\left(\boldsymbol{\ell}_2 \cdot \boldsymbol{z}_2\right)+i\left(\boldsymbol{\ell}_3 \cdot \boldsymbol{z}_3\right)} \frac{z_{23 \perp \nu}}{z_{23}^2}\left(\frac{x_1 x_2 z_{12 \perp \alpha}+x_1 x_3 z_{13 \perp \alpha}}{x_2+x_3}\right) \frac{Q}{Z} K_1(Q Z) \nonumber \\  =-\frac{(2 \pi)^4 \delta^2\left(\boldsymbol{\ell}_1+\boldsymbol{\ell}_2+\boldsymbol{\ell}_3\right) \ell_{1 \perp \alpha } \left(x_3 \ell_{2 \perp \nu}-x_2 \ell_{3 \perp \nu}\right)}{x_2 x_3\left(x_2+x_3\right)\left(\ell_1^2+x_1 \bar{x}_1 Q^2\right)\left(\frac{\ell_1^2}{x_1}+\frac{\ell_2^2}{x_2}+\frac{\ell_3^2}{x_3}+Q^2\right)} \; .
\label{Eq:FourTrasfBessel4}
\end{gather}

\bibliographystyle{jhep}

\providecommand{\href}[2]{#2}\begingroup\raggedright\endgroup

\end{document}